\newcommand*{\ATLASLATEXPATH}{}
\documentclass[cernpreprint, atlasdraft=true, texlive=2016, UKenglish]{\ATLASLATEXPATH atlasdoc}

\usepackage[subfig]{\ATLASLATEXPATH atlaspackage}
\usepackage{\ATLASLATEXPATH atlasbiblatex}
 
\usepackage{\ATLASLATEXPATH atlasphysics}
 
\graphicspath{{logos/}{figures/}}
 
\usepackage{ANA-HIGG-2018-29-PAPER-defs}
\usepackage{adjustbox}
\usepackage{multirow}
\usepackage{subfig}
\usepackage[inline]{enumitem}
\usepackage{harpoon}
\usepackage{hyperref}

\AtlasTitle{Measurements of the Higgs boson inclusive and differential fiducial cross sections in the 4$\ell$ decay channel at $\sqrt{s}$ = 13 TeV }
 
\AtlasAbstract{
Inclusive and differential fiducial cross sections of the Higgs boson are measured  in the $H \to ZZ^{*} \to 4\ell$ ($\ell = e,\mu$) decay channel. The results are based on proton-proton collision data produced at the Large Hadron Collider at a centre-of-mass energy of 13 TeV and recorded by the ATLAS detector from 2015 to 2018, equivalent to an integrated luminosity of 139~fb$^{-1}$. The inclusive fiducial cross section for the $H \to ZZ^{*} \to 4\ell$ process is measured to be $\sigma_\mathrm{fid} = 3.28 \pm 0.32$~fb, in agreement with the Standard Model prediction of $\sigma_\mathrm{fid, SM} = 3.41 \pm 0.18 $~fb. Differential fiducial cross sections are measured for a variety of observables which are sensitive to the production and decay of the Higgs boson. All measurements are in agreement with the Standard Model predictions.  The results are used to constrain anomalous Higgs boson interactions with Standard Model particles.}
 
\author{The ATLAS Collaboration}

\AtlasJournalRef{Eur. Phys. J. C 80 (2020) 942}
\AtlasDOI{10.1140/epjc/s10052-020-8223-0}
\AtlasJournal{EPJC}
\PreprintIdNumber{CERN-EP-2020-035}

\hypersetup{pdftitle={ATLAS document},pdfauthor={The ATLAS Collaboration}}
 
\begin{document}
 
\maketitle
 
\tableofcontents

\section{Introduction}
\label{sec:intro}
 
The observation of the Higgs boson by the ATLAS and CMS Collaborations~\cite{HIGG-2012-27, CMS-HIG-12-028} using data from proton--proton $(pp)$ collisions at the Large Hadron Collider (LHC) recorded in 2011 and 2012 at centre-of-mass energies of $\sqrt{s}=7$~\TeV\ and 8~\TeV, respectively, was a major step forward in the understanding of the electroweak (EW) symmetry breaking mechanism~\cite{Englert:1964et,Higgs:1964pj,Guralnik:1964eu}. 
Studies of the spin and parity of the Higgs boson, its coupling structure to other particles, and measurements of fiducial and differential cross sections have been performed ~\cite{HIGG-2013-17, CMS-HIG-14-018, HIGG-2014-06, CMS-HIG-14-009, HIGG-2015-07, HIGG-2016-25, HIGG-2016-22, HIGG-2016-29,HIGG-2017-02,HIGG-2014-14,HIGG-2016-21,CMS-HIG-16-041,CMS-HIG-16-041, CMS-HIG-17-011,CMS-HIG-17-028,HIGG-2017-14,STDM-2017-09,HIGG-2017-07,HIGG-2016-07,HIGG-2018-04,HIGG-2017-06,HIGG-2018-13,HIGG-2016-33,HIGG-2017-11}. These show no significant deviations from the Standard Model (SM) predictions for the Higgs boson with a mass of $125.09\pm 0.24$~\GeV~\cite{CMS-HIG-14-042}.

This paper presents updated  inclusive and differential cross-section measurements of the Higgs boson in the $H \to ZZ^{*} \to 4\ell$ decay channel (where $\ell = e\text{ or }\mu$). 
The full ATLAS Run 2 dataset, consisting of $pp$ collision data at $\sqrt{s} = 13$~\TeV\ taken between 2015 and 2018, is used for this analysis. The total integrated luminosity after imposing data quality requirements is \Lum, with a data-taking efficiency of 91.5\%.

All measurements are performed with the assumption that the mass of the Higgs boson is $125$~\GeV, and are compared with SM predictions. The signal is extracted from a binned likelihood fit to the four-lepton invariant mass, $m_{4\ell}$, distribution. All major background processes are estimated from data. In particular, the normalisation of the dominant non-resonant $ZZ^{*}$ background is now constrained from dedicated data sidebands rather than from simulation. Signal events are corrected for detector measurement inefficiency and resolution by unfolding using the detector response matrix in the likelihood fit, in place of a bin-by-bin correction. Compared with the previous published results~\cite{HIGG-2016-25}, this paper also benefits from the full LHC Run 2 integrated luminosity, improved event and electron reconstruction~\cite{ATL-PHYS-PUB-2017-022,EGAM-2018-01}, and improved lepton isolation to mitigate the impact of additional $pp$ interactions in the same or neighbouring bunch crossing (pile-up). The fiducial phase-space definition has also been updated with respect to the previous publication to harmonise the selection of the leptons.

The paper is organised as follows.
A brief introduction of the ATLAS detector is given in Section~\ref{sec:detector}, while
in Section~\ref{sec:samples}, the data and simulated signal and background samples are described.
The selection of the Higgs boson candidate events is detailed in Section~\ref{sec:selection}.
Section ~\ref{sec:fiducial} outlines the fiducial phase-space definition and the observables that are unfolded,
while the background modelling is described in Section~\ref{sec:bkg}.
The unfolding strategy is described in Section~\ref{sec:sig_extraction}.
The experimental and theoretical systematic uncertainties, detailed in Section~\ref{sec:systematics}, are taken into account for the statistical interpretation of the data.
The final results are presented in Section~\ref{sec:results} and their interpretation to constrain possible beyond the SM (BSM) contact interactions or non-SM values of the $b$- and $c$-quark Yukawa couplings are shown in Section~\ref{sec:interpretations}.
Concluding remarks are given in Section~\ref{sec:conclusion}.
More information about general aspects of the analysis is contained in the concurrent Ref.~\cite{STXS}, where, in particular, details of
the event selection and background estimation can be found.
 
\section{The ATLAS detector}
\label{sec:detector}
\newcommand{\AtlasCoordFootnote}{
ATLAS uses a right-handed coordinate system with its origin at the nominal interaction point (IP) in the centre of the detector and the $z$-axis along the beam pipe.
The $x$-axis points from the IP to the centre of the LHC ring,
and the $y$-axis points upwards. Cylindrical coordinates $(r,\phi)$ are used in the transverse plane,  $\phi$ being the azimuthal angle around the $z$-axis.
The pseudorapidity is defined in terms of the polar angle $\theta$ as $\eta = -\ln \tan(\theta/2)$ and the rapidity is defined as $y = \frac{1}{2}\ln\frac{E+{\mathbf{p}}_z}{E-{\mathbf {p}}_z}$. Angular distance is measured in units of $\Delta R \equiv \sqrt{(\Delta\eta)^{2} + (\Delta\phi)^{2}}$.}
 
The ATLAS detector~\cite{PERF-2007-01}  is a multipurpose particle detector
with a forward--backward symmetric cylindrical geometry\footnote{\AtlasCoordFootnote} and a near $4\pi$ coverage in solid angle. It consists of an inner tracking detector (ID) surrounded by a thin superconducting solenoid, which provides a \SI{2}{\tesla} axial magnetic field, electromagnetic (EM) and hadron calorimeters, and a muon spectrometer.
The inner tracking detector covers the pseudorapidity range $|\eta| < 2.5$.
It consists of a silicon pixel detector, including the newly installed insertable B-layer~\cite{ATLAS-TDR-2010-19, PIX-2018-001}, a silicon microstrip detector, and a straw-tube tracking detector featuring transition radiation to aid in the identification of electrons.
Lead/liquid-argon (LAr) sampling calorimeters provide electromagnetic energy measurements
with high granularity.
A steel/scintillator-tile hadron calorimeter covers the central pseudorapidity range ($|\eta| < 1.7$).
The endcap and forward regions are instrumented up to $|\eta| = 4.9$ with LAr calorimeters for both the EM and hadronic energy measurements.
The calorimeters are surrounded by the muon spectrometer, which has
three large air-core toroidal superconducting magnets with eight coils each.
The field integral of the toroid magnets ranges between \num{2.0} and \SI{6.0}{\tesla\metre} across most of the detector.
The muon spectrometer includes a system of precision tracking chambers and fast detectors for triggering with a coverage of $|\eta| < 2.7$.
Events are selected using a first-level trigger implemented in custom electronics, which reduces the event
rate to a maximum of 100~kHz using a subset of detector information. Software algorithms with access
to the full detector information are then used in the high-level trigger to  yield a recorded event rate of
about 1~kHz~\cite{TRIG-2016-01}.
 
 
\section{Theoretical predictions and event simulation}
\label{sec:samples}
 
The production of the SM Higgs boson via gluon--gluon fusion (\ggF), via vector-boson fusion (\VBF), with an associated vector boson (\VH, where $V$ is a $W$ or $Z$ boson), and with a top quark pair (\ttH) was modelled with the \progname{Powheg-Box}~v2 Monte Carlo (MC) event generator~\cite{Alioli:2010xd,Frixione:2007vw,Nason:2004rx,Bagnaschi:2011tu,Aglietti:2004nj,Hamilton:2015nsa,powheg5,Luisoni:2013kna}. Table~\ref{tab:MCSignal} summarises the predicted SM production cross sections and branching ratios for the \htollllbrief{} decay for $m_H=125$~\gev\ together with their theoretical accuracy.
 
\begin{table*}[htbp]
\centering
\caption{Predicted SM Higgs boson production cross sections ($\sigma$) for \ggF,
\VBF and five associated production modes
in $pp$ collisions for $m_H=125$~\gev\ at    $\sqrt{\mathrm{s}}=13~\tev$~\cite{Dittmaier:2011ti,Dittmaier:2012vm,Heinemeyer:2013tqa,Anastasiou:2015ema,Anastasiou:2016cez,Actis:2008ug,Anastasiou:2008tj,Grazzini:2013mca,Stewart:2011cf, Ciccolini:2007jr,Ciccolini:2007ec,Bolzoni:2010xr, Brein:2003wg,Denner:2011id,Altenkamp:2012sx,Beenakker:2002nc,Dawson:2003zu, Yu:2014cka,Frixione:2015zaa,Dawson:2003kb,Dittmaier:2003ej,Harlander:2003ai,Djouadi:1997yw,Djouadi:2006bz,Bredenstein:2006rh,Bredenstein:2006ha,Boselli:2015aha,Butterworth:2015oua,Dulat:2015mca,Harland-Lang:2014zoa,Ball:2014uwa}. For $bbH$ the accuracy of calculations in the 4- and 5-flavour schemes (FS)  is reported.
The quoted uncertainties correspond to the total theoretical systematic uncertainties calculated by adding in quadrature the uncertainties due to missing higher-order corrections and PDF$+\alpha_\mathrm{s}$.  The decay branching ratios (\BR) with the associated uncertainty for \htoZZ\ and \htollllbrief, with $\ell = e, \mu$, are also given.
\label{tab:MCSignal}}
\vspace{0.1cm}
{\renewcommand{\arraystretch}{1.1}
 
\begin{tabular}{ll|c|r}
\hline \hline
 
\multicolumn{2}{l|}{Production process}& Accuracy & \multicolumn{1}{c}{$\sigma$~[pb]}\\
\hline

\ggF& $\left(gg\to H \right)$& $\textrm{N}^{3}\textrm{LO}$ in QCD, NLO in EW &  $48.6 \pm 2.4\phantom{00}$    \\
 
\VBF& $\left(qq'\to Hqq' \right)$&  (approximate) NNLO in QCD, NLO in EW &
$3.78  \pm 0.08\phantom{0}$    \\
 
\WH&  $\left(q\bar{q'}\to WH \right)$  &  NNLO in QCD, NLO in EW &
$1.373 \pm 0.028$ \\
 
\ZH&  $\left(q\bar{q}/gg\to ZH \right)$ &  NNLO in QCD, NLO in EW &
$0.88 \pm 0.04\phantom{0}$    \\
 
\ttH&  $\left(q\bar{q}/gg\to t\bar{t}H \right)$ & NLO in QCD, NLO in EW &
$0.51 \pm 0.05\phantom{0}$ \\
 
\bbH&  $\left(q\bar{q}/gg\to b\bar{b}H \right)$ &  NNLO (NLO) in QCD for 5FS (4FS) & $0.49 \pm 0.12\phantom{0}$ \\
 
$tH$ &  $\left(q\bar{q}/gg\to tH\right)$ &  NLO in QCD &
$0.09 \pm 0.01\phantom{0}$ \\
\hline
\hline
\multicolumn{2}{l|}{Decay process}& NLO in QCD, NLO in EW  & \multicolumn{1}{c}{\BR~[$\cdot$ 10$^{-4}$]}\\
\hline
\multicolumn{2}{l|}{\htoZZ}       &  & $262  \pm 6\phantom{.000}$ \\
\multicolumn{2}{l|}{\htollllbrief}&  &$\phantom{00}1.240 \pm 0.027$ \\

\hline\hline
\end{tabular}
}
\end{table*}
 
For \ggF, the PDF4LHC next-to-next-to-leading-order (NNLO) set of parton distribution functions (PDF) was used, while for all other production modes, the PDF4LHC next-to-leading-order (NLO) set was used \cite{Butterworth:2015oua}.
The simulation of \ggF Higgs boson production  used the \progname{Powheg} method for merging the NLO Higgs + jet cross section with the parton shower and the \progname{MiNLO} method \cite{Hamilton:2012rf} to simultaneously achieve NLO accuracy for the inclusive Higgs boson production. In a second step, a reweighting procedure (NNLOPS) \cite{Hamilton:2013fea}, exploiting the Higgs boson rapidity distribution, was applied using the \progname{HNNLO} program \cite{Catani:2007vq,Grazzini:2008tf} to achieve NNLO accuracy in the strong coupling constant $\alpha_\mathrm{s}$.
 
The matrix elements of the \VBF, $q\bar{q}\rightarrow \VH$ and \ttH production mechanisms were calculated to NLO accuracy in QCD.
For \VH production, the \progname{MiNLO} method was used to merge 0- and 1-jet events \cite{Hamilton:2012rf,Luisoni:2013kna}.
The $gg\rightarrow ZH$ contribution was modelled at leading order (LO) in QCD.
 
The production of a Higgs boson in association with a bottom quark pair (\bbH) was simulated at NLO with \progname{MadGraph5\_aMC@NLO}~v2.3.3 \cite{Wiesemann:2014ioa}, using the CT10 NLO PDF~\cite{Lai:2010vv}. The production in association with a single top quark ($tH$+$X$ where $X$ is either $jb$ or $W$, defined in the following as $tH$) was
simulated at NLO with \progname{MadGraph5\_aMC@NLO}~v2.6.0 using the  NNPDF30 PDF set~\cite{Ball:2014uwa}.
 
For all production mechanisms the \progname{Pythia~8}~\cite{Sjostrand:2007gs} generator was used for the \htollllbrief{} decay as well as for the parton shower modelling. The AZNLO set of tuned parameters~\cite{STDM-2012-23} was used, except for $ttH$, where, like for the $t\bar{t}$ samples, the A14 tune~\cite{ATL-PHYS-PUB-2014-021} was employed.
The event generator was interfaced to \progname{EvtGen}~v1.2.0 \cite{Lange:2001uf} for simulation of the bottom and charm hadron decays. All signal samples were simulated for a Higgs boson mass $m_H=$~125~\GeV.
 
For additional cross-checks, the \ggF sample was also generated with \progname{MadGraph5\_aMC@NLO}. This simulation has NLO QCD accuracy for zero, one and two additional partons merged with the \progname{FxFx} merging scheme~\cite{Alwall:2014hca,Frederix:2012ps}, and top and bottom quark mass effects are taken into account~\cite{Frederix:2016cnl,Mantler:2015vba,Mantler:2012bj}. Higgs boson are decayed using \progname{Madspin}~\cite{Frixione:2007zp,Artoisenet:2012st}. Some final results are also compared with \ggF predictions calculated with \progname{RadISH}, which provides resummation at N$^{3}$LL+NNLO accuracy~\cite{Banfi:2012jm,Banfi:2015pju,Monni:2016ktx,Bizon:2017rah,Monni:2019yyr}, and uses \progname{MATRIX} for the fixed-order  calculation~\cite{Grazzini:2017mhc,Grazzini:2015wpa}.
Similarly, \ggF predictions are also obtained from \progname{NNLOJET} for distributions of Higgs plus one- or two-jet events~\cite{Chen:2014gva,Chen:2016zka,Chen:2019wxf}. Neither of these two predictions are included for the case in which there are zero jets. Additionally, final results for several of the variables that probe the kinematics of the Higgs boson decay products include comparisons with \progname{Hto4l} and \progname{Prophecy4f}. These two programs include the  full NLO electroweak corrections to the Higgs boson decay into four charged leptons~\cite{Boselli:2015aha,Boselli:2017pef,Bredenstein:2006rh,Bredenstein:2006nk,Bredenstein:2006ha,Altenkamp:2017ldc,Altenkamp:2017kxk,Altenkamp:2018bcs,Denner:2018opp}.
 
The samples are normalised to cross sections obtained from the best available predictions as provided in Refs.~\cite{Dittmaier:2011ti,Dittmaier:2012vm,Heinemeyer:2013tqa,YR4,Djouadi:1997yw,Djouadi:2006bz,Dulat:2015mca,Harland-Lang:2014zoa,Ball:2014uwa}.
The SM branching ratio prediction, taken from \progname{Prophecy4f}~\cite{Bredenstein:2006rh,Bredenstein:2006nk},  includes the full NLO EW corrections, and interference effects which result in a branching ratio that is 10\% higher for same-flavour final states ($4\mu$ and $4e$) than for different-flavour states ($2e2\mu$ and $2\mu2e$).
 
For the BSM interpretation, described in Section~\ref{sec:PseudoObs}, deviations from the SM are studied using a ggF sample generated with \progname{MadGraph5\_aMC@NLO} using the \progname{HPOprodMFV} UFO model \cite{Gainer:2018qjm} with \progname{FeynRules}~\cite{Alloul:2013bka}  at LO and the \progname{NNPDF23} PDF set. The sample was interfaced to \PYTHIAV{8} using the \progname{A14} parameter set~\cite{ATL-PHYS-PUB-2014-021}. For studies of the Yukawa couplings described in Section~\ref{sec:LightYukawa}, the gluon-initiated component of the prediction was calculated using \progname{RadISH}, while \progname{MadGraph5\_aMC@NLO} was used for the quark-initiated component with \progname{FxFx} merging for 0- and 1-jet final states.

The \zzstar\ continuum background from quark--antiquark annihilation was modelled using \progname{Sherpa}~2.2.2 \cite{Gleisberg:2008ta,Gleisberg:2008fv,Cascioli:2011va},
which provides a matrix element calculation accurate to NLO in $\alpha_\mathrm{s}$  for 0- and 1-jet final states, and LO accuracy for 2- and 3-jet final states.
The merging with the \progname{Sherpa} parton shower \cite{Schumann:2007mg} was performed using the ME+PS@NLO prescription~\cite{Hoeche:2012yf}.
The NLO EW corrections were applied as a function of the invariant mass of the $ZZ^*$ system $m_{\zzstar}$ \cite{Biedermann:2016yvs,Biedermann:2016lvg}.
This process was also simulated using two additional MC generators. The first is \progname{Powheg-Box}~v2 interfaced to \progname{Pythia~8} for parton showering and hadronisation, with \progname{EvtGen} for the simulation of bottom and charm hadron decays. The second is \progname{MadGraph5\_aMC@NLO} with \progname{FxFx} merging at NLO for 0- and 1-jet final states and interfaced to \progname{Pythia~8} for parton showering.
 
The gluon-induced \zzstar\ production was modelled by \progname{Sherpa} 2.2.2 \cite{Gleisberg:2008ta,Gleisberg:2008fv,Cascioli:2011va} at LO in QCD  for 0- and 1-jet final states.
The higher-order QCD effects for the $gg\rightarrow ZZ^{*}$ continuum production have been calculated for massless quark loops \cite{Caola:2015psa,Caola:2015rqy,Campbell:2016ivq} in the heavy top-quark
approximation \cite{Melnikov:2015laa}, including the $gg\rightarrow H^{*} \rightarrow ZZ$ processes \cite{Bonvini:2013jha,Li:2015jva}. The $gg \to ZZ$ simulation cross section is scaled by a $K$-factor of 1.7$\pm$1.0, defined as the ratio of the higher-order to leading-order cross section predictions.
Production of \zzstar\ via vector-boson scattering was simulated at LO in QCD with the \progname{Sherpa} 2.2.2 generator.
 
The \emph{WZ} background was modelled using \progname{Powheg-Box}~v2 interfaced to \progname{Pythia~8}
and \progname{EvtGen} v1.2.0 for the simulation of bottom and charm hadron decays. The
triboson backgrounds \emph{ZZZ}, \emph{WZZ}, and \emph{WWZ} with four or more prompt leptons (denoted by $VVV$ hereafter) were modelled
using \progname{Sherpa} 2.2.2.
The simulation of $t\bar{t}+Z$ events with both top quarks decaying semileptonically and the $Z$ boson decaying leptonically was performed with \progname{MadGraph5\_aMC@NLO} interfaced to \progname{Pythia~8}. The total cross section is normalised to the prediction of Ref.~\cite{Frixione:2015zaa}, which includes the two dominant terms at both the LO and
the NLO in a mixed perturbative expansion in the QCD and EW couplings. For modelling comparisons, \progname{Sherpa}~2.2.1 was used to simulate $t\bar{t}+Z$ events at LO.
The smaller $tWZ$, $t\bar{t}W^{+}W^{-}$, $t\bar{t}t$, $t\bar{t}t\bar{t}$ and $tZ$ background processes were simulated with \progname{MadGraph5\_aMC@NLO} interfaced to \progname{Pythia~8}.

The modelling of events containing $Z$ bosons with associated jets (\zjets)
was performed using the \progname{Sherpa}~2.2.1 generator. Matrix elements were calculated for up to two partons at NLO and four partons at LO using
\progname{Comix} \cite{Gleisberg:2008fv} and \progname{OpenLoops} \cite{Cascioli:2011va},
and merged with the \progname{Sherpa} parton shower \cite{Schumann:2007mg} using
the \progname{ME+PS@NLO} prescription \cite{Hoeche:2012yf}. The NNPDF3.0 NNLO PDF set was used in conjunction
with a dedicated set of tuned parton shower parameters.
 
The $t\bar{t}$ background was modelled using \progname{Powheg-Box}~v2 interfaced to \progname{Pythia~8}
for parton showering, hadronisation, and the underlying event, and to \progname{EvtGen} v1.2.0 for heavy-flavour hadron decays. For this sample, the A14 tune was used~\cite{ATL-PHYS-PUB-2016-020}.
Simulated \zjets\ and  $t\bar{t}$  background samples are normalised to the data-driven estimates described in Section \ref{sec:bkg}.

Generated events were processed through the ATLAS detector simulation~\cite{SOFT-2010-01} within the \progname{Geant4} framework \cite{GEANT4} and reconstructed in the same way as collision data.
Additional $pp$ interactions in the same and nearby
bunch crossings are included in the simulation.  The pile-up was modelled by overlaying the original hard-scattering event with simulated inelastic $pp$ events generated with
\textsc{Pythia~8}~\cite{Sjostrand:2007gs}
using the NNPDF2.3LO set of PDFs~\cite{Ball:2012cx} and the A3 tune~\cite{ATL-PHYS-PUB-2016-017}.
 
\section{Event reconstruction and selection}
\label{sec:selection}
The details of the selection and reconstruction of Higgs boson candidate events are provided in  Ref.~\cite{STXS}, while a brief description is provided here.
Single-lepton, dilepton, and trilepton triggers are employed and ensure a signal selection efficiency above 98\%. 
Data events are subjected to quality requirements and  are required to have at least one vertex with two associated ID tracks with transverse momentum $\pt>500$~\MeV. The primary interaction vertex is selected as the one with the largest $\sum{\pt^2}$ of all associated tracks.
 
The lepton identification
requirements follow the inclusive event selection described in Ref.~\cite{STXS}. All muons are required to satisfy $\pt > 5$~\GeV\ and $\abseta < 2.7$, except those that are reconstructed with ID tracks matched to energy deposits in the calorimeter (calorimeter-tagged), which must satisfy $\pt > 15$~\GeV\ and $\abseta < 0.1$.  No more than one calorimeter-tagged or stand-alone muon is allowed per event, where stand-alone muons have not been matched to an ID track. Electrons are required to satisfy $\et > 7$~\GeV\ and $\abseta < 2.47$.
Jets are reconstructed using the anti-$k_{t}$ algorithm with a radius parameter $R = 0.4$ and applied to Particle Flow objects~\cite{PERF-2015-09}. Jets are required to have $\pt > 30$~\GeV\ and $\abseta < 4.5$. Jets within $\abseta < 2.5$ are identified as containing a $b$-hadron using the MV2c10 $b$-tagging algorithm at the 70\% efficiency working point~\cite{ATL-PHYS-PUB-2015-022, PERF-2016-05}. If a jet overlaps geometrically with a reconstructed muon (electron) within a cone of radial size $\Delta R = 0.1 (0.2)$, the jet is removed.

Same-flavour opposite-charge (SFOC) lepton pairs are selected to form Higgs boson candidates. The SFOC lepton pair with mass \monetwo~closest to the $Z$ boson mass is called the leading pair, while the other becomes the subleading pair, with mass \mthreefour. If multiple combinations of SFOC pairs exist, the Higgs boson candidate with \monetwo~closest to the $Z$ boson mass is chosen.
The three leading leptons of each Higgs boson candidate are required to satisfy $\pt > 20$, 15, 10~\GeV. Higgs boson candidate events are subjected to further selection requirements on the dilepton masses,  lepton separation, $J/\psi$ veto, impact parameter significance ($d_0/\sigma(d_0)$), and  vertex quality, as outlined in Table~\ref{tab:evtSel}.
In addition, isolation requirements are imposed on the leptons to suppress the $t\bar{t}$ and \zjets\ reducible backgrounds.
If an extra prompt lepton with $\pt > 12$~\GeV\ passing all identification and isolation requirements detailed previously is present in the event, the final Higgs boson candidate is chosen using a method based on the matrix element (ME).
The matrix element is calculated at LO using \progname{MadGraph5\_aMC@NLO} and the quadruplet with the highest ME value is chosen. This increases the probability of selecting the correct Higgs boson candidate in cases where the extra lepton comes from the decay of a vector boson or top quark in \VH-leptonic  or \ttH/$tH$ production.
The four-lepton mass resolution is improved by accounting for reconstructed final-state radiation (FSR) photons in the $Z$ boson decay.
After selection criteria are applied, events are divided into bins for each variable of interest for the differential cross-section measurements.
Finally, all measurements presented in this paper are performed within a four-lepton mass window of $105 < m_{4\ell} < 160$~\GeV.
The signal selection efficiency is about 31\%, 21\%, 17\%, and 16\% for the $4\mu$, $2e2\mu$, $2\mu2e$, and $4e$ final states, respectively. Here, the first lepton pair refers to the lepton pair with an invariant mass closest to the $Z$ boson mass.
 
\begin{table}
\centering
\caption{A summary of event selection requirements for leptons and Higgs boson candidates outlined in Section~\ref{sec:selection}. SFOC lepton pairs are same-flavour opposite-charge lepton pairs. For the mass requirement of the subleading lepton pair, $m_{\textrm{threshold}}$ is 12~\GeV\ for $m_{4\ell}<140$~\GeV, and rises linearly until reaching 50~\GeV\ for $m_{4\ell} = 190$~\GeV.   \label{tab:evtSel} }
\vspace{0.2cm}
\begin{tabular}{ll}
\hline\hline
\multicolumn{2}{c}{\textbf{Leptons and jets}} \\
\hline
Muons & $\pt>5$~\GeV, $\abseta<2.7$  \\
Electrons & $\et>7$~\GeV, $\abseta<2.47$  \\
Jets &  $\pt > 30$~\gev, $\abseta< 4.5$ \\
\hline
\multicolumn{2}{c}{\textbf{Lepton selection and pairing}} \\
\hline
Lepton kinematics & $\pt > 20, 15, 10$~\GeV  \\
Leading pair ($m_{12}$) & SFOC lepton pair with smallest $|m_Z-m_{\ell\ell}|$  \\
Subleading pair ($m_{34}$) & Remaining SFOC lepton pair with  smallest $|m_Z-m_{\ell\ell}|$  \\
\hline
\multicolumn{2}{c}{\textbf{Event selection (at most one Higgs boson candidate per channel)}} \\
\hline
Mass requirements & 50~\GeV$ < m_{12} < 106$~\GeV \, and  $m_{\textrm{threshold}} < m_{34} < 115$~\GeV  \\
Lepton separation: & $\Delta R(\ell_{i},\ell_{j}) > 0.1$  \\
Lepton/Jet separation & $\Delta R(\mu_{i}(e_i), \textrm{jet}) > 0.1 (0.2)$ \\
 
$J/\psi$~veto & $m(\ell_{i},\ell_{j}) > 5$~\GeV\  for all SFOC lepton pairs  \\
Impact parameter & $|d_{0}|/\sigma(d_{0})$ < 5 (3) for electrons (muons) \\
Mass window &  $105~\GeV < m_{4\ell} < 160$~\GeV  \\
Vertex selection: & $\chi^{2}/N_{\mathrm{dof}}$ < 6 (9) for $4\mu$ (other channels) \\
If extra lepton with $\pt > 12$~\GeV\ & Quadruplet with largest matrix element (ME) value \\
\hline\hline
\end{tabular}
\end{table}

\FloatBarrier
 

\section{Fiducial phase space and unfolded observables}
\label{sec:fiducial}
 
The fiducial cross sections are defined using simulation at particle level and the selection requirements outlined in Table~\ref{tab:FiducialRegion}. In order to minimise model-dependent acceptance extrapolations, these are chosen to closely match the selection requirements of the detector-level analysis after the event reconstruction.
 
The fiducial selection is applied to final-state
electrons and muons that do not originate from hadrons or $\tau$-lepton decays, after `dressing' them, i.e., the
four-momenta of photons within a cone of size $\Delta R = 0.1$ around the lepton are added to the lepton's four-momentum. The photons which originate from hadron decays are excluded.
Particle-level jets are reconstructed from final-state neutral and charged particles using the anti-$k_t$ algorithm  with radius parameter $R=0.4$. Electrons, muons, neutrinos (if they are not from hadron decays) and photons from Higgs decays as well as those used to dress leptons are excluded from the jet clustering. A jet is labelled as a $b$-jet if there is a $b$-hadron with $\pt > 5$~\GeV\ within a cone of size $\Delta R = 0.3$ around the jet axis. Jets are removed if they are within a cone of size $\Delta R = 0.1$ around a selected lepton.
 
Quadruplet selection using the selected dressed leptons follows the same procedure as for reconstructed events.
In the case of \VH or \STXSttH production, additional leptons not originating from a Higgs boson decay can induce a `lepton mispairing' when assigning them to the leading and subleading $Z$ bosons. To improve the lepton pairing efficiency,   the matrix-element-based pairing method as described in Section~\ref{sec:selection} is employed. The variables used in the differential cross-section measurement are calculated using the dressed leptons of the quadruplets.
 
The acceptance of the fiducial selection, defined as the ratio of the
number of events passing the particle-level selection to the number of events generated in a given bin or final state  (with respect to the full phase space of
$H \rightarrow\ ZZ^* \rightarrow 2\ell 2\ell'$, where $\ell,\ell' = e~\mathrm{or}~\mu$), is about 49\% for each final state for a SM Higgs boson with $\mH = 125$~\gev. The ratio of the
number of events passing the selection after detector simulation and event reconstruction
to those passing the particle-level selection is about 45\%.
About 1.6\% of the events which pass the detector-level selection fail the particle-level selection. This is mostly due to resolution effects for muons. For electrons channels, the difference in the reconstructed and fiducial phase space definition, has an additional comparable contribution.

\begin{table}[th]
\centering
\caption{List of event selection requirements which define the fiducial phase space for the cross-section measurement. SFOC lepton pairs are same-flavour
opposite-charge lepton pairs. \label{tab:FiducialRegion}}
\vspace{0.2cm}
\begin{tabular}{ll}
\hline\hline
\multicolumn{2}{c}{\textbf{Leptons and jets}} \\
\hline
Leptons & $\pt>5$ \GeV, $|\eta|<2.7$  \\
Jets &  \pt~$>$ 30 \gev, $|y|<$ \,4.4 \\
\hline
\multicolumn{2}{c}{\textbf{Lepton selection and pairing}} \\
\hline
Lepton kinematics & $\pt > 20, 15, 10 $ \GeV  \\
Leading pair ($m_{12}$) & SFOC lepton pair with smallest $|m_Z-m_{\ell\ell}|$  \\
Subleading pair ($m_{34}$) & remaining SFOC lepton pair with  smallest $|m_Z-m_{\ell\ell}|$  \\
\hline
\multicolumn{2}{c}{\textbf{Event selection (at most one quadruplet per event)}} \\
\hline
Mass requirements & 50 \GeV$ < m_{12} < 106$ \GeV \, and  12 \GeV $< m_{34} < 115$ \GeV  \\
Lepton separation & $\Delta R(\ell_{i},\ell_{j}) > 0.1$ \\
Lepton/Jet separation & $\Delta R(\ell_{i}, \textrm{jet}) > 0.1$ \\
$J/\psi$~veto & $m(\ell_{i},\ell_{j}) > 5$ \GeV \,  for all SFOC lepton pairs  \\
Mass window &  105~\GeV $< m_{4\ell} < 160 $ \GeV  \\
If extra lepton with $\pt>12$ GeV&  Quadruplet with  largest matrix element value \\
\hline\hline
\end{tabular}
 
\end{table}
Within the fiducial phase space defined above, differential cross sections are measured for variables which are sensitive to both the production and decay of the Higgs boson. For example, the transverse momentum distribution of the Higgs boson provides a test of perturbative QCD calculations, is sensitive to the structure of the Higgs boson interactions and is sensitive to charm and bottom Yukawa couplings. The rapidity of the Higgs boson is sensitive to the choice of parton distribution functions for the colliding protons, and is also influenced by QCD radiative corrections. The invariant masses of the leading and subleading lepton pair are sensitive to higher-order electroweak corrections to the Higgs boson decay, and are sensitive to BSM contributions. These two variables and the angular variables of the Higgs boson decay are also of interest due to their sensitivity to the spin and parity of the Higgs boson, as well as to same-flavour pair final-state interference and EW corrections. Variables related to jets probe QCD radiation effects and the Higgs boson production. The jet multiplicity is sensitive to different production mechanisms and provides sensitivity to the theoretical modelling of high-$\pt$ quark and gluon emission. The transverse momentum of the jets directly probes the quark and gluon radiation. The invariant mass of the two leading jets is also sensitive to the production mechanisms of the Higgs boson, while the signed angle in the transverse plane  of the two leading jets is a test of the spin and parity of the Higgs boson. Jet-related variables, in particular double differential variables, also probe the effects of QCD resummation. Additional variables which combine the properties related to the kinematics of the Higgs boson and the jets are also considered. A summary of all the variables and their descriptions is given in Table~\ref{tab:varDefinition}.

\begin{table}[th]
\centering
\caption{Definitions of observables for which differential cross sections are measured. The angular variables are defined as in Ref.~\cite{Gao_2010}.
In addition to the single observables listed, the following double differential observables are built using variables defined below: $m_{12}$ vs.\ $m_{34}$, \ptH~vs.\ \njets, \ptH~vs.\ \ljpt, \ptH~vs.\ \ptHj, \ptH~vs.\ $|y_{4\ell}|$, \ptHj\ vs.\ $m_{4\ell j}$, \ljpt~vs.\ \sjpt, and \ljpt~vs.\ \ljy\ (where \ljy| is the rapidity of the leading jet). Jet-related variables are inclusive, while for the jet multiplicity the results are provided in both the inclusive and exclusive jet bins. $\Delta\phi_{jj}$ is defined as $\phi^{\textrm{lead.\ jet}} - \phi^{\textrm{sublead.\ jet}}$ if $\eta^{\textrm{lead.\ jet}} > \eta^{\textrm{sublead.\ jet}}$ or as $ \phi^{\textrm{sublead.\ jet}} - \phi^{\textrm{lead.\ jet}} $  if $ \eta^{\textrm{sublead.\ jet}} > \eta^{\textrm{lead.\ jet}}$. If $\Delta\phi_{jj} < 0$, $2\pi$ is added to the value.
\label{tab:varDefinition}}
\vspace{0.2cm}
\begin{tabular}{c | p{11cm}}
\hline\hline
\multicolumn{2}{c}{\textbf{Higgs boson kinematic-related variables}} \\
\hline
\ptH, $|y_{4\ell}|$ & Transverse momentum and rapidity of the four-lepton system \\
$m_{12}$, $m_{34}$     & Invariant mass of the leading and subleading lepton pair     \\
$|\cos\theta^{\ast}|$ & Magnitude of the cosine of the decay angle of the leading lepton pair in the four-lepton rest frame relative to  the beam axis \\
$\cos\theta_1$, $\cos\theta_2$ & Production angles of the anti-leptons from the two $Z$ bosons, where the angle is relative to the $Z$ vector. \\ 
$\phi$, $\phi_1$ & Two azimuthal angles between the three planes constructed from the $Z$~bosons and leptons in the Higgs boson rest frame.\\
\hline
\multicolumn{2}{c}{\textbf{Jet-related variables}} \\
\hline
$N_{\textrm{jets}}$, $N_{b\textrm{-jets}}$   & Jet and $b$-jet multiplicity      \\
\ljpt, \sjpt\ & Transverse momentum of the leading and subleading jet, for events with at least one and two jets, respectively. Here, the leading jet refers to the jet with the highest $\pt$ in the event, while subleading refers to the jet with the second-highest $\pt$. \\
$m_{jj}$, $|\Delta\eta_{jj}|$, $\Delta\phi_{jj}$
& Invariant mass, difference in pseudorapidity, and signed difference in $\phi$ of the leading and subleading jets for events with at least two jets \\
\hline
\multicolumn{2}{c}{\textbf{Higgs boson and jet-related variables}} \\
\hline
\ptHj, $m_{4\ell j}$ & Transverse momentum and invariant mass of the four-lepton system and leading jet, for events with at least one jet \\
\ptHjj, $m_{4\ell jj}$& Transverse momentum and invariant mass of the four-lepton system and leading and subleading jets, for events with at least two jets \\
\hline\hline
\end{tabular}
 
\end{table}
 
\section{Background estimation}
\label{sec:bkg}

Non-resonant SM  $(Z^{(\ast)}/\gamma^{\ast})(Z^{(\ast)}/\gamma^{\ast})$ production via $q\bar{q}$ annihilation and gluon--gluon fusion, referred to as $ZZ^{\ast}$, can result in four prompt leptons in the final state and constitutes the largest background for this analysis. While for previous analyses \cite{HIGG-2016-25, HIGG-2016-22} both the shape and the normalisation of this background were exclusively estimated with simulation, in this paper the normalisation is constrained with a data-driven technique. The systematic uncertainty is  reduced because both the theoretical and luminosity uncertainties no longer contribute to the normalisation uncertainty.
The normalisation of the non-resonant $ZZ^*$ component, which dominates outside the Higgs boson peak region, is obtained from data by extending the mass interval considered from 115--130~\GeV\ to 105--160~\GeV. The increased mass interval allows an estimation of this process with minimal impact on the expected sensitivity for the signal process.
This contribution is determined as part of the $4\ell$ mass fit (discussed in Section~\ref{sec:sig_extraction}) in the full four-lepton mass region, with the shape of the background taken from simulation.

The $ZZ^*$ normalisation is estimated separately in each bin of each differential observable, where a different $ZZ^{\ast}$ scaling factor is used for each observable bin.	In phase-space regions where the $ZZ^*$ component in the $m_{4\ell}$ sidebands is too low to provide a reliable estimate of its contribution, the estimate is evaluated simultaneously  for several differential bins.\footnote{The same normalisation factor is used for neighbouring bins until the increase in uncertainty on the expected cross section in each measured bin is less than 5\% of the total uncertainty.}

Other background processes, such as \zjets, \ttbar, and $WZ$, contain at least one jet, photon or lepton from a hadron decay that is misidentified as a prompt lepton. These reducible backgrounds are significantly smaller than the non-resonant $ZZ^*$  background and are estimated using data where possible, following
slightly different approaches for the \llmumu\ and \llee\ final states~\cite{HIGG-2016-22, HIGG-2016-25,STXS}.
 
In the \llmumu\ final states, the normalisations for the \zjets\ and \ttbar\ backgrounds are determined by performing fits to the invariant mass of the leading lepton pair in dedicated independent control regions which target each background process for each bin of the differential observables. 
Depending on the background process being targeted, the control regions are formed by relaxing the $\chi^2$ requirement on the four-lepton vertex fit, and by inverting or relaxing isolation and/or impact-parameter requirements on the subleading muon pair. Additional control regions ($e\mu \mu\mu$ and \llmumuSS) are used to improve the background estimate by reducing the statistical uncertainty of the fitted normalisation. Transfer factors to extrapolate from the control regions to the signal region are obtained separately for \ttbar\ and \zjets\ using simulation. This method is performed in each differential bin. The $m_{4\ell}$ shape for both processes in each bin is obtained from simulation. 
 
The \llee\ control-region selection requires the electrons in the subleading lepton pair to have the same charge, and relaxes the identification, impact parameter and isolation requirements on the electron candidate with the lowest transverse energy. This electron candidate, denoted by $X$,  can be a light-flavour jet, an electron from photon conversion or an electron from heavy-flavour hadron decay.
The heavy-flavour background is completely determined from simulation, whereas the light-flavour and photon conversion background is obtained with the sPlot~method \cite{splot}. This is based on a fit to the number of hits in the innermost ID layer in the data control region. Transfer factors to extrapolate from the \llee\ control region to the signal region for the light-flavour jets and converted photons, obtained from simulated samples, are corrected using a $Z+X$ data control region. The corrected transfer factors are then used to extrapolate the extracted yields to the signal region. Both the extraction of the global yield in the control region and the extrapolation to the signal mass region are performed in bins of the transverse momentum of the electron candidate and the jet multiplicity. In order to extract the shape of the backgrounds from light-flavour jets and photon conversions for each observable, a similar method is used, except that the extraction and extrapolation is performed only as a function of the transverse momentum of the electron candidate, ignoring the binning in jet multiplicity.
 
Additional contributions from rare processes, such as $tXX$ ($t\bar{t}Z$, $t\bar{t}W$, $tWZ$ and other rare top-associated processes) and $VVV$ are estimated from simulation.
 
\section{Signal extraction and unfolding}
\label{sec:sig_extraction}

To extract the number of signal events in each bin of a differential distribution (or for each decay final state for the inclusive fiducial cross section), invariant mass templates for the Higgs boson signal and the background processes are fitted to the $m_{4\ell}$ distribution in data.
Compared to the previous analysis~\cite{HIGG-2016-25}, the non-resonant $ZZ^{\ast}$ background is fitted simultaneously with the signal and constrained by extending the $m_{4\ell}$ fit range from 115--130~\GeV\ to 105--160~\GeV.
 
For the total and fiducial cross sections in different final states, the same normalisation factor is used for the $ZZ^{\ast}$ contribution. For the differential cross-section measurements,  multiple $ZZ^{\ast}$ normalisation factors are introduced in the model, as described in Section~\ref{sec:bkg}.
The reducible background, composed of \zjets, \ttbar, and $WZ$ processes, is estimated from dedicated control regions as described in Section~\ref{sec:bkg} and its overall normalisation and shape can vary within the associated systematic uncertainties.
Finally, for the differential distributions, no splitting into decay final states is performed, and the SM $\zzstar \rightarrow 4\ell$ decay fractions are assumed.
 
The number of expected events $N_{i}$ in each observable reconstruction bin $i$, expressed as a function of  $m_{4\ell}$, is given by
 
\begin{align*}
N_{i}(m_{4\ell})=\sum_{j} r_{ij}\cdot(1+f_{i}^{\mathrm{non fid}})\cdot\sigma^{\mathrm{fid}}_{j}\cdot\mathcal{P}_i(m_{4\ell})\cdot\mathcal{L}+N^{\mathrm{bkg}}_{i}(m_{4\ell})
\end{align*}
with
\begin{align}
\label{eq:fidXS}
\sigma_{j}^{\mathrm{fid}}= \sigma_j \cdot A_j \cdot \mathcal{B}
\end{align}
where $A_j$ is the acceptance in the fiducial phase space and  $\sigma_j$ the total cross section in fiducial bin $j$, $\mathcal{L}$ is the integrated luminosity, $\mathcal{B}$ is the branching ratio and $N^{\mathrm{bkg}}_{i}(m_{4\ell})$ is the background contribution. The index $j$ runs over all observable bins in the fiducial phase space.
The term $\mathcal{P}_i(m_{4\ell})$ is the $m_{4\ell}$ signal shape containing the fraction of events as a function of $m_{4\ell}$ expected in each reconstruction bin, taken from MC simulation.
The term  $r_{ij}$ represents the detector response matrix,
created with simulated signal samples and averaged across the different production modes using the expected SM cross-sections \cite{YR4}. These factors correspond to the probability that an event generated within the fiducial volume in the observable bin $j$ is reconstructed in bin $i$.

The normalisation, $f_{i}^{\mathrm{non fid}}$, represents the fraction of events which are outside of the fiducial region but are reconstructed within the signal region.
This ranges from 1.1\% to 1.7\% depending on the bin of the unfolded observable or final state.

The detector response matrix accounts for bin-to-bin migrations in the unfolding of the signal.
It was chosen over the bin-by-bin correction factor technique used in the previous analyses~\cite{HIGG-2016-25,HIGG-2016-22} due to its lower model dependence.
Biases introduced via the unfolding method are minimised when using the response matrix; however, matrix unfolding can amplify small fluctuations in data when the response matrix is characterised by a large condition number.\footnote{The condition number is defined as the ratio of the maximum and minimum singular values of the matrix. Values close to 1 signify a well-conditioned matrix with low sensitivity to statistical fluctuations on the input.}
The binning choice made for all observables ensures a statistical significance of more than $2\sigma$ for the signal process. The binning is also chosen to minimise migrations between bins. In general, the bin width is more than twice the experimental resolution. As a result, the response matrices for all the variables considered  are well-conditioned, with a condition number less than 2.5.
The fluctuations of the unfolded distribution can be further reduced using regularisation techniques.
Unfolding tests done with toy data sets indicate that while regularisation provides a modest reduction of the statistical uncertainty, this reduction is counterbalanced by the bias introduced by this technique.
Therefore, no regularisation of the unfolding was applied. Two of the jet-related variables are also provided in Appendix A using a regularised unfolding method, and are compatible with the matrix-unfolded results presented here.
 
\begin{figure}[!htbp]
\centering
\subfloat[]{\includegraphics[width=0.45\linewidth]{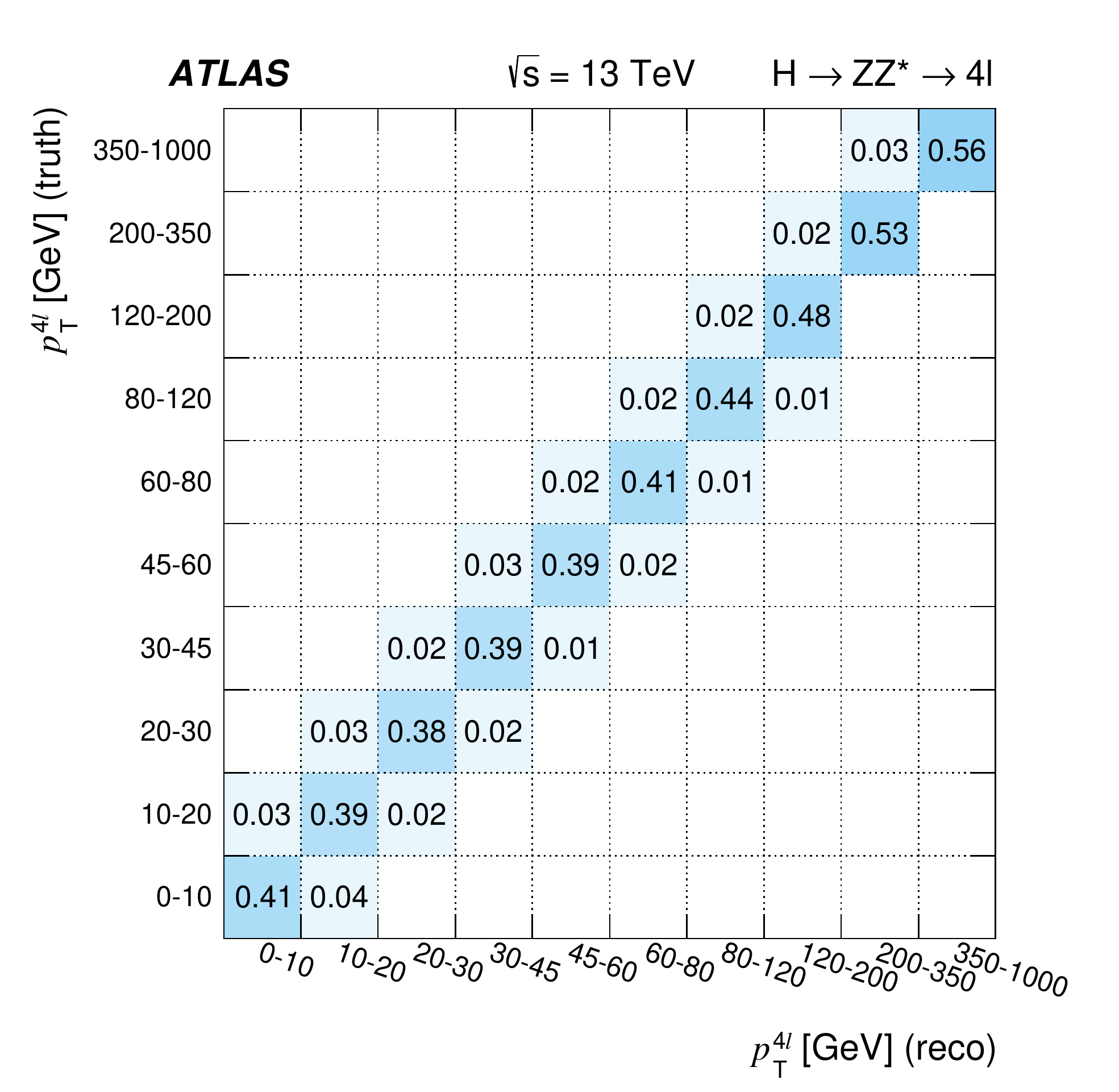}\label{fig:rspmtx_pt}}
\subfloat[]{\includegraphics[width=0.45\linewidth]{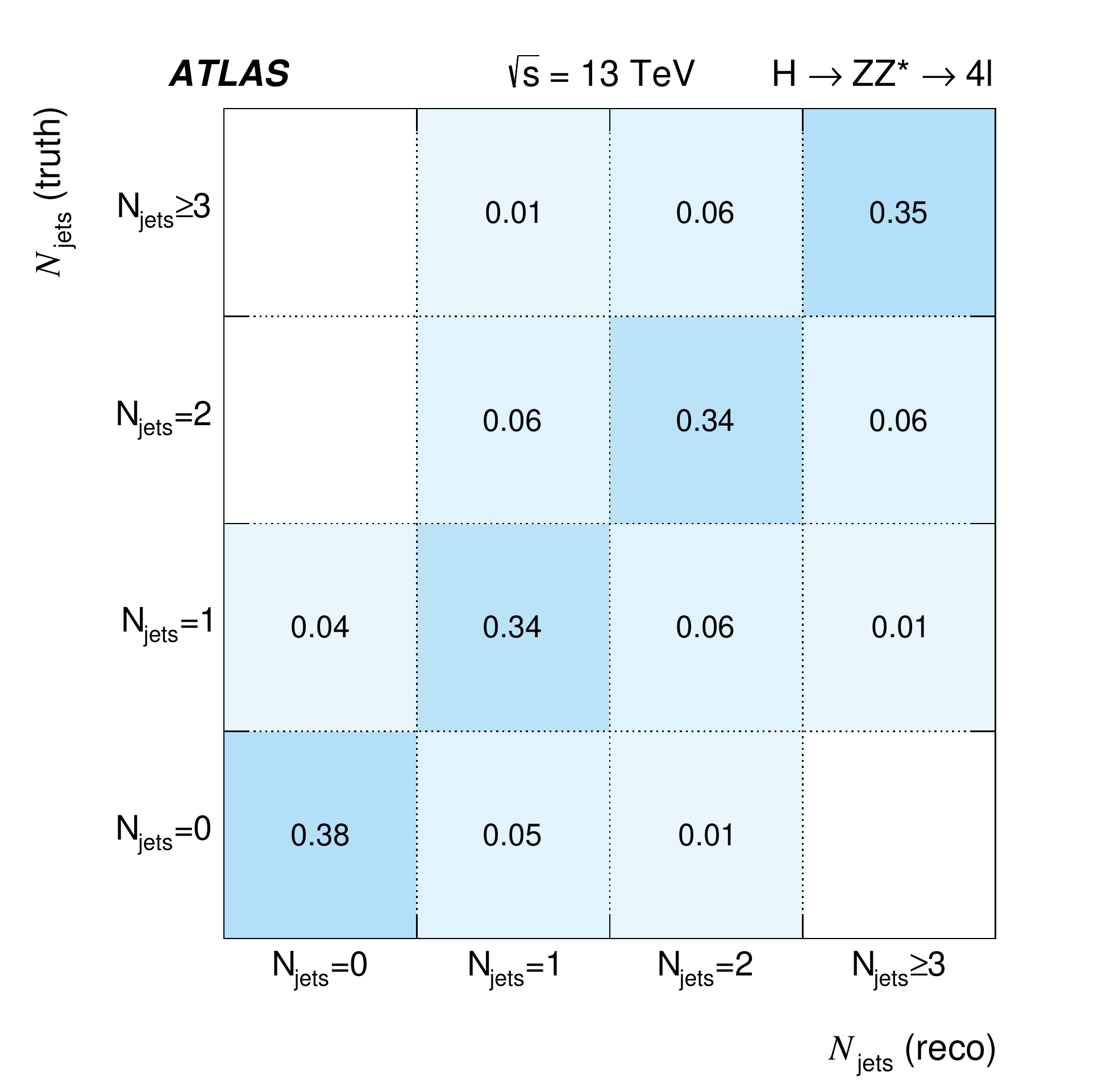}\label{fig:crspmtx_njets}}\\
\subfloat[]{\includegraphics[width=0.45\linewidth]{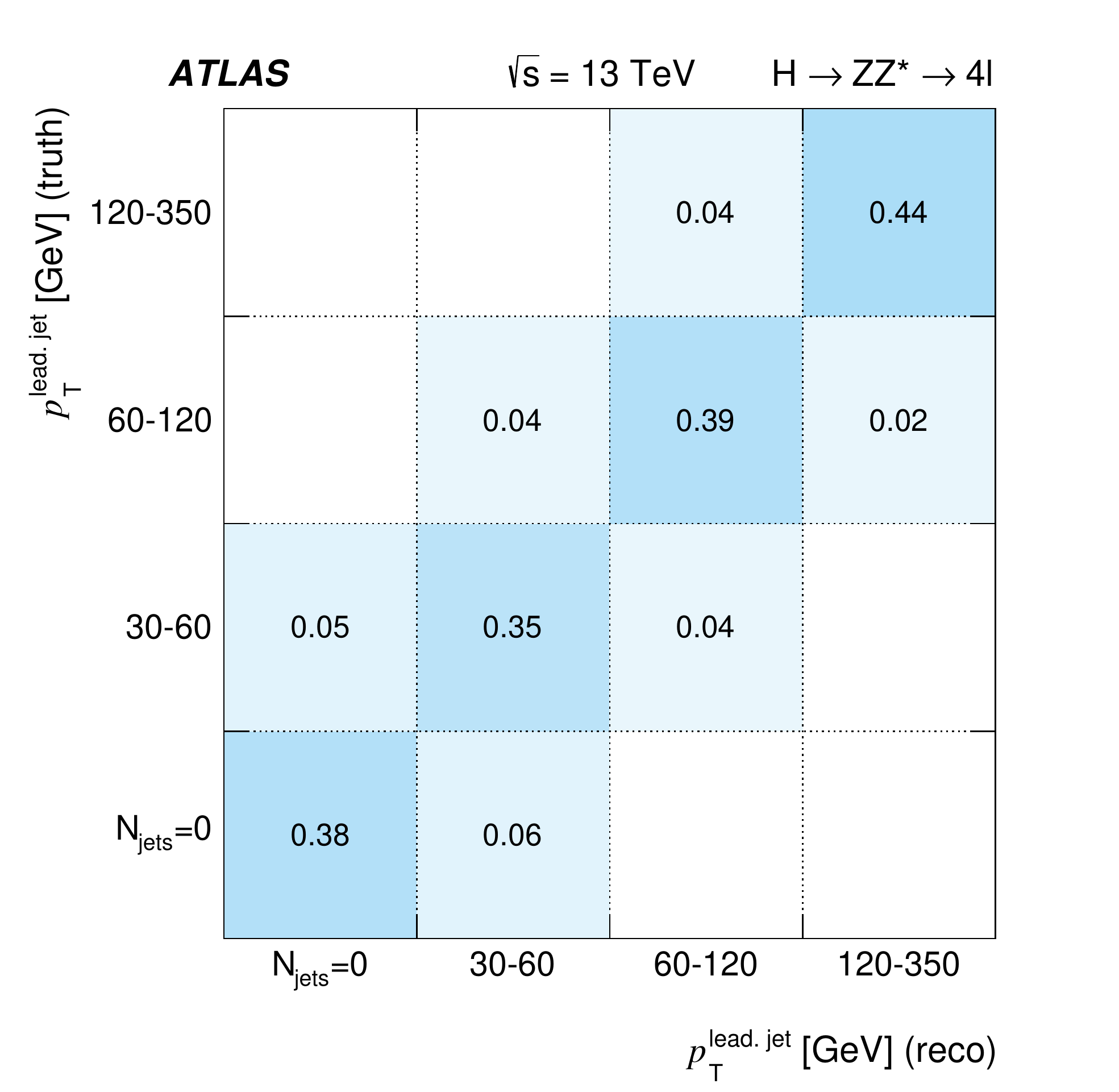}\label{fig:rspmtx_ljpt}}
\subfloat[]{\includegraphics[width=0.45\linewidth]{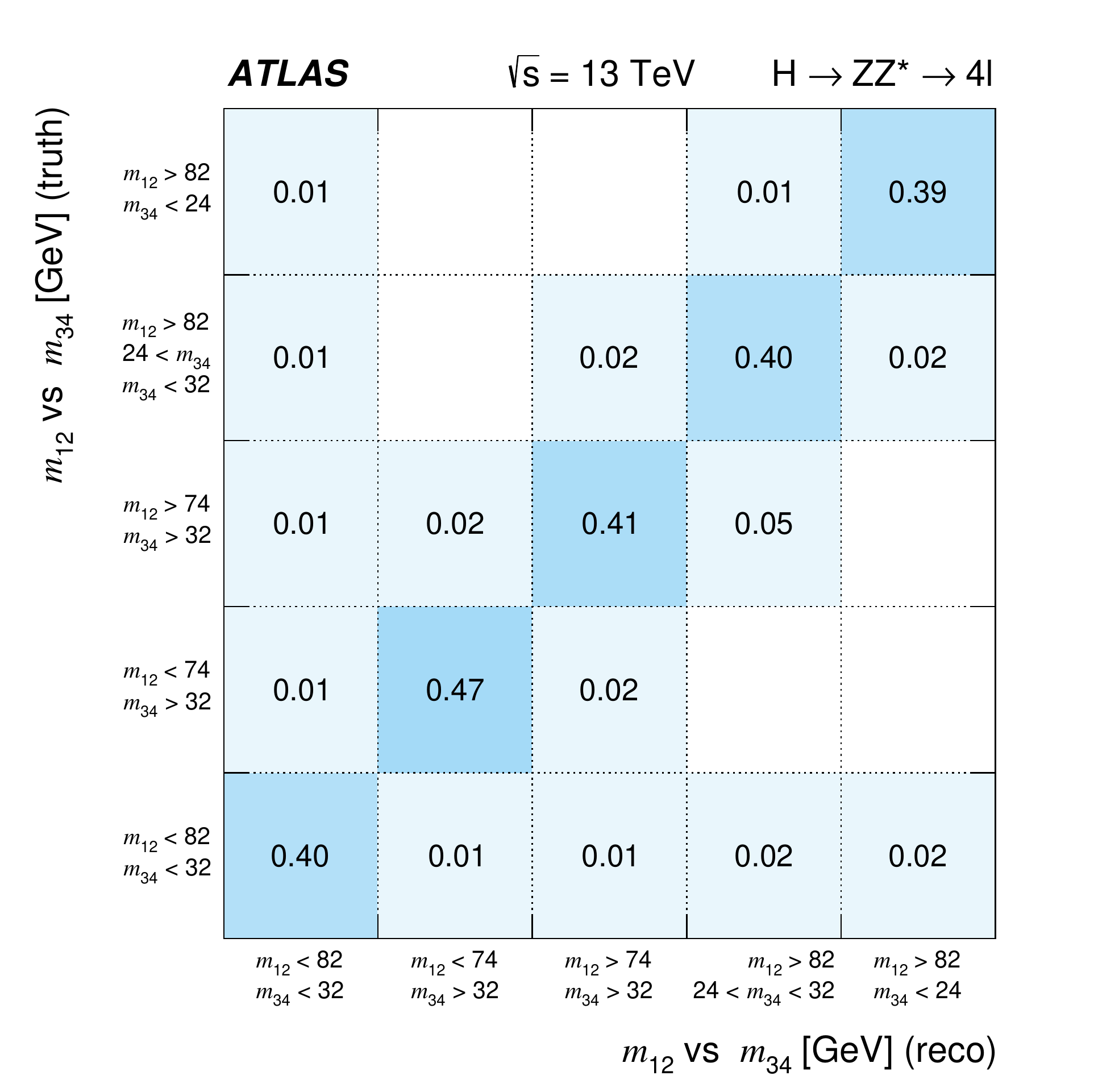}\label{fig:rspmtx_m12m34}}
\caption{Response matrices, derived using simulation, for  (a) the transverse momentum of the four-lepton system \ptH, (b) the number of jets \njets, (c) the transverse momentum of the leading jet \ljpt, and (d) the mass of the leading versus subleading lepton pair $m_{12}~\mathrm{vs.}~m_{34}$. Only reconstructed events that were matched to generator-level (`truth') events are included. Bins below 0.005 are omitted for clarity.
\label{fig:resp_matrix}}
\end{figure}

Figure~\ref{fig:resp_matrix} shows the response matrix for the \ptH\, \njets\, \ljpt, and $m_{12}~\mathrm{vs.}~m_{34}$  observables.
For  \ptH, the purity of the bins ranges from 87\% at low \ptH, where the bins are narrow, to 97\% at high  \ptH, where wider bins are defined. The purity is defined as the percentage of reconstructed events which match the particle-level events in that bin. For the \njets\ observable, the migrations are more relevant due to the relatively worse jet energy resolution and the presence of pile-up jets in the reconstructed events. This brings the purity for the for \njets\ $\ge 3$ bin down to 68\%. The \ljpt\ migrations are similarly larger, with the lowest purity value of 67\% occurring in the lowest \ljpt\ bin. The $m_{12}~\mathrm{vs.}~m_{34}$ observable, like \ptH, has a higher purity. All bins have a purity of around 90\% except the first bin, which has a purity of 78\%.

 
 
\section{Systematic uncertainties}
\label{sec:systematics}
 
The systematic uncertainties include experimental uncertainties, such as those in object reconstruction, identification, isolation, resolution, and trigger efficiencies, as well as theoretical uncertainties related to the modelling of the signal and background processes.
More detail is provided in Ref.~\cite{STXS}, while a brief overview of the dominant sources of uncertainty is provided here.
The impacts of the experimental and theoretical uncertainties on the measurements are summarised in Table~\ref{tab:RankingSummarXS}.
\subsection{Experimental uncertainties}
\label{subsec:systematics_exp}

The uncertainty in the predicted yields due to pile-up modelling ranges between 1\% and 2\%.
The uncertainty in the integrated luminosity is 1.7\% and affects the signal yields and simulated background estimates when not constrained by the sidebands.
 
The electron (muon) reconstruction and identification efficiency uncertainties are approximately $1.0$--$2.0 \%$ ($< 1.0\%$). The uncertainty in the expected yields due to the muon and electron isolation efficiencies is also considered, and is approximately 1\%. Lepton energy momentum scale and resolution uncertainties have negligible impacts on the presented results.
 
The impact of uncertainties in the jet energy scale and resolution (of between 1\% and 3\%) is only relevant for the jet-related differential cross-section measurements, where their impact is typically between 3\% and 5\%, and is negligible in the other measurements.
The uncertainty in the performance of the $b$-tagging algorithm is at the level of a few percent over most of the jet \pt\ range~\cite{PERF-2016-05}.
 
The impact of the  precision  of the Higgs boson mass measurement, $m_{H} = 125.09 \pm 0.24$~\GeV~\cite{HIGG-2014-14}, on the signal acceptance due to the signal region mass-window requirement is negligible.
 
For the data-driven measurement of the reducible background, three sources of uncertainty are considered: statistical uncertainty, overall systematic uncertainty for each of \llmumu\ and \llee, and a shape systematic uncertainty which varies with the differential variable. 
Impacts from these sources of uncertainty range from less than 1\% to a maximum of around 3\%.
The inclusive reducible background estimate has a relatively small (3\%) statistical uncertainty, which has minimal impact on the cross section.

\begin{table*}
\centering
\caption{Fractional uncertainties for the inclusive fiducial and total cross sections, and ranges of systematic uncertainties for the differential measurements. The columns `$e/\mu$' and `Jets' represent the experimental uncertainties in lepton and jet reconstruction and identification, respectively.
The \zjets, $t\bar{t}$, $tXX$ (Other Bkg.) column includes uncertainties related to the estimation of these background sources.
The $ZZ^*$ theory ($ZZ^*$ th.) uncertainties include the PDF and scale variations. Signal theory (Sig th.) uncertainties include PDF choice, QCD scale, and shower modelling of the signal. Finally, the column labelled `Comp.' contains uncertainties related to production mode composition and unfolding bias which affect the response matrices. The uncertainties have been rounded to the nearest 0.5\%, except for the luminosity uncertainty, which has been measured to be 1.7\%.
\label{tab:RankingSummarXS} }
\vspace{0.2cm}
\footnotesize
\renewcommand{\arraystretch}{1.3}
\begin{tabular}{ l c c c c c c c c c c c }
\hline
Observable  & \multicolumn{1}{c}{Stat.} & \multicolumn{1}{c}{Syst.}  & \multicolumn{7}{c}{Dominant systematic components [\%] }  \\
&  \multicolumn{1}{c}{unc. [\%]} & \multicolumn{1}{c}{unc. [\%]} & \multicolumn{1}{c}{Lumi.} & \multicolumn{1}{c}{$e/\mu$} & \multicolumn{1}{c}{Jets} & \multicolumn{1}{c}{Other Bkg.} & \multicolumn{1}{c}{\zzstar\ Th.} & \multicolumn{1}{c}{Sig. Th.} & \multicolumn{1}{c}{Comp.} \\
\hline

$\sigma_\mathrm{comb}$      &  9    &  3  &  1.7  &  2  &  $<0.5$  &  $<0.5$  &  1    &  1.5  &  $<0.5$  \\
 
$\sigma_\mathrm{4\mu}$      &  15   &  4  &  1.7  &  3  &  $<0.5$  &  $<0.5$  &  1.5  &  1    &  $<0.5$  \\
$\sigma_\mathrm{4e}$        &  26   &  8  &  1.7  &  7  &  $<0.5$  &  $<0.5$  &  1.5  &  1.5  &  $<0.5$  \\
$\sigma_\mathrm{2\mu2e}$    &  20   &  7  &  1.7  &  5  &  $<0.5$  &  $<0.5$  &  2    &  1.5  &  $<0.5$  \\
$\sigma_\mathrm{2e2\mu}$    &  15   &  3  &  1.7  &  2  &  $<0.5$  &  $<0.5$  &  1    &  1.5  &  $<0.5$  \\

d$\sigma$ / d\ptH           &  20--46   &  2--8   &  1.7  &  1--3  &  1--2
&  $<0.5$  &  1--6   &  1--2  &  $<1$  \\
d$\sigma$ / d\monetwo       &  12--42   &  3--6   &  1.7  &  2--3  & $<1$
&  $<0.5$  &  1--2   &  1--2  &  $<1$  \\
d$\sigma$ / d\mthreefour    &  20--82   &  3--12  &  1.7  &  2--3  & $<1$
&   1--2    &  1--8   &  1--3  &  $<1$  \\
d$\sigma$ / d\yH            &  22--81   &  3--6   &  1.7  &  2--3  & $<1$
&  $<0.5$  &  1--5   &  1--3  &  $<1$  \\
d$\sigma$ / d\costhetastar  &  23--113  &  3--6   &  1.7  &  2--3  & $<1$
&   1--2    &  1--7   &  1--3  &  $<0.5$  \\
d$\sigma$ / d\costhetaone   &  23--44   &  3--6   &  1.7  &  2--3  & $<1$
&  $<0.5$  &  1--3   &  1--2  &  $<1$  \\
d$\sigma$ / d\costhetatwo   &  22--39   &  3--6   &  1.7  &  2--3  & $<1$
&  $<0.5$  &  1--3   &  1--3  &  $<1$  \\
d$\sigma$ / d$\phi$         &  20--29   &  2--5   &  1.7  &  2--3  & $<1$
&  $<0.5$  &  1--3   &  1--2  &  $<0.5$  \\
d$\sigma$ / d\phione        &  22--33   &  3--6   &  1.7  &  2--3  & $<1$
&  $<0.5$  &  1--2   &  1--3  &  $<0.5$  \\
d$\sigma$ / d\njets         &  15--37   &  6--14  &  1.7  &  1--3  & 4--10
&  $<0.5$  &  1--4   &  3--7  &  1--4  \\
d$\sigma$ / d\nbjets        &  15--67   &  6--15  &  1.7  &  1--3  & 4--5
&   1--3    &  1--2   &  3--9  &  1--4  \\
d$\sigma$ / d\ljpt          &  15--34   &  3--13  &  1.7  &  1--3  & 4--10
&  $<0.5$  &  1--2   &  1--5  & $<0.5$ \\
d$\sigma$ / d\sjpt          &  11--67   &  5--22  &  1.7  &  1--3  & 2--12
&  $<1$  &  1--3   &  2--15 &  1--5  \\
d$\sigma$ / d\mjj           &  11--50   &  5--18  &  1.7  &  1--3  & 1--11
&  $<0.5$  &  1--3   &  2--15 &  1--2  \\
d$\sigma$ / d$\eta_{jj}$    &  11--57   &  5--17  &  1.7  &  1--3  & 2--10
&  $<0.5$  &  1--2   &  2--14 &  1--4  \\
d$\sigma$ / d$\phi_{jj}$    &  11--50   &  4--18  &  1.7  &  1--3  & 2--9
&  $<0.5$  &  1--3   &  2--14 &  1--6  \\
d$\sigma$ / d\mflj          &  15--66   &  4--19  &  1.7  &  1--3  & 3--9
&  $<0.5$  &  1--6   &  3--14 &  1--8  \\
d$\sigma$ / d\mfljj         &  11--182  &  5--67  &  1.7  &  1--3  & 4--24
&  $<0.5$  &  1--5   &  2--35 &  1--9  \\
d$\sigma$ / d\ptHj          &  15--76   &  6--13  &  1.7  &  1--3  & 2--8
&  $<1$  &  1--5   &  3--9  &  1--3  \\
d$\sigma$ / d\ptHjj         &  11--76   &  5--27  &  1.7  &  2--3  & 2--9
&   1--2    &  1--4   &  3--17 &  1--12  \\
 
d$^2\sigma$ / d\monetwo d\mthreefour &  16--65  &  3--11  &  1.7  &  2--3  & $<1$
&   1--2    &  1--9   &  1--3 &  1--2   \\
d$^2\sigma$ / d\ptH d\yH    &  23--63   &  2--13   &  1.7  &  1--3  & 1--2
&  $<1$  &  1--6   &  1--5  &  1--2   \\
d$^2\sigma$ / d\ptH d\njets &  23--93   &  4--193  &  1.7  &  2--14 & 2--25
&   1--3    &  1--7   &  1--12 &  1--92  \\
d$^2\sigma$ / d\ptHj d\mflj &  15--41   &  4--12   &  1.7  &  1--3  & 2--8
&  $<0.5$  &  1--5   &  2--9  &  $<1$  \\
d$^2\sigma$ / d\ptH d\ptHj  &  15--53   &  3--10  &  1.7  &  1--3  & 2--8
&  $<1$  &  1--2   &  2--6  &  1--2   \\
d$^2\sigma$ / d\ptH d\ljpt  &  15--84   &  3--21  &  1.7  &  1--3  & 2--18
&   1--10   &  1--3   &  2--9  &  1--3   \\
d$^2\sigma$ / d\ljpt d\ljy  &  15--38   &  3--11  &  1.7  &  1--3  & 2--9
&  $<0.5$  &  1--2   &  1--4  &  1--2   \\
d$^2\sigma$ / d\ljpt d\sjpt &  15--63   &  5--22  &  1.7  &  1--3  & 4--15
&  $<0.5$  &  1--4   &  3--11 &  1--7  \\
 
\hline
\hline
\end{tabular}
\end{table*}

\subsection{Theoretical uncertainties}
\label{subsec:systematics_theory_couplings}
 
Sources of theoretical uncertainty include missing higher-order corrections, parton shower and underlying event modelling, and PDF$+\alpha_\mathrm{s}$ uncertainties, and these all affect modelling of the signal and background processes.
For measurements of the cross section, the impact of these theory systematic uncertainties on the signal comes from their effects on the response matrix.
 
The prediction of the \ggF process in different \njets~categories and migration effects on the \njets~\ggF cross sections are large sources of theoretical uncertainty, which are accounted for using the approach detailed in Ref.~\cite{YR4}. The QCD scale uncertainty from the factorisation and renormalisation scales, resummation scales, and migrations between $N$-jet phase-space bins are considered~\cite{Stewart:2013faa, Liu:2013hba, Boughezal:2013oha, Stewart:2011cf, Gangal:2013nxa}.
The impact of QCD scale variations on the Higgs boson \pT\ distribution as well as the uncertainty of the \pT\ distribution in the 0-jet bins are also taken into account.
Higher-order impacts on the \pT\ distribution predictions due to treating the top quark mass as infinite in the heavy-quark loop are accounted for by comparing these predictions with finite-mass calculations.
For the \VBF production mode, the uncertainty due to missing higher orders in QCD are considered, including migration effects in number of jets, transverse momentum of the Higgs boson, transverse momentum of the Higgs boson and leading dijet system, and the invariant mass of the two leading jets as outlined in the scheme presented  in Ref.  \cite{bendavid2018les}.

For production modes other than ggF and VBF, the effects of QCD scale uncertainties are estimated by considering all configurations of renormalisation and factorisation scales varied by a factor of two. In each experimental bin, the largest difference between all the variations and the nominal configuration  is assigned as uncertainty.

The effects of parton shower and multiple-parton interaction modelling uncertainties on the acceptance are estimated using tune eigenvector variations as well as comparisons between acceptances calculated with \progname{Pythia 8} and \progname{Herwig 7} parton showering algorithms.

PDF uncertainty impacts are estimated using the eigenvector variations of the \progname{PDF4LHC\_NLO\_30} Hessian PDF set, following the \progname{PDF4LHC} recommendations \cite{Butterworth:2015oua}.

For the cross sections extrapolated to the full phase space, an additional uncertainty (2.2\%) related to the $H\to ZZ^*$ branching ratio \cite{Bredenstein:2006rh,Bredenstein:2006ha} is included in the measurement.

Since the \zzstar\ process normalisation is  constrained by performing a simultaneous fit of sideband regions enriched in this contribution
together with the signal region, most of the  theoretical uncertainty in the normalisation for this background vanishes.\footnote{Except in cases where the cross-section bins are merged into a single \zzstar\ bin, where the relative normalisation uncertainties are included.}
The uncertainties due to missing higher-order effects in QCD are estimated by varying the
factorisation and renormalisation QCD scales by a factor of two; the impact of the PDF uncertainty
is estimated using the MC replicas of the \progname{NNPDF3.0} PDF set. Uncertainties due to  the parton shower modelling for the \zzstar\ process are considered as well.
The impact of these uncertainties is below 2\% for all the fiducial differential cross sections.
In addition, the $m_{4\ell}$ shape obtained from \progname{Sherpa} is compared with that obtained from \progname{Powheg} and \progname{MadGraph5\_aMC@NLO} and the difference is taken as an additional source of systematic uncertainty. In each $m_{4\ell}$ bin, the largest difference between \progname{Sherpa} and \progname{Powheg} or \progname{MadGraph5\_aMC@NLO} is used, and the systematic uncertainty is determined by interpolating between these shapes. Typically, \progname{Sherpa} and \progname{Powheg} have the largest difference in the predicted $m_{4\ell}$ shape, with the impact linearly varying from approximately $\pm 10\%$ at low $m_{4\ell}$ to $\mp 2\%$ at high $m_{4\ell}$.
 
The uncertainty in the gluon-induced  \zzstar\  process is taken into account as well by changing the relative composition between the quark-initiated and gluon-initiated \zzstar\ components according to the theoretical uncertainty in the predicted cross sections.

Finally, unfolding-related uncertainties arise from uncertainties in the production mode composition that affect the response matrices, as well as from uncertainties in the bias introduced by the unfolding method. For the former, an uncertainty is assessed by varying the production cross sections within their measured uncertainties taken from Ref.~\cite{HIGG-2016-22}, and has an impact of less than 1\%. In the latter case, the uncertainty in the bias is obtained independently per bin by comparing the unfolded cross section from simulation with that expected when varying the underlying true cross sections of the simulated data sample within the expected statistical error. The impact of this uncertainty is typically negligible in distributions such as \ptH, where the response matrix is largely diagonal, but can be of the order of 10\% in distributions with larger bin migrations, such as \njets.
 
 
\clearpage
\section{Results}
\label{sec:results}
 
Results are presented for the full set of inclusive and differential variables outlined in Section~\ref{sec:fiducial}. Section~\ref{sec:yields} presents the data yields from the full Run 2 data set. Section~\ref{sec:thestat} provides details of the statistical procedure used for the extraction of the measurements. Cross-section results, and comparisons with SM predictions, are provided in Sections~\ref{sec:fidResults} and~\ref{sec:diffXSresults}.
 
\subsection{Measured data yields}
\label{sec:yields}

The observed number of events in each of the four decay final states, and the expected signal and background yields before fitting to data (pre-fit), are presented in Table~\ref{tab:recoyields}. These events have passed the event selection and fall  in a narrow window around the Higgs boson mass peak (115 $< \mfourl < $130~\GeV).

\begin{table*}[htbb!]
\centering
\caption{Expected (pre-fit) and observed numbers of events in the four decay final states after the event selection, in the mass range \hbox{115~\gev$< \mfourl <$ 130~\gev}.
The sum of the expected number of SM Higgs boson events and the estimated background yields is compared with the data. Combined statistical and systematic
uncertainties are included for the predictions (see Section~\ref{sec:systematics}).
\label{tab:recoyields}}
\vspace{0.2cm}
\begin{tabular}{*{6}{c}}
\hline \hline
\noalign{\vspace{0.05cm}}
Final    & \multicolumn{1}{c}{Signal}   & \multicolumn{1}{c}{\zzstar}    & \multicolumn{1}{c}{Other}		 & \multicolumn{1}{c}{Total}  &      \multicolumn{1}{c}{Observed}\\
state    &  							   & \multicolumn{1}{c}{background}	& \multicolumn{1}{c}{backgrounds}& \multicolumn{1}{c}{expected}& 								 \\\hline
 
$4\mu$  & $  78    \pm  5 $ &	$  38.0  \pm  2.1 $ &	$  2.85  \pm  0.18 $ &	$  119   \pm  5~~   $ &	115~~ \\
$2e2\mu$  & $  53.0  \pm  3.1 $ &	$  26.1  \pm  1.4 $ &	$  2.98  \pm  0.19 $ &	$  82.0  \pm  3.4 $ &	96 \\
$2\mu2e$  & $  40.1  \pm  2.9 $ &	$  17.3  \pm  1.3 $ &	$  3.6   \pm  0.5  $ &	$  61.0  \pm  3.2 $ &	57 \\
$4e$  & $  35.3  \pm  2.6 $ &	$  15.0  \pm  1.5 $ &	$  2.91  \pm  0.33 $ &	$  53.2  \pm  3.1 $ &	42 \\
 
\noalign{\vspace{0.05cm}}
\hline
Total  & $  206  \pm  13 $ &	$  96  \pm  6 $ &	$  12.2  \pm  1.0~~ $ &	$  315  \pm  14 $ &	310~~ \\

\hline \hline
 
\end{tabular}
\end{table*}

\begin{figure}[!htbp]
\centering
\includegraphics[width=0.55\linewidth]{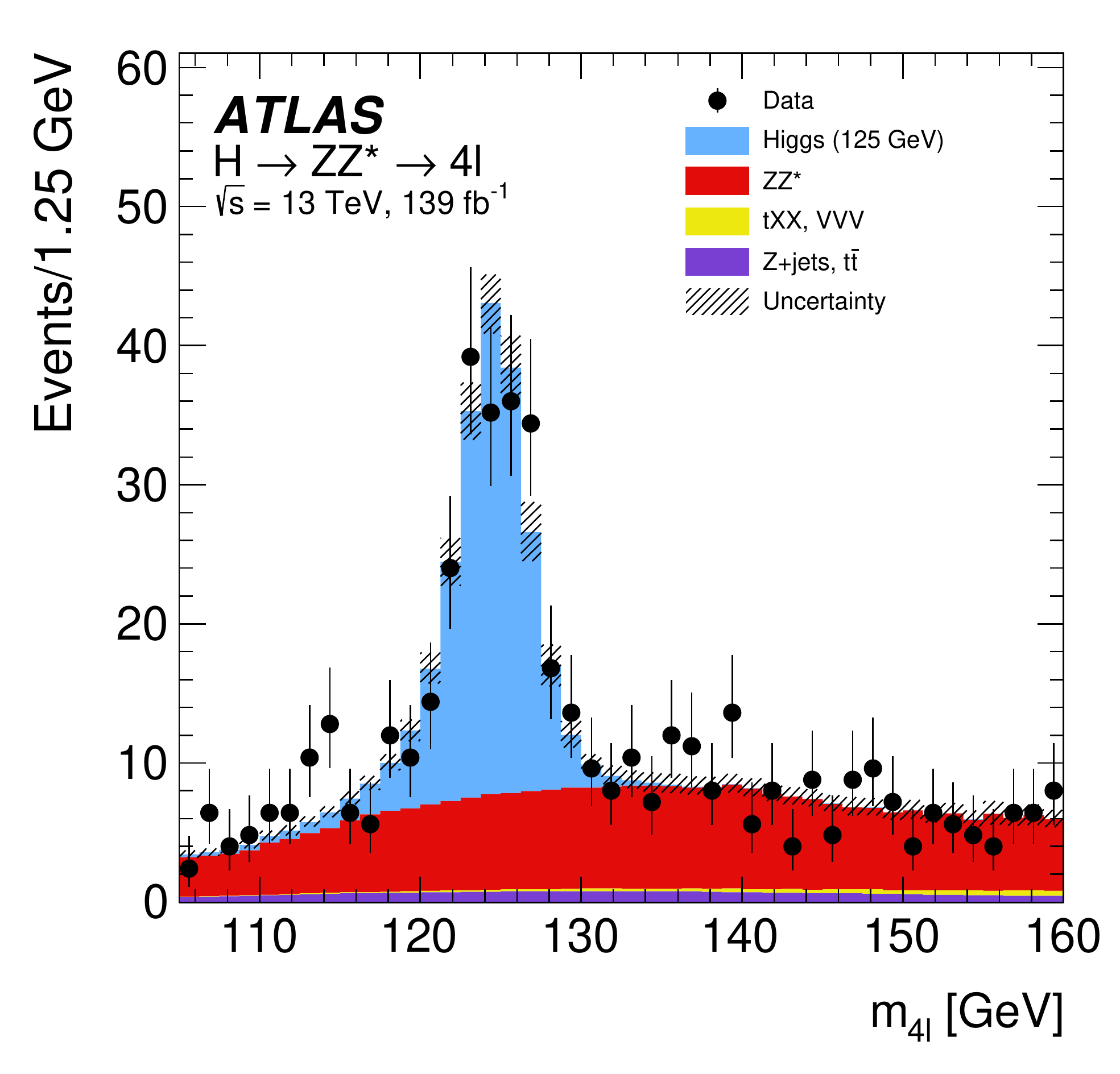}
 
\caption{The observed and expected (pre-fit) inclusive four-lepton invariant mass distributions for the selected Higgs boson candidates, shown
for an integrated luminosity of \Lum\ and at $\sqrt{\mathrm{s}}=$~13~TeV. The uncertainty in the prediction is shown by the hatched band, 
which includes the theoretical uncertainties of the SM cross section for the signal and the \zzstar\ background.
}
\label{fig:recoyields}
\end{figure}

\begin{figure}[!htbp]
\centering
\subfloat[]{ \includegraphics[width=0.48\linewidth]{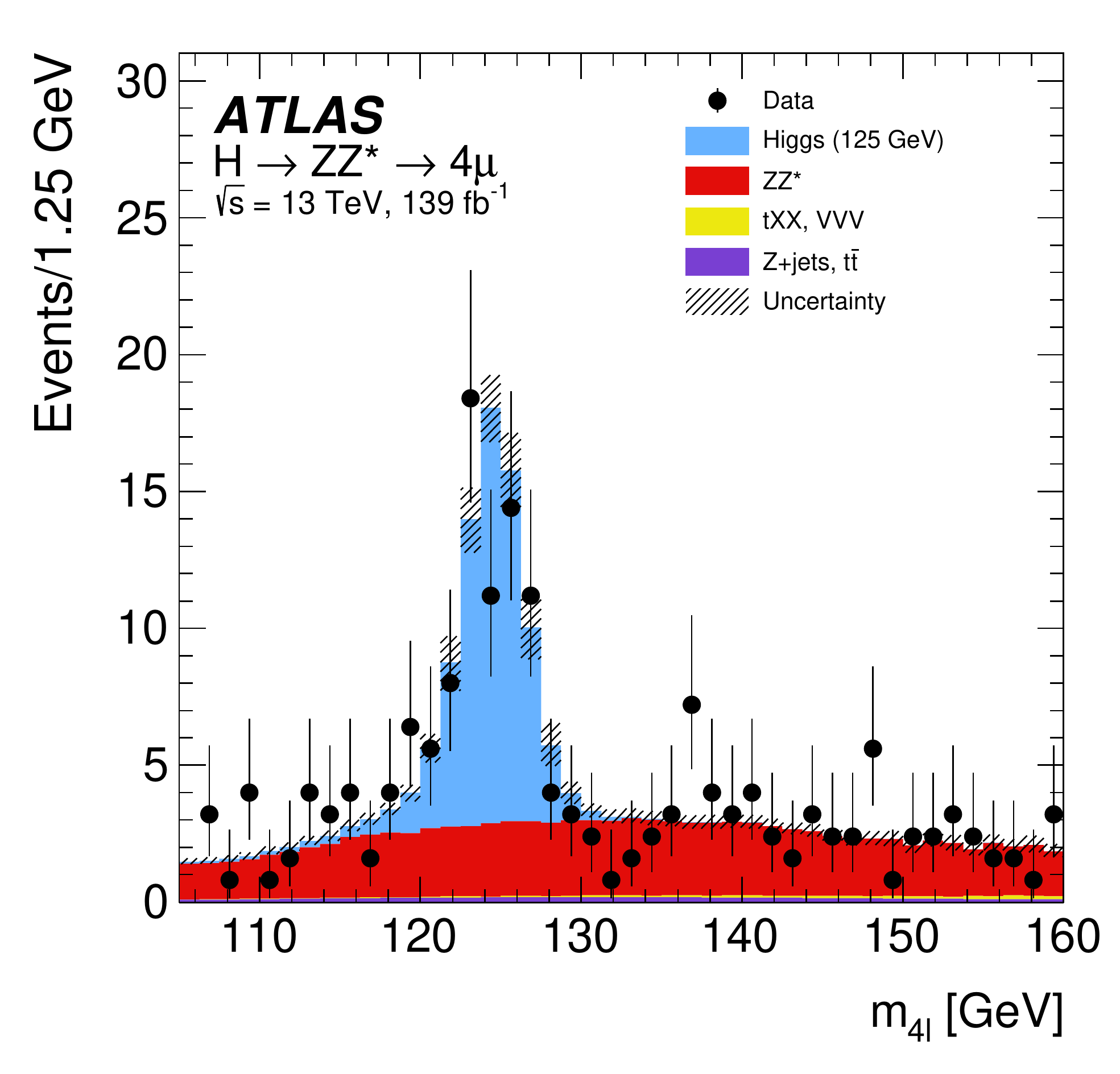}\label{fig:m4l4m}}
\subfloat[]{ \includegraphics[width=0.48\linewidth]{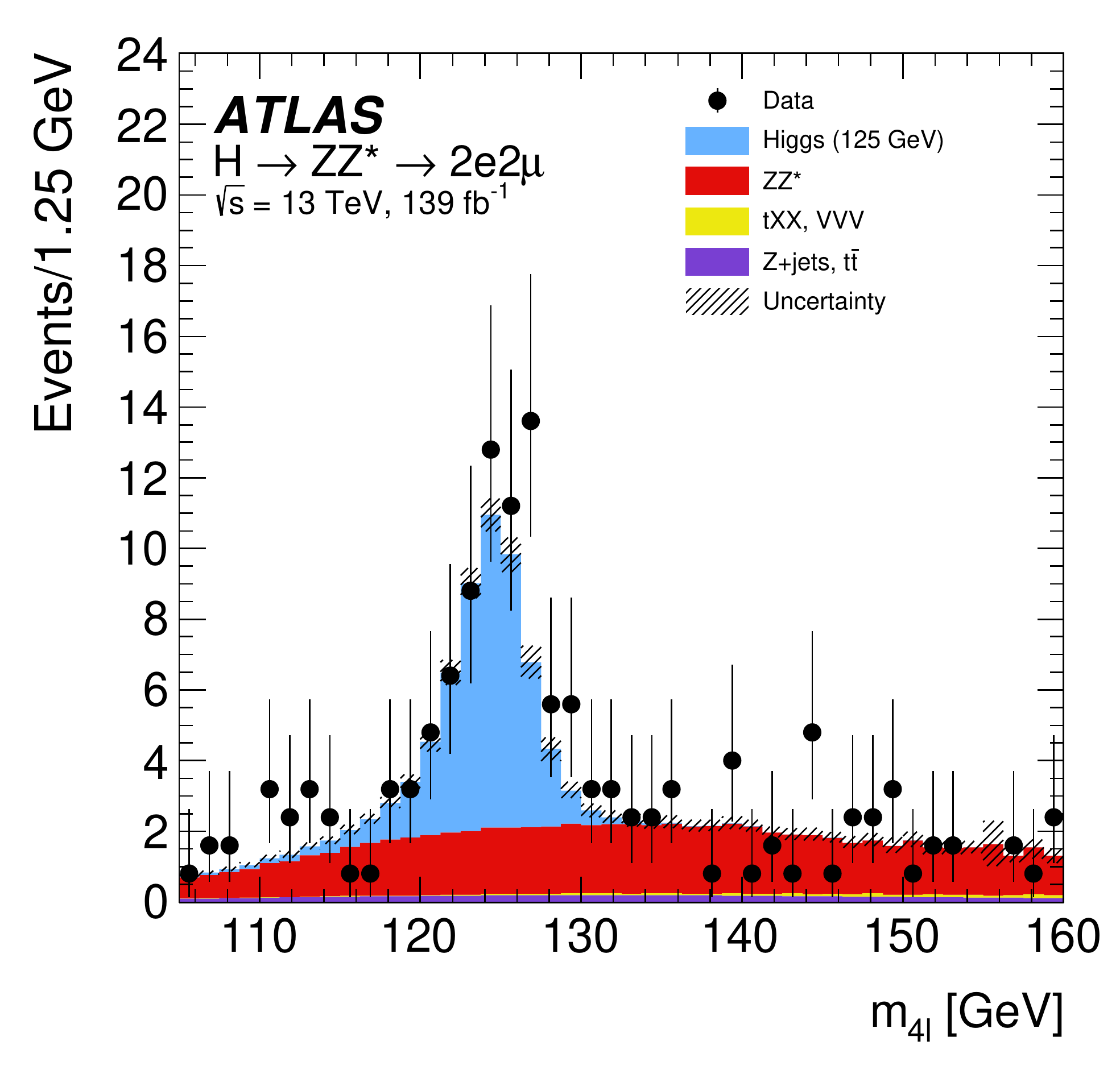}\label{fig:m4l2e2m}} \\
\subfloat[]{ \includegraphics[width=0.48\linewidth]{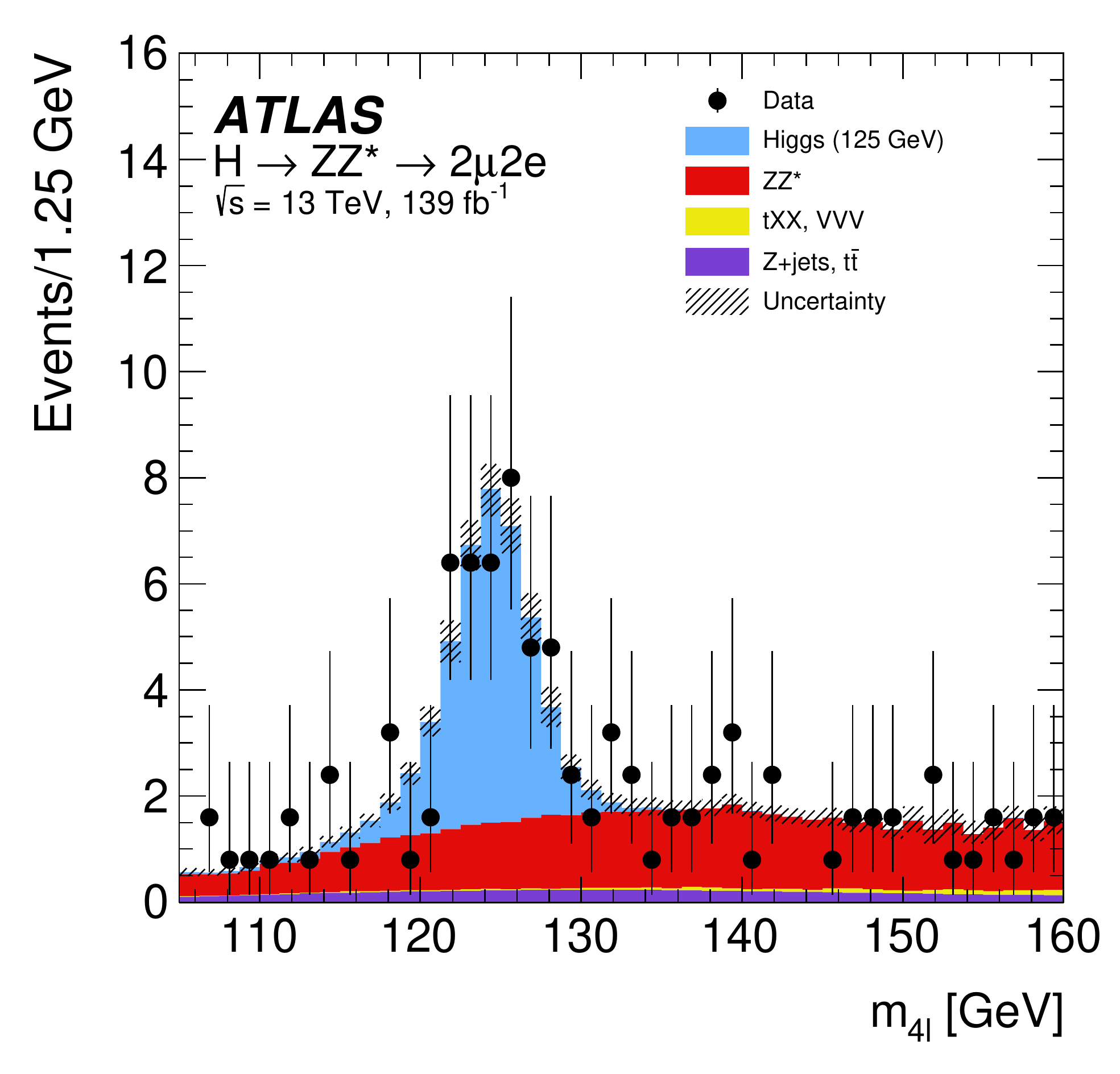}\label{fig:m4l2m2e}}
\subfloat[]{ \includegraphics[width=0.48\linewidth]{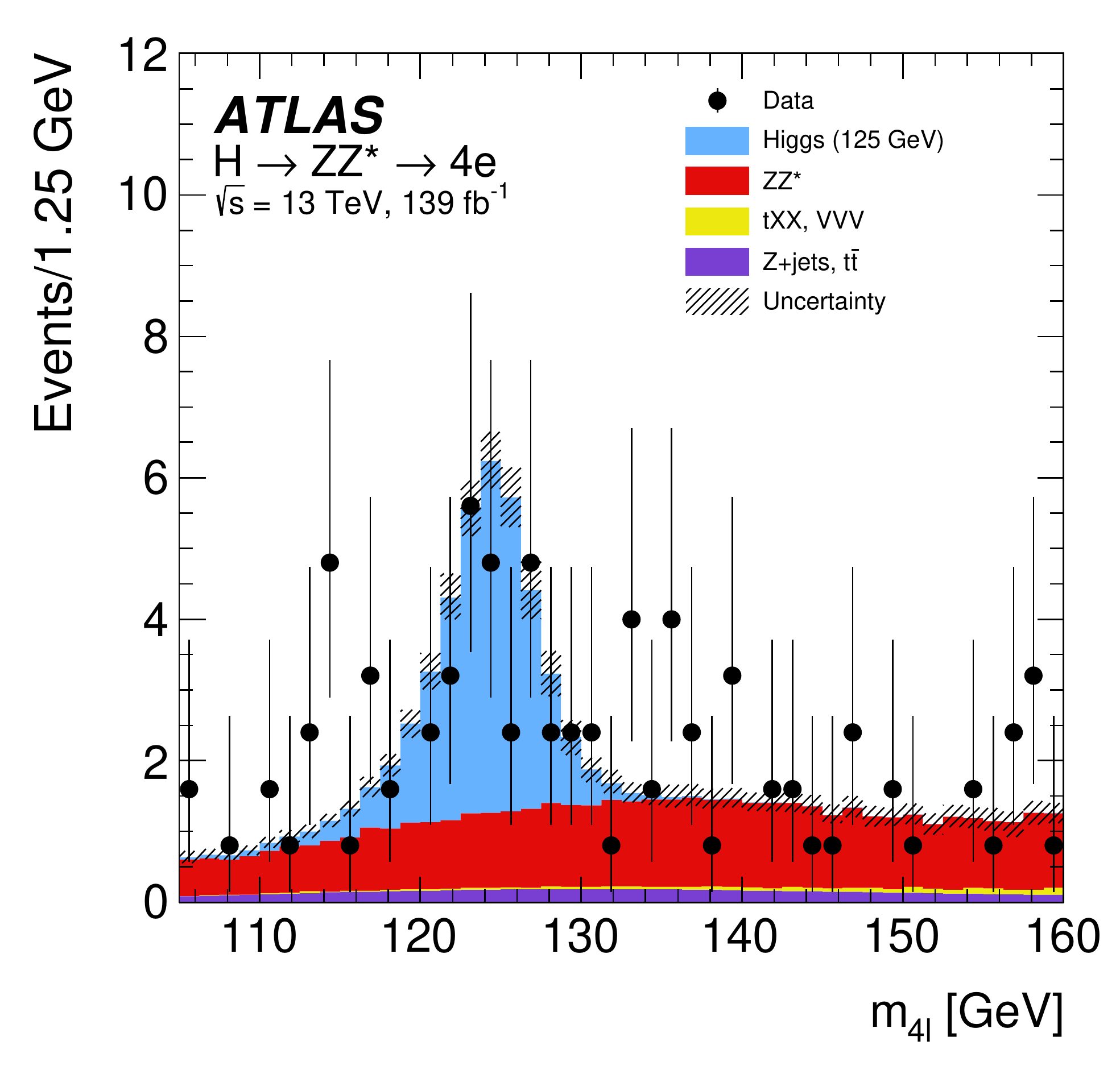}\label{fig:m4l4e}}
\caption{The observed and expected (pre-fit) four-lepton invariant mass distribution for the selected Higgs boson candidates, for the different decay final states \protect\subref{fig:m4l4m} $4\mu$, \protect\subref{fig:m4l2e2m} $2e2\mu$, \protect\subref{fig:m4l2m2e} $2\mu2e$, \protect\subref{fig:m4l4e} $4e$. The uncertainty in the prediction is shown by the hatched band, 
which includes the theoretical uncertainties of the SM cross section for the signal and the \zzstar\ background.}
\label{fig:recoyields2}
\end{figure}

Figures~\ref{fig:recoyields} and \ref{fig:recoyields2} show the expected and observed four-lepton invariant mass distributions, inclusively and per final state respectively.
The $m_{4\ell}$ distribution shows two clear peaks corresponding to $Z \to 4 \ell$
production and the Higgs boson signal with a mass near 125~\GeV.
 
\begin{figure}[!htbp]
\begin{center}
\subfloat[]{\includegraphics[width=0.46\textwidth]{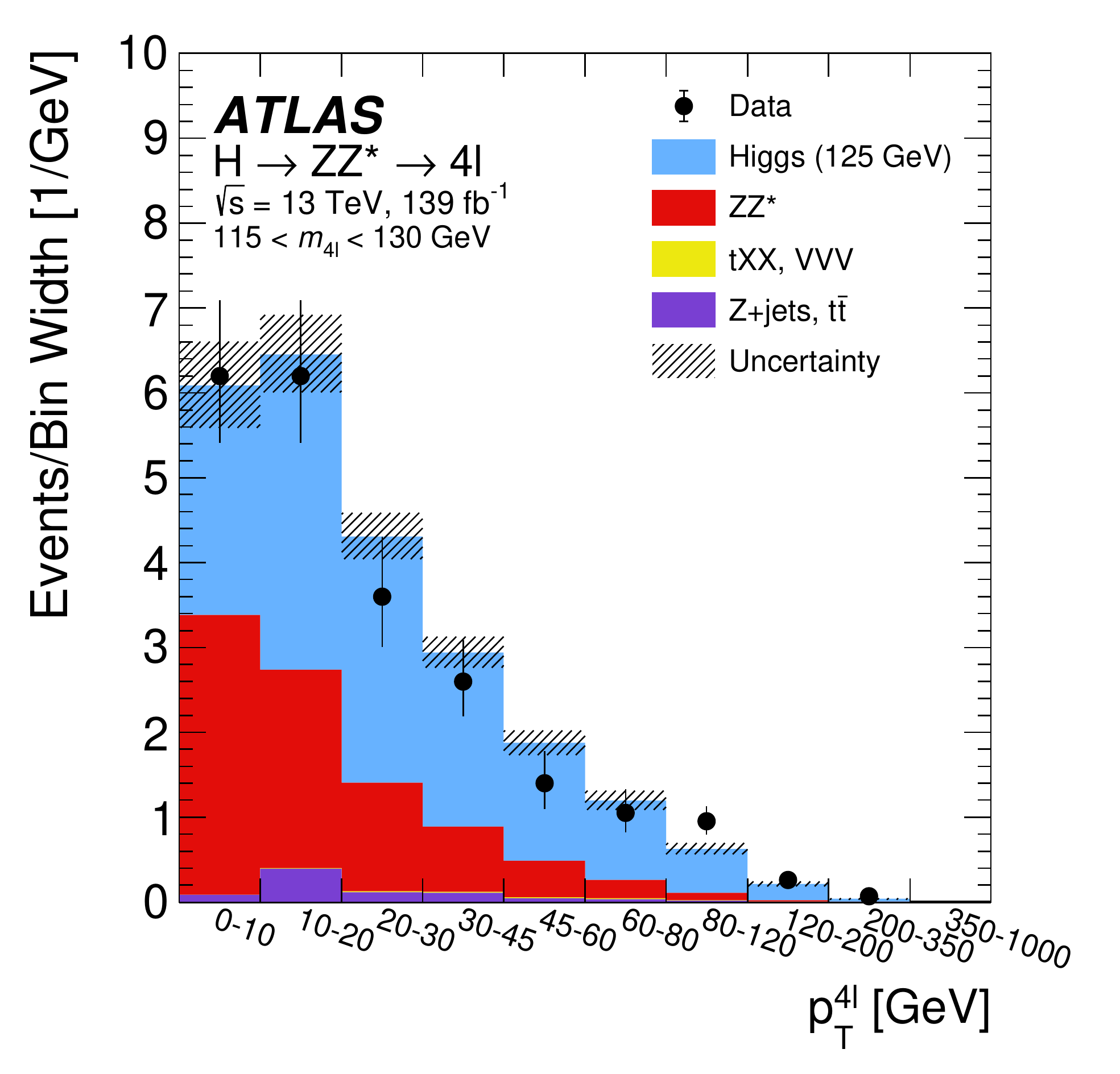}\label{fig:pt4lreco}}\\
\subfloat[]{\includegraphics[width=0.46\textwidth]{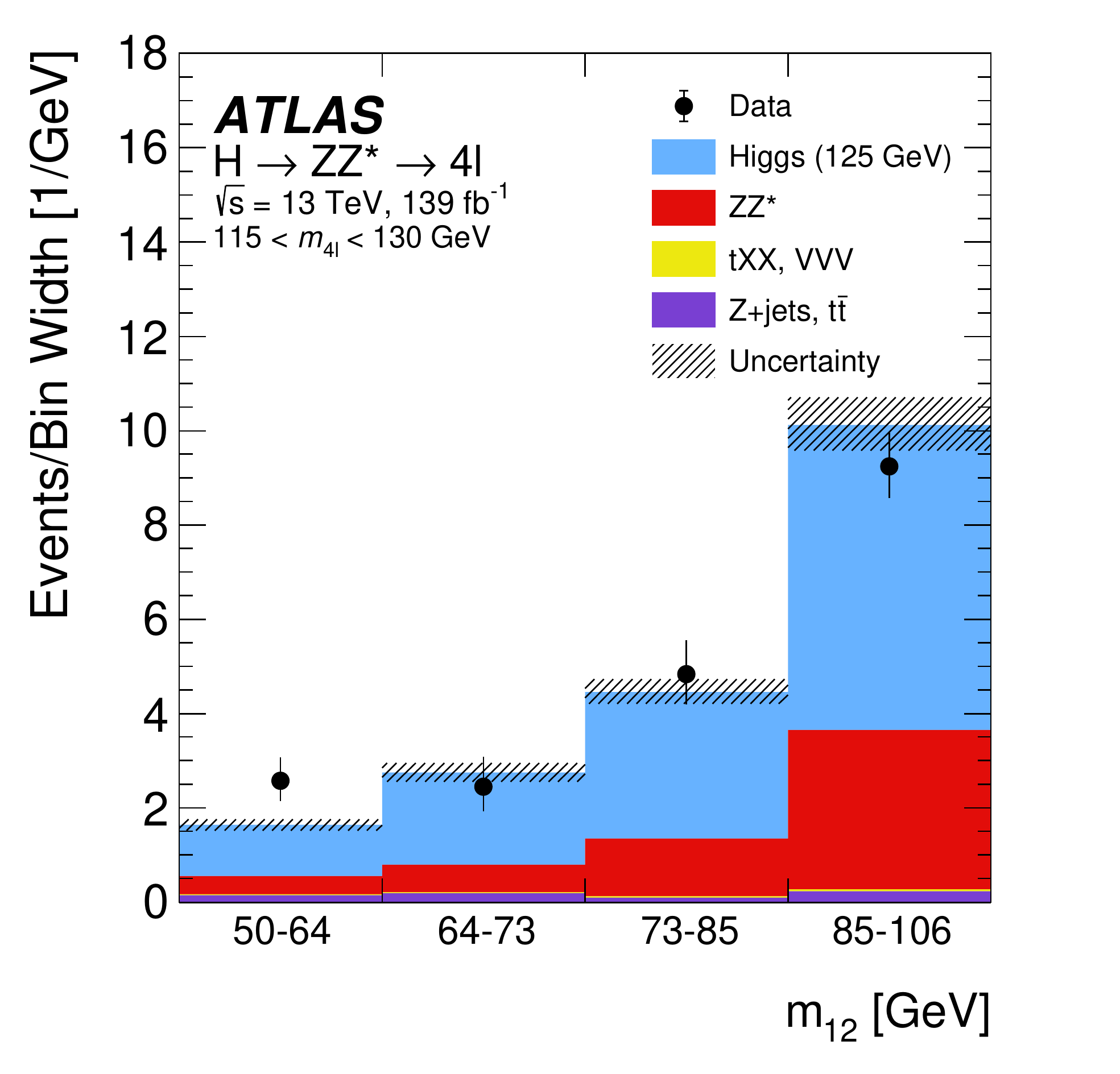}\label{fig:m12reco}}
\subfloat[]{\includegraphics[width=0.46\textwidth]{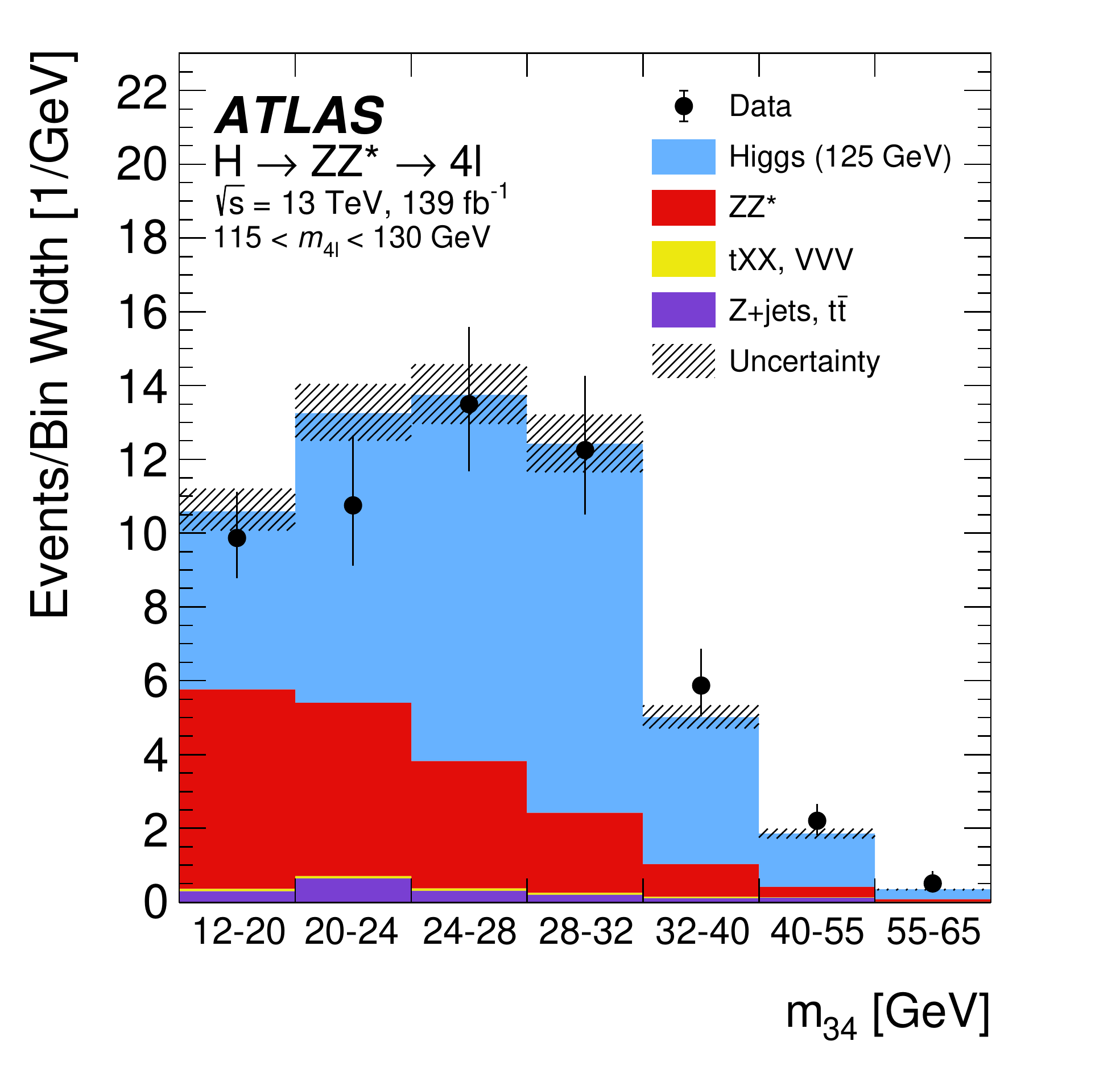}\label{fig:m34reco}}
\end{center}
\vspace{-0.3cm}
\caption{The observed and expected (pre-fit) distributions of \protect\subref{fig:pt4lreco}  \ptH, \protect\subref{fig:m12reco}  \monetwo, and \protect\subref{fig:m34reco}  \mthreefour\
in the mass region $115 < \mfourl < 130$~\GeV, for an integrated luminosity of \Lum~collected at $\sqrt{\mathrm{s}}=$~13~\TeV. A SM Higgs boson signal with a
mass $m_{H}$ = 125~\gev\ is assumed. The uncertainty in the prediction is shown by the hatched band, 
which includes the theoretical uncertainties of the SM cross section for the signal and the \zzstar\ background.}
\label{fig:categoryVars1}
\end{figure}

\begin{figure}[!htbp]
\begin{center}
\subfloat[]{\includegraphics[width=0.5\textwidth]{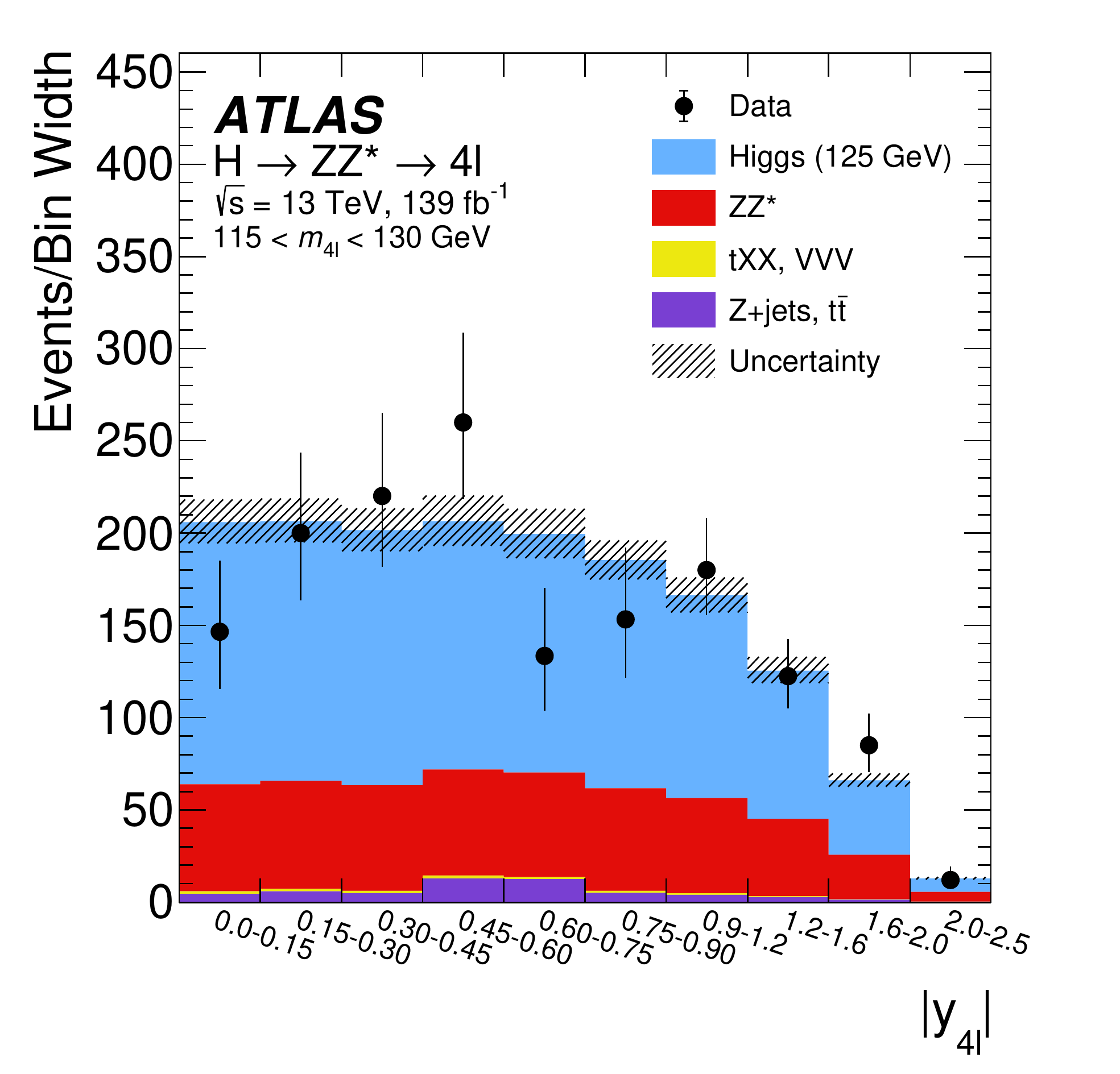}\label{fig:y4lreco}}
\subfloat[]{\includegraphics[width=0.5\textwidth]{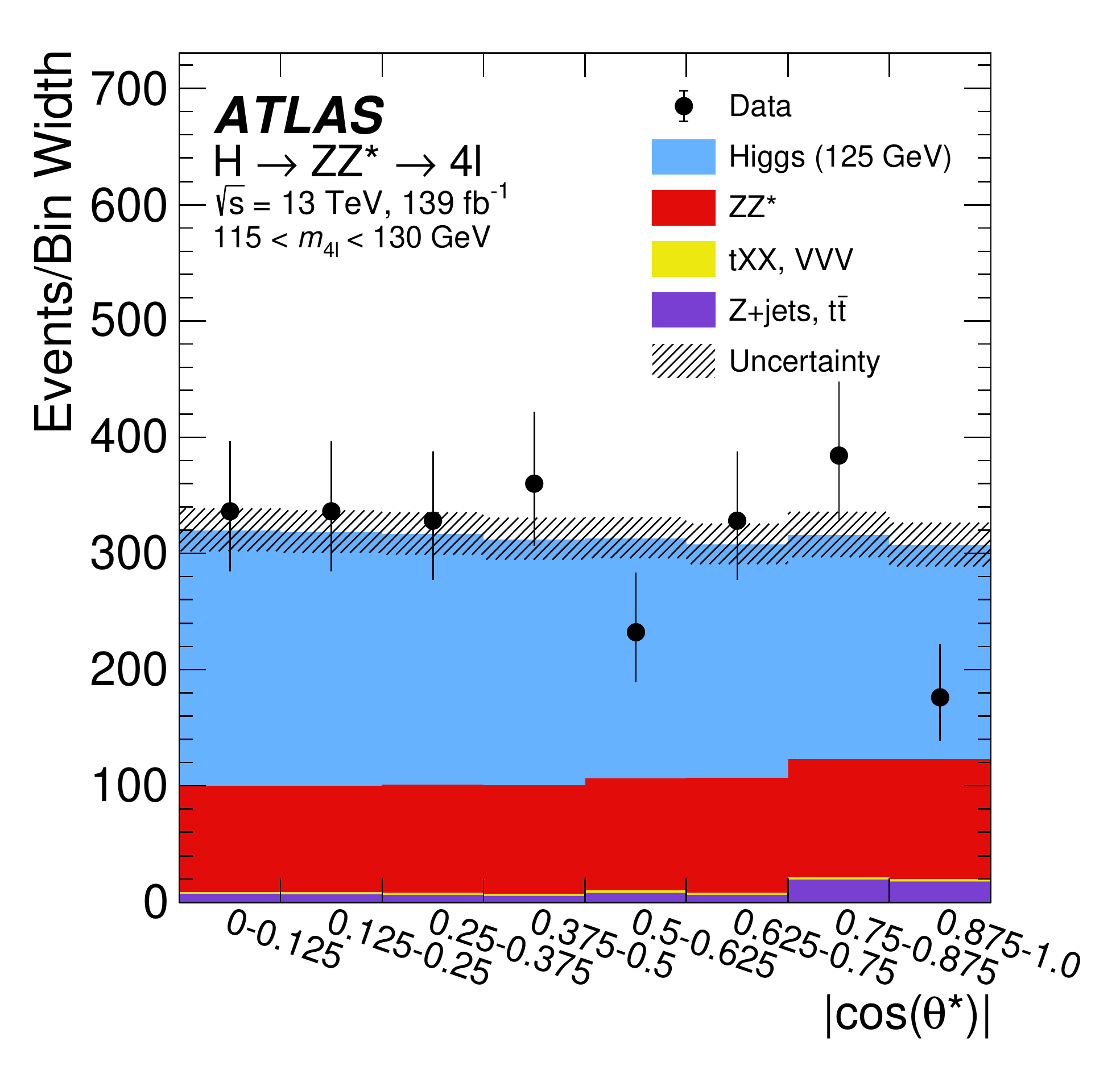}\label{fig:ctsreco}}\\
\end{center}
\vspace{-0.3cm}
\caption{The observed and expected (pre-fit) distributions of \protect\subref{fig:y4lreco} \yH\ and \protect\subref{fig:ctsreco} \costhetastar\
in the mass region $115 < \mfourl < 130$~\GeV, for an integrated luminosity of \Lum~collected at $\sqrt{\mathrm{s}}=$~13~\TeV. A SM Higgs boson signal with a
mass $m_{H}$ = 125~\gev\ is assumed. The uncertainty in the prediction is shown by the hatched band, 
which includes the theoretical uncertainties of the SM cross section for the signal and the \zzstar\ background.}
\label{fig:categoryVars2}
\end{figure}

\begin{figure}[!htbp]
\begin{center}
\subfloat[]{\includegraphics[width=0.5\textwidth]{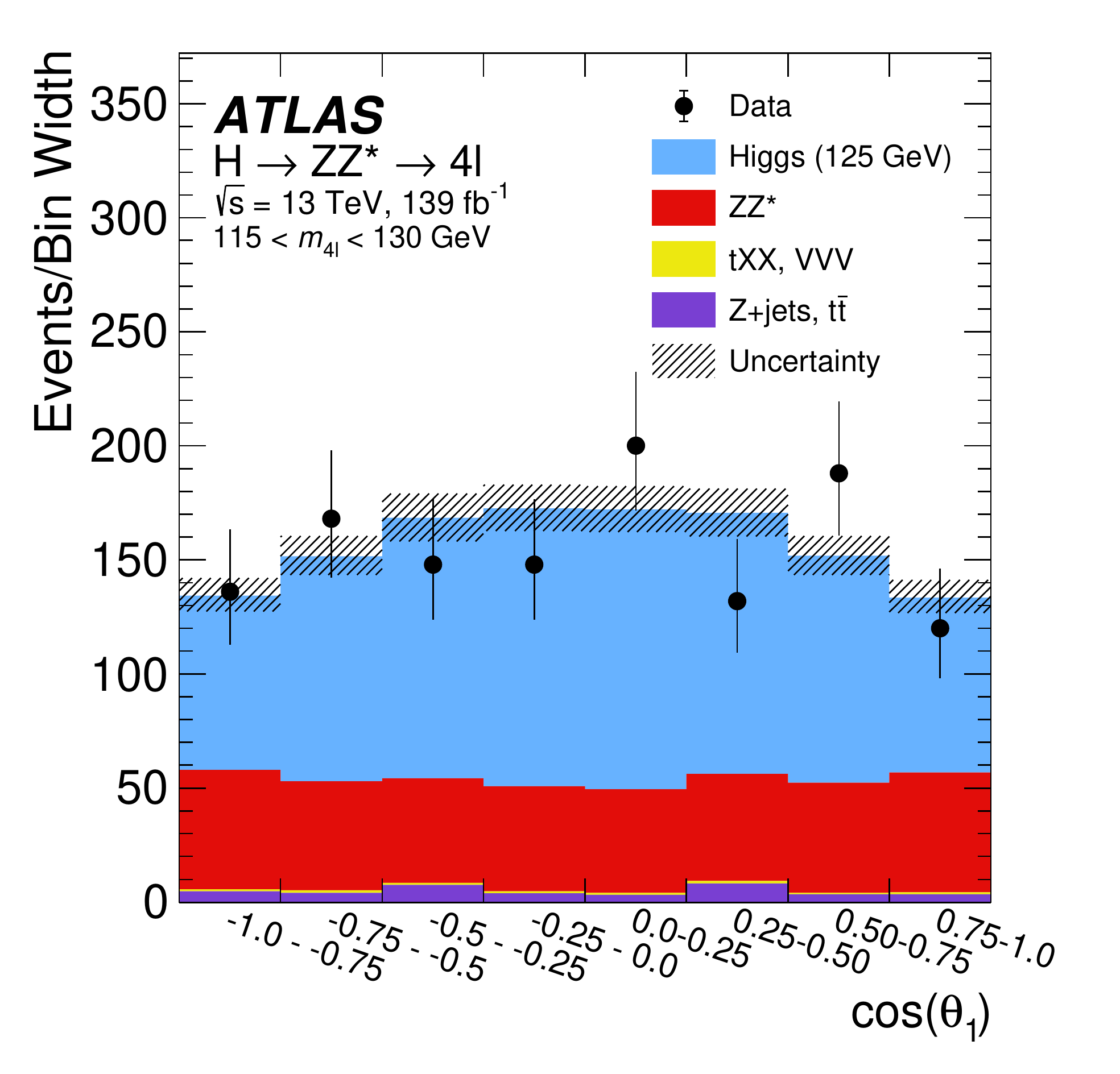}\label{fig:ct1reco}}
\subfloat[]{\includegraphics[width=0.5\textwidth]{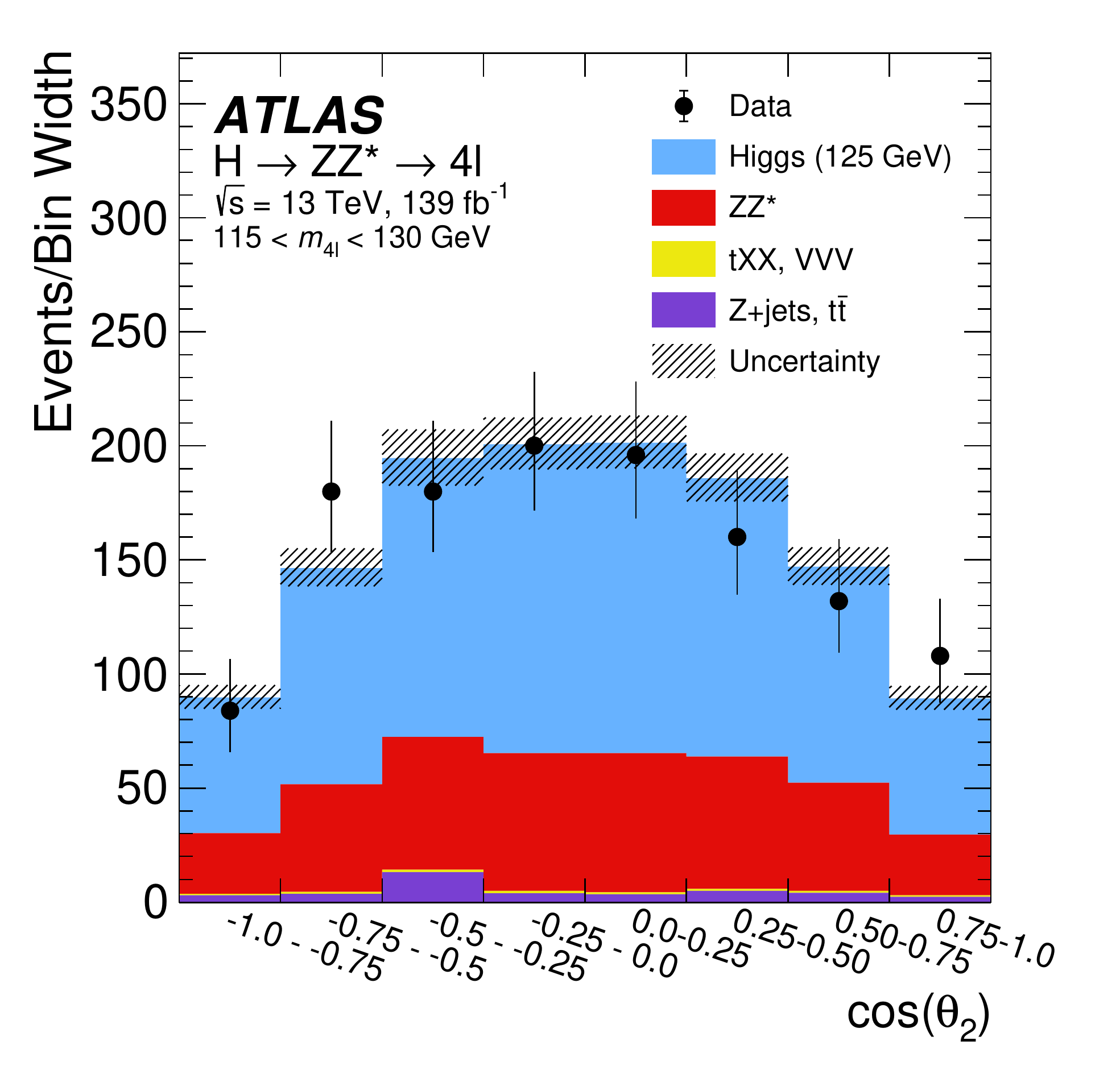}\label{fig:ct2reco}}\\
\subfloat[]{\includegraphics[width=0.5\textwidth]{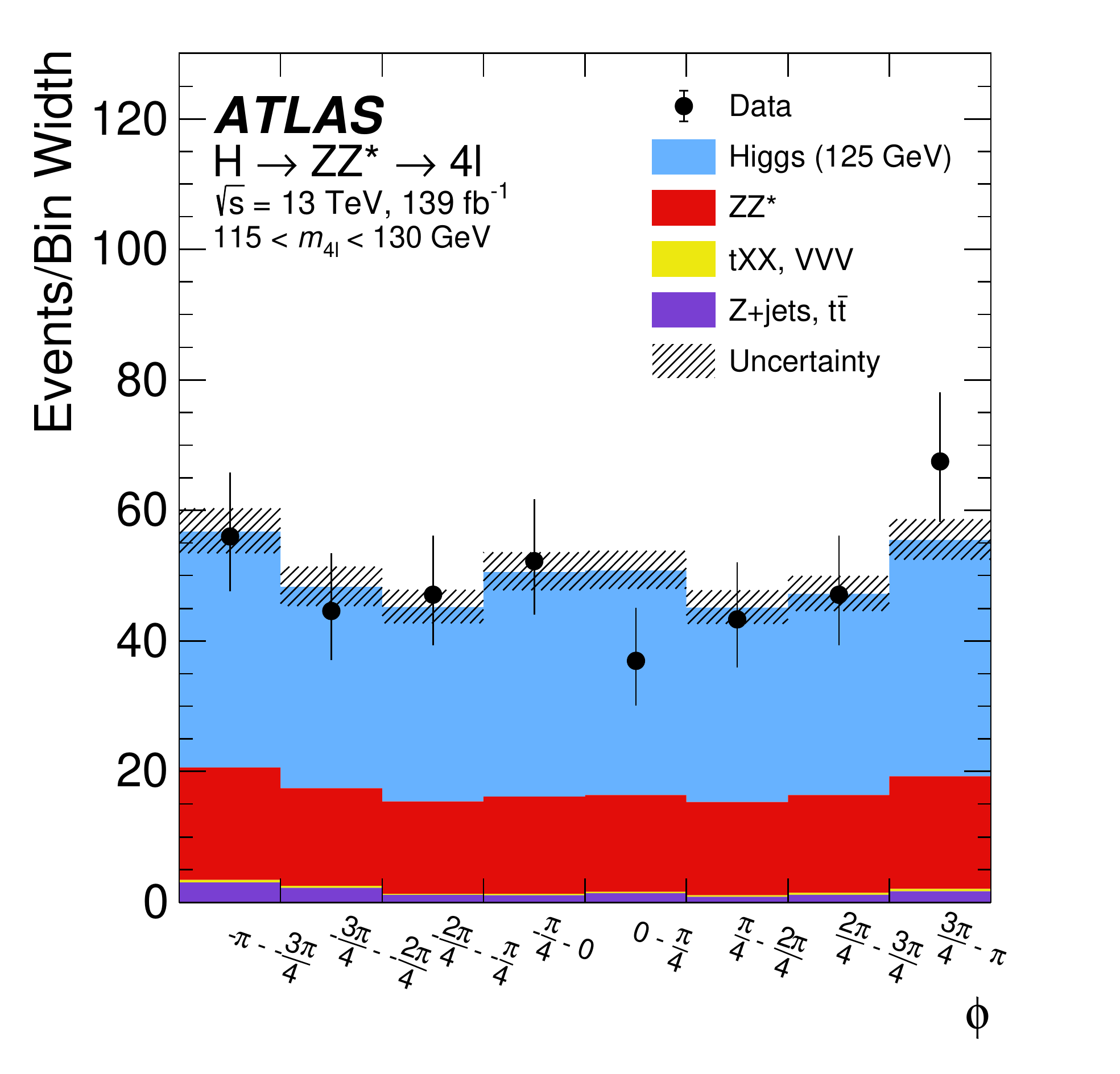}\label{fig:phireco}}
\subfloat[]{\includegraphics[width=0.5\textwidth]{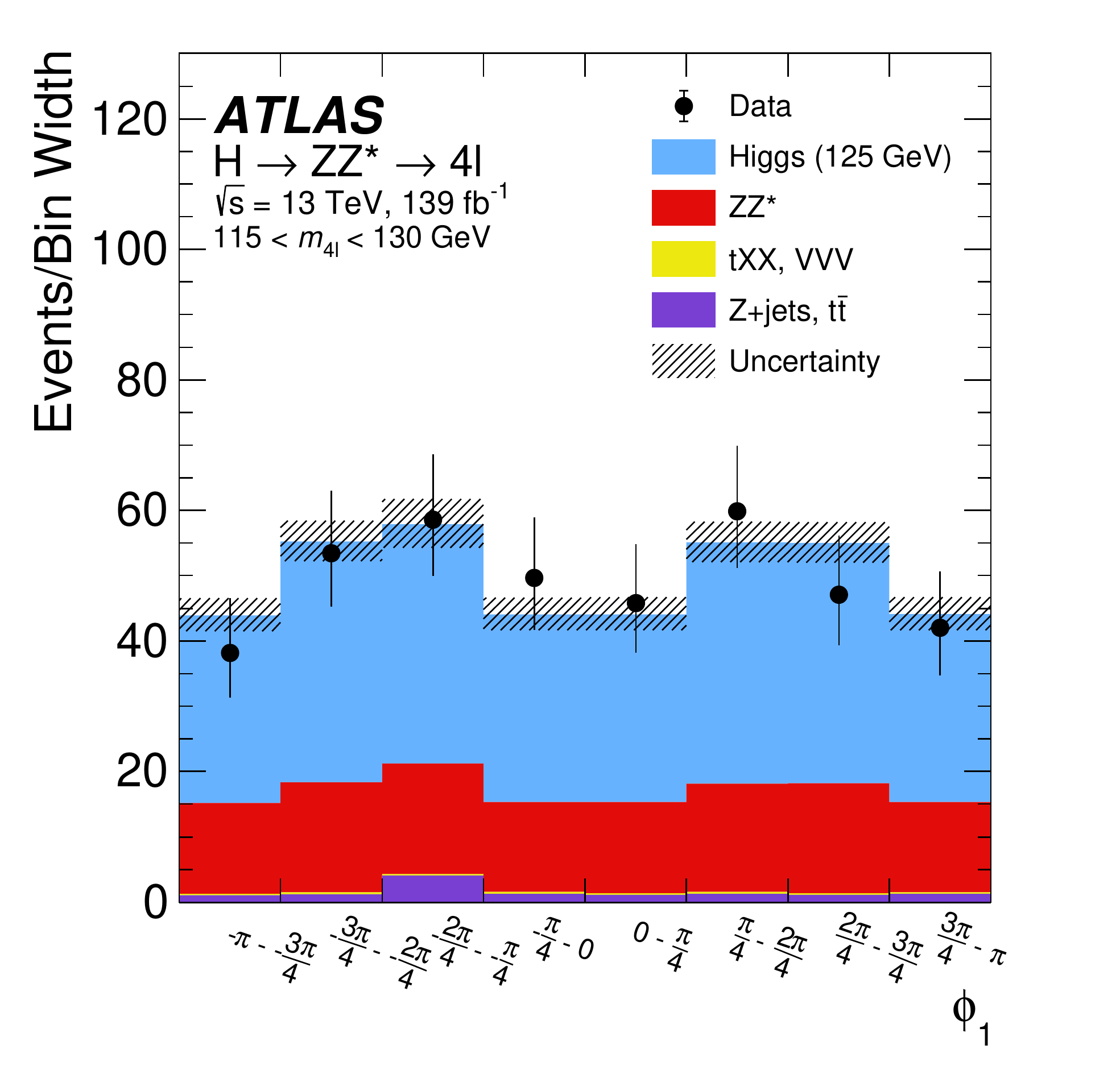}\label{fig:phi1reco}}
\end{center}
\vspace{-0.3cm}
\caption{The observed and expected (pre-fit) distributions of \protect\subref{fig:ct1reco} \costhetaone, \protect\subref{fig:ct2reco} \costhetatwo, \protect\subref{fig:phireco} \phiZZ, and \protect\subref{fig:phi1reco} \phione\
in the mass region $115 < \mfourl < 130$~\GeV, for an integrated luminosity of \Lum~collected at $\sqrt{\mathrm{s}}=$~13~\TeV. A SM Higgs boson signal with a
mass $m_{H}$ = 125~\gev\ is assumed. The uncertainty in the prediction is shown by the hatched band, 
which includes the theoretical uncertainties of the SM cross section for the signal and the \zzstar\ background.}
\label{fig:categoryVars2p5}
\end{figure}

\begin{figure}[!htbp]
\begin{center}
\subfloat[]{\includegraphics[width=0.5\textwidth]{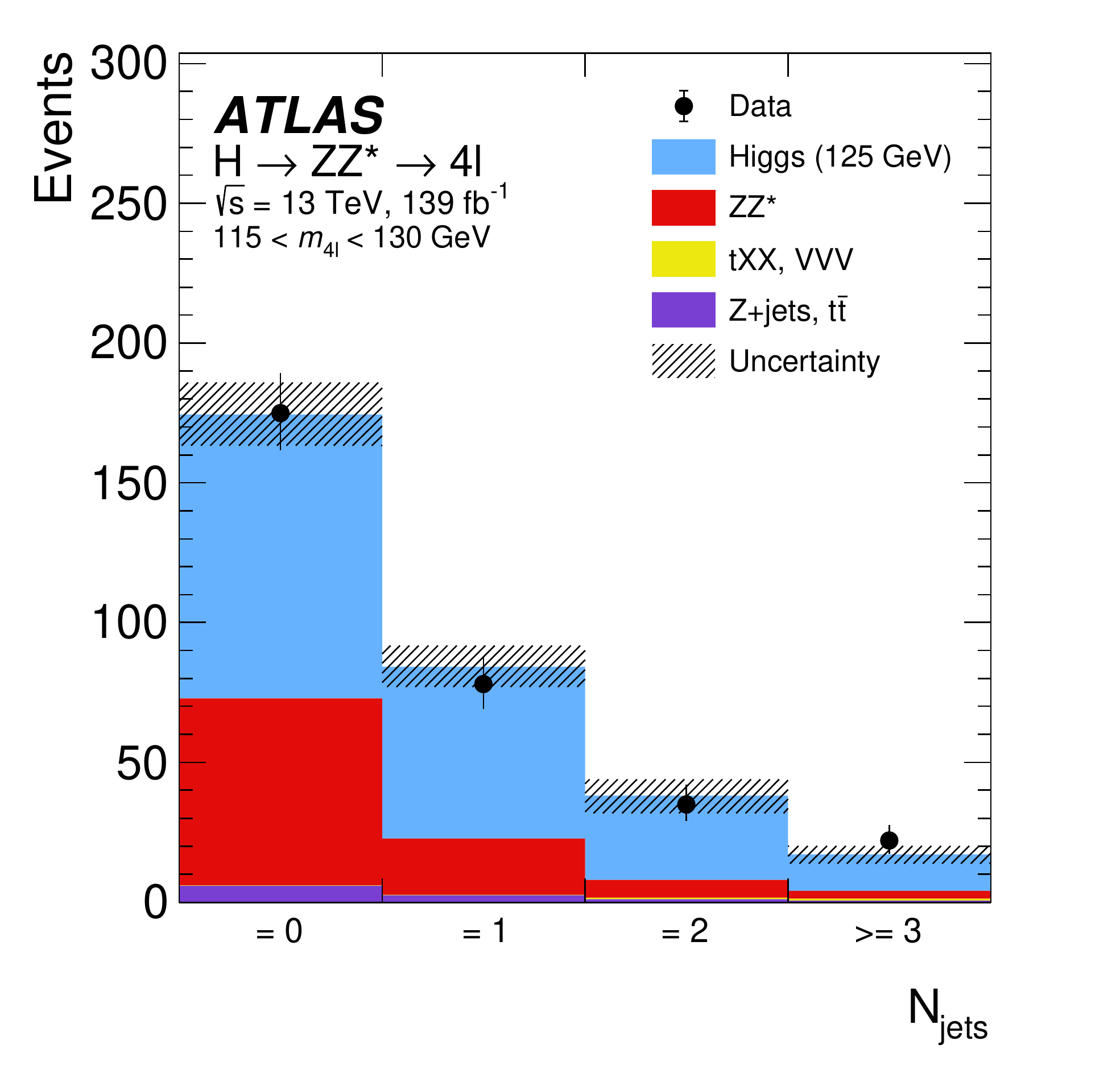}\label{fig:njetsreco}}
\subfloat[]{\includegraphics[width=0.5\textwidth]{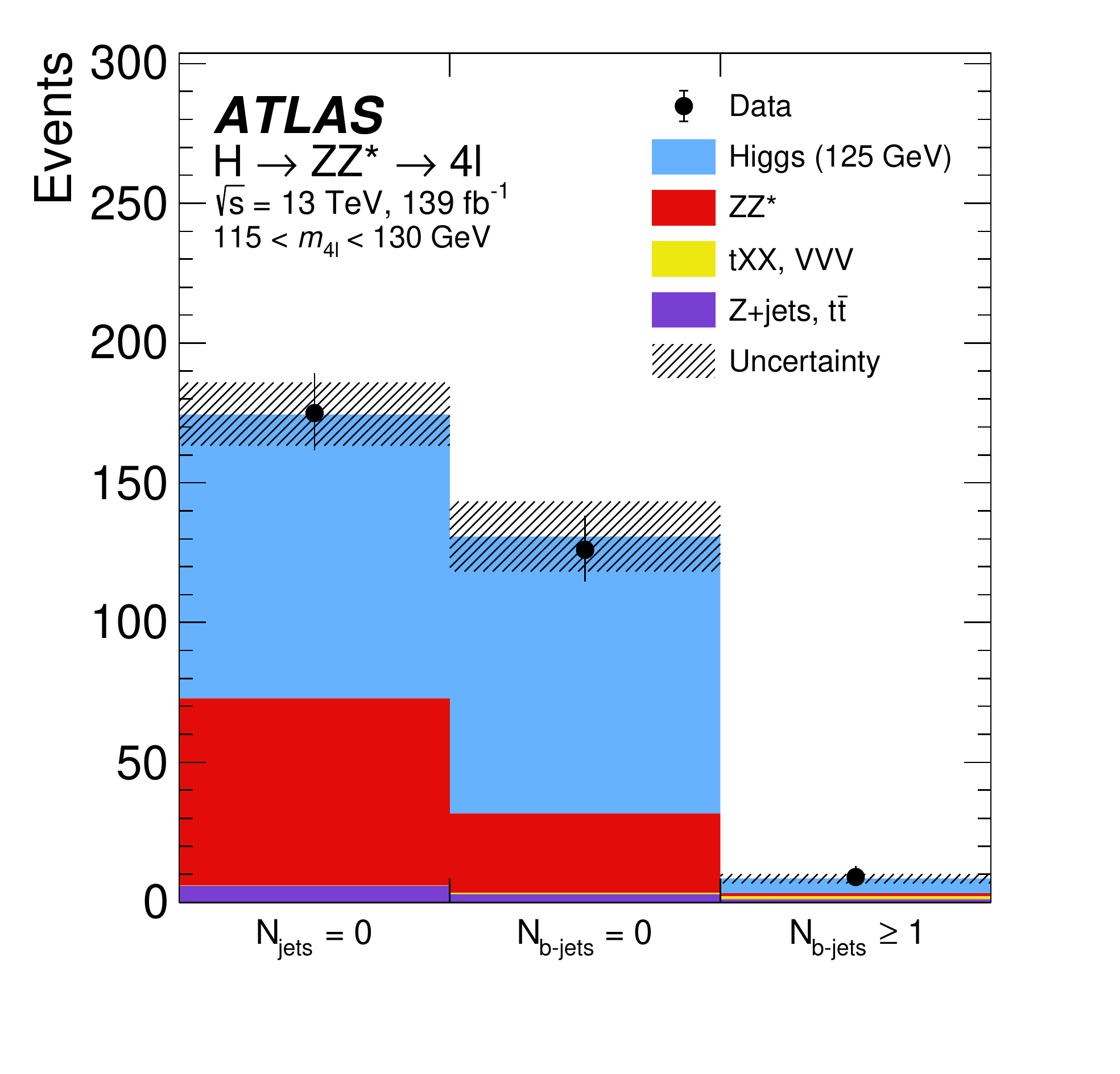}\label{fig:nbjetsreco}}\\
\subfloat[]{\includegraphics[width=0.5\textwidth]{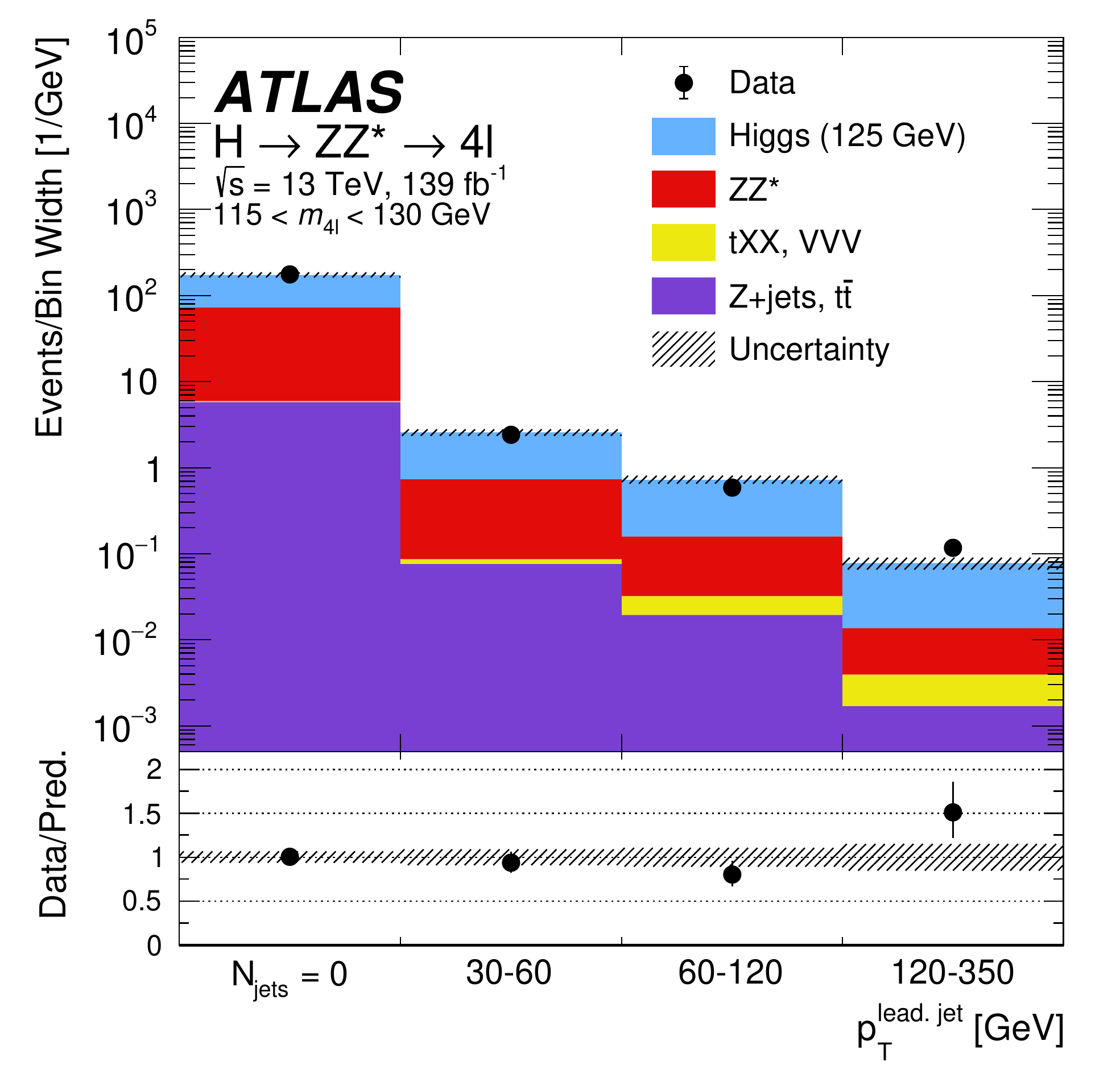}\label{fig:ljptreco}}
\subfloat[]{\includegraphics[width=0.5\textwidth]{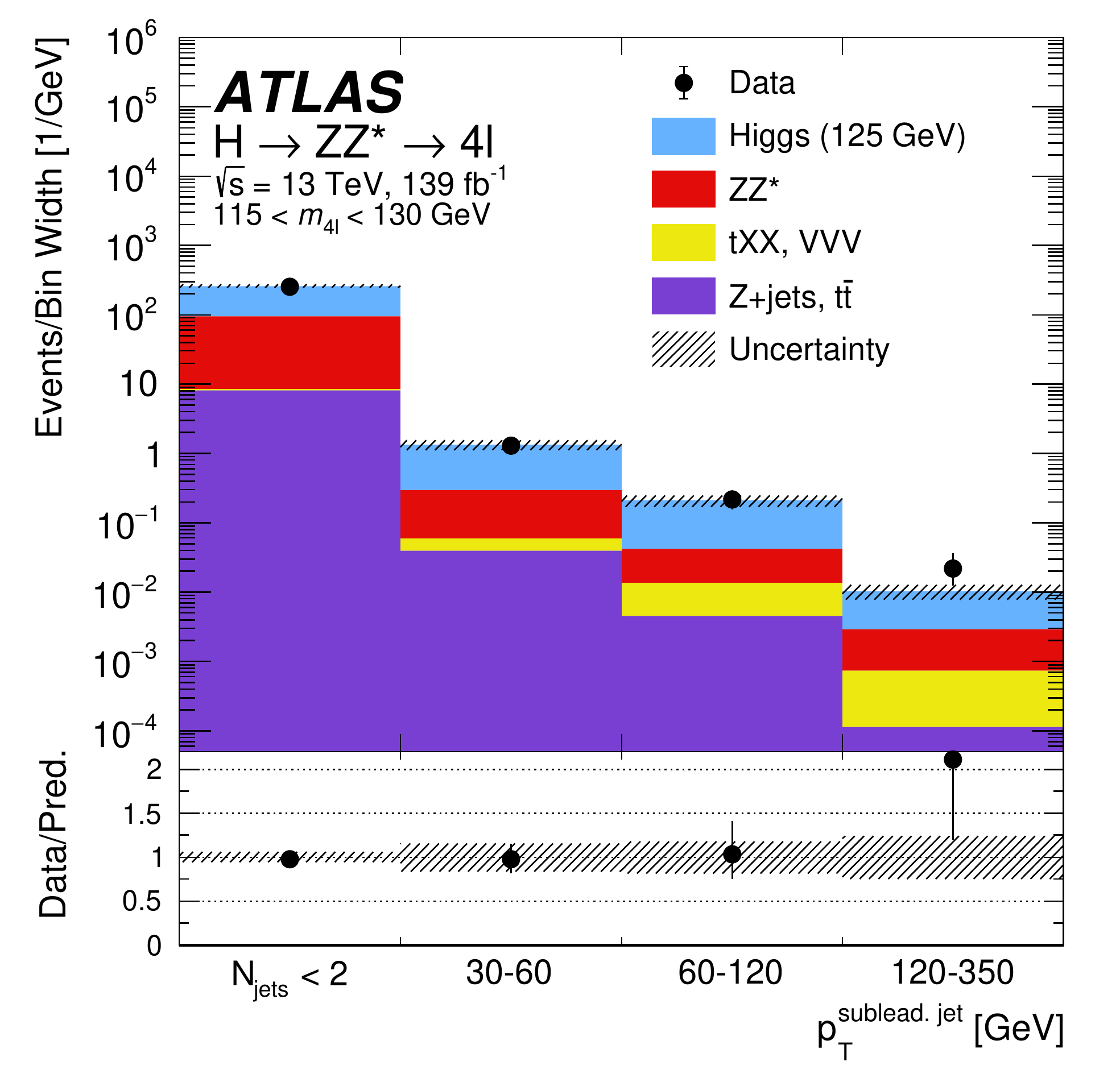}\label{fig:sjptreco}}\\
\end{center}
\vspace{-0.3cm}
\caption{The observed and expected (pre-fit) distributions of \protect\subref{fig:njetsreco} \njets, \protect\subref{fig:nbjetsreco} \nbjets, \protect\subref{fig:ljptreco} \ljpt, and \protect\subref{fig:sjptreco} \sjpt\
in the mass region $115 < \mfourl < 130$~\GeV, for an integrated luminosity of \Lum~collected at $\sqrt{\mathrm{s}}=$~13~\TeV. A SM Higgs boson signal with a
mass $m_{H}$ = 125~\gev\ is assumed.
In distribution \protect\subref{fig:ljptreco}, the first bin contains events with zero jets, while in distribution \protect\subref{fig:sjptreco}, the first bin contains events with fewer than two jets. In both \protect\subref{fig:ljptreco} and \protect\subref{fig:sjptreco}, all bins except the first are divided by the bin width.
The uncertainty in the prediction is shown by the hatched band, 
which includes the theoretical uncertainties of the SM cross section for the signal and the \zzstar\ background.}
\label{fig:categoryVars3}
\end{figure}

\begin{figure}[!htbp]
\begin{center}
\subfloat[]{\includegraphics[width=0.5\textwidth]{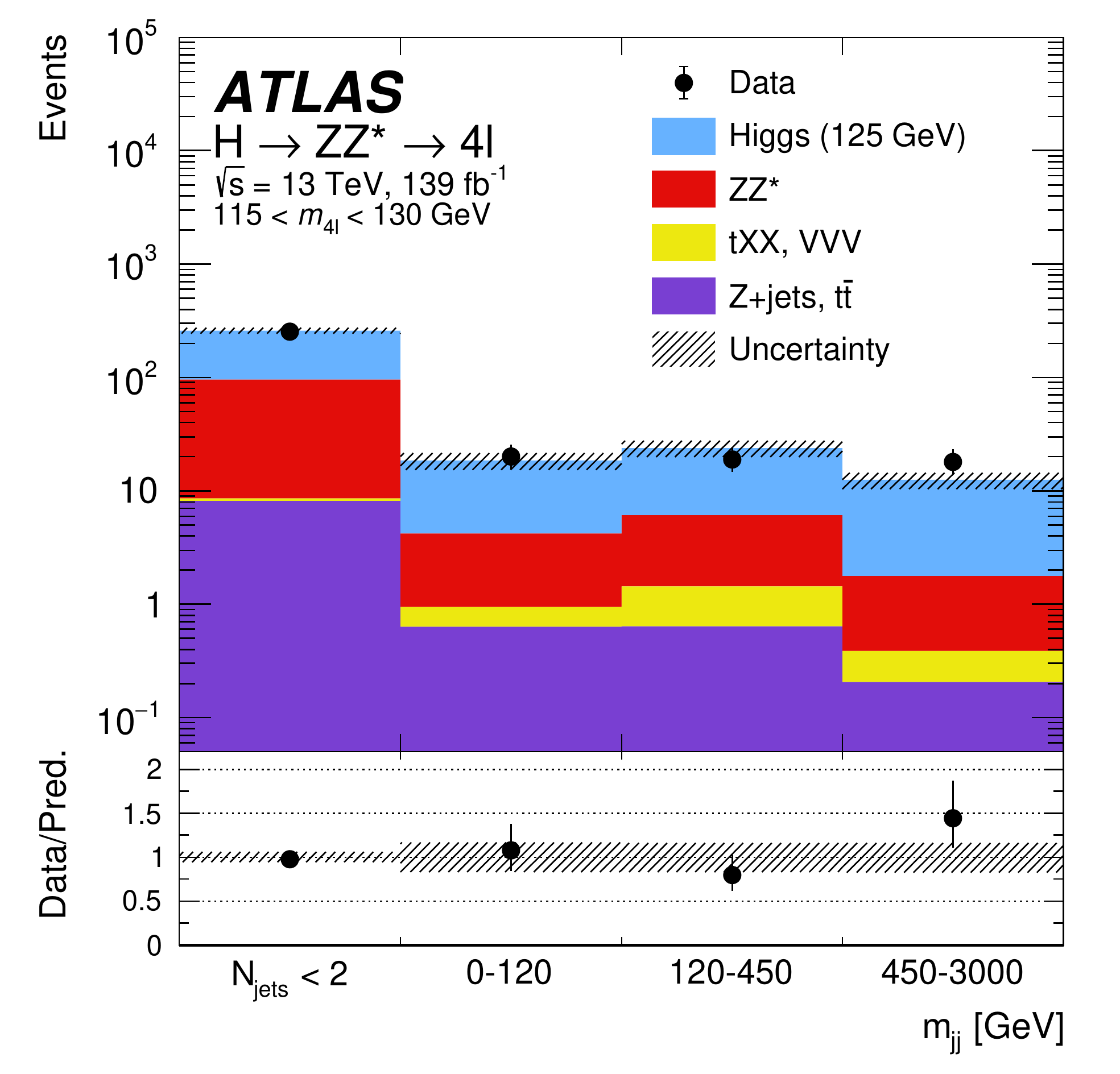}\label{fig:mjjreco}}
\subfloat[]{\includegraphics[width=0.5\textwidth]{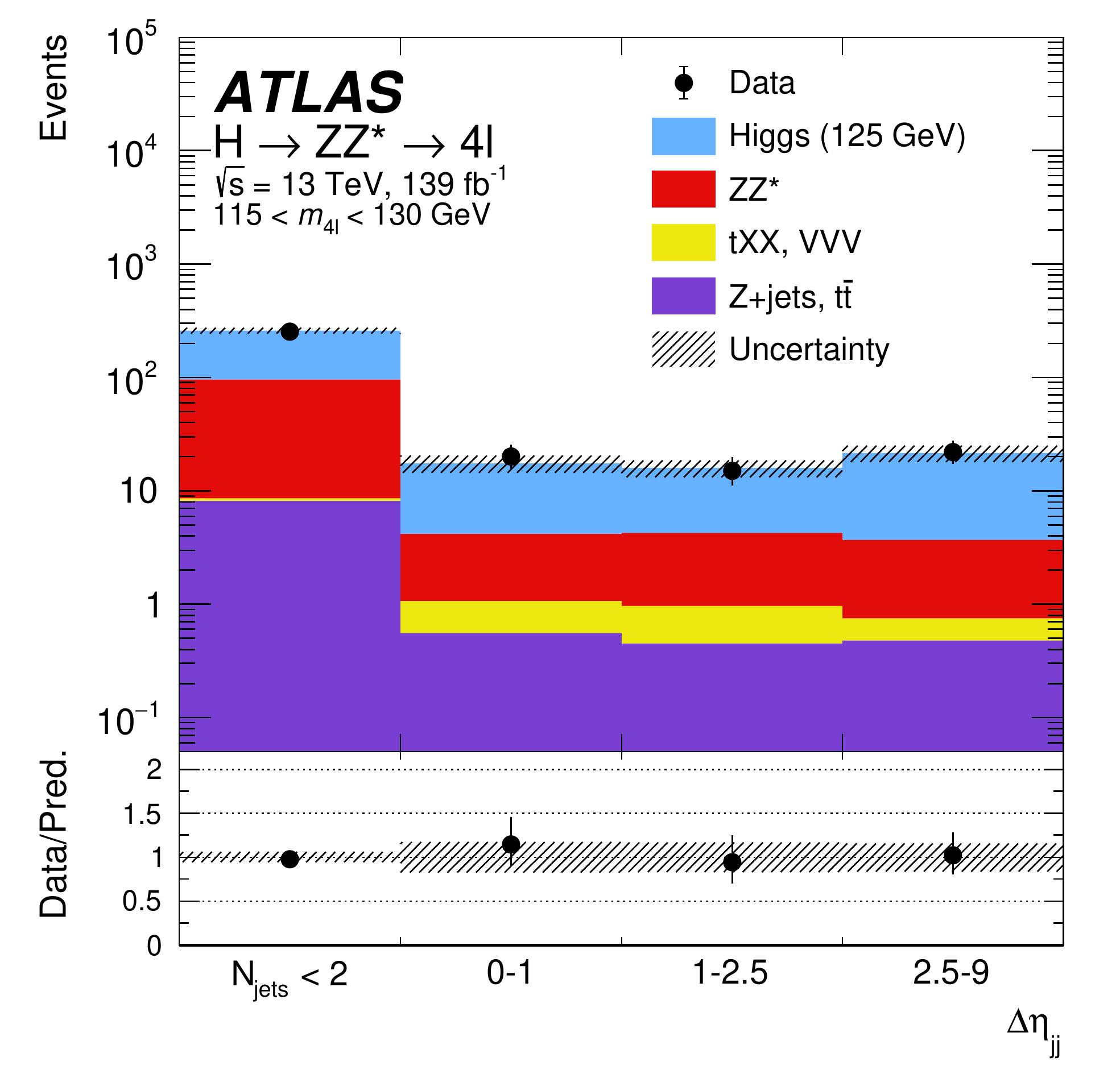}\label{fig:etajjreco}}\\
\subfloat[]{\includegraphics[width=0.5\textwidth]{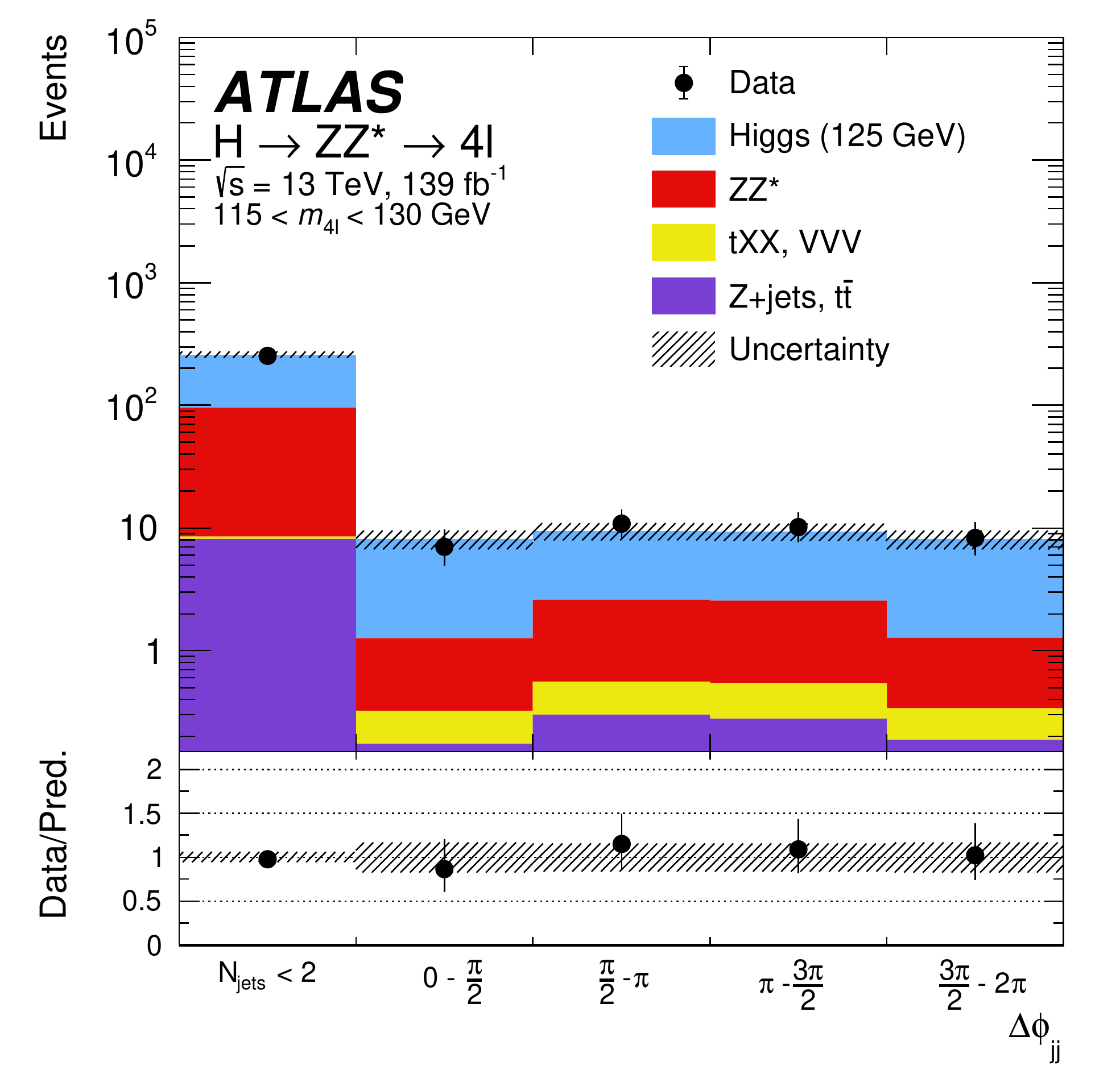}\label{fig:phijjreco}}
\end{center}
\vspace{-0.3cm}
\caption{The observed and expected (pre-fit) distributions of
\protect\subref{fig:mjjreco} \mjj, \protect\subref{fig:etajjreco} \detajj, and \protect\subref{fig:phijjreco} \dphijj\
in the mass region $115 < \mfourl < 130$~\GeV, for an integrated luminosity of \Lum~collected at $\sqrt{\mathrm{s}}=$~13~\TeV. A SM Higgs boson signal with a
mass $m_{H}$ = 125~\gev\ is assumed.
In all distributions, the first bin contains events with fewer than two jets.
The uncertainty in the prediction is shown by the hatched band, 
which includes the theoretical uncertainties of the SM cross section for the signal and the \zzstar\ background.}
\label{fig:categoryVars3p5}
\end{figure}

\begin{figure}[!htbp]
\begin{center}
\subfloat[]{\includegraphics[width=0.5\textwidth]{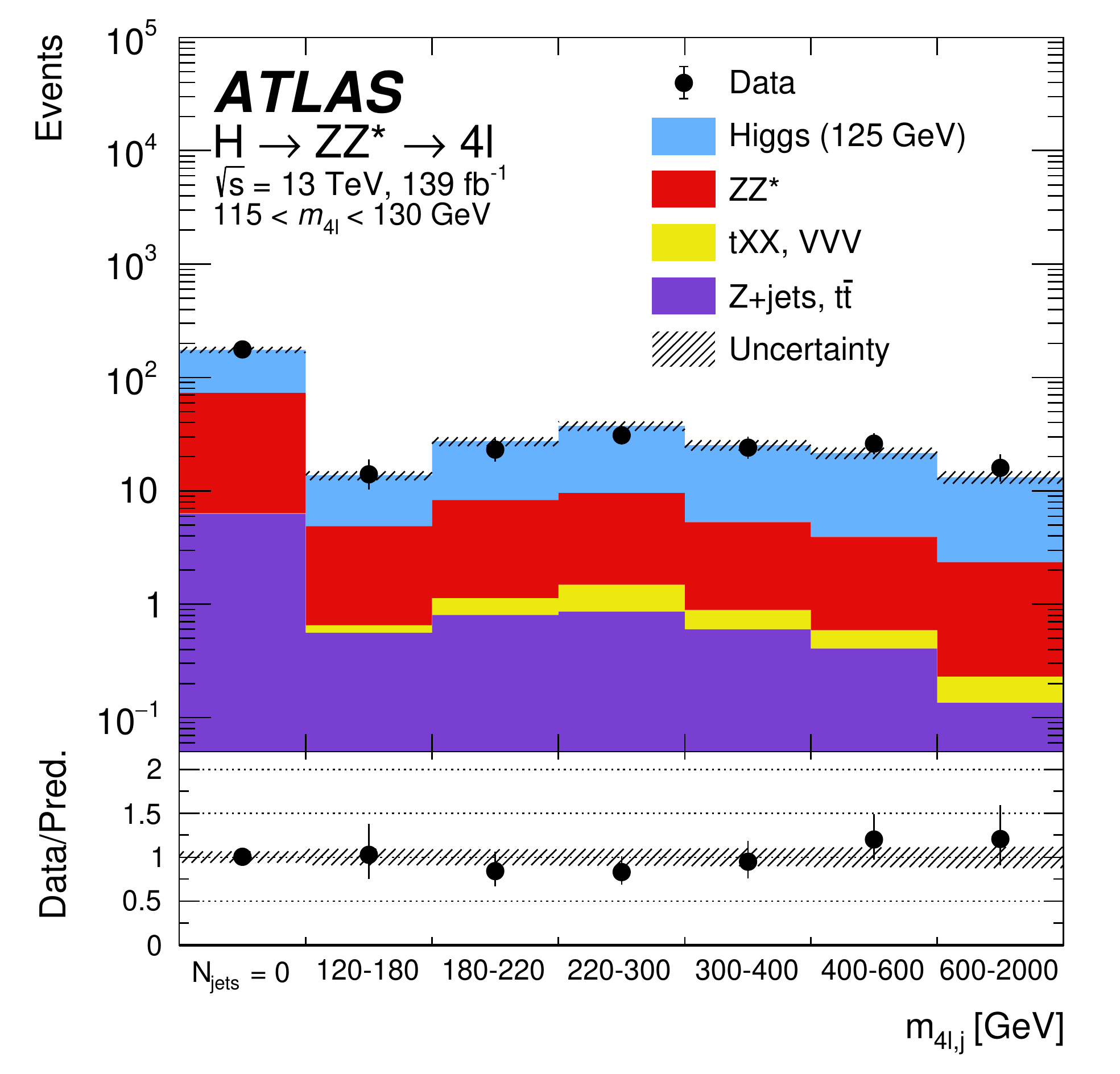}\label{fig:m4ljreco}}
\subfloat[]{\includegraphics[width=0.5\textwidth]{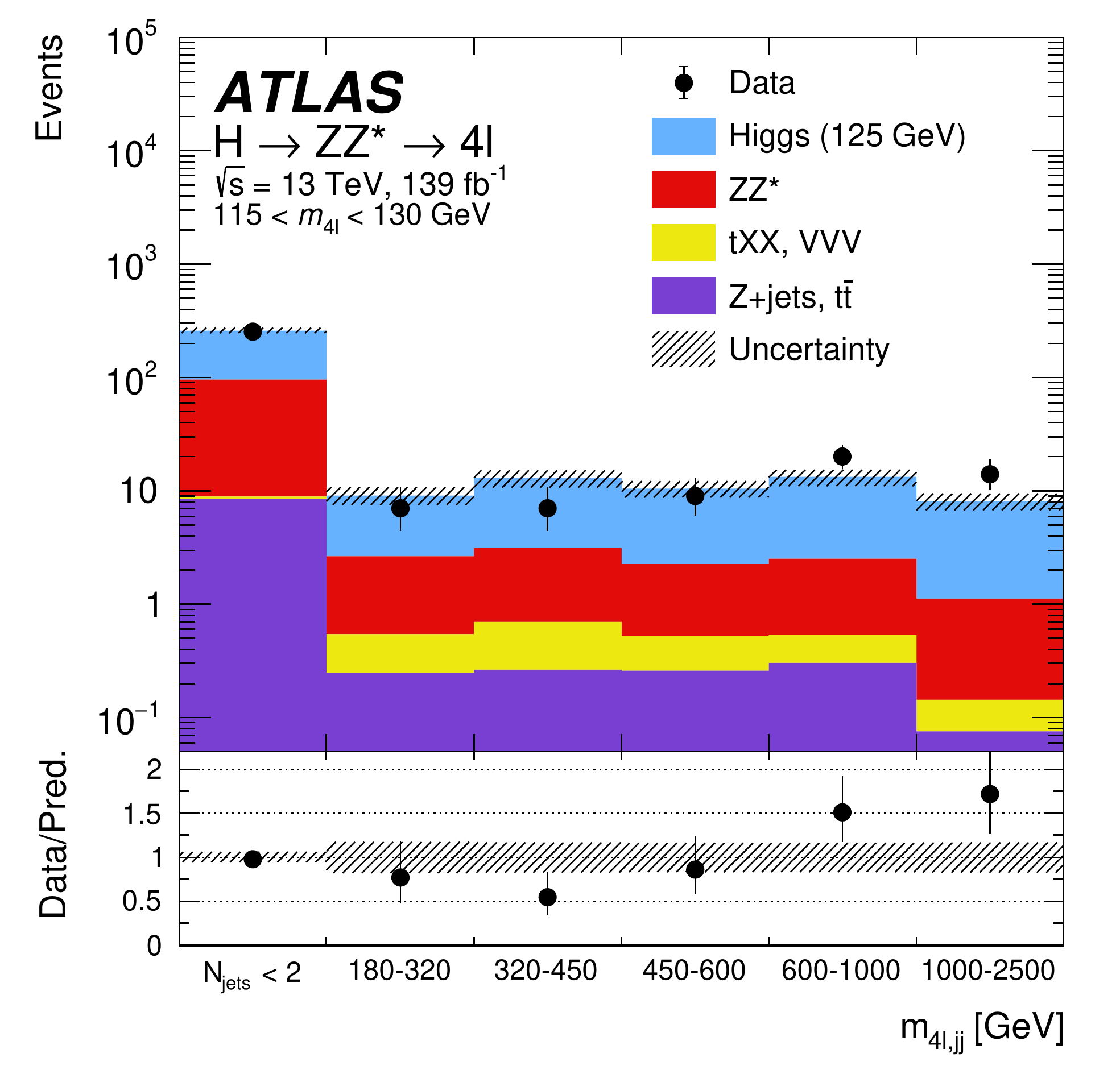}\label{fig:m4ljjreco}}\\
\subfloat[]{\includegraphics[width=0.5\textwidth]{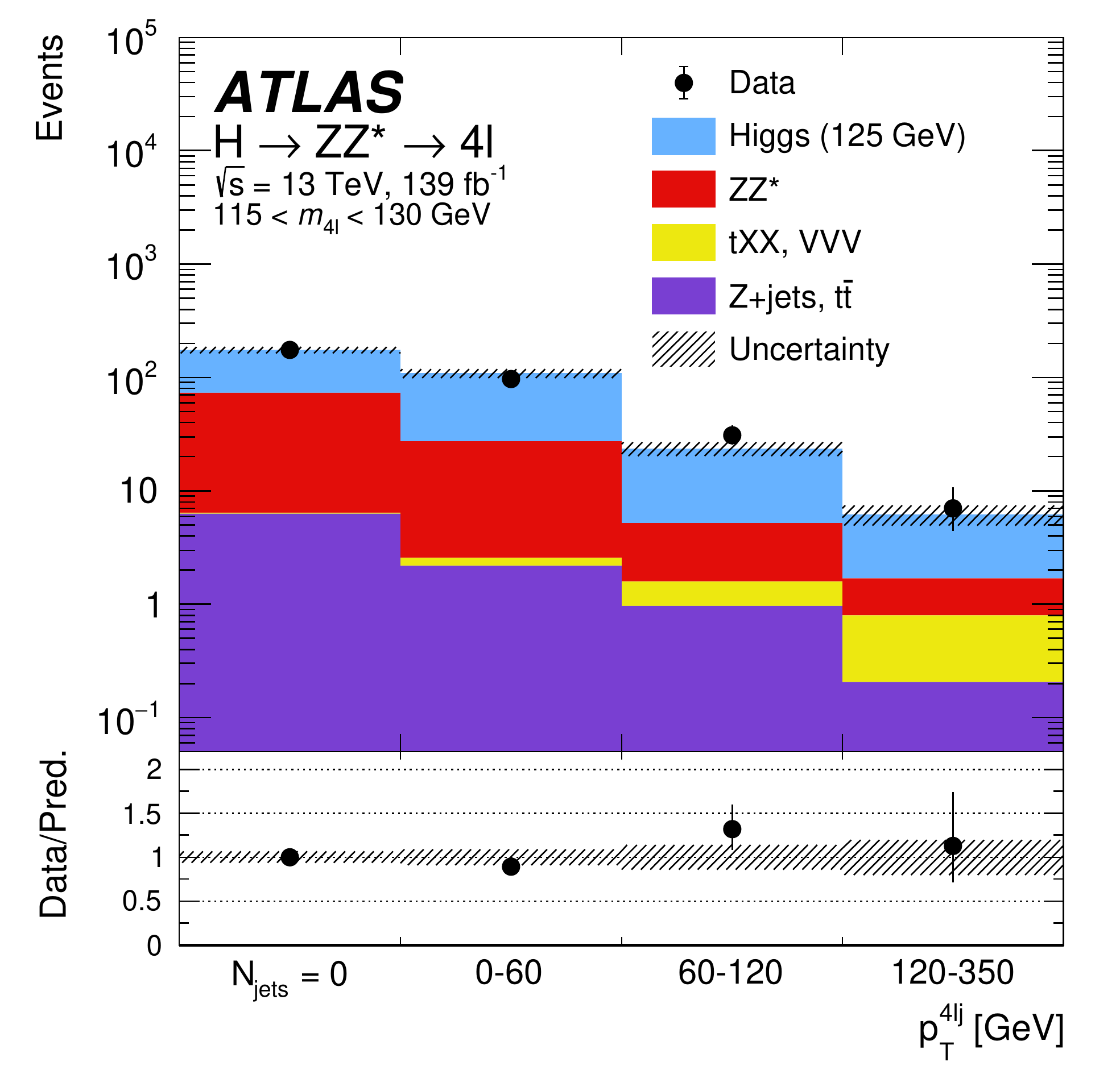}\label{fig:pt4ljreco}}
\subfloat[]{\includegraphics[width=0.5\textwidth]{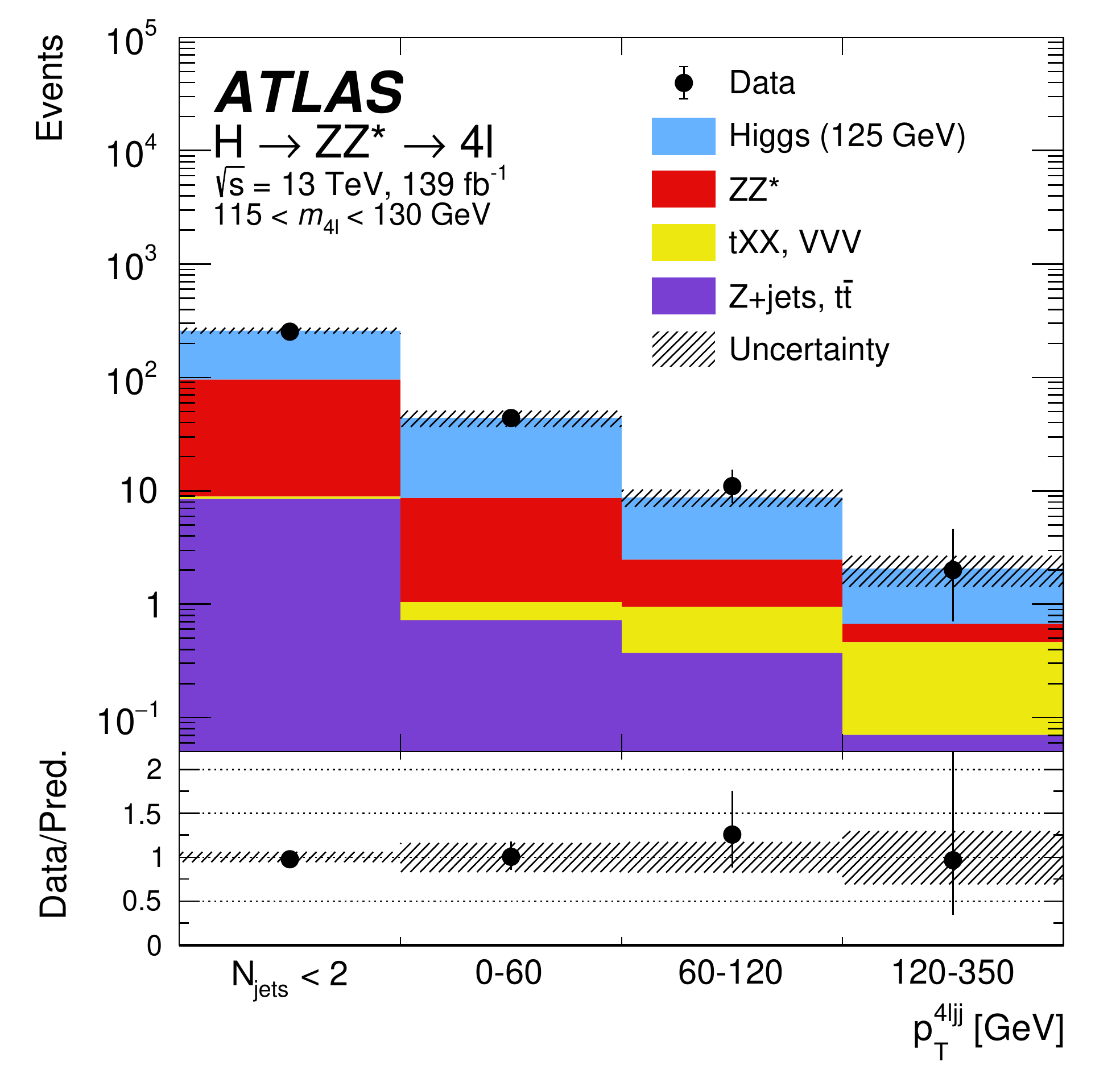}\label{fig:pt4ljjreco}}
\end{center}
\vspace{-0.3cm}
\caption{The observed and expected (pre-fit) distributions of
\protect\subref{fig:m4ljreco} \mflj,
\protect\subref{fig:m4ljjreco}  \mfljj,
\protect\subref{fig:pt4ljreco}  \ptHj, and
\protect\subref{fig:pt4ljjreco}  \ptHjj\
in the mass region $115 < \mfourl < 130$~\GeV, for an integrated luminosity of \Lum~collected at $\sqrt{\mathrm{s}}=$~13~\TeV. A SM Higgs boson signal with a
mass $m_{H}$ = 125~\gev\ is assumed.
The first bin in \protect\subref{fig:m4ljreco} and \protect\subref{fig:pt4ljreco} contains events with no jets, while the first bin in \protect\subref{fig:m4ljjreco} and \protect\subref{fig:pt4ljjreco} contains events with fewer than two jets.
The uncertainty in the prediction is shown by the hatched band, which includes the theoretical uncertainties of the SM cross section for the signal and the \zzstar\ background.}
\label{fig:categoryVars4}
\end{figure}

\begin{figure}[!htbp]
\begin{center}
\subfloat[]{\includegraphics[width=0.8\textwidth]{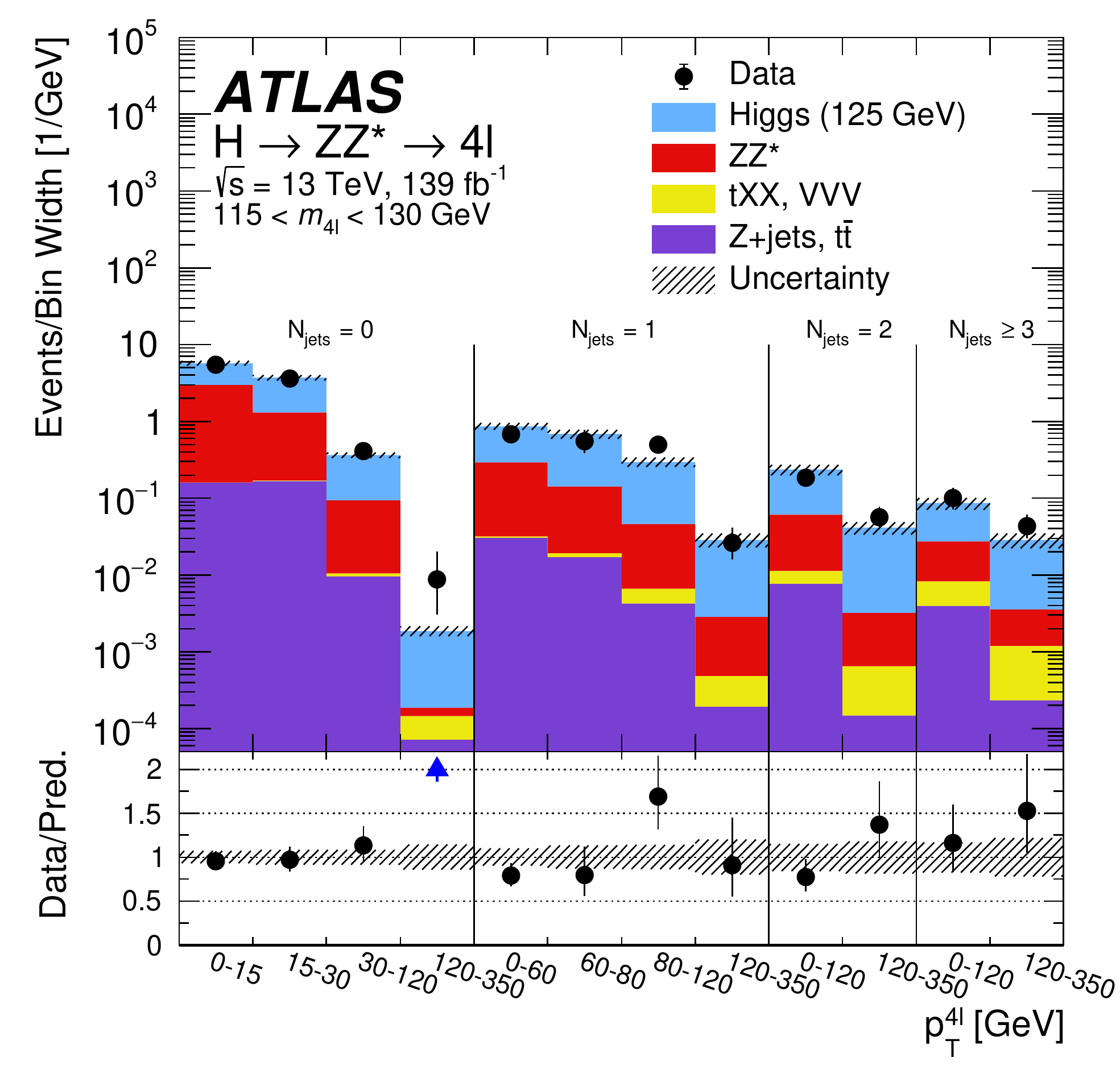}}
\end{center}
\vspace{-0.3cm}
\caption{The observed and expected (pre-fit) distributions of \ptH\ in \njets\ bins in the mass region $115 < \mfourl < 130$~\GeV, for an integrated luminosity of \Lum~collected at $\sqrt{\mathrm{s}}=$~13~\TeV. A SM Higgs boson signal with a
mass $m_{H}$ = 125~\gev\ is assumed. The uncertainty in the prediction is shown by the hatched band, which includes the theoretical uncertainties of the SM cross section for the signal and the \zzstar\ background.}
\label{fig:categoryVars5}
\end{figure}

\begin{figure}[!htbp]
\begin{center}
\subfloat[]{\includegraphics[width=0.8\textwidth]{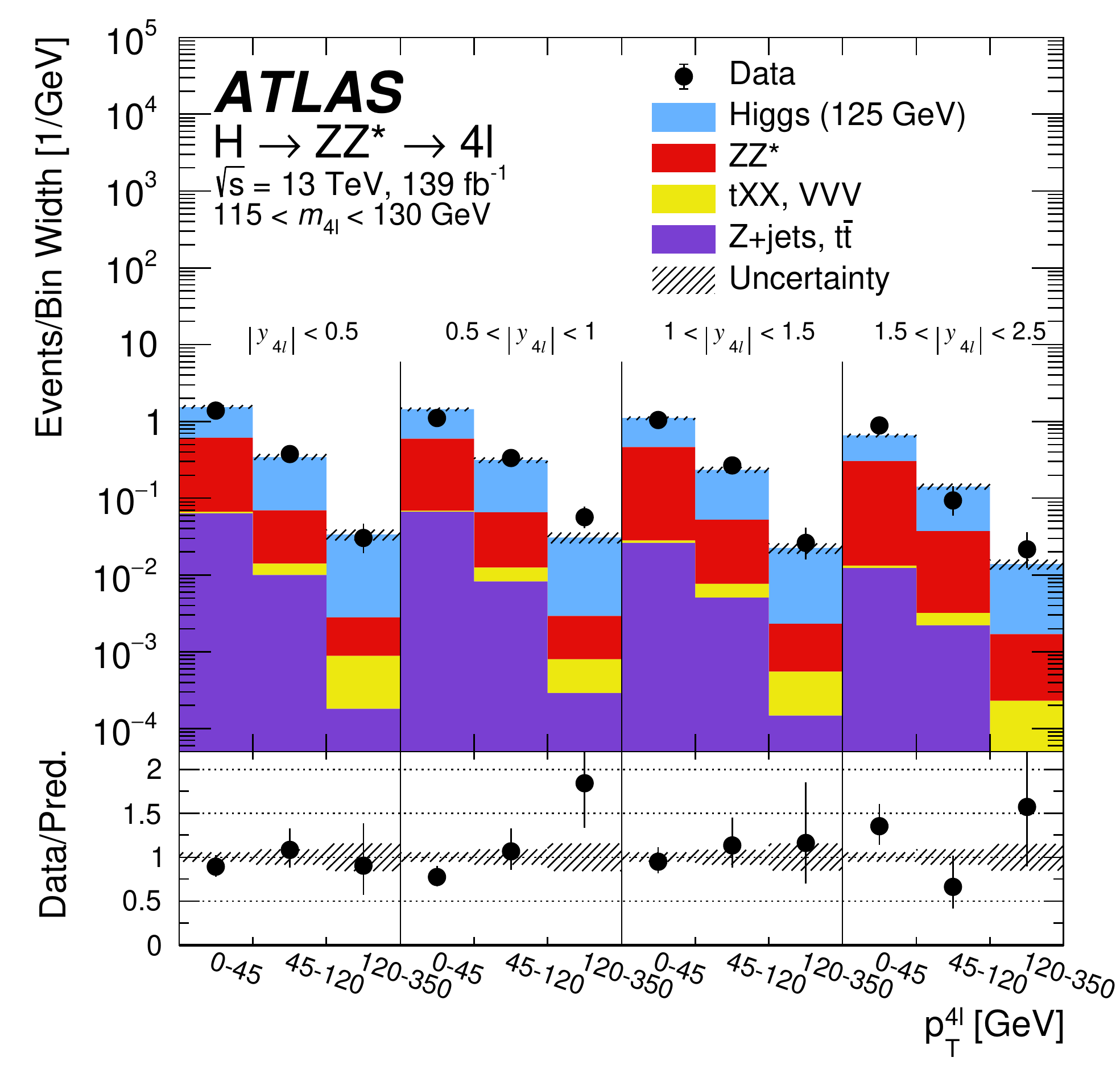}}
\end{center}
\vspace{-0.3cm}
\caption{The observed and expected (pre-fit) distributions of \ptH\ in \yH\ bins in the mass region $115 < \mfourl < 130$~\GeV, for an integrated luminosity of \Lum~collected at $\sqrt{\mathrm{s}}=$~13~\TeV. A SM Higgs boson signal with a
mass $m_{H}$ = 125~\gev\ is assumed. The uncertainty in the prediction is shown by the hatched band, which includes the theoretical uncertainties of the SM cross section for the signal and the \zzstar\ background.}
\label{fig:categoryVars6}
\end{figure}

\begin{figure}[!htbp]
\begin{center}
\subfloat[]{\includegraphics[width=0.9\textwidth]{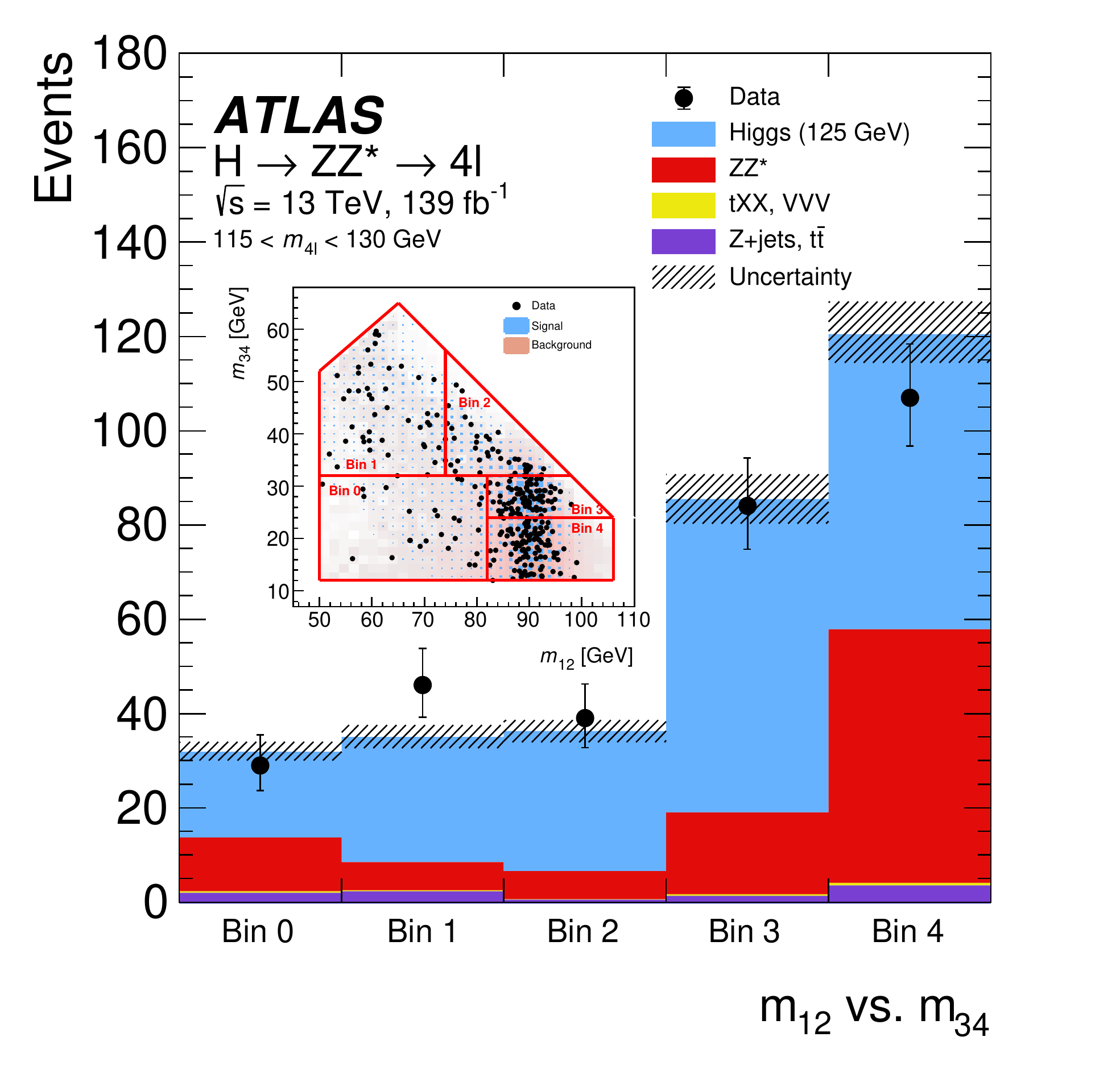}}
\end{center}
\vspace{-0.3cm}
\caption{The observed and expected (pre-fit) distribution in bins of
the leading vs.\ subleading $Z$ boson mass, \monetwo\ vs.\ \mthreefour. The same distribution in the 2D plane is provided in the inset plot, where the black dots depict data and the blue and pink shaded areas represent simulated signal and background, respectively. The red lines depict the bin boundaries, chosen as described in Section~\protect\ref{sec:sig_extraction}. These distributions correspond to
the mass region $115 < \mfourl < 130$~\gev\ for an integrated luminosity of \Lum~collected at $\sqrt{\mathrm{s}}=$~13~\TeV. A SM Higgs boson signal with a
mass $m_{H}$ = 125~\gev\ is assumed. The uncertainty in the prediction is shown by the hatched band, which includes the theoretical uncertainties of the SM cross section for the signal and the \zzstar\ background.}
\label{fig:categoryVars7}
\end{figure}

\begin{figure}[!htbp]
\begin{center}
\subfloat[]{\includegraphics[width=0.9\textwidth]{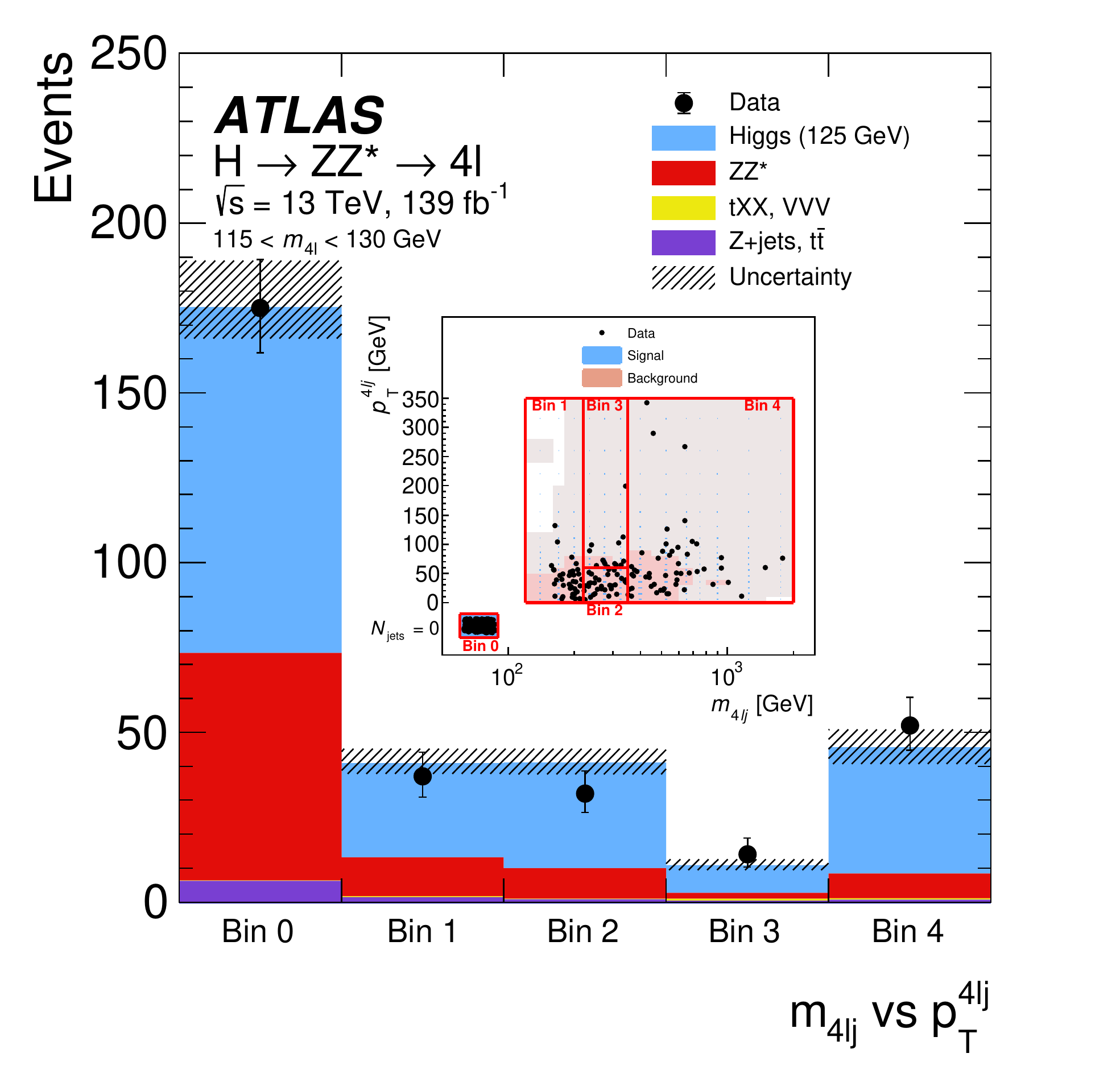}}
\end{center}
\vspace{-0.3cm}
\caption{The observed and expected (pre-fit) distribution in bins of
the transverse momentum of the four-lepton plus leading-jet system vs.\ the invariant mass of the four-lepton plus leading-jet system, \ptHj\ vs.\ \mflj. The same distribution in the 2D plane is provided in the inset plot, where the black dots depict data and the blue and pink shaded areas represent simulated signal and background, respectively. The red lines depict the bin boundaries, chosen as described in Section~\protect\ref{sec:sig_extraction}. These distributions correspond to
the mass region $115 < \mfourl < 130$~\gev\ for an integrated luminosity of \Lum~collected at $\sqrt{\mathrm{s}}=$~13~\TeV. A SM Higgs boson signal with a
mass $m_{H}$ = 125~\gev\ is assumed. The uncertainty in the prediction is shown by the hatched band, which includes the theoretical uncertainties of the SM cross section for the signal and the \zzstar\ background.}
\label{fig:categoryVars8}
\end{figure}

\begin{figure}[!htbp]
\begin{center}
\subfloat[]{\includegraphics[width=0.9\textwidth]{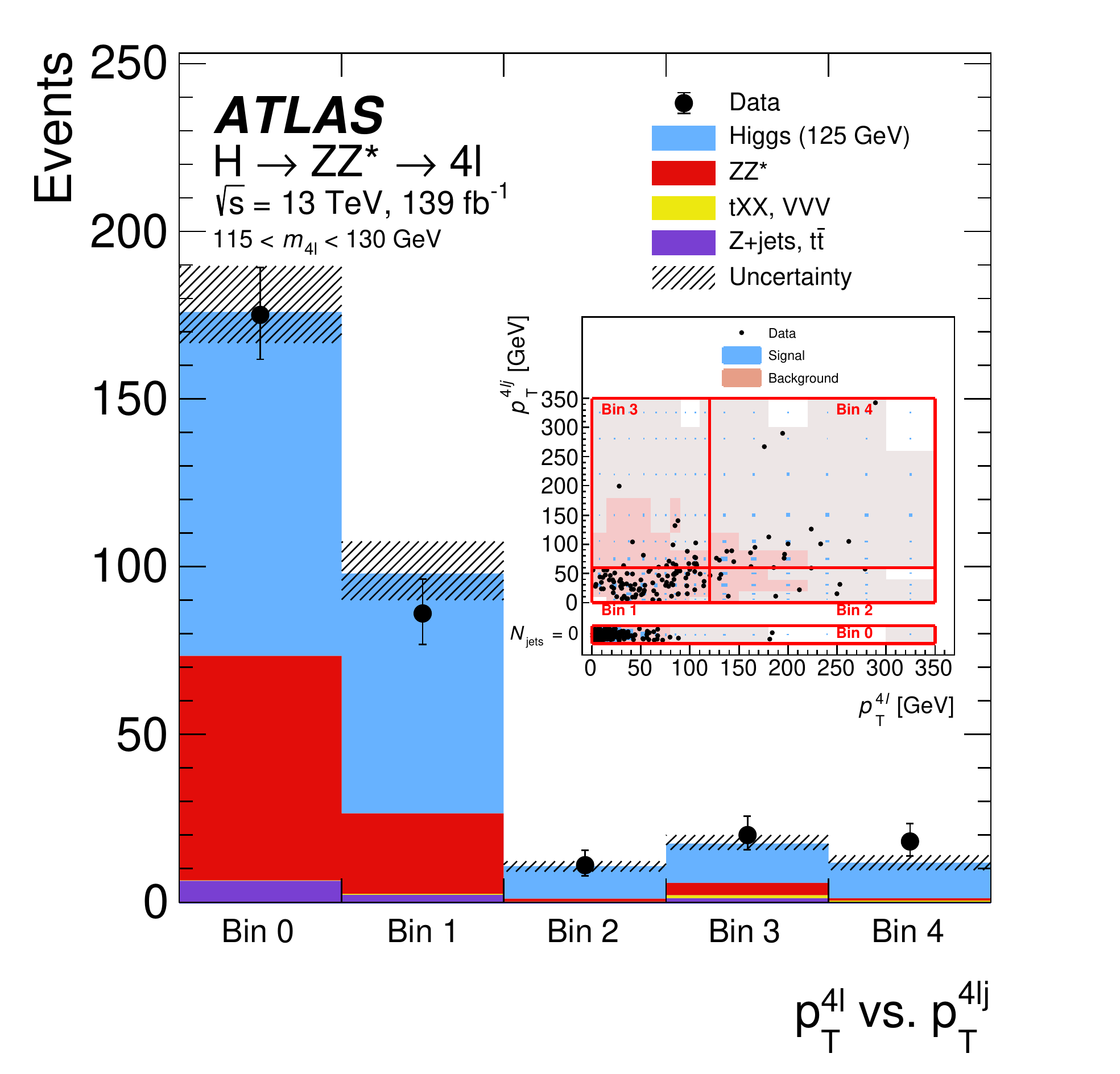}  } \\
\end{center}
\vspace{-0.3cm}
\caption{The observed and expected (pre-fit) distribution in bins of
the transverse momentum of the four-lepton system vs.\ the transverse momentum of the four-lepton plus leading-jet system, \ptH\ vs.\ \ptHj.	The same distribution in the 2D plane is shown in the inset plot, where the black dots depict data and the blue and pink shaded areas represent simulated signal and background, respectively. The red lines depict the bin boundaries, chosen as described in Section~\protect\ref{sec:sig_extraction}. These distributions correspond to
the mass region $115 < \mfourl < 130$~\gev\ for an integrated luminosity of \Lum~collected at $\sqrt{\mathrm{s}}=$~13~\TeV. A SM Higgs boson signal with a
mass $m_{H}$ = 125~\gev\ is assumed. The uncertainty in the prediction is shown by the hatched band, which includes the theoretical uncertainties of the SM cross section for the signal and the \zzstar\ background.}
\label{fig:categoryVars9}
\end{figure}

\begin{figure}[!htbp]
\begin{center}
\subfloat[]{\includegraphics[width=0.9\textwidth]{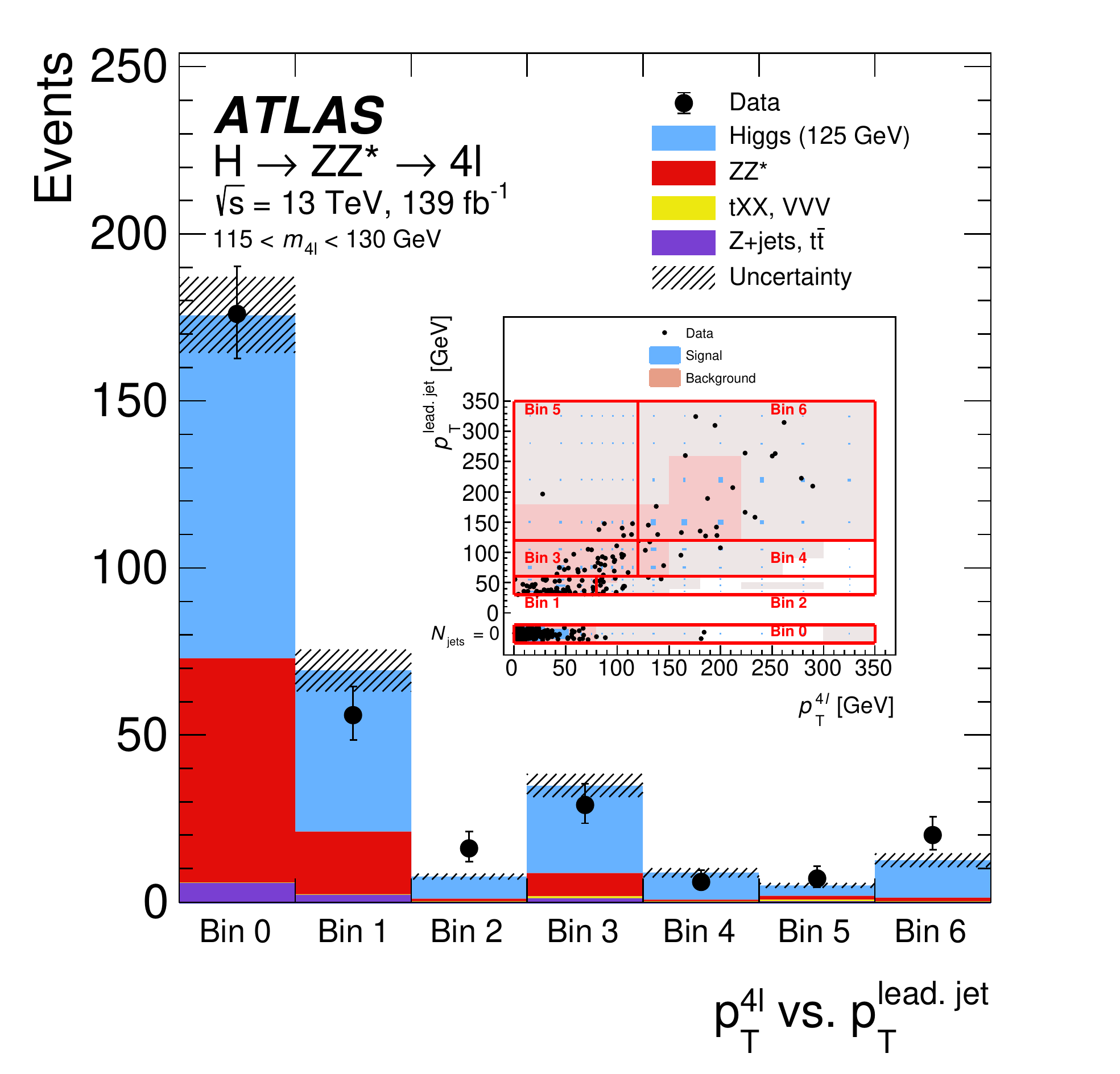}  }
\end{center}
\vspace{-0.3cm}
\caption{The observed and expected (pre-fit) distribution in bins of
the transverse momentum of the four-lepton system vs.\ the transverse momentum of the leading jet, \ptH\  vs.\ \ljpt.
The same distribution in the 2D plane is provided in the inset plot, where the black dots depict data and the blue and pink shaded areas represent simulated signal and background, respectively. The red lines depict the bin boundaries, chosen as described in Section~\protect\ref{sec:sig_extraction}. These distributions correspond to
the mass region $115 < \mfourl < 130$~\gev\ for an integrated luminosity of \Lum~collected at $\sqrt{\mathrm{s}}=$~13~\TeV. A SM Higgs boson signal with a
mass $m_{H}$ = 125~\gev\ is assumed. The uncertainty in the prediction is shown by the hatched band, which includes the theoretical uncertainties of the SM cross section for the signal and the \zzstar\ background.}
\label{fig:categoryVars10}
\end{figure}

\begin{figure}[!htbp]
\begin{center}
\subfloat[]{\includegraphics[width=0.9\textwidth]{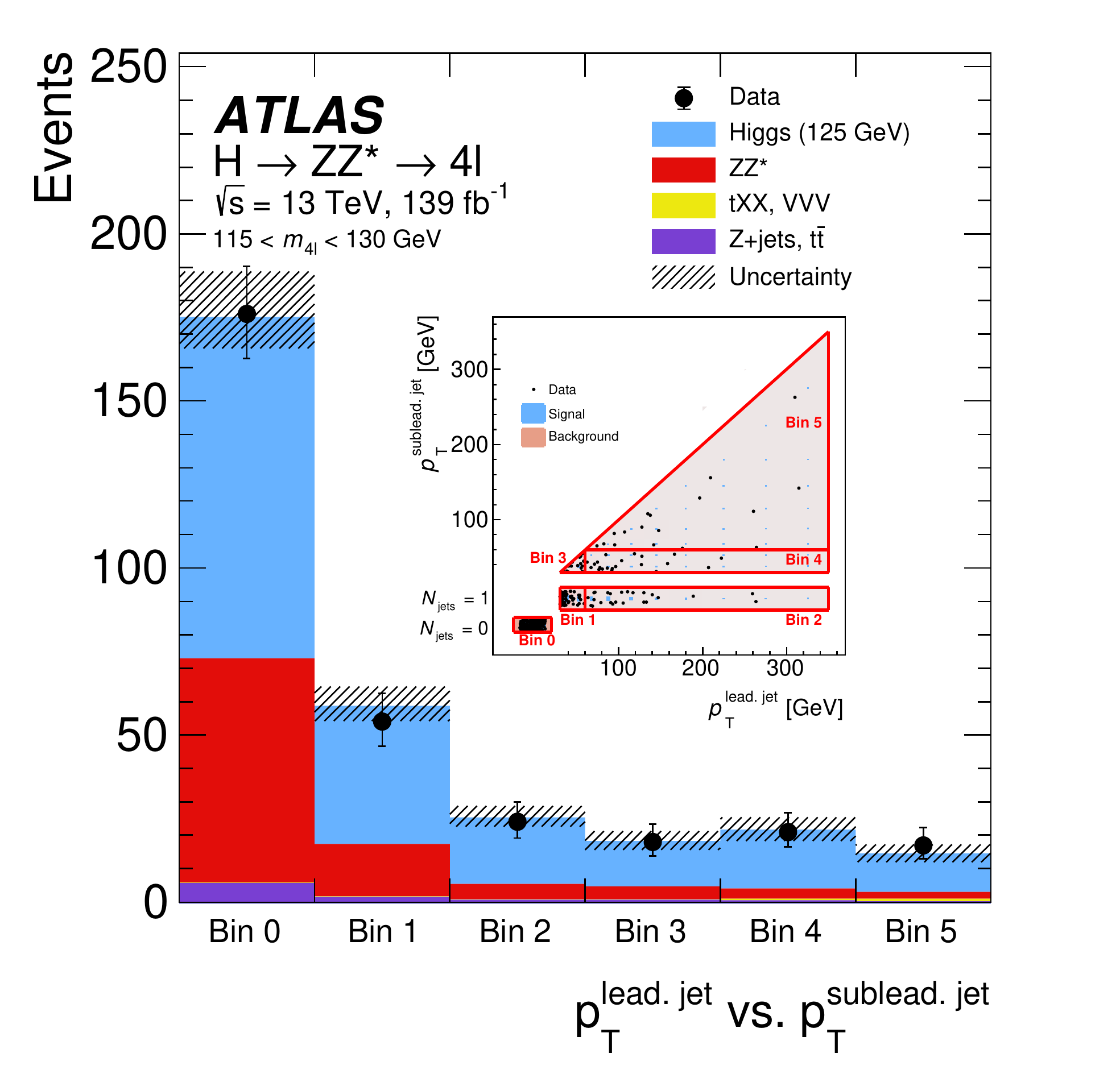}}
\end{center}
\vspace{-0.3cm}
\caption{The observed and expected (pre-fit) distribution in bins of
the transverse momentum of the leading vs.\ subleading jet, \ljpt\ vs.\ \sjpt. The same distribution in the 2D plane is provided in the inset plot, where the black dots depict data and the blue and pink shaded areas represent simulated signal and background, respectively. The red lines depict the bin boundaries, chosen as described in Section~\protect\ref{sec:sig_extraction}. These distributions correspond to
the mass region $115 < \mfourl < 130$~\gev\ for an integrated luminosity of \Lum~collected at $\sqrt{\mathrm{s}}=$~13~\TeV. A SM Higgs boson signal with a
mass $m_{H}$ = 125~\gev\ is assumed. The uncertainty in the prediction is shown by the hatched band, which includes the theoretical uncertainties of the SM cross section for the signal and the \zzstar\ background. \ljpt\ and \sjpt\ are required to have \pt greater than 30~\gev. }
\label{fig:categoryVars11}
\end{figure}

\begin{figure}[!htbp]
\begin{center}
\subfloat[]{\includegraphics[width=0.9\textwidth]{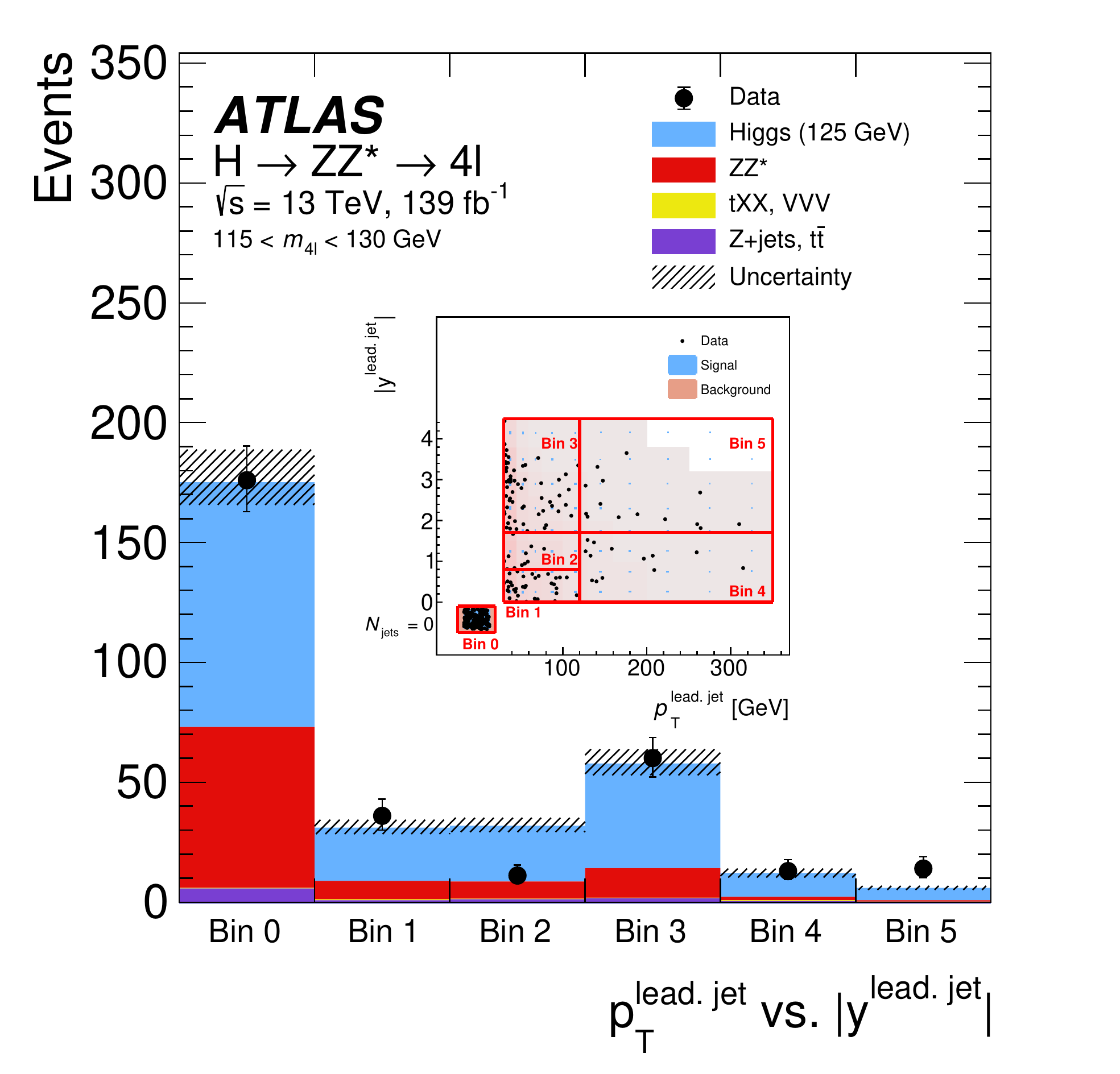}}
\end{center}
\vspace{-0.3cm}
\caption{The observed and expected (pre-fit) distribution in bins of
the transverse momentum vs.\ the rapidity of the leading jet, \ljpt\ vs.\ \ljy. The same distribution in the 2D plane is provided in the inset plot, where the black dots depict data and the blue and pink shaded areas represent simulated signal and background, respectively. The red lines depict the bin boundaries, chosen as described in Section~\protect\ref{sec:sig_extraction}. These distributions correspond to
the mass region $115 < \mfourl < 130$~\gev\ for an integrated luminosity of \Lum~collected at $\sqrt{\mathrm{s}}=$~13~\TeV. A SM Higgs boson signal with a
mass $m_{H}$ = 125~\gev\ is assumed. The uncertainty in the prediction is shown by the hatched band, which includes the theoretical uncertainties of the SM cross section for the signal and the \zzstar\ background.}
\label{fig:categoryVars12}
\end{figure}

\clearpage
The observed and expected distributions of one-dimensional observables are
shown in Figures~\ref{fig:categoryVars1}--\ref{fig:categoryVars4}. In addition, the observed and expected distributions for the two-dimensional observables are shown in Figures~\ref{fig:categoryVars5}--\ref{fig:categoryVars12}. 
All these figures show events selected within an  \mfourl\ mass range of 115--130~\GeV.
Further details of the compatibility with the SM are reported in Section \ref{sec:diffXSresults}.

\FloatBarrier
\subsection{Statistical analysis}
\label{sec:thestat}
 
The inclusive fiducial and  differential cross sections are measured using a binned profile-likelihood-ratio fit~\cite{Cowan:2010js}, taking into account all bins of a given distribution. The likelihood function includes the shape and normalisation uncertainties of the signal and background predictions as nuisance parameters, as outlined in Section~\ref{sec:systematics}.  The cross sections are extracted by minimising two times
the negative logarithm of the profile likelihood ratio, $-2 \ln{\Lambda}$. In the asymptotic approximation, i.e.\ the large sample limit, $-2 \ln{\Lambda}$ behaves as a $
\chi^2$ distribution with one degree of freedom. The compatibility of a measured cross section and its theoretical prediction is tested by computing a $p$-value
based on the difference between the value of $-2 \ln{\Lambda}$ at the best-fit value and the value obtained by fixing the cross section in each bin to that
predicted by theory. These $p$-values do not include the uncertainties in the theoretical predictions. 
For all measured observables the asymptotic approximation is
validated with pseudo-experiments, and where the number of observed events is less than three, the uncertainties are corrected to the values obtained with the pseudo-experiments.
 
For the fiducial and differential cross-section measurements, the fitted \mfourl\ distribution in each final state or differential bin is used to extract the measured cross section following Eq.~(\ref{eq:fidXS}).
The fiducial cross sections of the four final
states can either be summed to obtain an inclusive fiducial cross section, or they can be combined assuming the SM $\zzstar \rightarrow 4\ell$ relative branching
ratios. The latter combination is more model dependent, but benefits from a smaller statistical uncertainty.

 
\subsection{Inclusive fiducial cross-section measurements}
\label{sec:fidResults}

\begin{figure}[!htbp]
\centering
\subfloat[]{\includegraphics[width=0.56\linewidth]{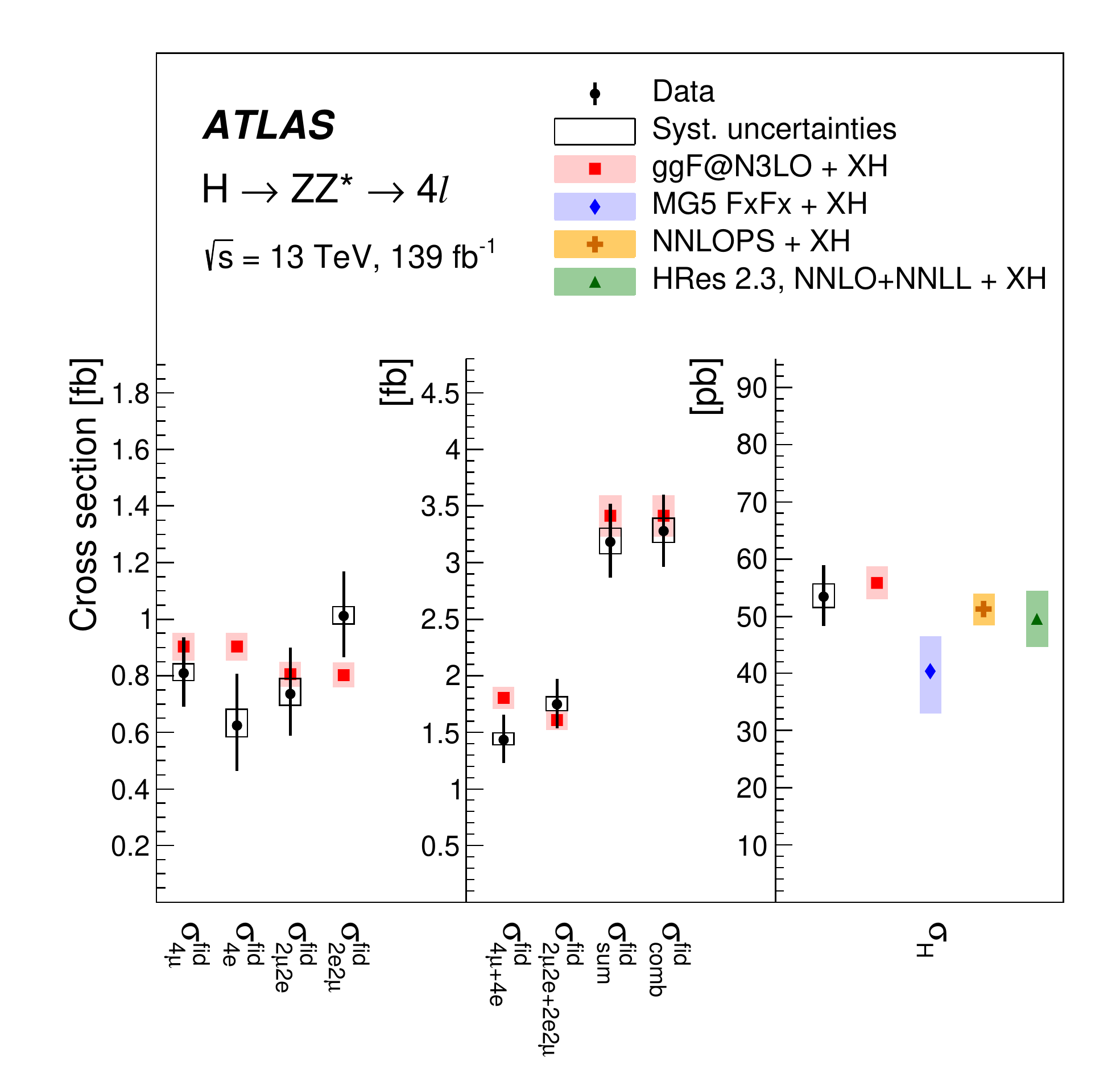}\label{fig:summary_fid}}\\
\subfloat[]{\includegraphics[width=0.44\linewidth]{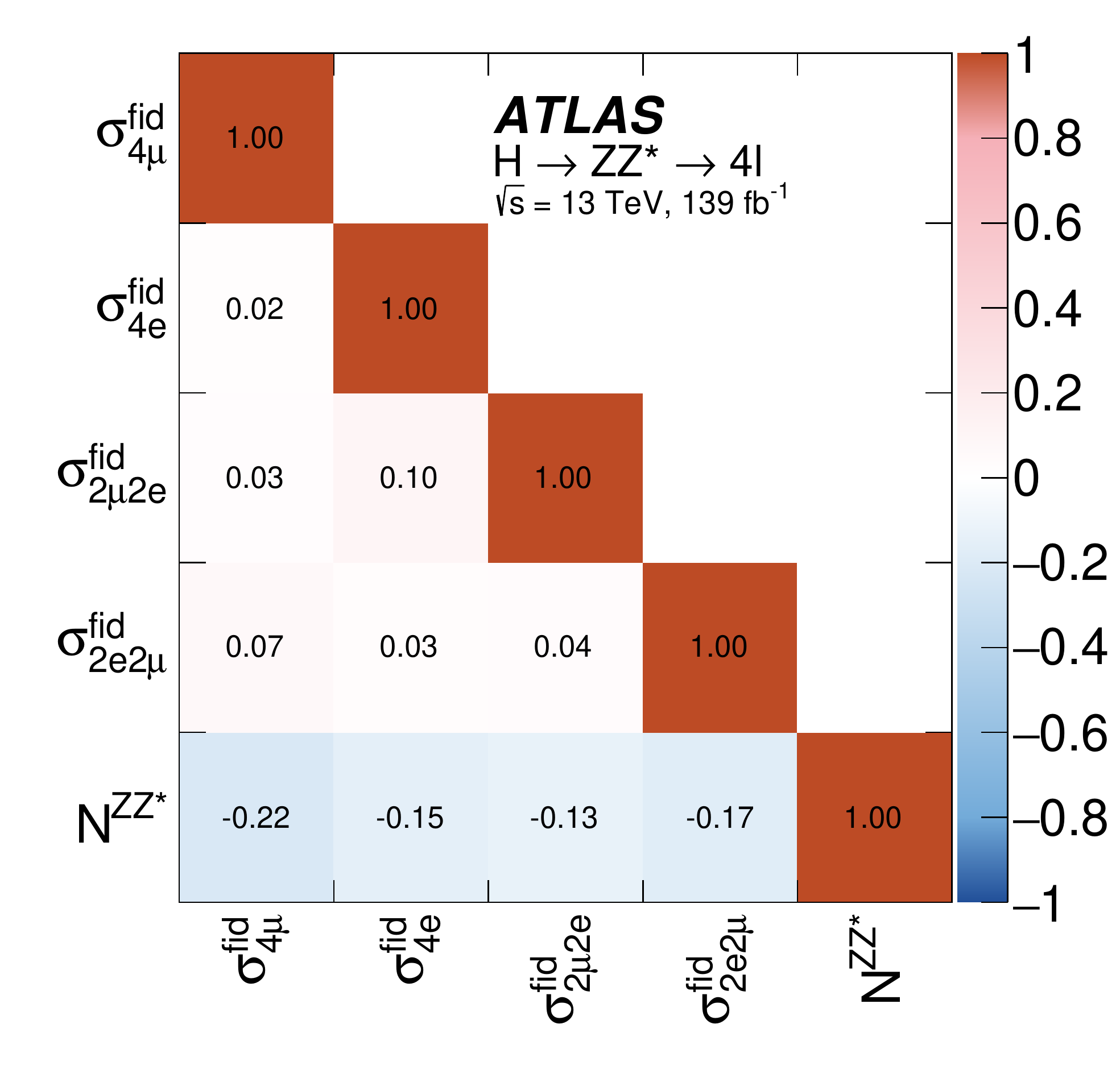}\label{fig:corrMatrix_fid}}
\caption{\protect\subref{fig:summary_fid} The fiducial cross sections (left two panels) and total cross section (right panel) of Higgs boson production measured in the 4$\ell$ final state. The
fiducial cross sections are shown separately for each decay final state, and for same- and different-flavour decays. The inclusive fiducial cross section is measured
as the sum of all final states, as well as by combining the per-final-state measurements assuming SM $\zzstar \rightarrow 4\ell$ relative branching ratios. The total SM
prediction is accurate to N$^3$LO in QCD for the ggF process. The cross sections for all other Higgs boson production modes $XH$ are added. For the fiducial cross-section predictions, the SM cross sections are multiplied by the 
acceptances determined using the NNLOPS sample for ggF and the samples discussed in Section~\ref{sec:samples} for the other production modes. For the total cross section, the predictions by the generators \progname{NNLOPS} , \progname{Hres}, and \progname{MadGraph5\_aMC@NLO}-FxFx are also shown. The error bars on the data points show the total uncertainties, while the
systematic uncertainties are indicated by the boxes.
The shaded bands around the theoretical predictions indicate the PDF and scale uncertainties, calculated as described in Section~\ref{subsec:systematics_theory_couplings}. \protect\subref{fig:corrMatrix_fid} The correlation between the fiducial cross sections for the four individual decay final states and the \zzstar\ normalisation factor. \label{fig:unfolded_fid}}
\end{figure}

The fiducial production cross sections of the \hfourl\ process are presented in Table~\ref{tab:FiducialResults} and Figure~\ref{fig:unfolded_fid}. The left panel in Figure~\ref{fig:summary_fid} shows the fiducial cross sections for the four individual decay final states: $4\mu$, $4e$ decays (hereafter referred to as same flavour), and  $2\mu2e$, $2e2\mu$ decays (hereafter referred to as different flavour). The middle panel shows the cross sections for same- and different-flavour decays, which can provide a probe of same-flavour interference effects, as well as the inclusive fiducial cross sections obtained by either summing all 4$\ell$ decay final states or combining them assuming relative SM branching ratios.
 
The data are compared with the SM prediction after accounting for the fiducial acceptance as determined from the SM Higgs boson simulated samples (see
Section~\ref{sec:samples}).
 
The combined inclusive fiducial cross section is extrapolated to the full phase space, as shown in the right panel of Figure~\ref{fig:unfolded_fid}, using the fiducial acceptance as well as the branching ratios, with the uncertainties described in Section~\ref{sec:systematics}. The total cross section is also compared with the cross sections predicted by \progname{NNLOPS}, \progname{MadGraph5\_aMC@NLO}-FxFx (\progname{MG5}-FxFx) and \progname{Hres 2.3} \cite{Grazzini:2013mca, deFlorian:2012mx} for ggF, while for all other production modes the predictions described in Section~\ref{sec:samples} are used. For ggF, all generators predict cross sections that are lower than the N$^3$LO calculation. The $p$-values, calculated as described in Section~\ref{sec:thestat}, are shown in Table~\ref{tab:FiducialResults}. The probability of compatibility of the measured fiducial cross section ($\sigma_\mathrm{comb}$) and the Standard Model expectation is at the level of 67\%.
 
\renewcommand{\arraystretch}{1.2}
\begin{table*}[ht]
\centering
\caption{The fiducial and total cross sections of Higgs boson production measured in the 4$\ell$ final state.  The fiducial cross sections are given separately for
each decay final state, and for same- and different-flavour decays. The inclusive fiducial cross section is measured as the sum of all final states ($\sigma_\mathrm{sum}$), as well as by
combining the per-final-state measurements assuming SM $\zzstar \rightarrow 4\ell$ relative branching ratios ($\sigma_\mathrm{comb}$). For the total cross section ($\sigma_\mathrm{tot}$), the Higgs boson branching ratio at $m_{H} =$ 125~\GeV~ is assumed.
The total SM prediction is accurate to N$^3$LO in QCD and NLO EW for the ggF process. For the fiducial cross-section predictions, the SM cross sections are multiplied
by the acceptances determined using the NNLOPS sample for ggF. For all the other production modes, the cross sections from the samples discussed in Section~\ref{sec:samples} are added.
The $p$-values indicating the probability of compatibility of the measurement and the SM prediction are shown as well. They do not include the systematic uncertainty in the
theoretical predictions.  \label{tab:FiducialResults} }
\vspace{0.2cm}
\begin{tabular}{ lcllS[table-format=1.3,table-figures-uncertainty=1]cr }
\hline
\hline
Cross section [fb] & \multicolumn{3}{c}{Data ( $\pm$ (stat.) $\pm$ (syst.)})  & {Standard Model prediction} & $p$-value [\%]\\
\hline
$\sigma_{4\mu}$          & 0.81 & $\pm 0.12$ & $\pm 0.03$ &  0.90 \pm 0.05 & 46 \\
$\sigma_{4e}$            & 0.62 & $\pm 0.17$ & $\pm 0.05$ &  0.90 \pm 0.05 & 14 \\
$\sigma_{2\mu2e}$        & 0.74 & $\pm 0.15$ & $\pm 0.05$ &  0.80 \pm 0.04 & 67 \\
$\sigma_{2e2\mu}$        & 1.01 & $\pm 0.15$ & $\pm 0.03$ &  0.80 \pm 0.04 & 15 \\
\hline
$\sigma_{4\mu+4e}$       & 1.43 & $\pm 0.21$ & $\pm 0.05$ &  1.81 \pm 0.10 & 10 \\
$\sigma_{2\mu2e+2e2\mu}$ & 1.75 & $\pm 0.21$ & $\pm 0.06$ &  1.61 \pm 0.09 & 51 \\
\hline
$\sigma_\mathrm{sum}$    & 3.18 & $\pm 0.31$ & $\pm 0.11$ &  3.41 \pm 0.18 & 49 \\
$\sigma_\mathrm{comb}$   & 3.28 & $\pm 0.30$ & $\pm 0.11$ &  3.41 \pm 0.18 & 67 \\
\hline
\hline
$\sigma_\mathrm{tot}$ [pb] & 53.5~~~~ & $\pm 4.9$ & $\pm 2.1$ & 55.7 \pm 2.8 & 66 \\[0.1cm]
\hline
\hline
 
\end{tabular}
\end{table*}
 
\FloatBarrier
\subsection{Differential cross-section measurements}
\label{sec:diffXSresults}
 
\begin{figure}[!htbp]
\centering
\subfloat[]{\includegraphics[width=0.48\linewidth]{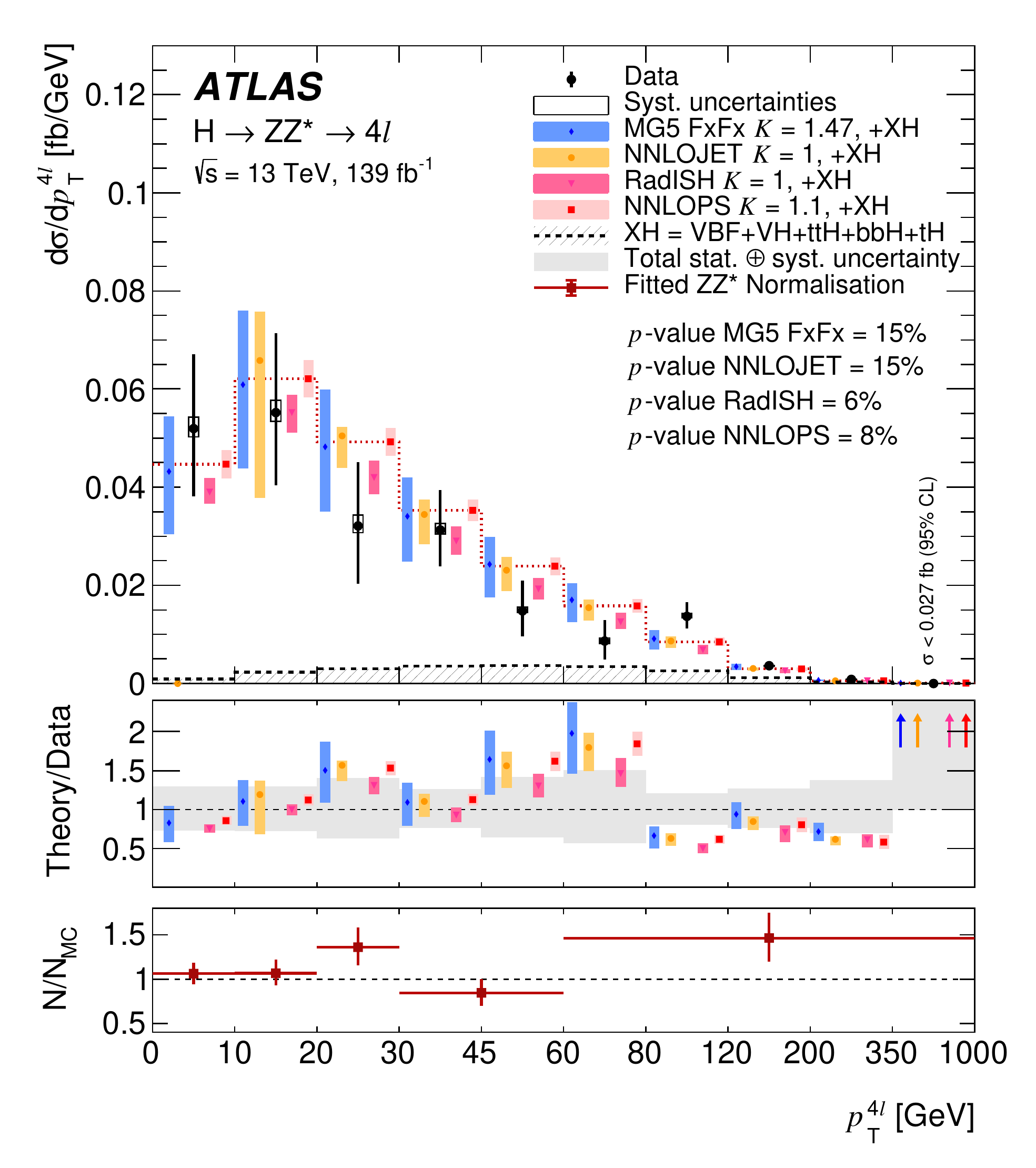}\label{fig:summary_pt}}
\subfloat[]{\includegraphics[width=0.5\linewidth]{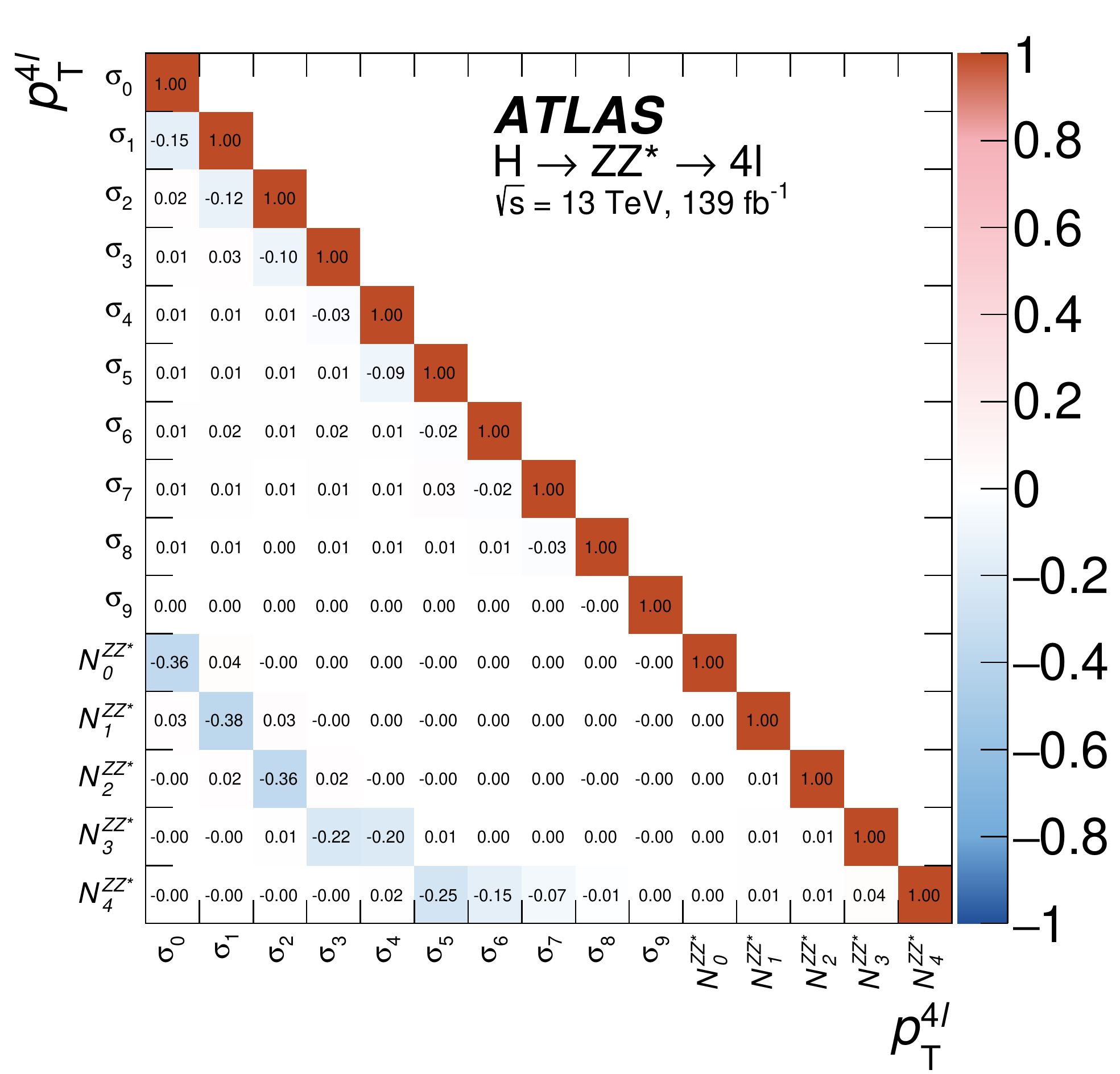}\label{fig:corrMatrix_pt}}
\caption{\protect\subref{fig:summary_pt} Differential fiducial cross section for the transverse momentum \ptH\ of the Higgs boson, along with~\protect\subref{fig:corrMatrix_pt}  the corresponding correlation matrix between the measured cross sections and the $ZZ^{\ast}$ background normalisation factors. 
The measured cross sections
are compared with ggF predictions by  \progname{MadGraph5\_aMC@NLO}-FxFx,
\progname{NNLOJET},
\progname{RadISH}, and
\progname{NNLOPS},
where
\progname{MadGraph5\_aMC@NLO}-FxFx and
\progname{NNLOPS} are
normalised to the N$^3$LO total cross section with the
listed $K$-factors
while the normalisations for \progname{NNLOJET} and \progname{RadISH} are to their respective predicted cross sections.
MC-based predictions for all other Higgs boson production modes $XH$ are normalised to the SM predictions. The error bars on the data points show the total uncertainties, while
the systematic uncertainties are indicated by the boxes. The shaded bands on the expected cross sections indicate the PDF and scale systematic uncertainties, calculated as described in Section~\ref{subsec:systematics_theory_couplings}. This includes the uncertainties related to the $XH$ production modes.
The $p$-values indicating the probability of compatibility of the measurement and the SM prediction are shown as well. They do not include the systematic uncertainty in the theoretical
predictions. The central panel of~\protect\subref{fig:summary_pt} shows the ratio of different predictions to the data, and the grey area represents the total uncertainty of the measurement. The bottom panel of~\protect\subref{fig:summary_pt} shows the ratios of the fitted values of the $ZZ^{\ast}$ normalisation factors to the predictions from MC simulation discussed in Section~\ref{sec:samples}. As indicated by the horizontal error bars, the $ZZ^{\ast}$ normalisation is estimated in each of the first three \ptH\ bins separately, while the next two bins share a common estimation factor, as do the last five bins. \label{fig:pt_unfolded} }
\end{figure}

\begin{figure}[!htbp]
\centering
\subfloat[]{\includegraphics[width=0.45\linewidth]{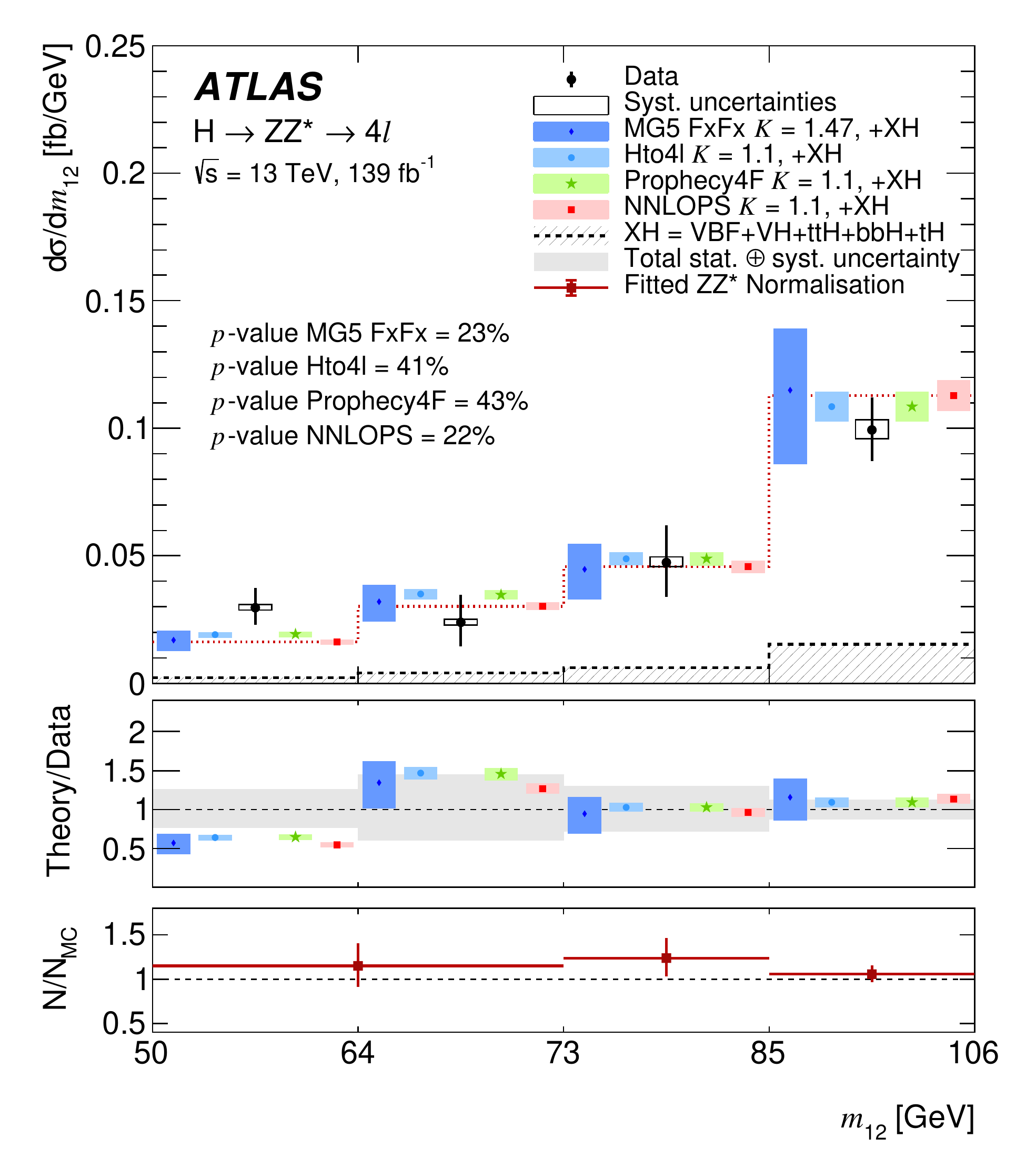}\label{fig:summary_m12}}
\subfloat[]{\includegraphics[width=0.48\linewidth]{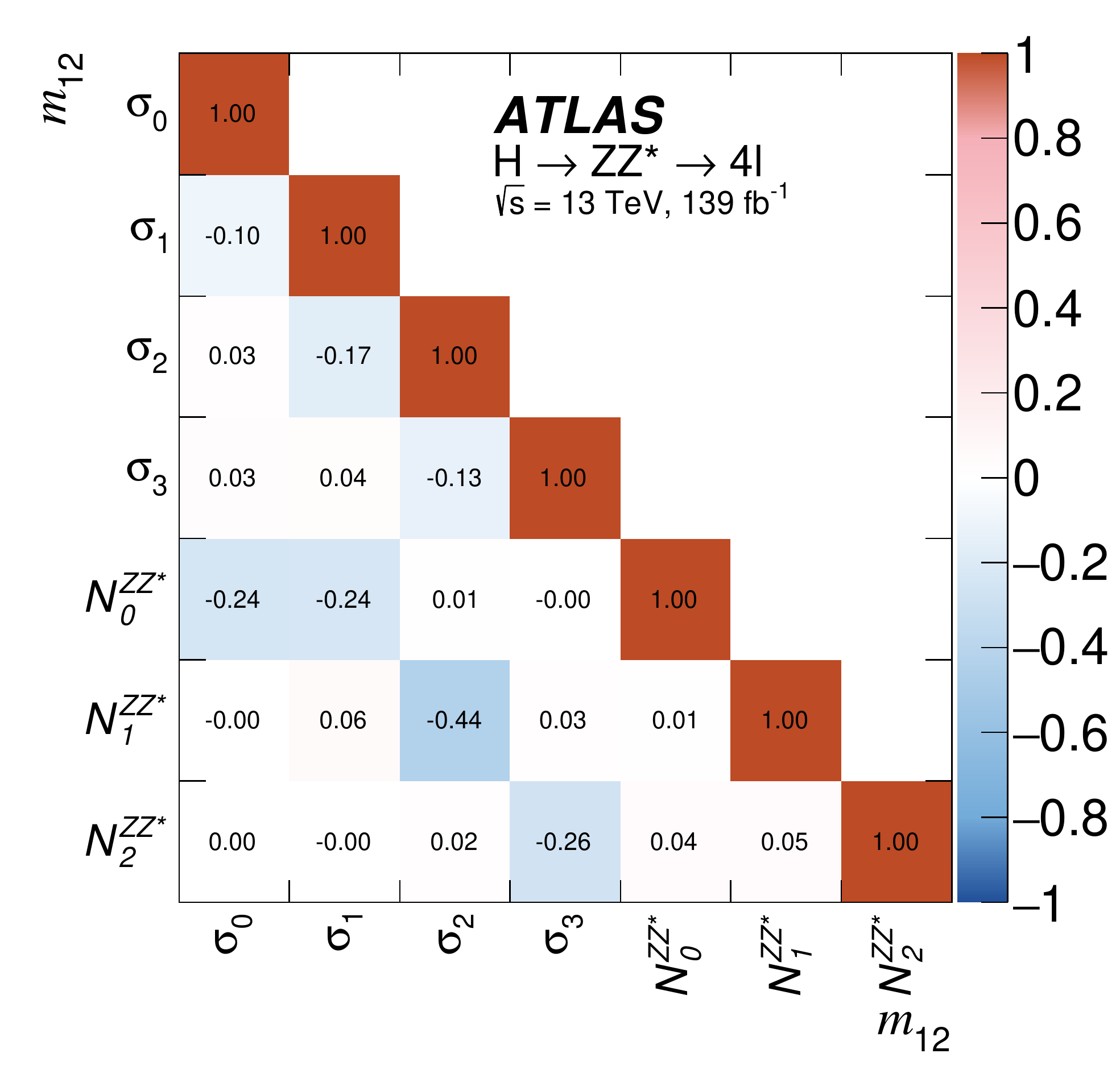}\label{fig:corrMatrix_m12}}\\
\subfloat[]{\includegraphics[width=0.45\linewidth]{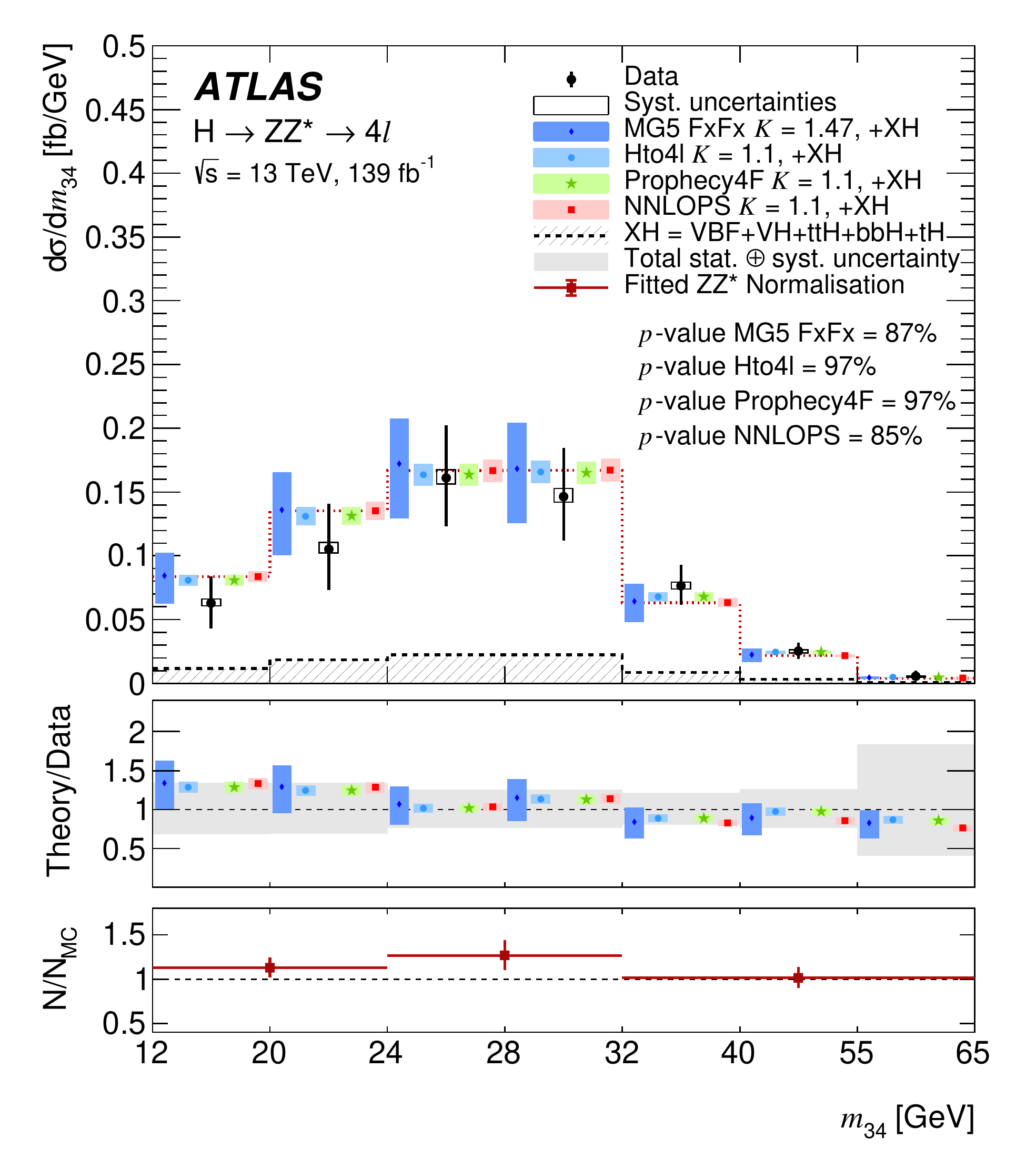}\label{fig:summary_m34}}
\subfloat[]{\includegraphics[width=0.48\linewidth]{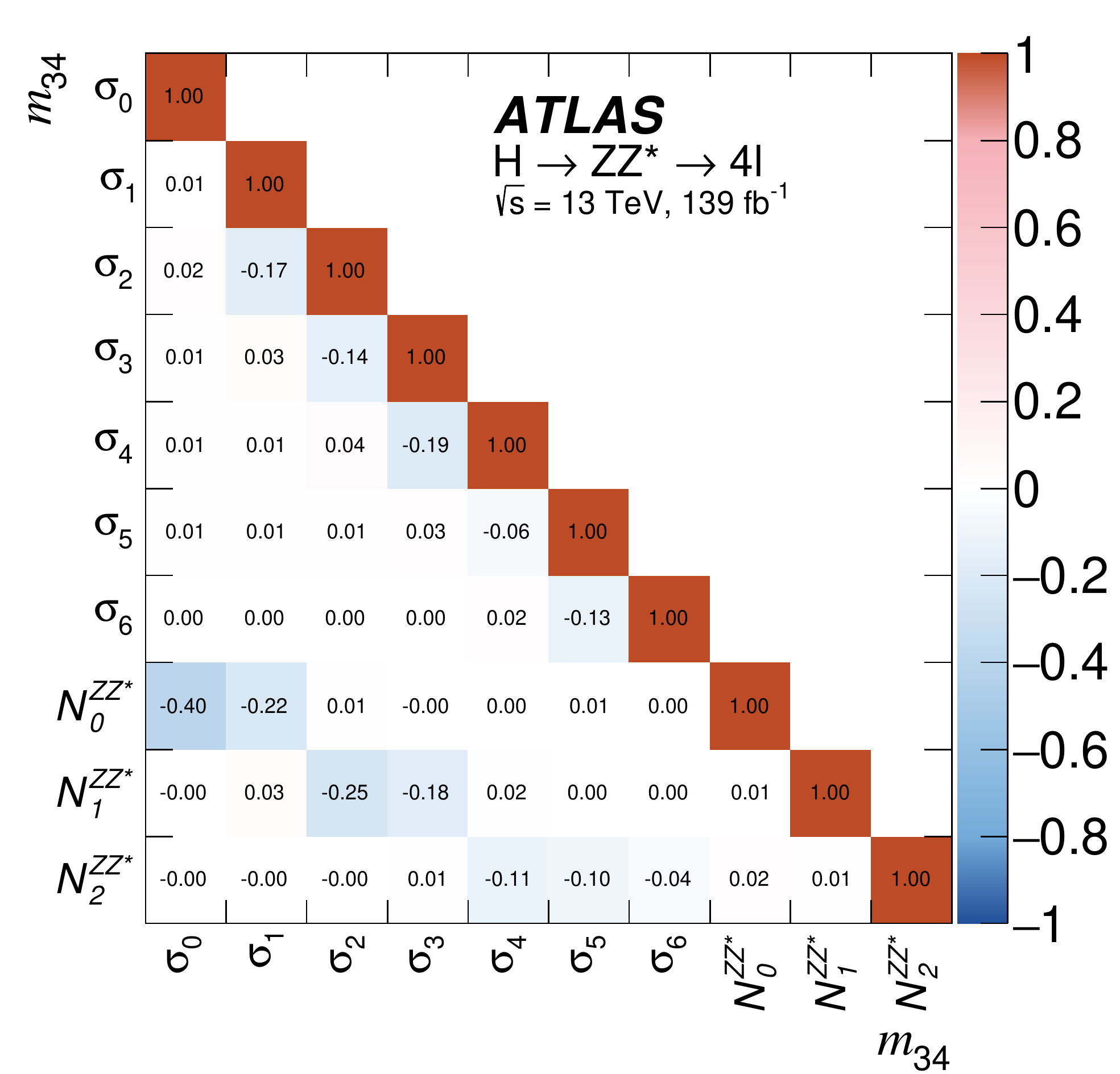}\label{fig:corrMatrix_m34}}
\caption{Differential fiducial cross sections for~\protect\subref{fig:summary_m12} the invariant mass \monetwo~of the leading $Z$ boson and~\protect\subref{fig:summary_m34} the invariant mass \mthreefour\ of the subleading $Z$ boson, along with the corresponding correlation matrices between the measured cross sections and the $ZZ^{\ast}$ background normalisation factors (\protect\subref{fig:corrMatrix_m12} and~\protect\subref{fig:corrMatrix_m34}). 
\label{fig:m12_m34_unfolded} }
\end{figure}

\begin{figure}[!htbp]
\centering
\subfloat[]{\includegraphics[width=0.45\linewidth]{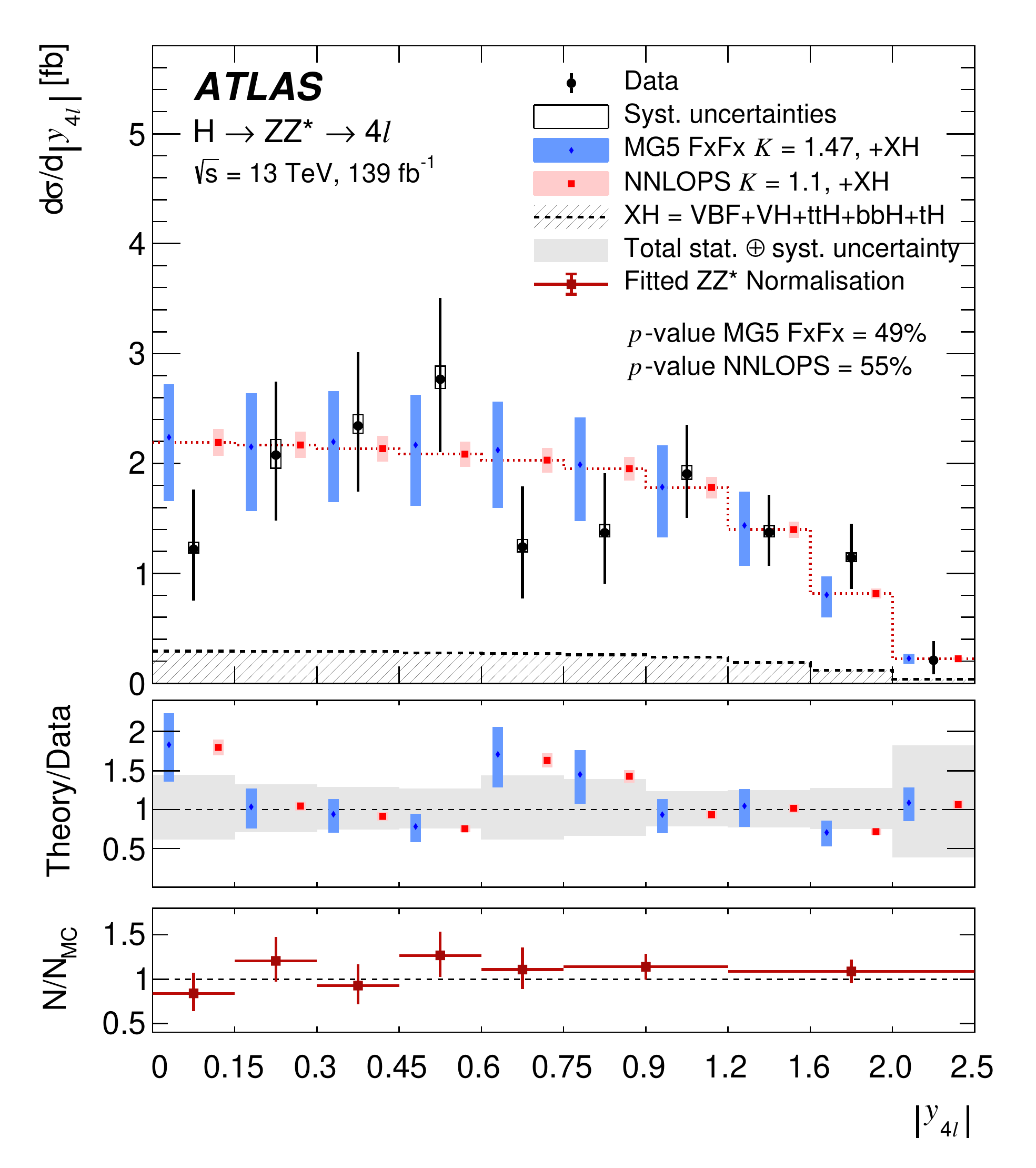}\label{fig:summary_y}}
\subfloat[]{\includegraphics[width=0.48\linewidth]{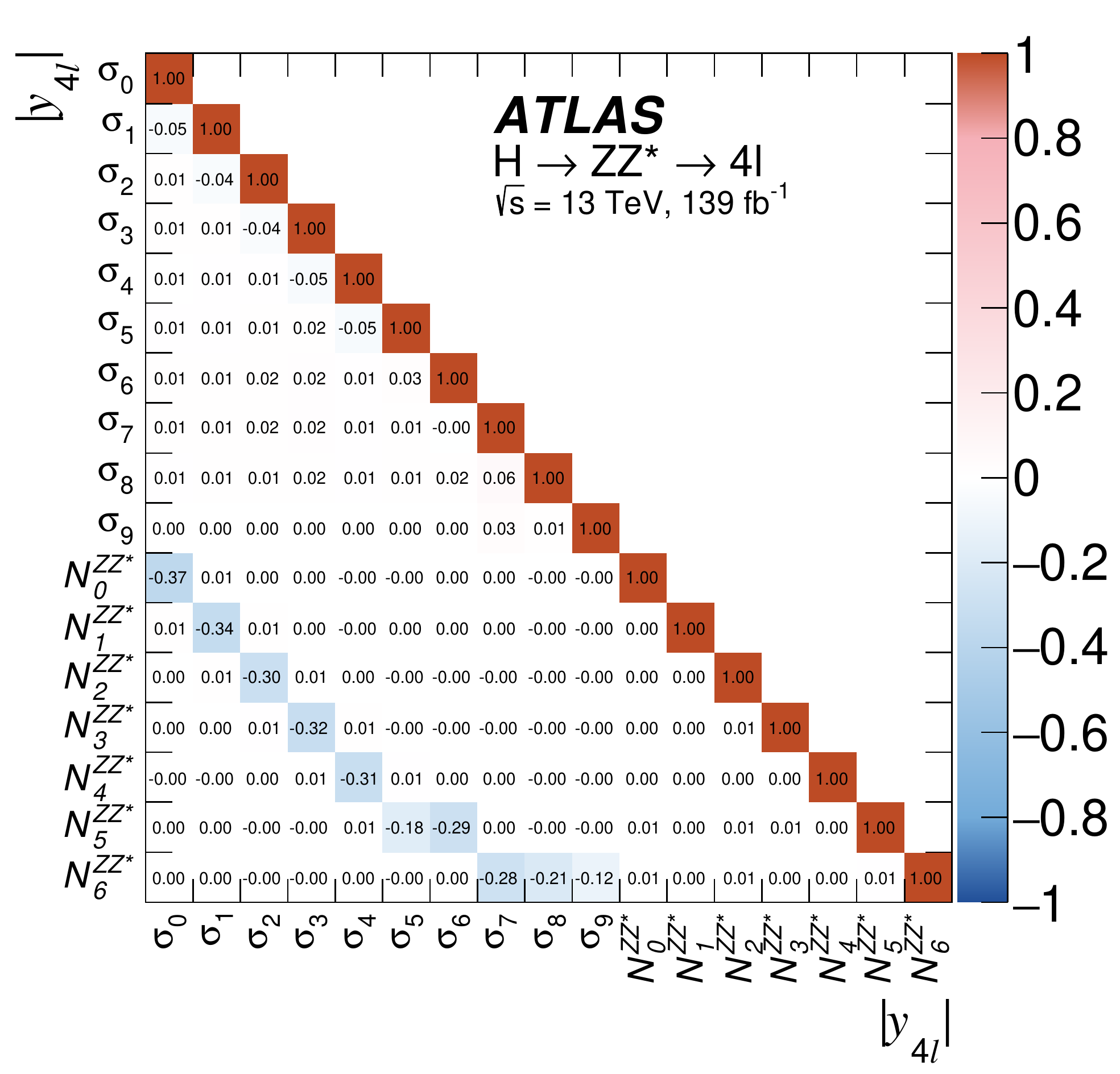}\label{fig:corrMatrix_y}}	\\
\subfloat[]{\includegraphics[width=0.45\linewidth]{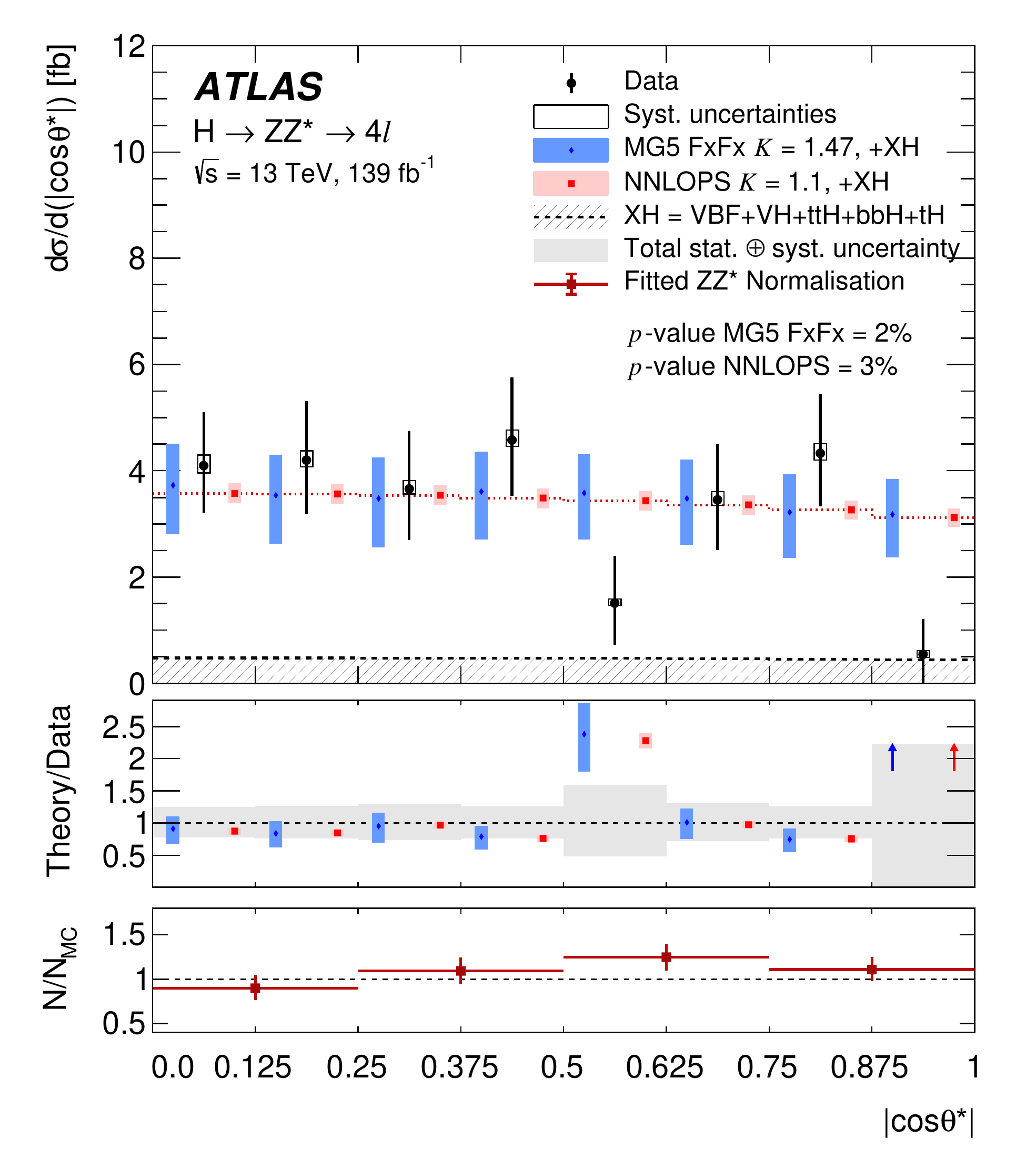}\label{fig:summary_cts}}
\subfloat[]{\includegraphics[width=0.48\linewidth]{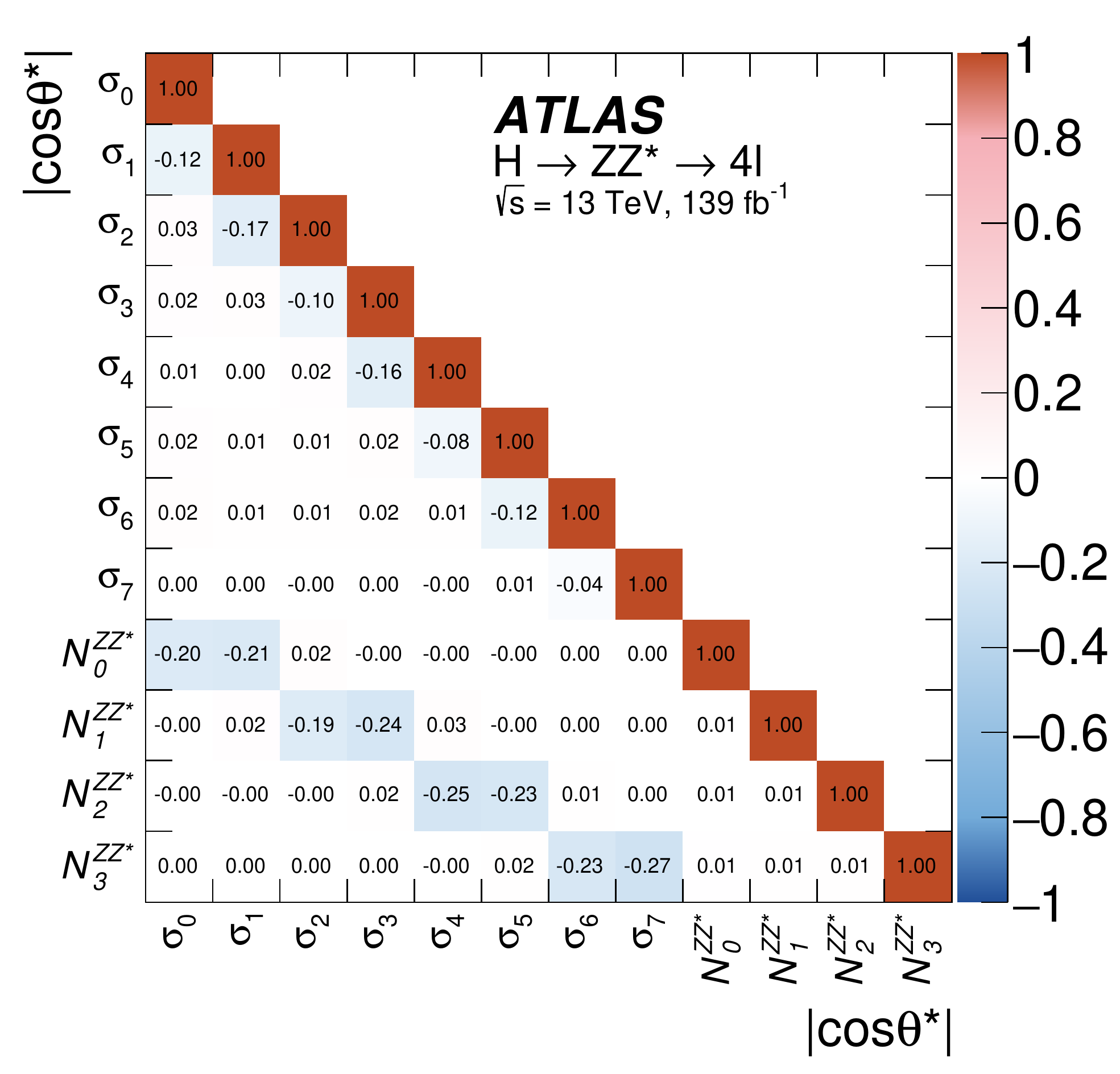}\label{fig:corrMatrix_cts}}
\caption{Differential fiducial cross sections for \protect\subref{fig:summary_y} the rapidity, \yH, of the Higgs boson and \protect\subref{fig:summary_cts} the production angle, \costhetastar, of the leading $Z$ boson. The corresponding correlation matrices between the measured cross sections and the $ZZ^{\ast}$ background normalisation factors are also shown (\protect\subref{fig:corrMatrix_y} and \protect\subref{fig:corrMatrix_cts}). 
\label{fig:y_cts_unfolded} }
\end{figure}

\begin{figure}[!htbp]
\centering
\subfloat[]{\includegraphics[width=0.45\linewidth]{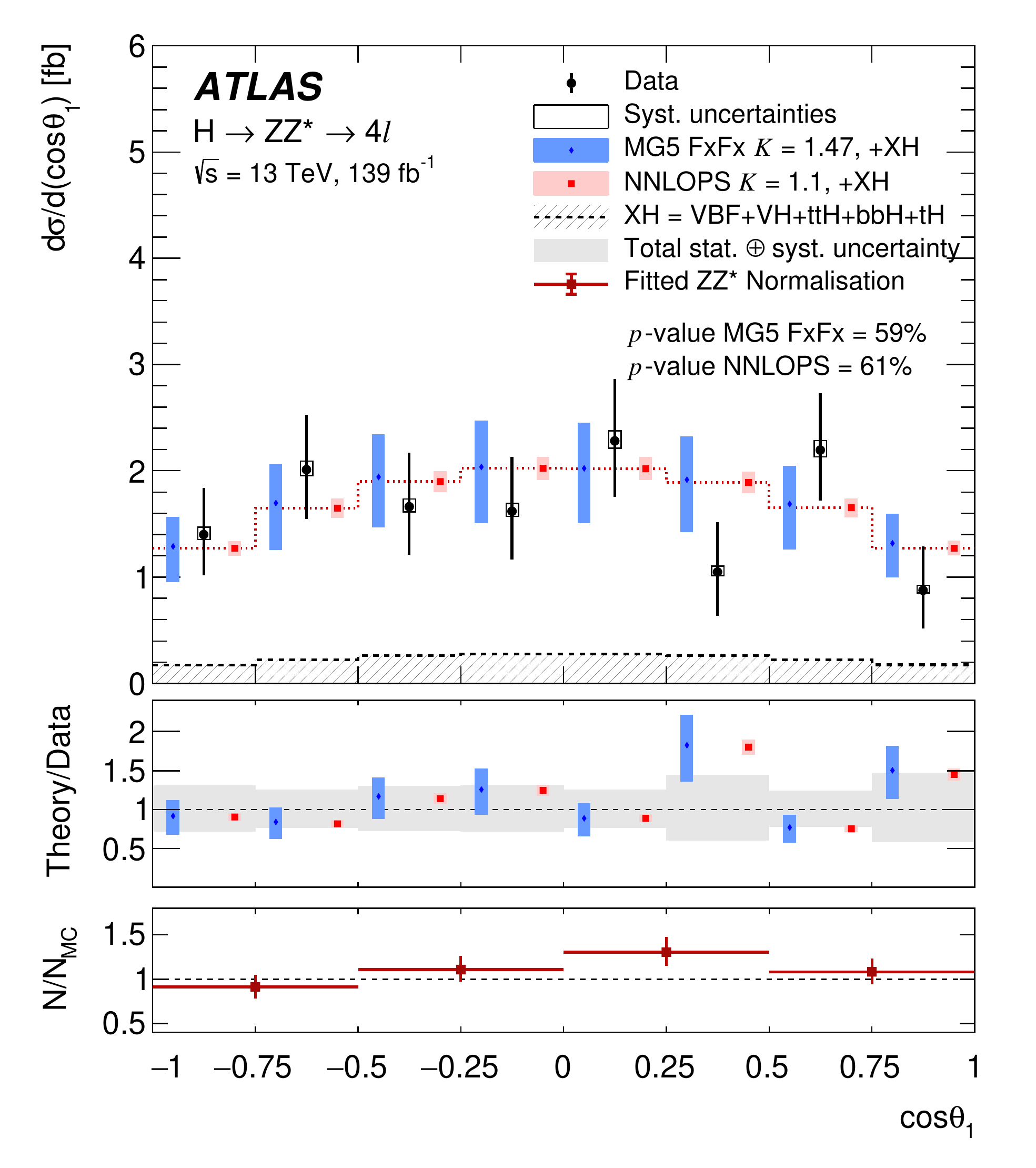}\label{fig:summary_ct1}}
\subfloat[]{\includegraphics[width=0.48\linewidth]{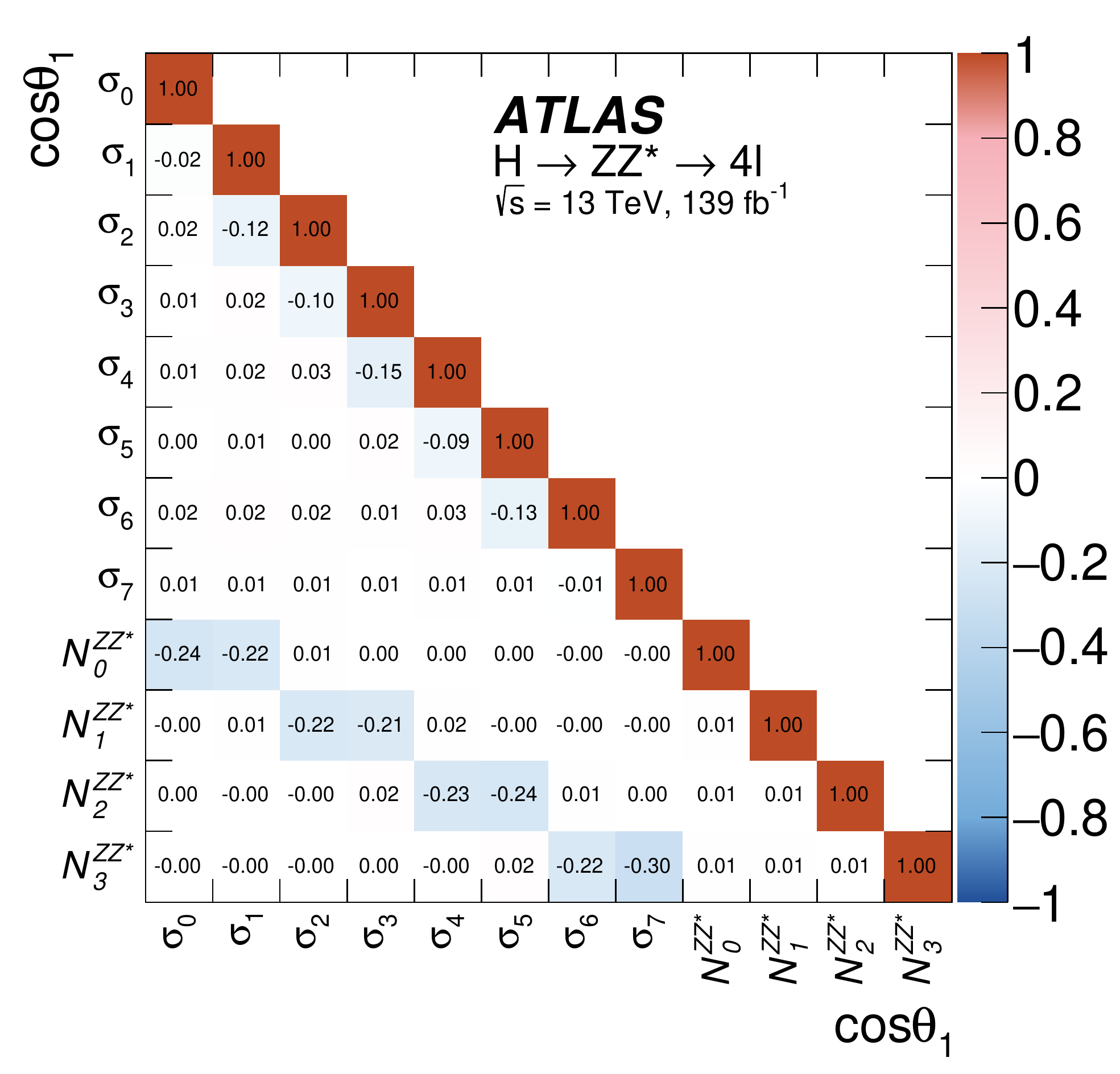}\label{fig:corrMatrix_ct1}}\\
\subfloat[]{\includegraphics[width=0.45\linewidth]{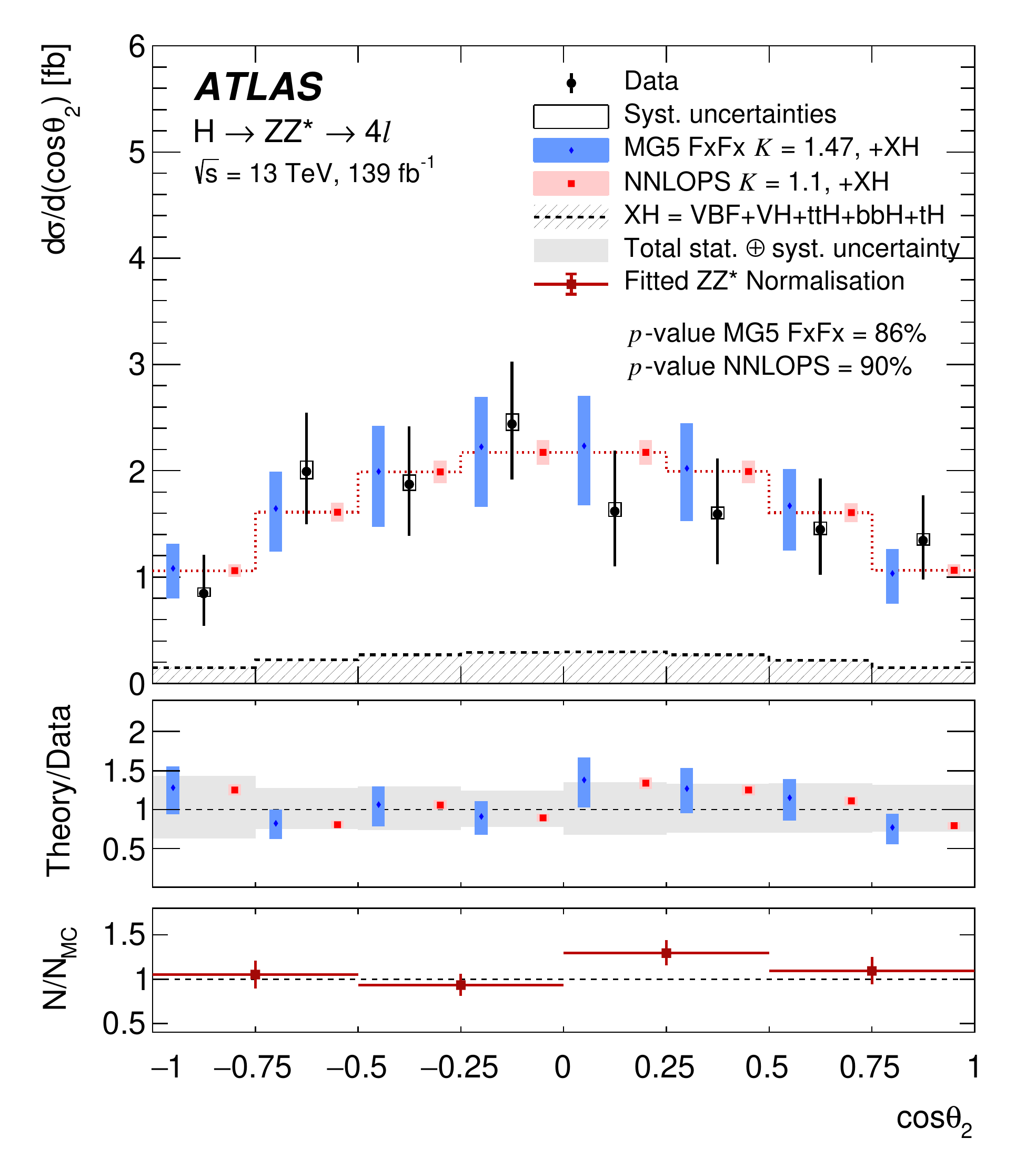}\label{fig:summary_ct2}}
\subfloat[]{\includegraphics[width=0.48\linewidth]{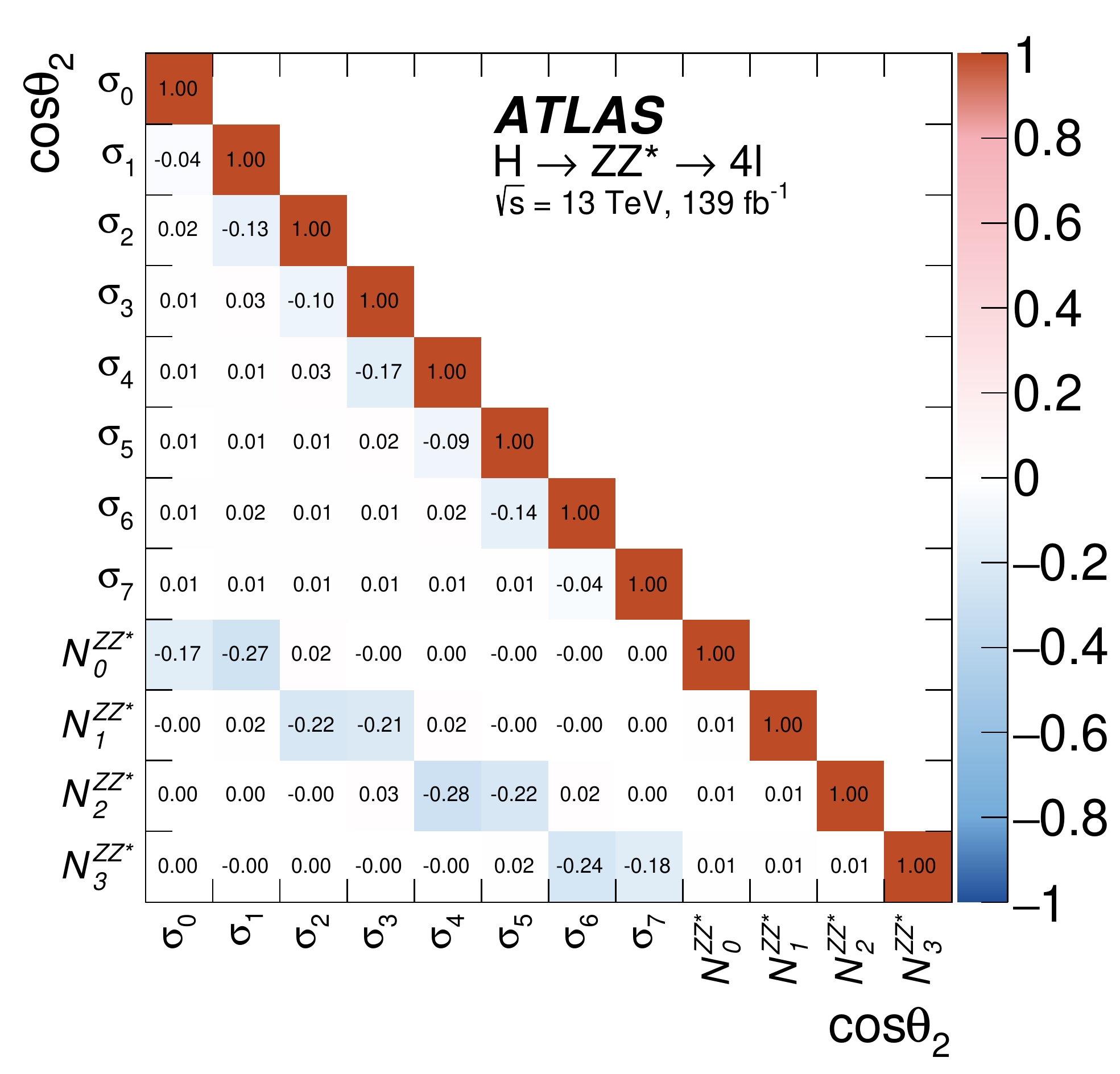}\label{fig:corrMatrix_ct2}}
\caption{Differential fiducial cross sections for~\protect\subref{fig:summary_ct1}  production angle, \costhetaone, of the anti-lepton from the leading $Z$ boson and~\protect\subref{fig:summary_ct2}  the production angle, \costhetatwo, of the anti-lepton from the subleading $Z$ boson. The corresponding correlation matrices between the measured cross sections and the $ZZ^{\ast}$ background normalisation factors are also shown (\protect\subref{fig:corrMatrix_ct1} and~\protect\subref{fig:corrMatrix_ct2}). 
\label{fig:ct1_ct2_unfolded} }
\end{figure}

\begin{figure}[!htbp]
\centering
\subfloat[]{\includegraphics[width=0.45\linewidth]{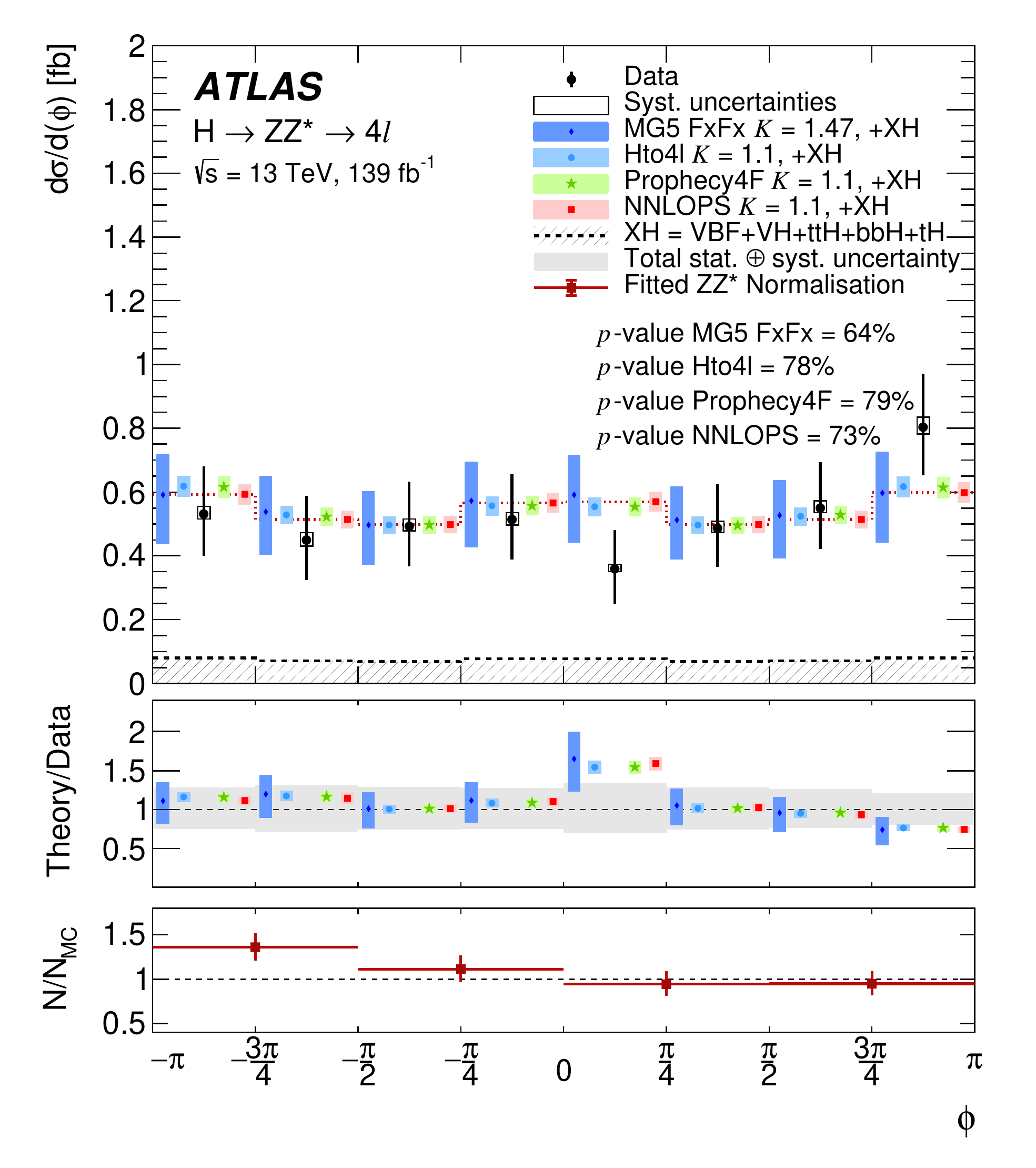}\label{fig:summary_phi}}
\subfloat[]{\includegraphics[width=0.48\linewidth]{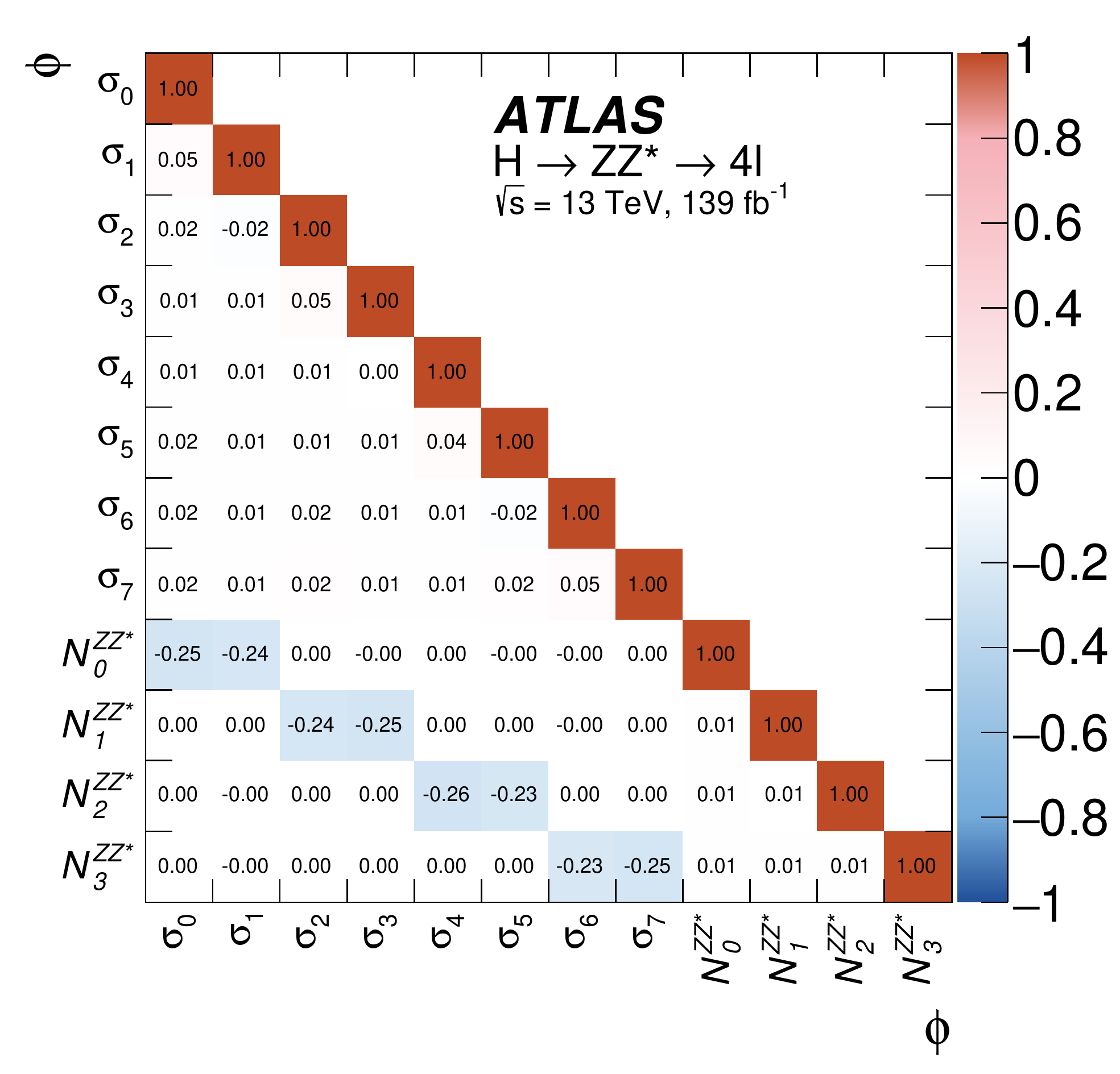}\label{fig:corrMatrix_phi}}\\
\subfloat[]{\includegraphics[width=0.45\linewidth]{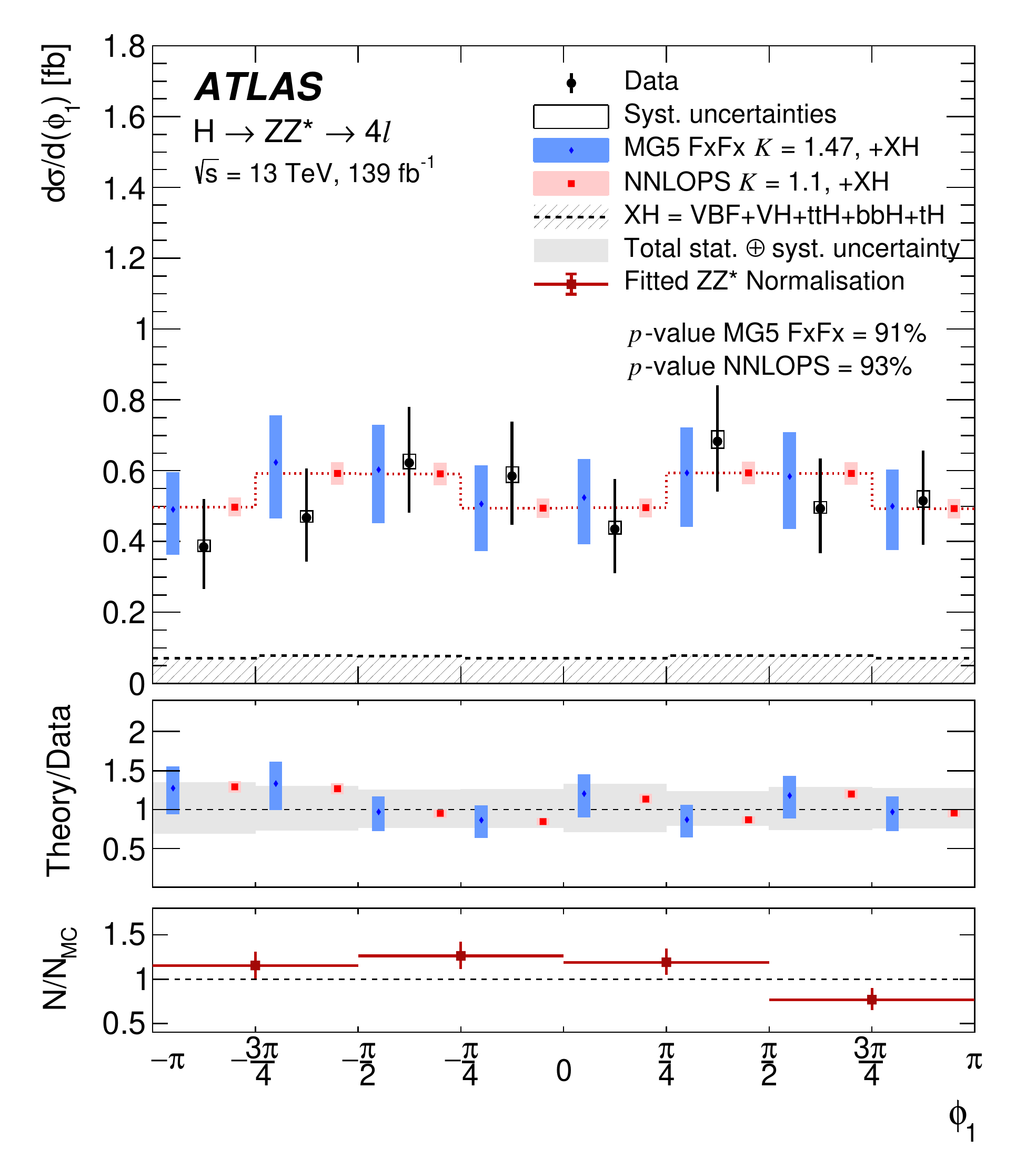}\label{fig:summary_phi1}}
\subfloat[]{\includegraphics[width=0.48\linewidth]{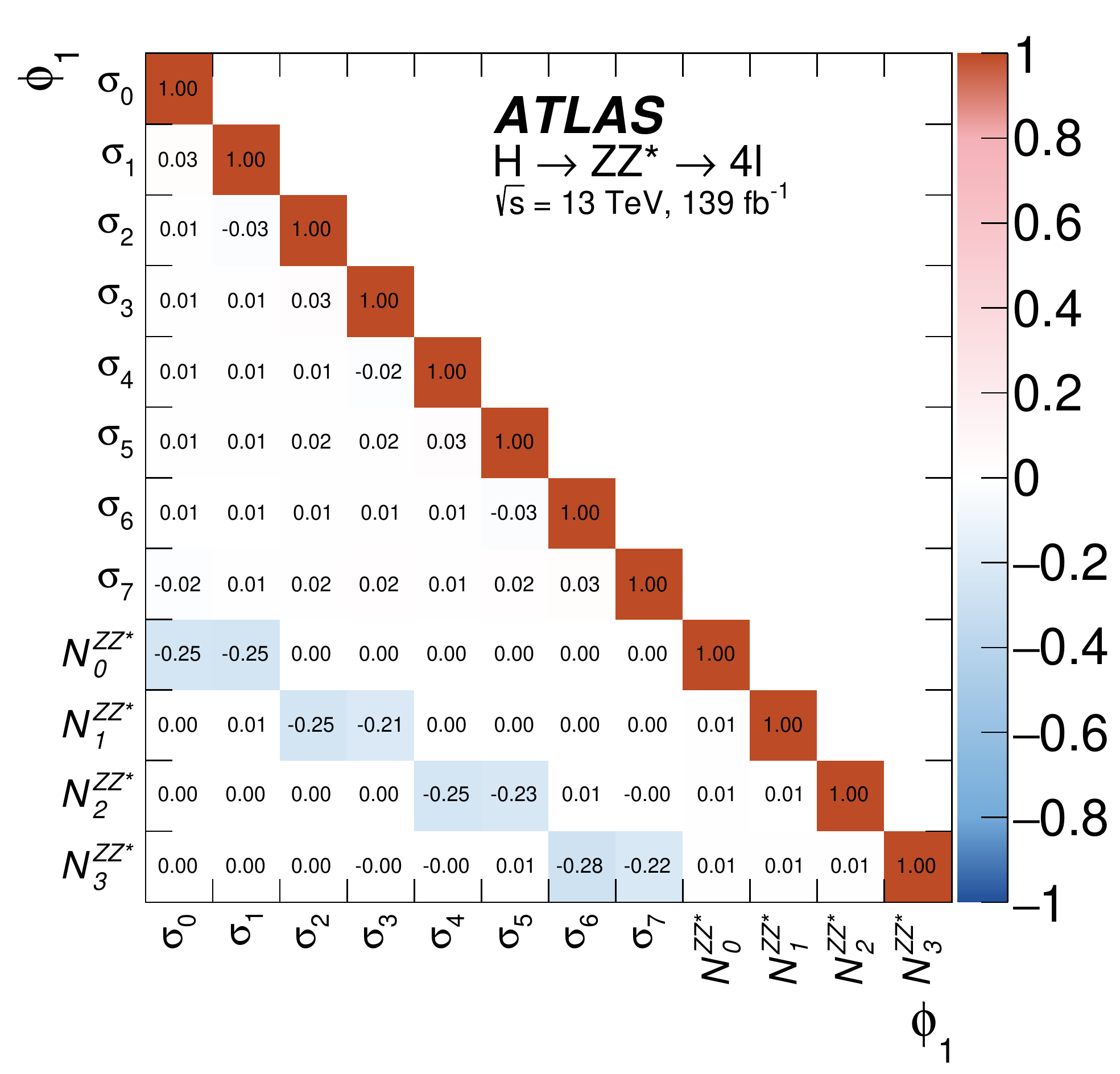}\label{fig:corrMatrix_phi1}}
\caption{Differential fiducial cross sections for~\protect\subref{fig:summary_phi} the azimuthal angle, \phiZZ, between the decay planes of the two reconstructed $Z$ bosons and~\protect\subref{fig:summary_phi1} the azimuthal angle, \phione, between the decay plane of the leading $Z$ boson and the plane formed by its four-momentum and the $z$-axis. The corresponding correlation matrices between the measured cross sections and the $ZZ^{\ast}$ background normalisation factors are also shown (\protect\subref{fig:corrMatrix_phi} and~\protect\subref{fig:corrMatrix_phi1}). 
\label{fig:phi_phi1_unfolded} }
\end{figure}

The measured differential production cross sections for the transverse momentum \ptH\ of the Higgs boson are shown in Figure~\ref{fig:pt_unfolded}, while the measured differential cross sections with respect to the masses of the leading and subleading $Z$ bosons resulting from the Higgs boson decay, \monetwo~and \mthreefour, are provided in Figure~\ref{fig:m12_m34_unfolded}.  Figures~\ref{fig:y_cts_unfolded}, \ref{fig:ct1_ct2_unfolded}, and \ref{fig:phi_phi1_unfolded} show the measured differential production cross sections with respect to angular variables, \yH, \costhetastar, \costhetaone, \costhetatwo, \phiZZ, and \phione, that probe the kinematics of the Higgs boson decay products.
 
\begin{figure}[!htbp]
\centering
\subfloat[]{\includegraphics[width=0.45\linewidth]{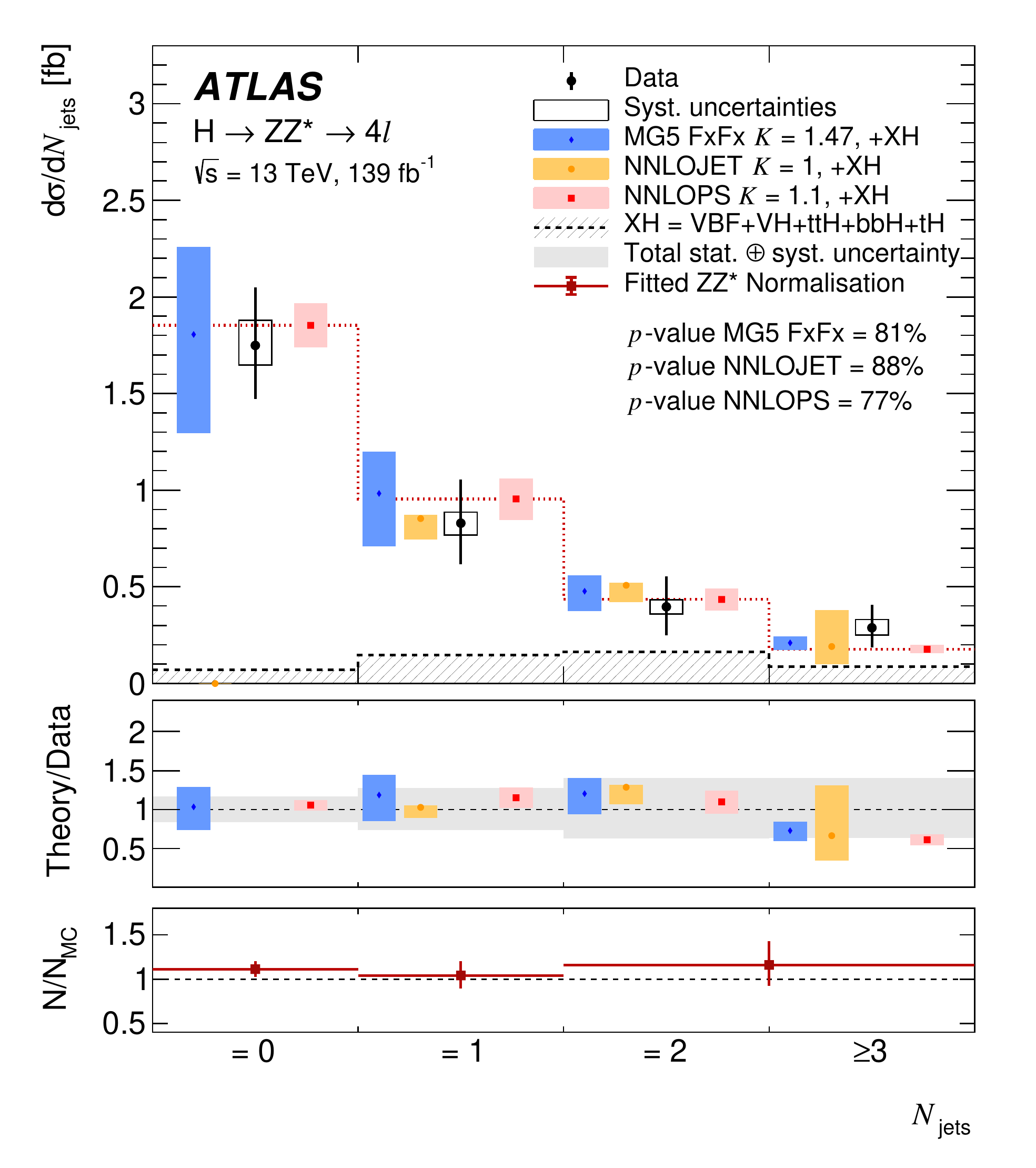}\label{fig:summary_njet}}
\subfloat[]{\includegraphics[width=0.45\linewidth]{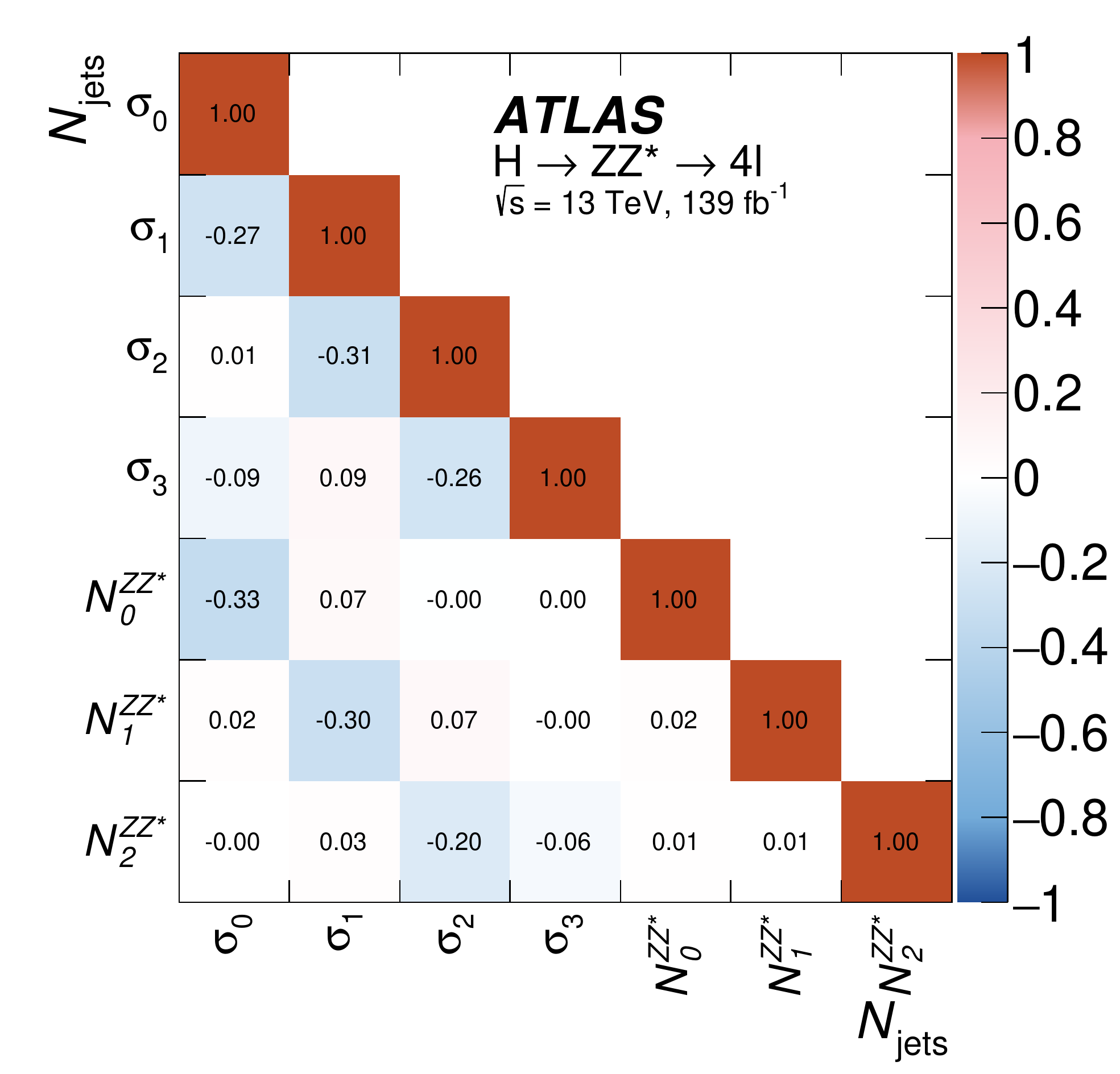}\label{fig:corrMatrix_njet}}\\
\subfloat[]{\includegraphics[width=0.45\linewidth]{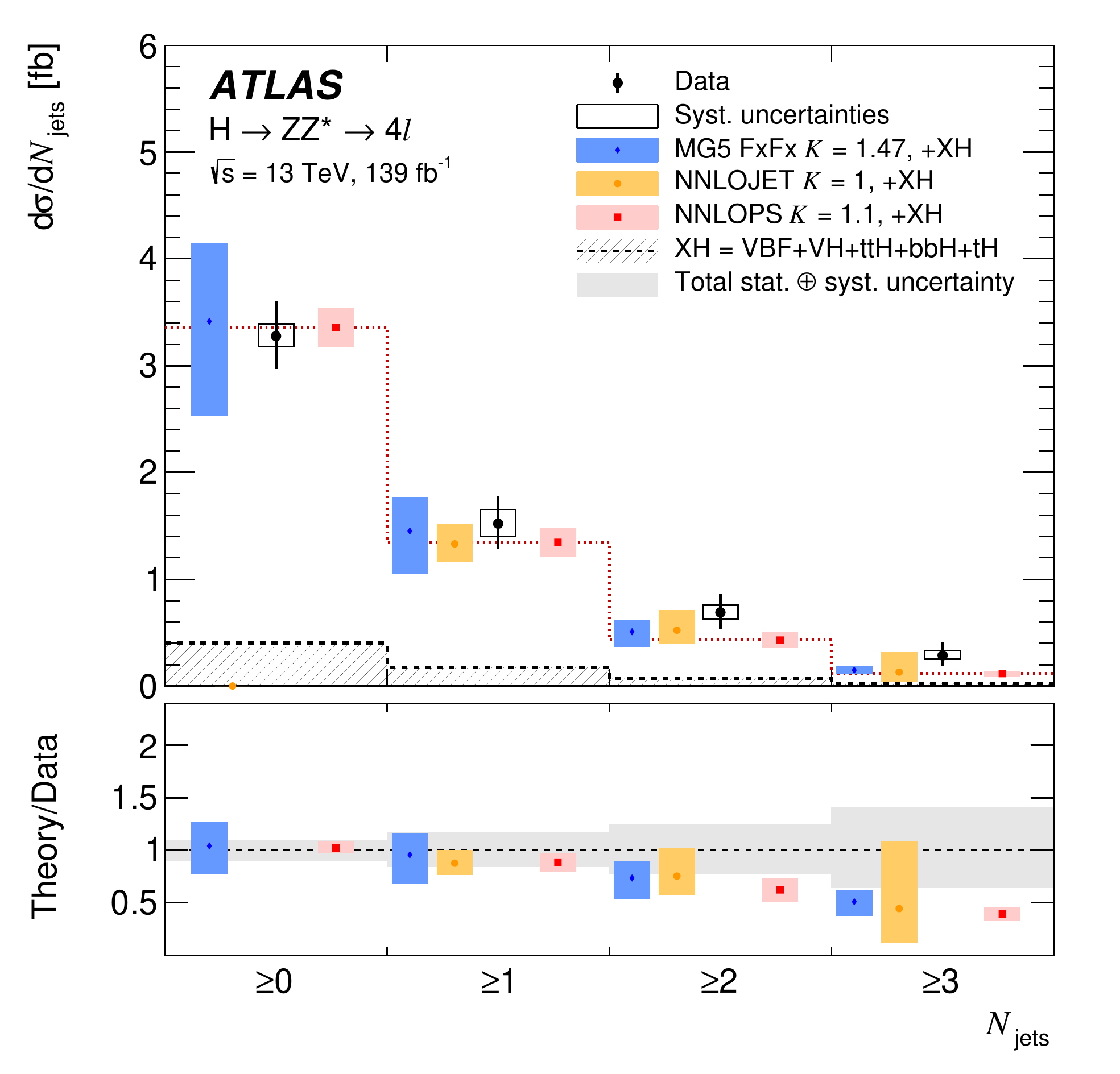}\label{fig:summary_njet_incl}}
\caption{Differential fiducial cross sections for~\protect\subref{fig:summary_njet} the jet multiplicity, \njets, in the selected events, and~\protect\subref{fig:summary_njet_incl}, the inclusive jet multiplicity. In the \njets\ distribution in~\protect\subref{fig:summary_njet}, the first three bins are exclusive in number of jets, while the fourth is inclusive. The corresponding correlation matrix between the measured cross sections and the $ZZ^{\ast}$ background normalisation factors is also shown in \protect\subref{fig:corrMatrix_njet}. In the \njets\ distribution in~\protect\subref{fig:summary_njet_incl}, all bins are inclusive, with the first bin including all events, the second including all events with at least one jet, and so on. 
\label{fig:njet_unfolded} }
\end{figure}

\begin{figure}[!htbp]
\centering
\subfloat[]{\includegraphics[width=0.45\linewidth]{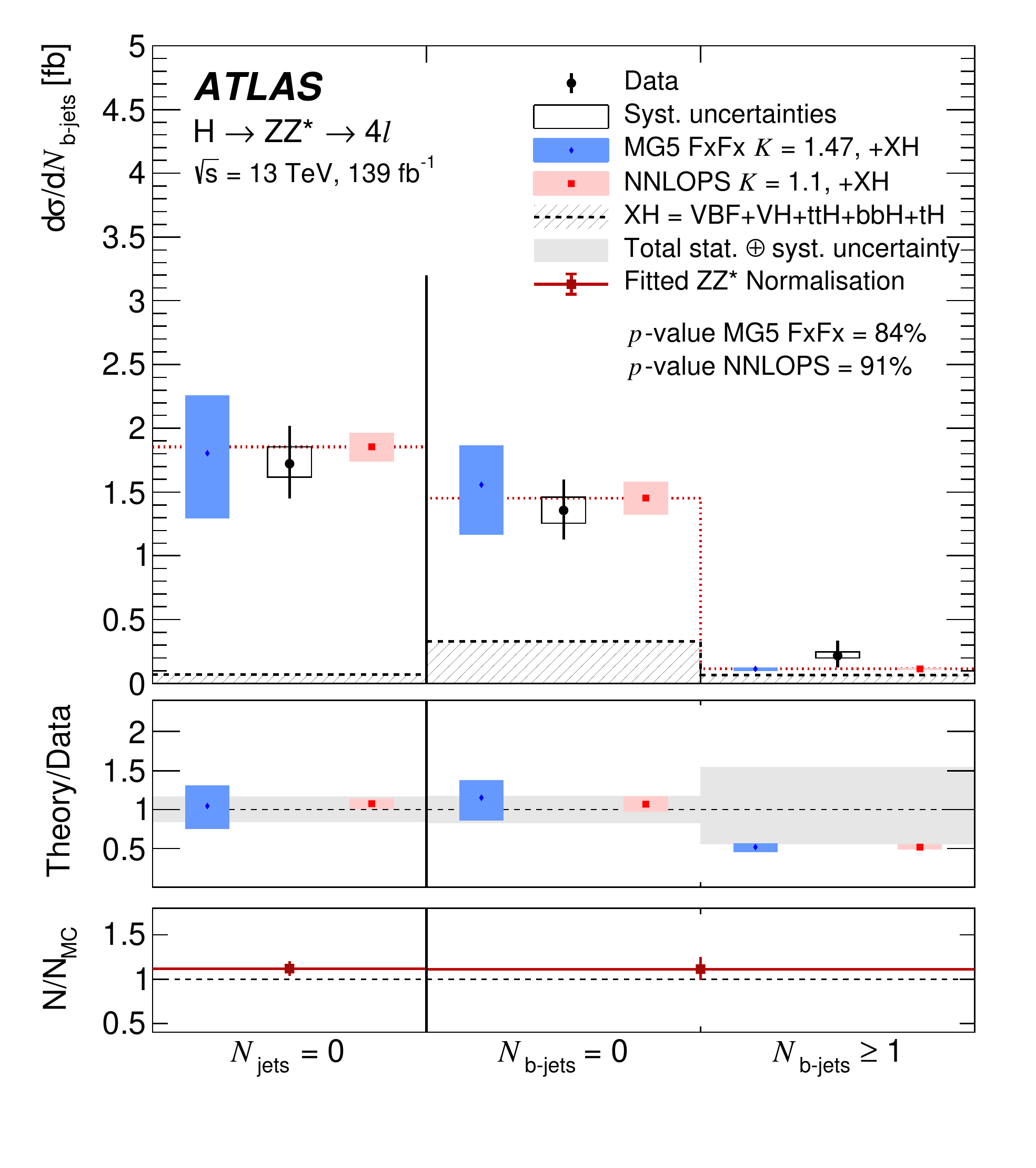}\label{fig:summary_nbjet}}
\subfloat[]{\includegraphics[width=0.45\linewidth]{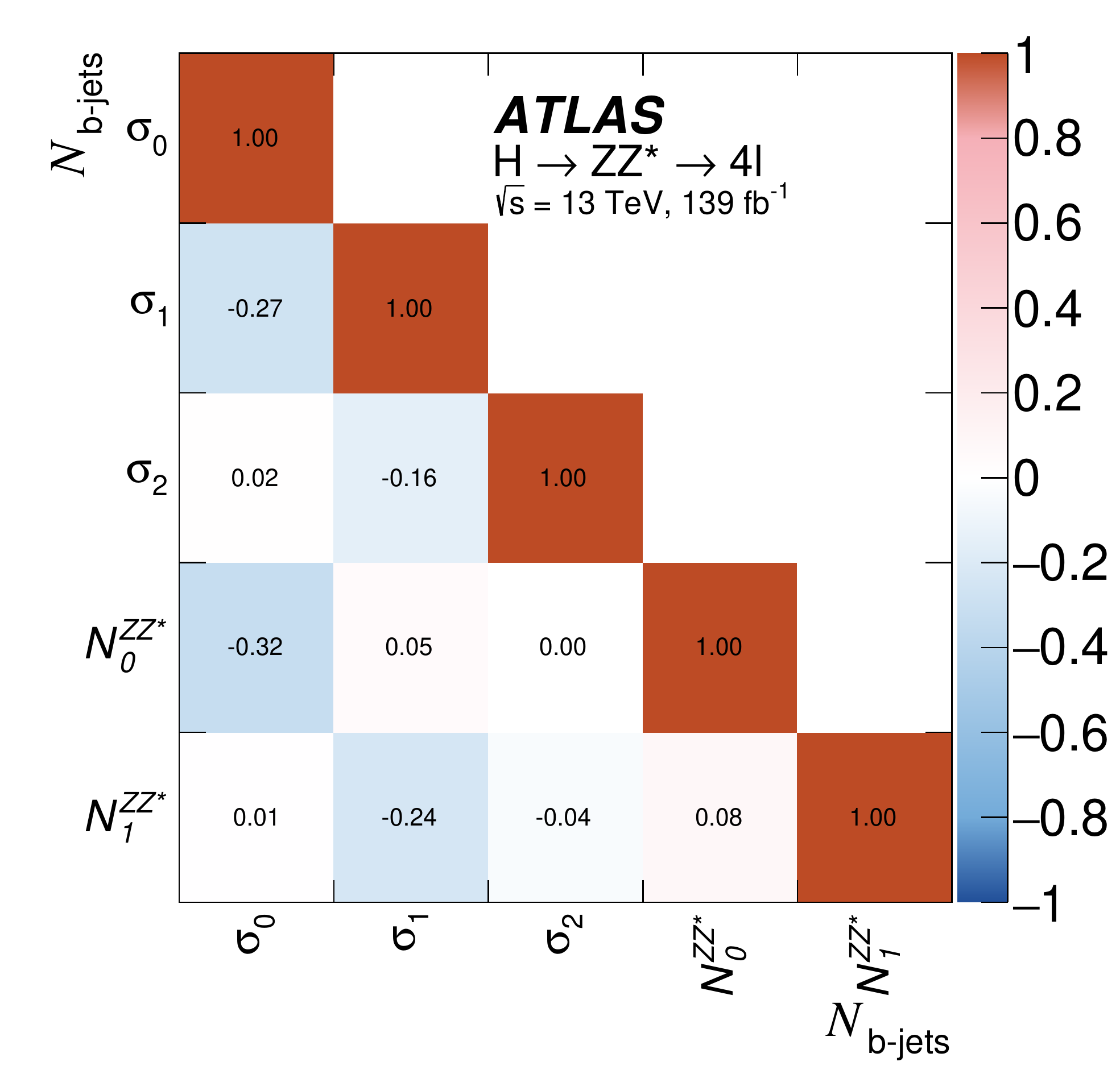}\label{fig:corrMatrix_nbjet}}
\caption{~\protect\subref{fig:summary_nbjet} Differential fiducial cross section as function of  the $b$-jet multiplicity, \nbjets. Three bins are considered. The first bin is filled with events which do not have any jets, the second is filled with events with at least one jet but no $b$-tagged jets, while the third includes all events with at least one $b$-tagged jet. The corresponding correlation matrices between the measured cross sections and the $ZZ^{\ast}$ background normalisation factors are also shown in~\protect\subref{fig:corrMatrix_nbjet}). 
\label{fig:nbjet_unfolded} }
\end{figure}

\begin{figure}[!htbp]
\centering
\subfloat[]{\includegraphics[width=0.45\linewidth]{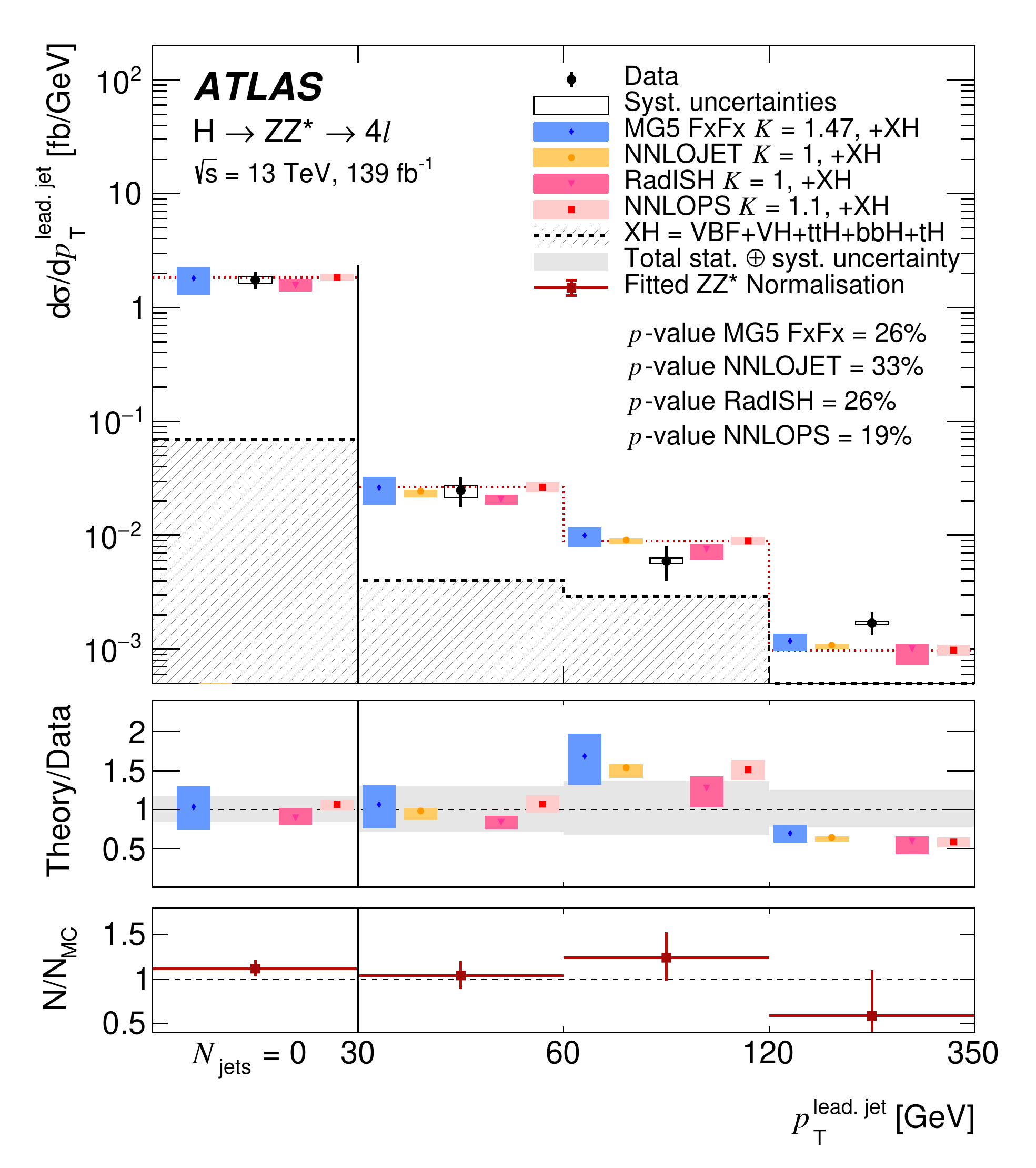}\label{fig:summary_ljpt}}
\subfloat[]{\includegraphics[width=0.45\linewidth]{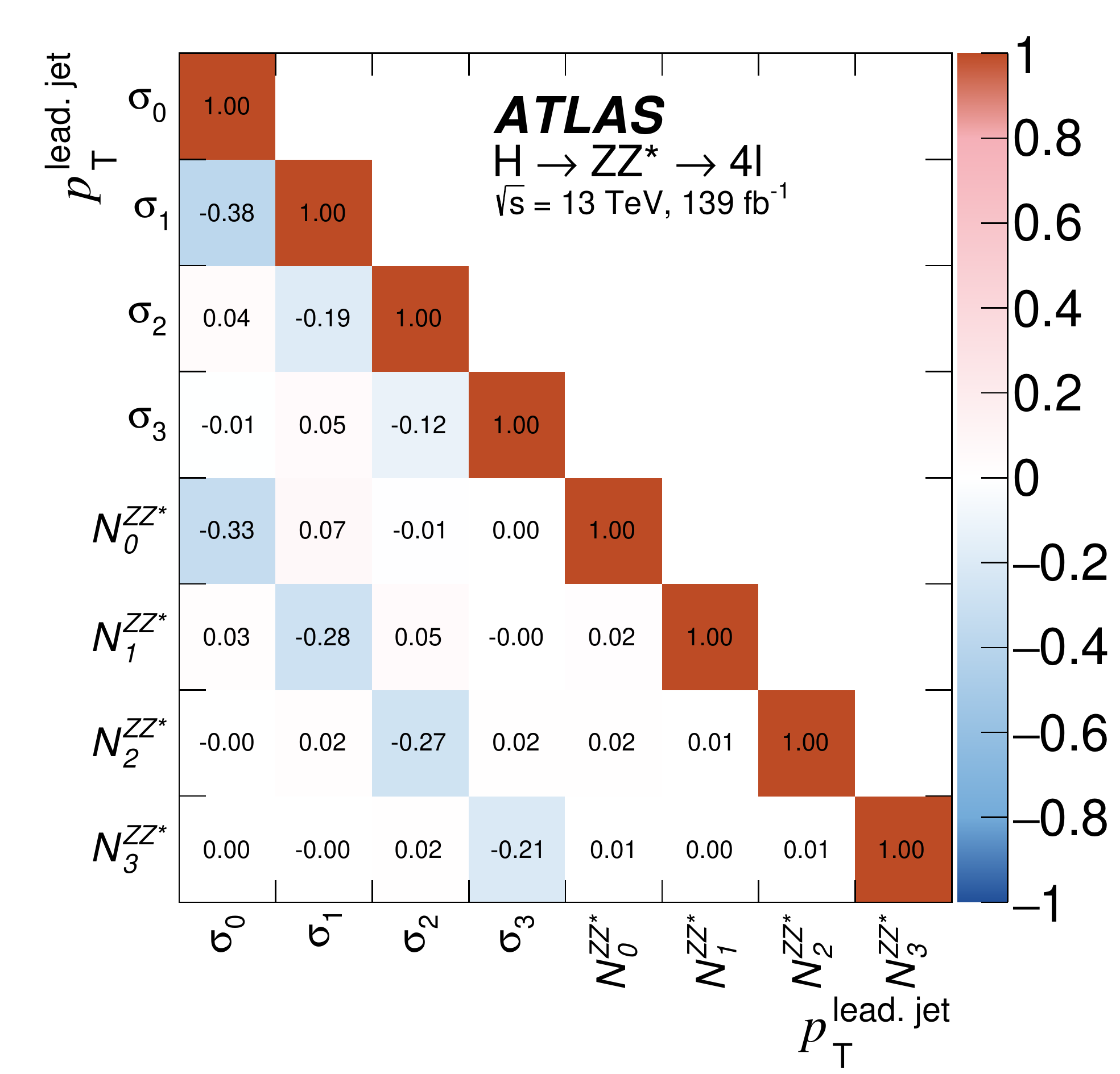}\label{fig:corrMatrix_ljpt}}\\
\subfloat[]{\includegraphics[width=0.45\linewidth]{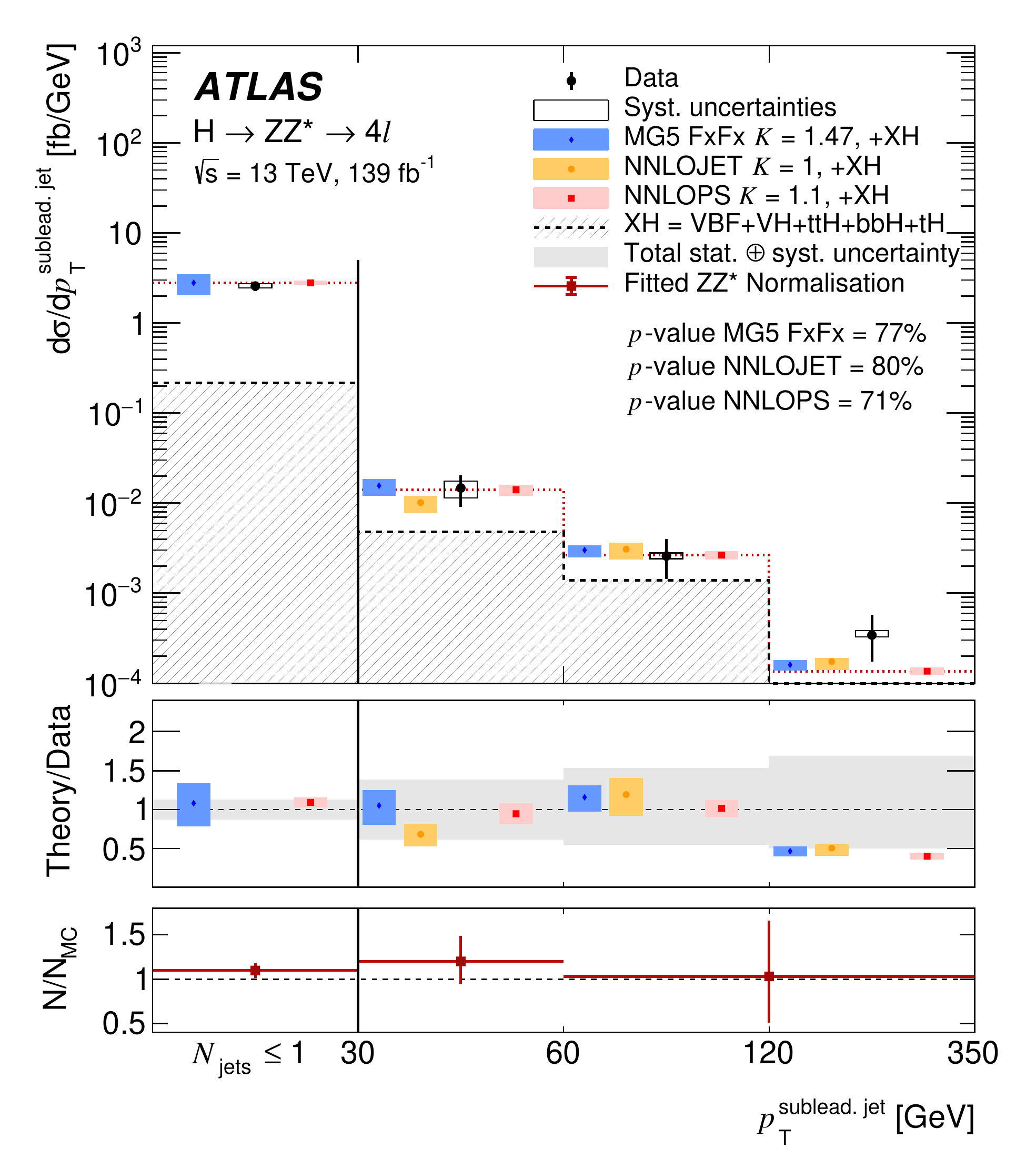}\label{fig:summary_sjpt}}
\subfloat[]{\includegraphics[width=0.45\linewidth]{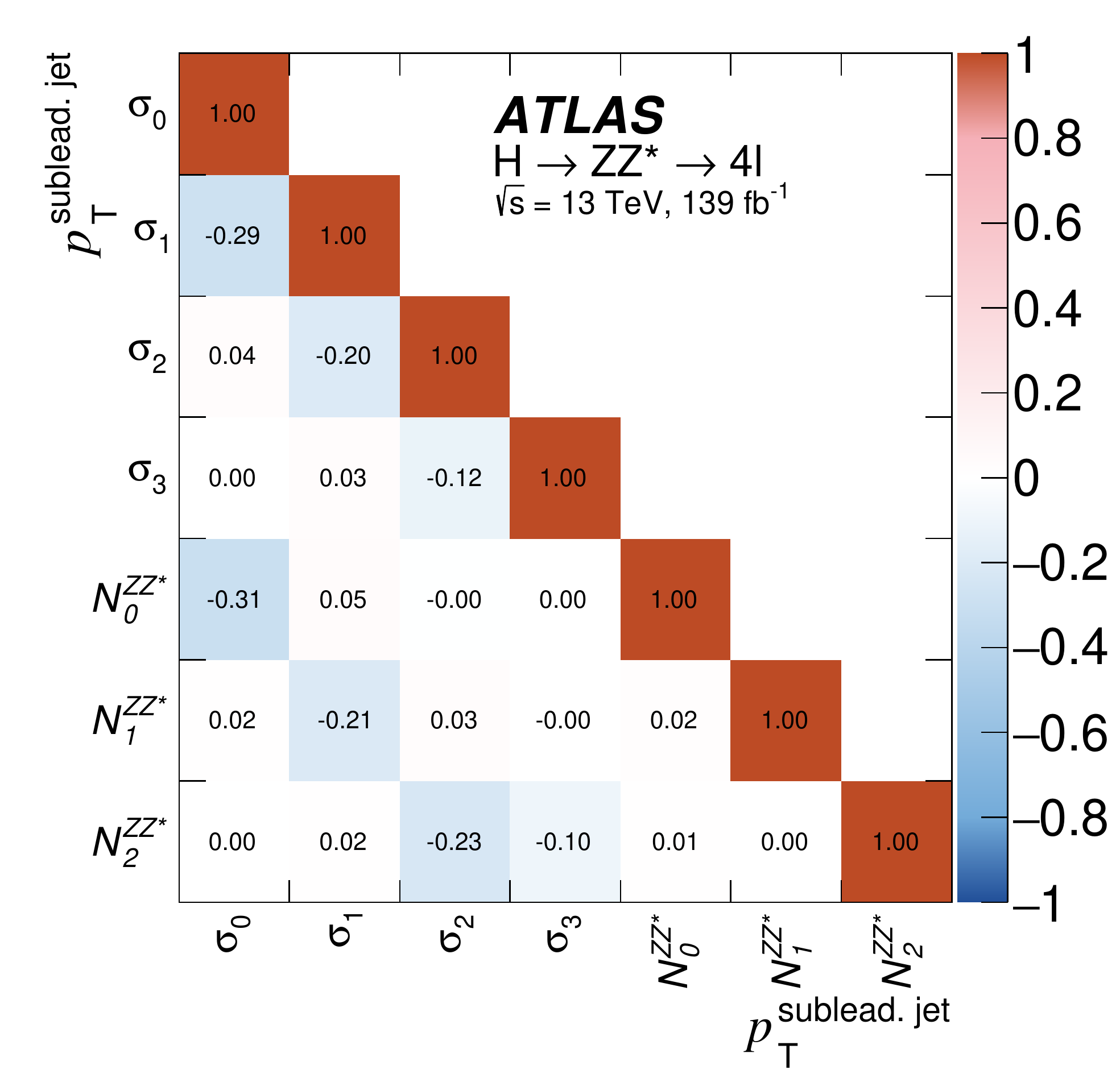}\label{fig:corrMatrix_sjpt}}
\caption{Differential fiducial cross sections for~\protect\subref{fig:summary_ljpt} the transverse momentum of the leading jet, \ljpt, in events with at least one jet, and~\protect\subref{fig:summary_sjpt} the transverse momentum of the subleading jet, \sjpt, in events with at least two jets. Leading and subleading jets refer to the jets with the highest and second-highest transverse momenta. The first bin contains events which do not pass the jet requirements. The corresponding correlation matrices between the measured cross sections and the $ZZ^{\ast}$ background normalisation factors are also shown (\protect\subref{fig:corrMatrix_ljpt} and ~\protect\subref{fig:corrMatrix_sjpt}). 
\label{fig:ljpt_sjpt_unfolded} }
\end{figure}

\begin{figure}[!htbp]
\centering
\subfloat[]{\includegraphics[width=0.45\linewidth]{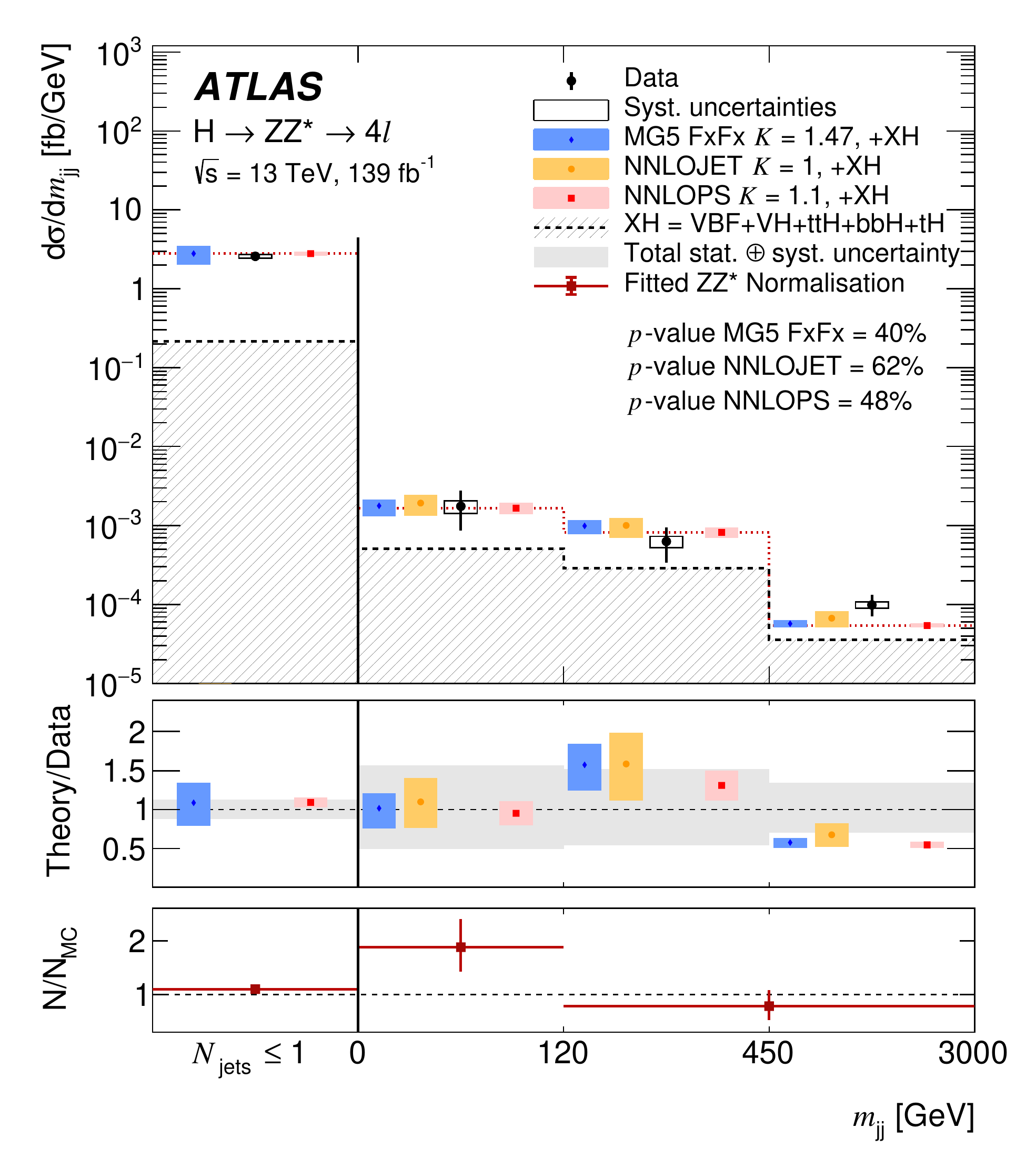}\label{fig:summary_mjj}}
\subfloat[]{\includegraphics[width=0.48\linewidth]{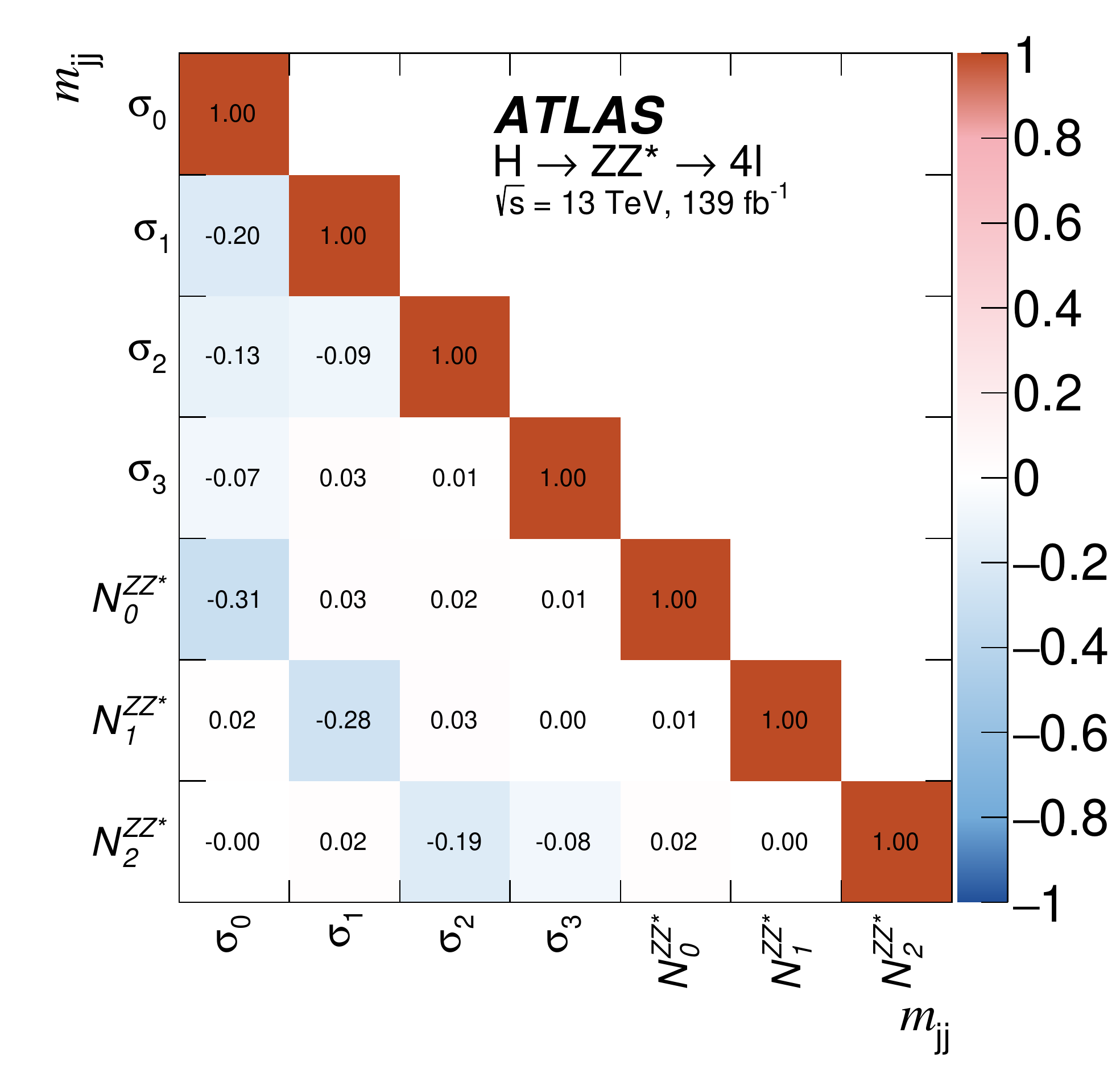}\label{fig:corrMatrix_mjj}}\\
\caption{Differential fiducial cross sections for~\protect\subref{fig:summary_mjj} the invariant mass of the two highest-$\pt$ jets, \mjj, in events with at least two jets. The corresponding correlation matrix between the measured cross sections and the $ZZ^{\ast}$ background normalisation factors is also provided (\protect\subref{fig:corrMatrix_mjj}). 
\label{fig:mjj_unfolded} }
\end{figure}

\begin{figure}[!htbp]
\centering
\subfloat[]{\includegraphics[width=0.45\linewidth]{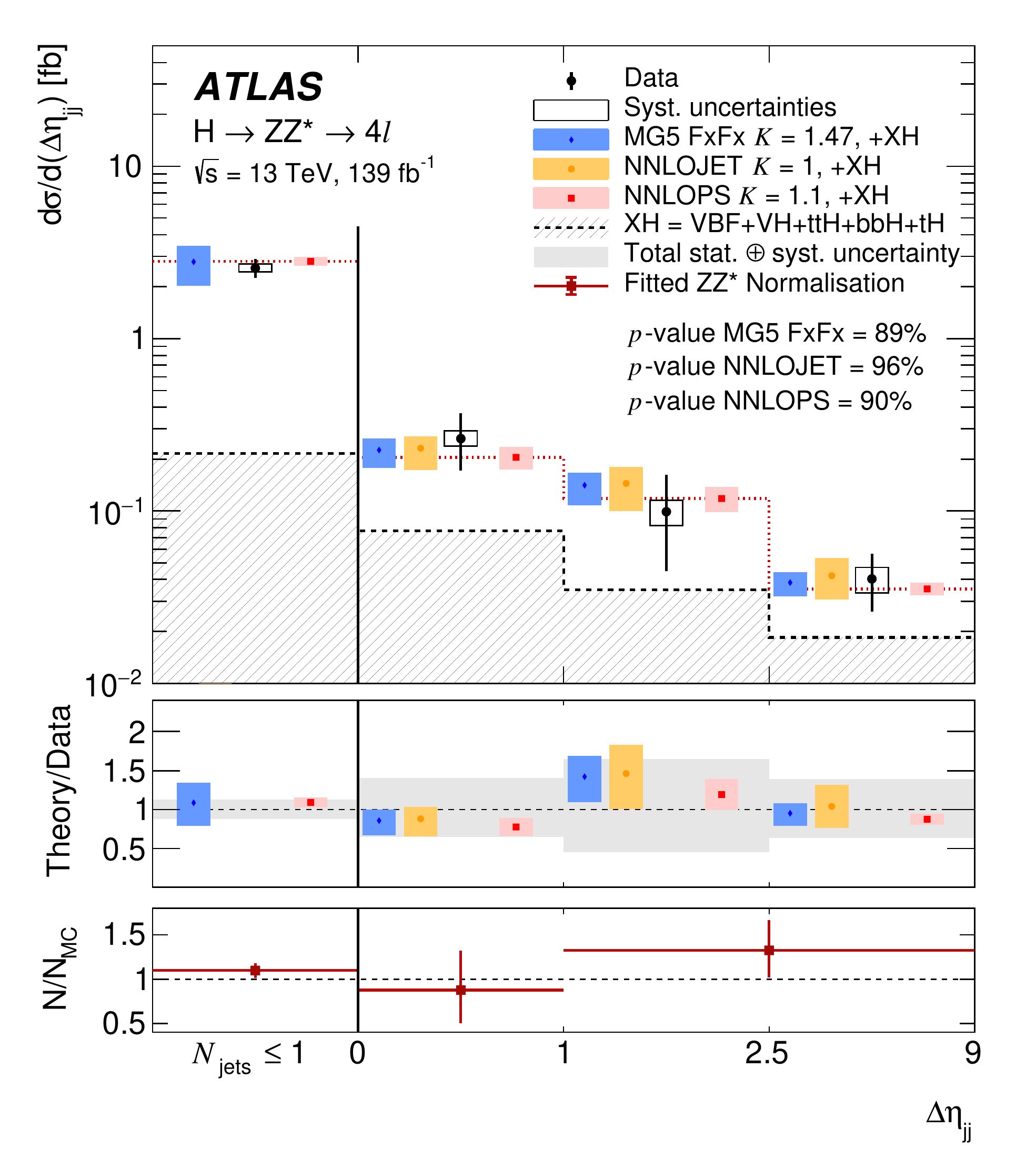}\label{fig:summary_etajj}}
\subfloat[]{\includegraphics[width=0.48\linewidth]{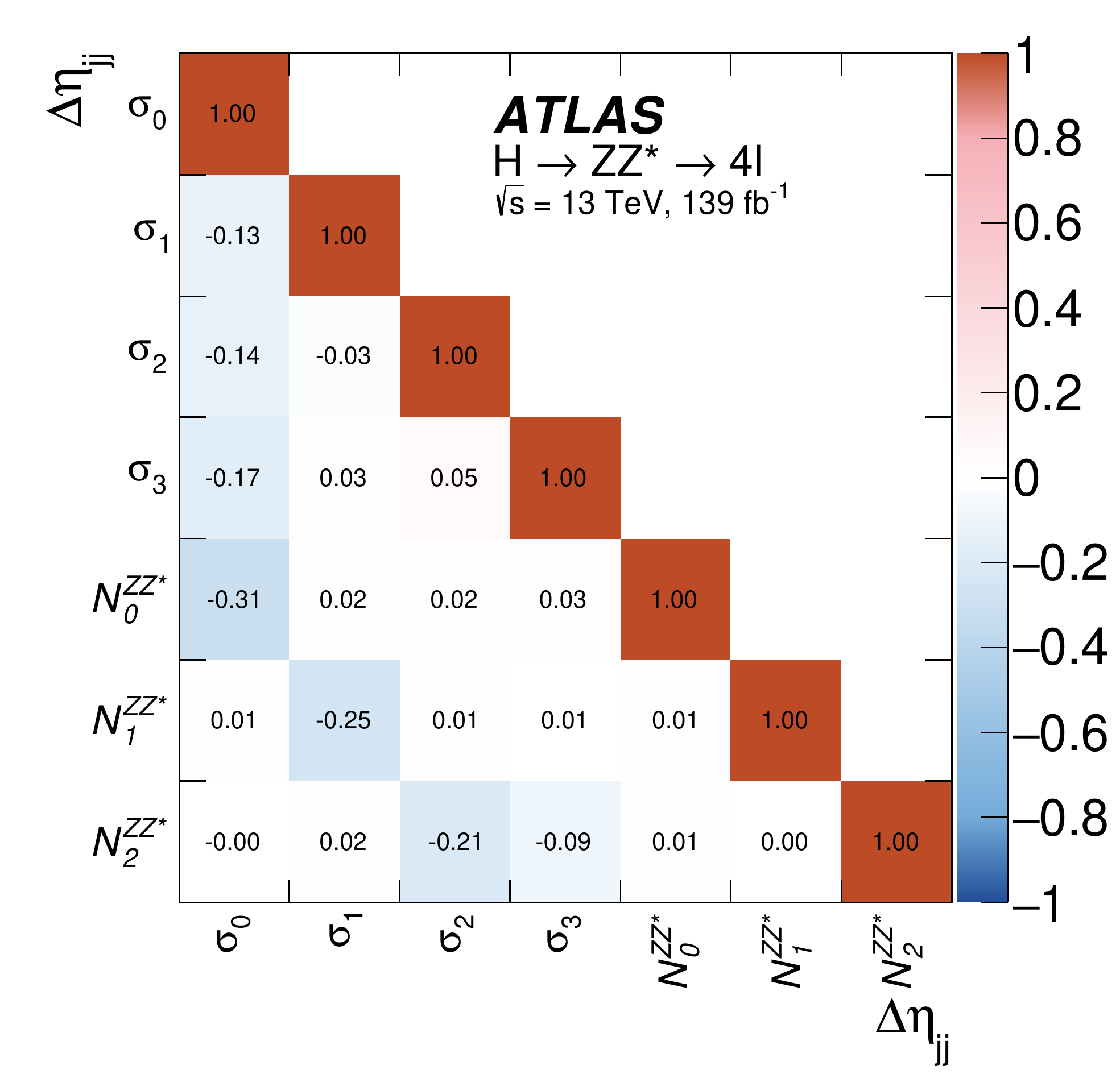}\label{fig:corrMatrix_etajj}}\\
\subfloat[]{\includegraphics[width=0.45\linewidth]{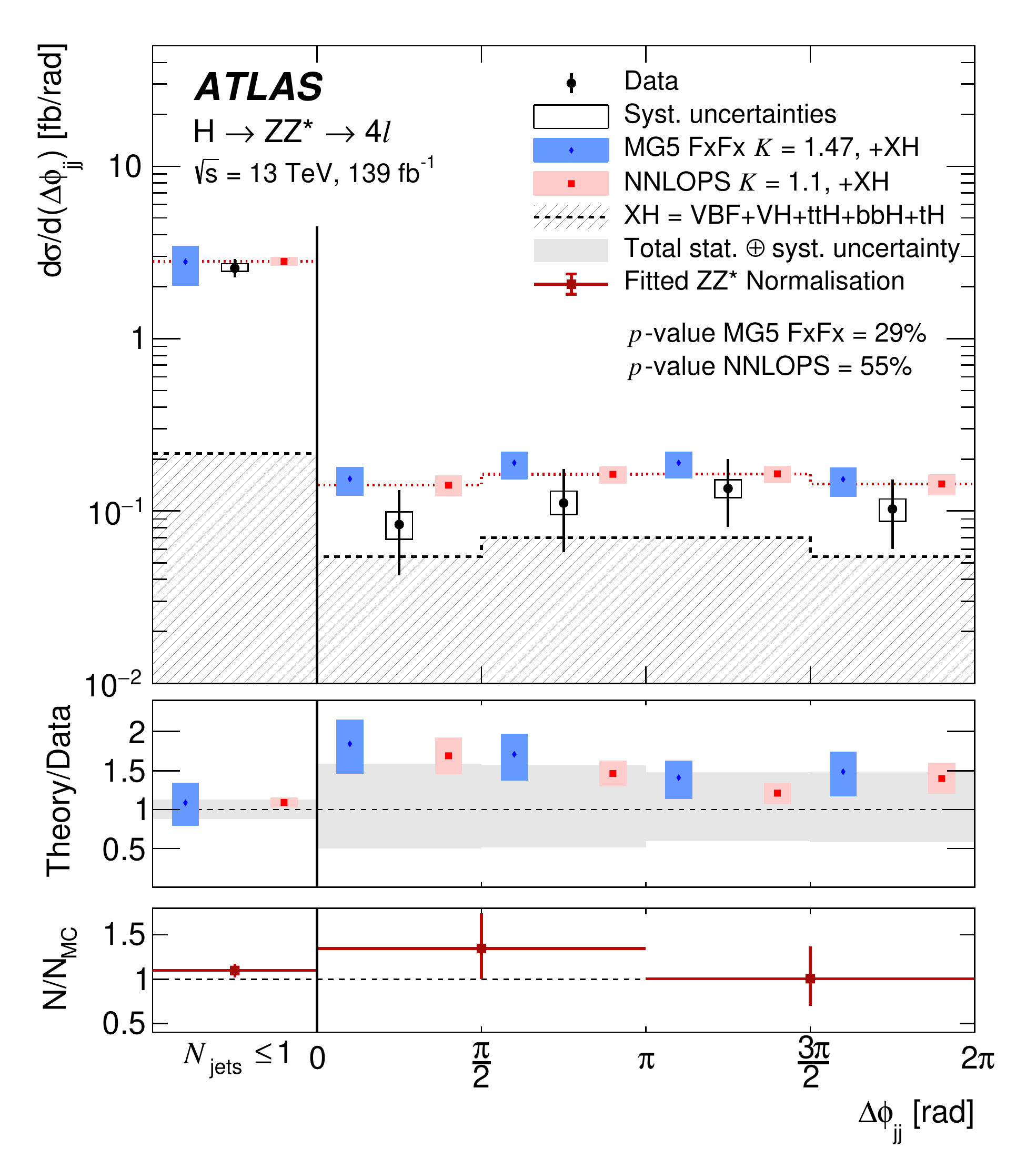}\label{fig:summary_phijj}}
\subfloat[]{\includegraphics[width=0.48\linewidth]{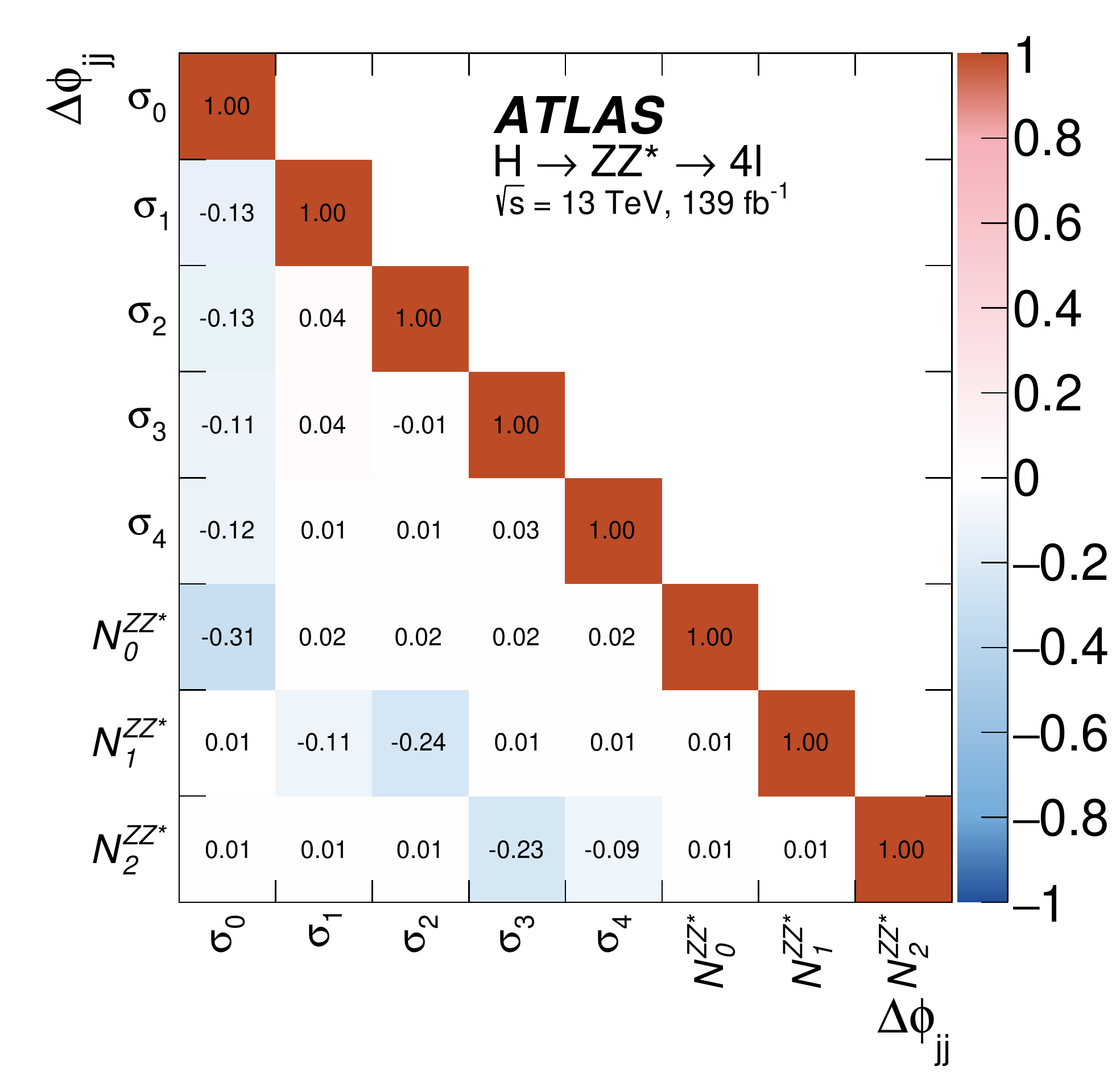}\label{fig:corrMatrix_phijj}}
\caption{Differential fiducial cross sections for~\protect\subref{fig:summary_etajj} the distance between these two jets in pseudorapidity, \detajj, and~\protect\subref{fig:summary_phijj} the distance between the two jets in $\phi$, \dphijj. The first bin contains events with fewer than two jets that pass the jet selection requirements. Finally, the corresponding correlation matrices between the measured cross sections and the $ZZ^{\ast}$ background normalisation factors are provided (\protect\subref{fig:corrMatrix_etajj} and~\protect\subref{fig:corrMatrix_phijj}). 
\label{fig:etajj_phijj_unfolded} }
\end{figure}

Differential production cross-section measurements with respect to variables that probe the jet activity in reconstructed Higgs boson events follow in Figures~\ref{fig:njet_unfolded}--\ref{fig:etajj_phijj_unfolded}. These include the exclusive and inclusive jet multiplicities, \njets, the $b$-jet multiplicity, \nbjets, variables measuring the transverse momentum of the jets, \ljpt\ and \sjpt, as well as variables that probe the kinematics of pairs of jets in events with at least two jets, \mjj, \detajj, and \dphijj.
 
\begin{figure}[!htbp]
\centering
\subfloat[]{\includegraphics[width=0.45\linewidth]{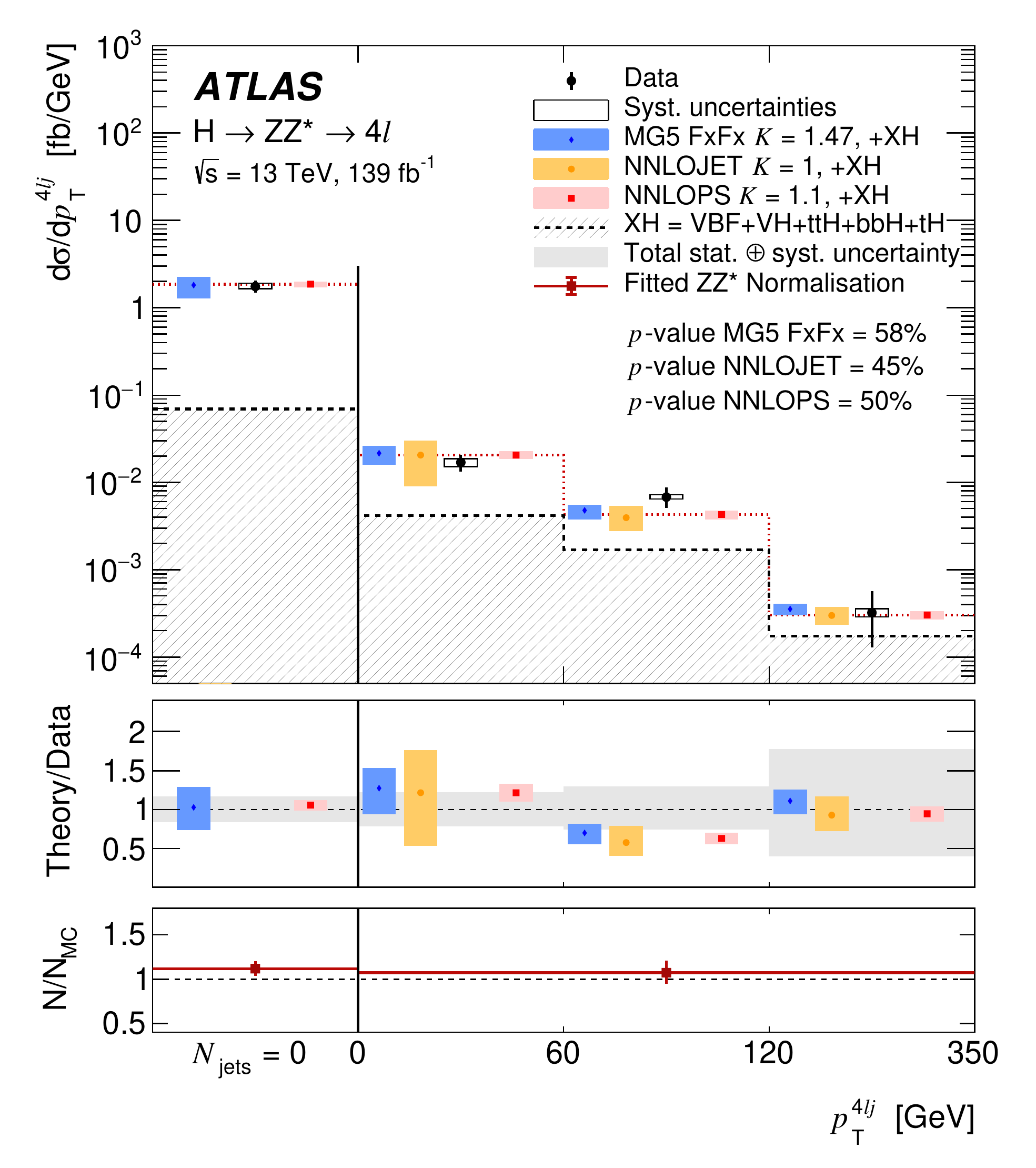}\label{fig:summary_pt4lj}}
\subfloat[]{\includegraphics[width=0.45\linewidth]{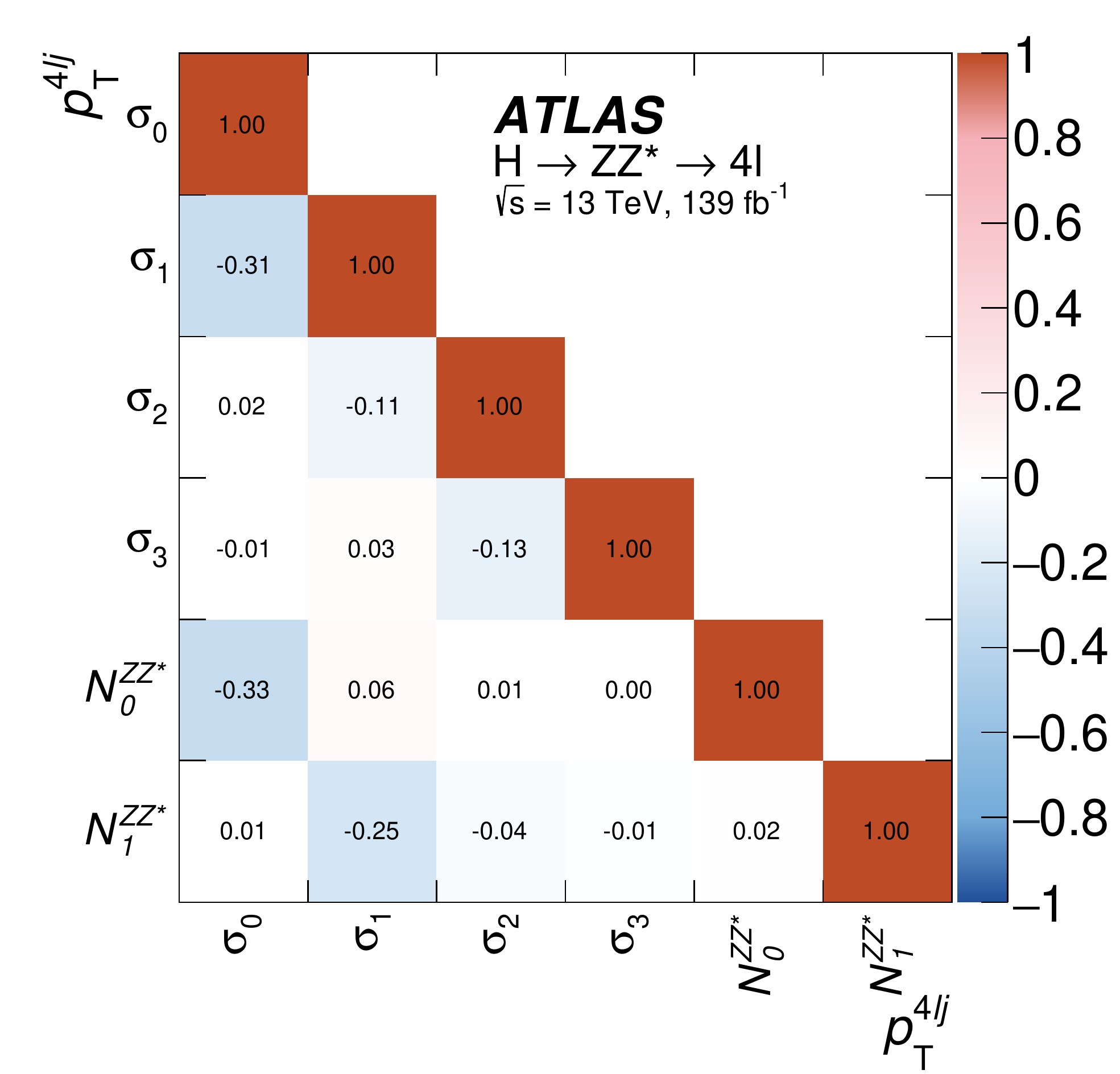}\label{fig:corrMatrix_pt4lj}}\\
\subfloat[]{\includegraphics[width=0.45\linewidth]{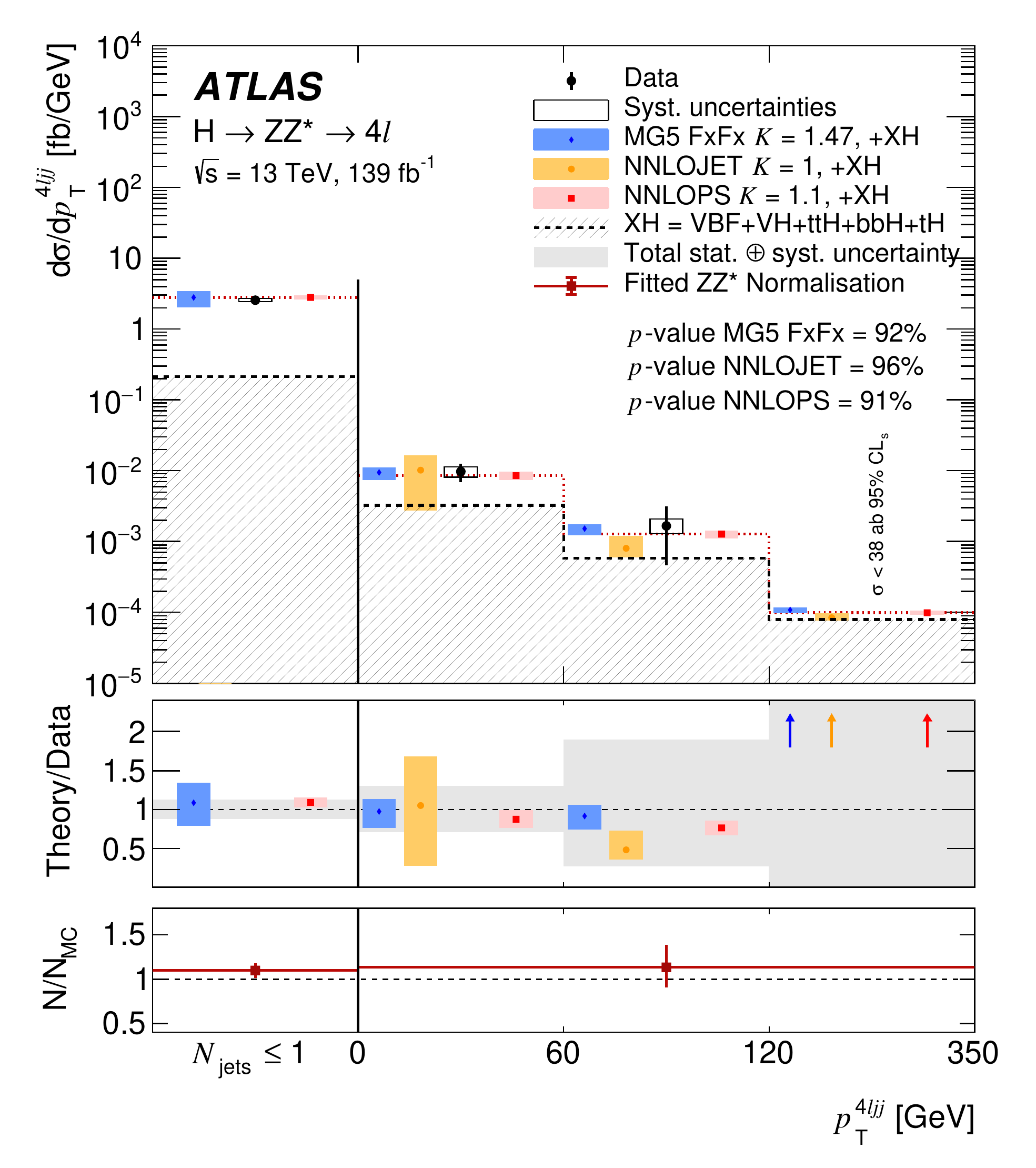}\label{fig:summary_pt4ljj}}
\subfloat[]{\includegraphics[width=0.45\linewidth]{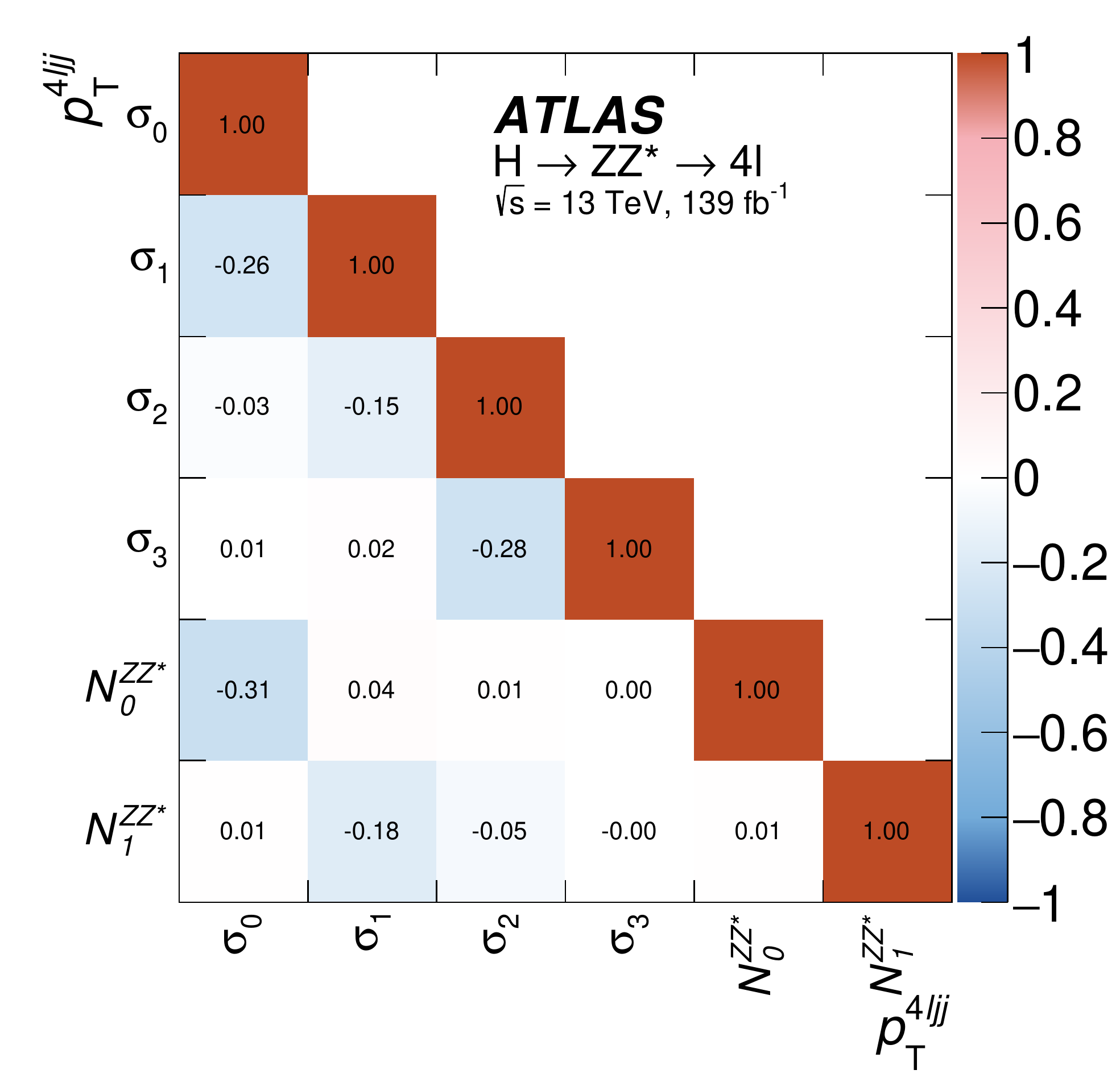}\label{fig:corrMatrix_pt4ljj}}
\caption{Differential fiducial cross sections for \protect\subref{fig:summary_pt4lj} the transverse momentum of the four-lepton plus jet system, in events with at least one jet, and \protect\subref{fig:summary_pt4ljj}  the transverse momentum of the four-lepton plus dijet system, in events with at least two jets. The corresponding correlation matrices between the measured cross sections and the $ZZ^{\ast}$ background normalisation factors are also shown (\protect\subref{fig:corrMatrix_pt4lj}  and \protect\subref{fig:corrMatrix_pt4ljj}). 
\label{fig:pt4lj_pt4ljj_unfolded} }
\end{figure}

\begin{figure}[!htbp]
\centering
\subfloat[]{\includegraphics[width=0.45\linewidth]{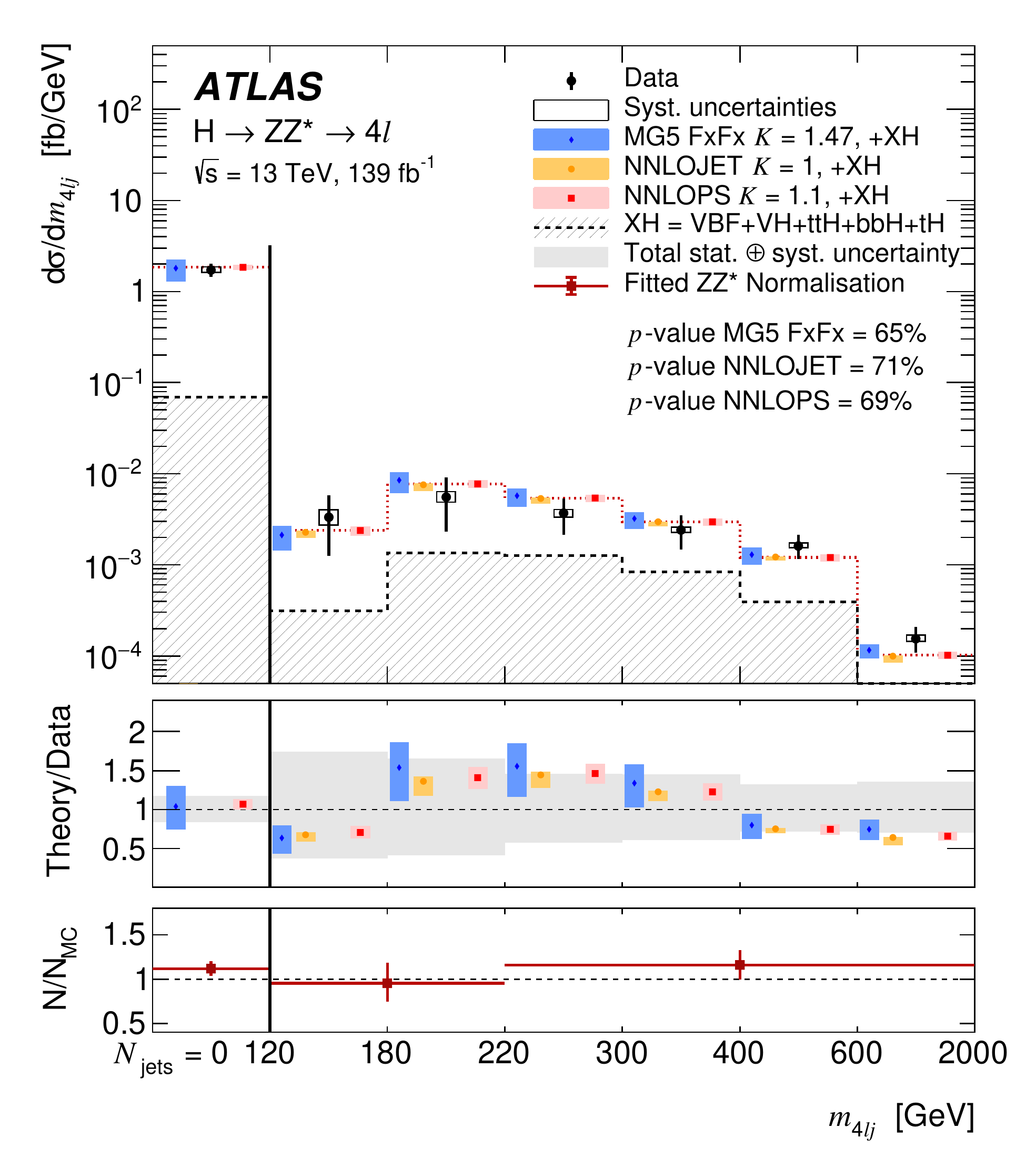}\label{fig:summary_m4lj}}
\subfloat[]{\includegraphics[width=0.48\linewidth]{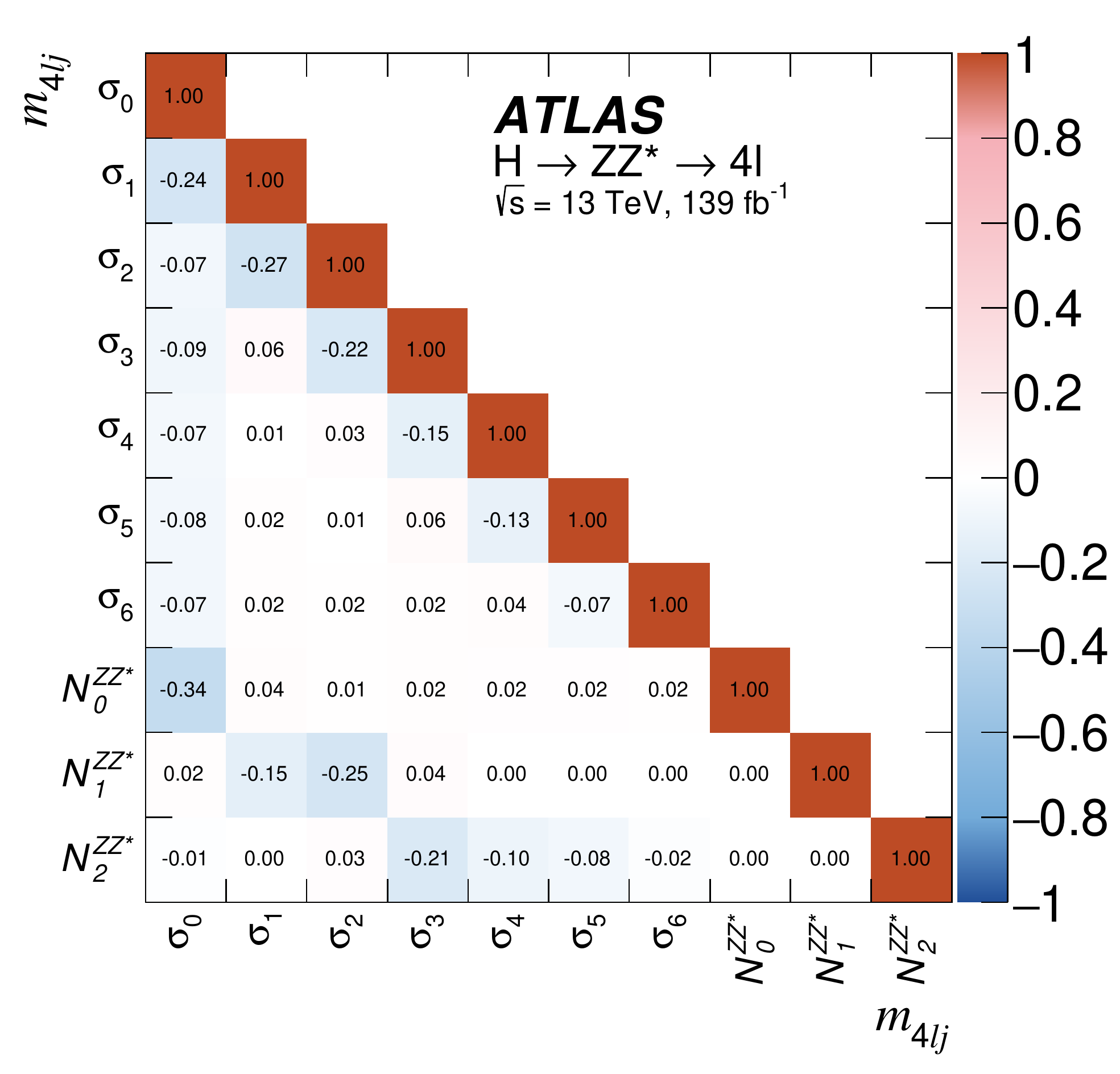}\label{fig:corrMatrix_m4lj}}\\
\subfloat[]{\includegraphics[width=0.45\linewidth]{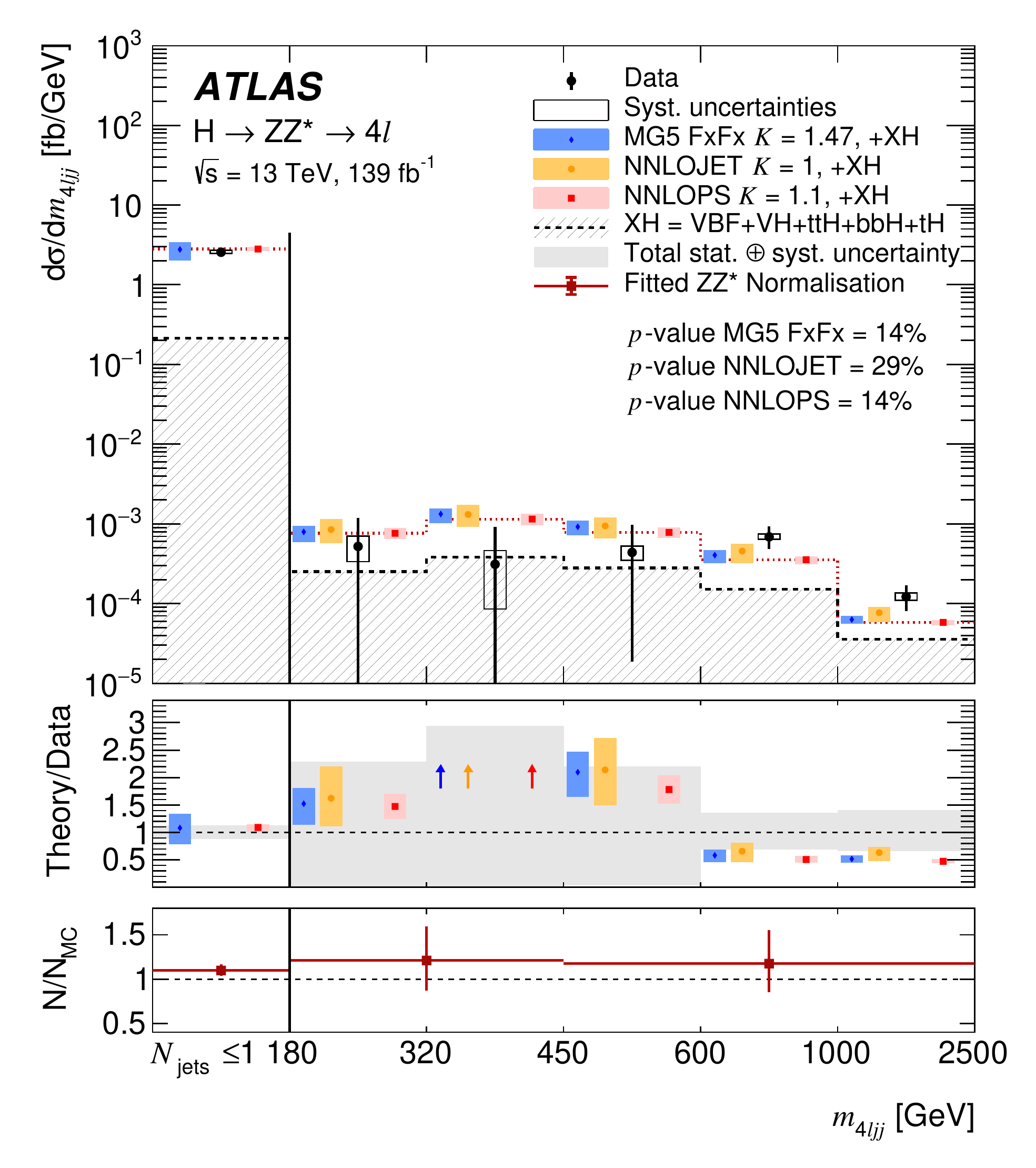}\label{fig:summary_m4ljj}}
\subfloat[]{\includegraphics[width=0.48\linewidth]{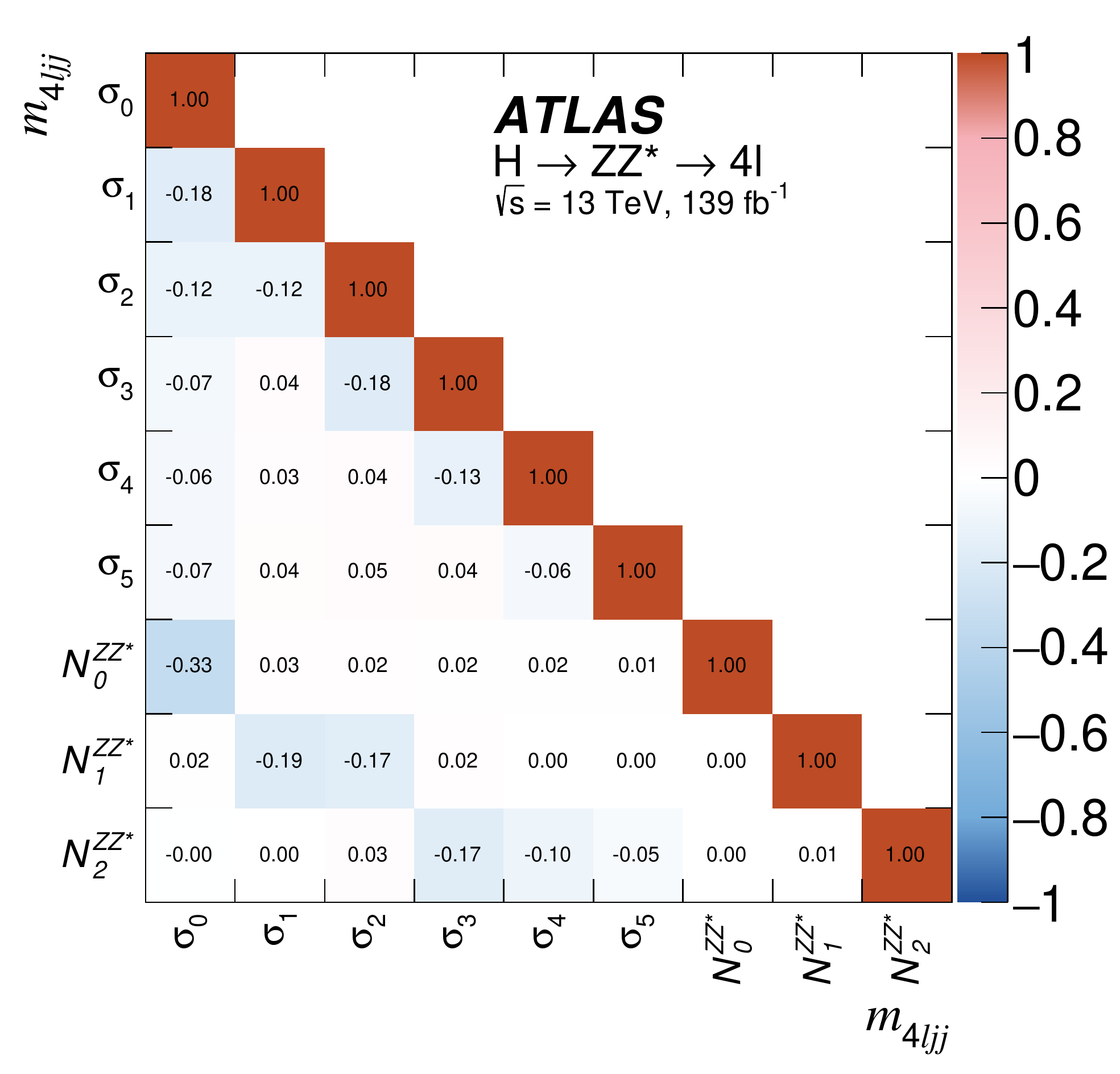}\label{fig:corrMatrix_m4ljj}}
\caption{Differential fiducial cross sections for \protect\subref{fig:summary_m4lj} the invariant mass of the four-lepton plus jet system, in events with at least one jet, and \protect\subref{fig:summary_m4ljj} the invariant mass of the four-lepton plus dijet system, in events with at least two jets. The corresponding correlation matrices between the measured cross sections and the $ZZ^{\ast}$ background normalisation factors are also shown (\protect\subref{fig:corrMatrix_m4lj} and \protect\subref{fig:corrMatrix_m4ljj}). 
\label{fig:m4lj_m4ljj_unfolded} }
\end{figure}

In addition,  differential cross-section measurements are provided for observables aimed at studying the relationship between the reconstructed Higgs boson and accompanying jets. These are presented in Figures~\ref{fig:pt4lj_pt4ljj_unfolded}--\ref{fig:m4lj_m4ljj_unfolded}.
 
\begin{figure}[!htbp]
\centering
\subfloat[]{\includegraphics[width=0.45\linewidth]{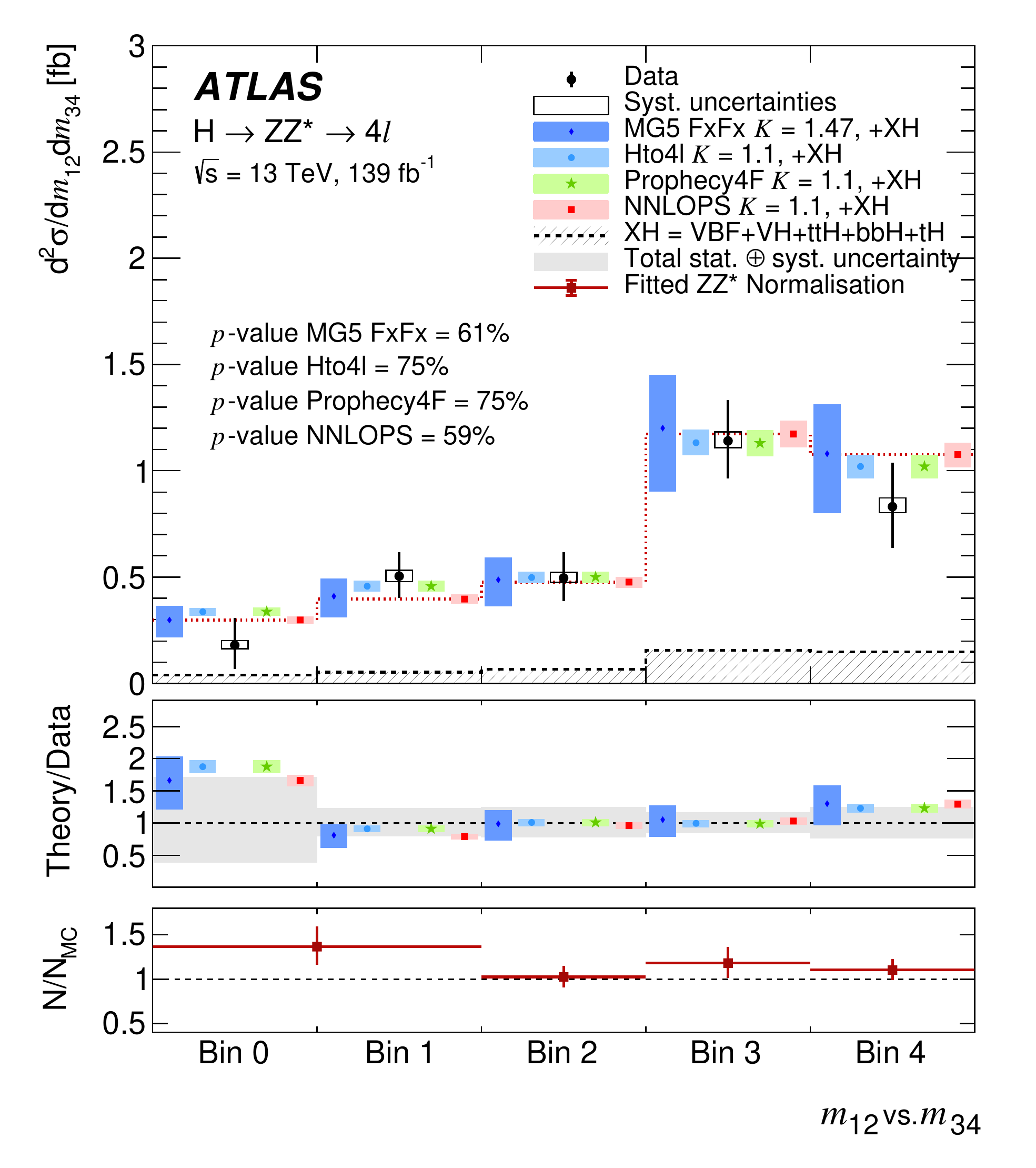}\label{fig:summary_m12m34}}
\subfloat[]{\includegraphics[width=0.48\linewidth]{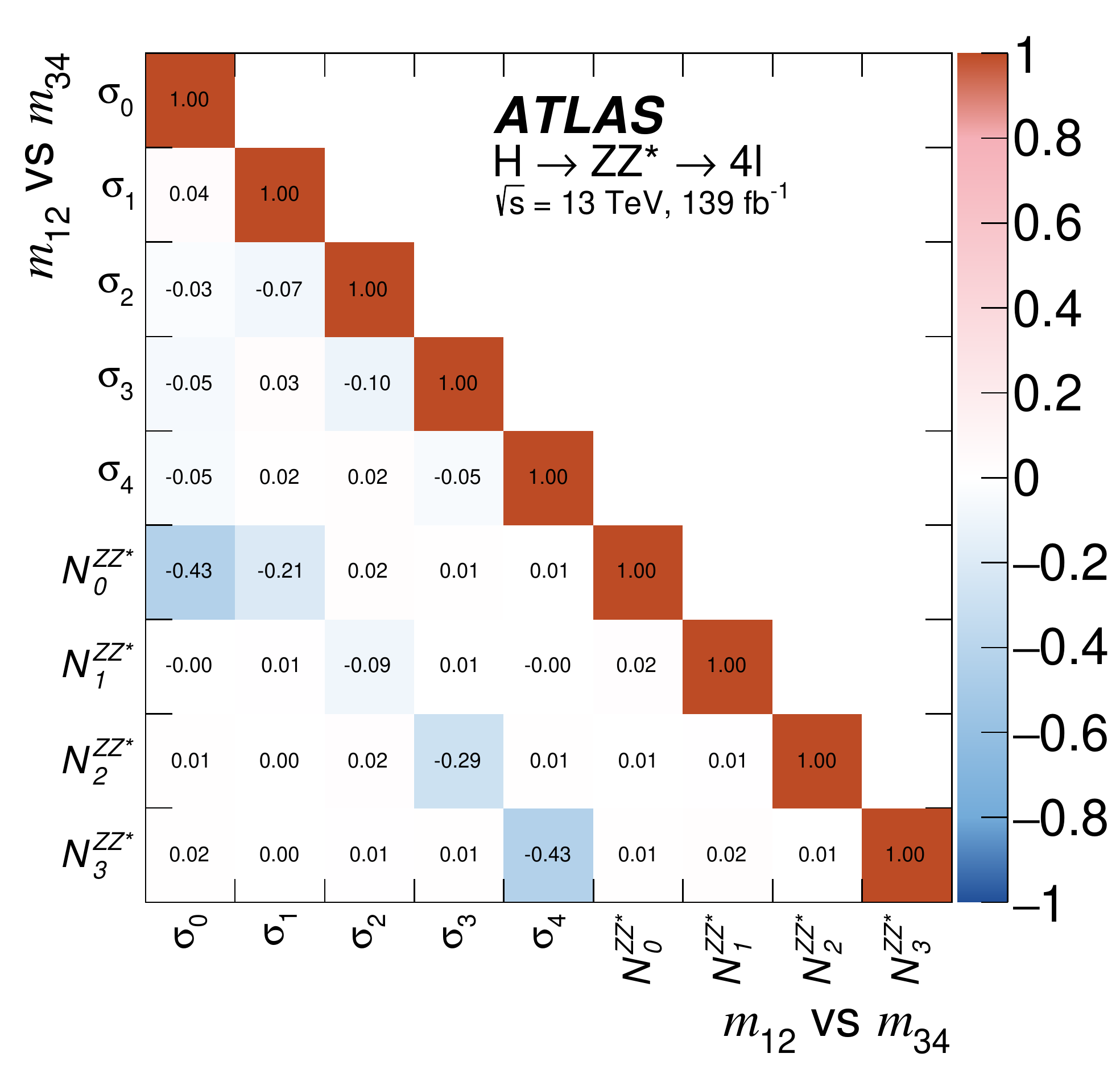}\label{fig:corrMatrix_m12m34}}\\
\caption{\protect\subref{fig:summary_m12m34} Differential fiducial cross section for the leading vs.\ subleading $Z$ boson mass, \monetwo\ vs.\ \mthreefour, and \protect\subref{fig:corrMatrix_m12m34} the corresponding correlation matrix between the measured cross sections and the $ZZ^{\ast}$ background normalisation factors. The bin boundaries are defined in Figure~\ref{fig:categoryVars7}.
\label{fig:m12m34_unfolded} }
\end{figure}

\begin{figure}[!htbp]
\centering
\subfloat[]{\includegraphics[width=0.45\linewidth]{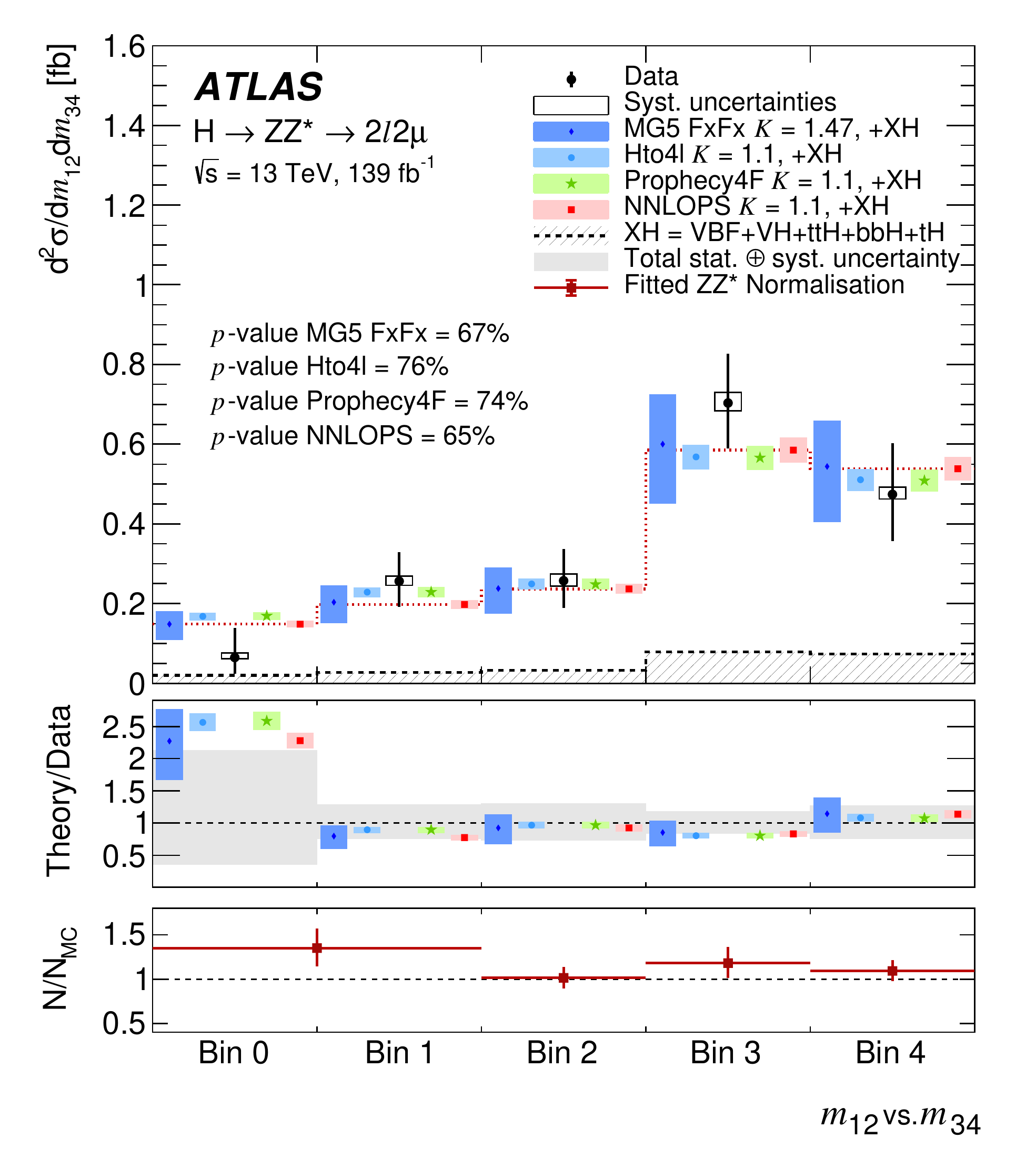}\label{fig:summary_m12m34_2l2mu}}
\subfloat[]{\includegraphics[width=0.45\linewidth]{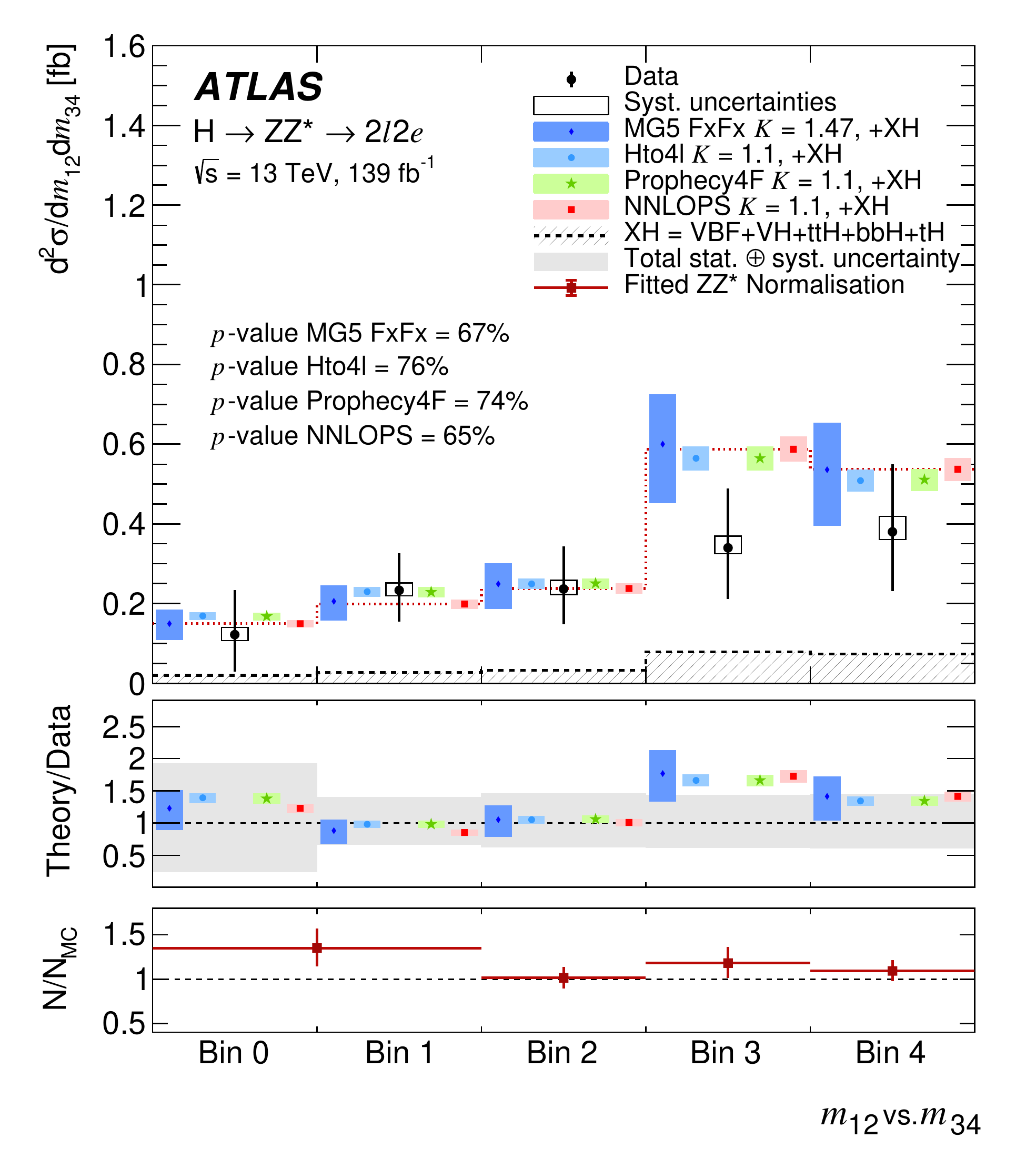}\label{fig:summary_m12m34_2l2e}} \\
\subfloat[]{\includegraphics[width=0.48\linewidth]{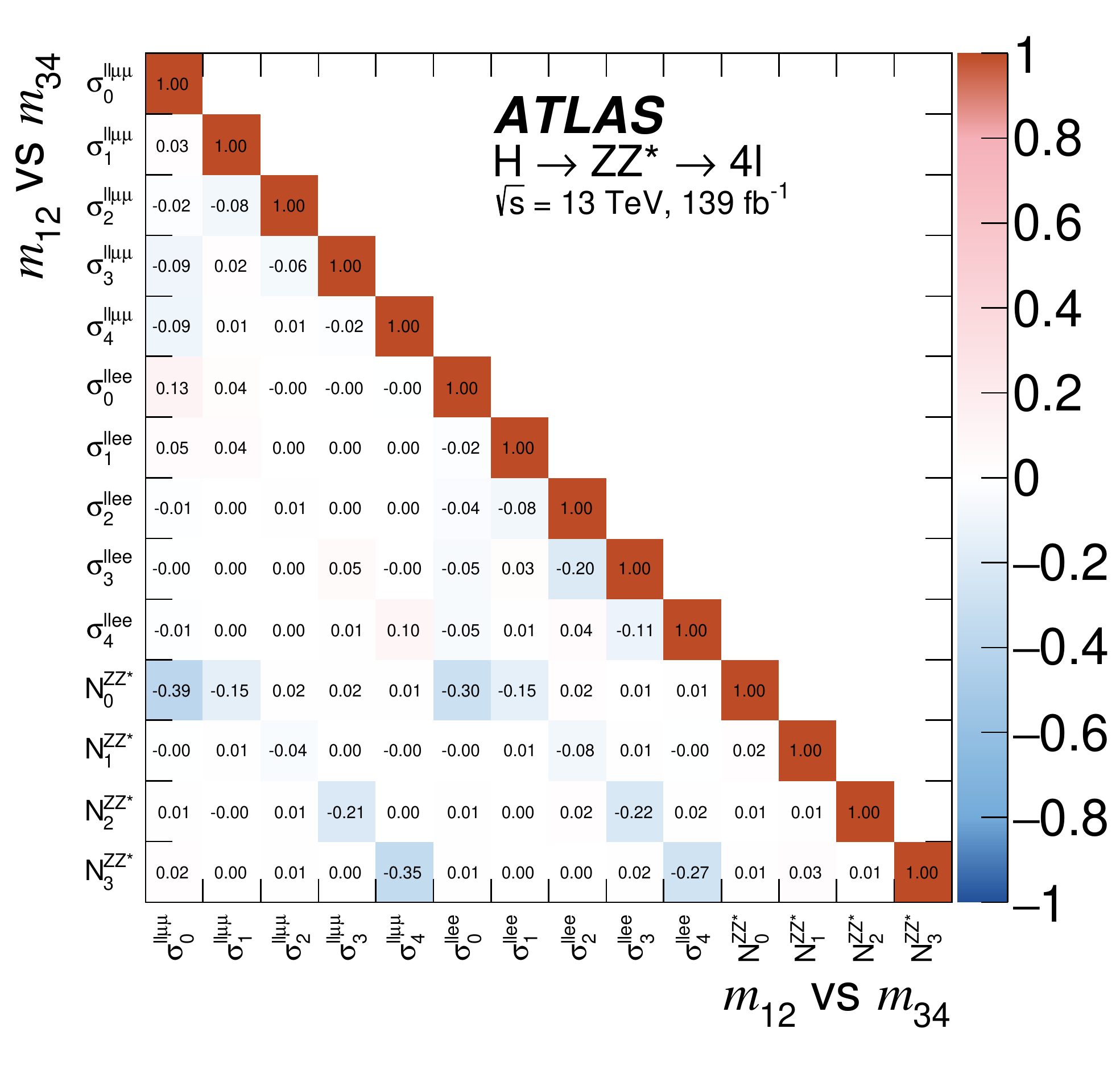}\label{fig:corrMatrix_m12m34_2l2e}}
\caption{Differential fiducial cross sections for the leading vs.\ subleading $Z$ boson mass, $m_{12}$ vs.\ $m_{34}$, in \protect\subref{fig:summary_m12m34_2l2mu} $\ell\ell\mu\mu$ and \protect\subref{fig:summary_m12m34_2l2e} $\ell\ell ee$  final states, along with \protect\subref{fig:corrMatrix_m12m34_2l2e} their corresponding correlation matrix between the measured cross sections and the $ZZ^{\ast}$ background normalisation factors. The bin boundaries are defined in Figure~\ref{fig:categoryVars7}.
\label{fig:m12m34_2l2l_unfolded} }
\end{figure}

\begin{figure}[!htbp]
\centering
\subfloat[]{\includegraphics[width=0.45\linewidth]{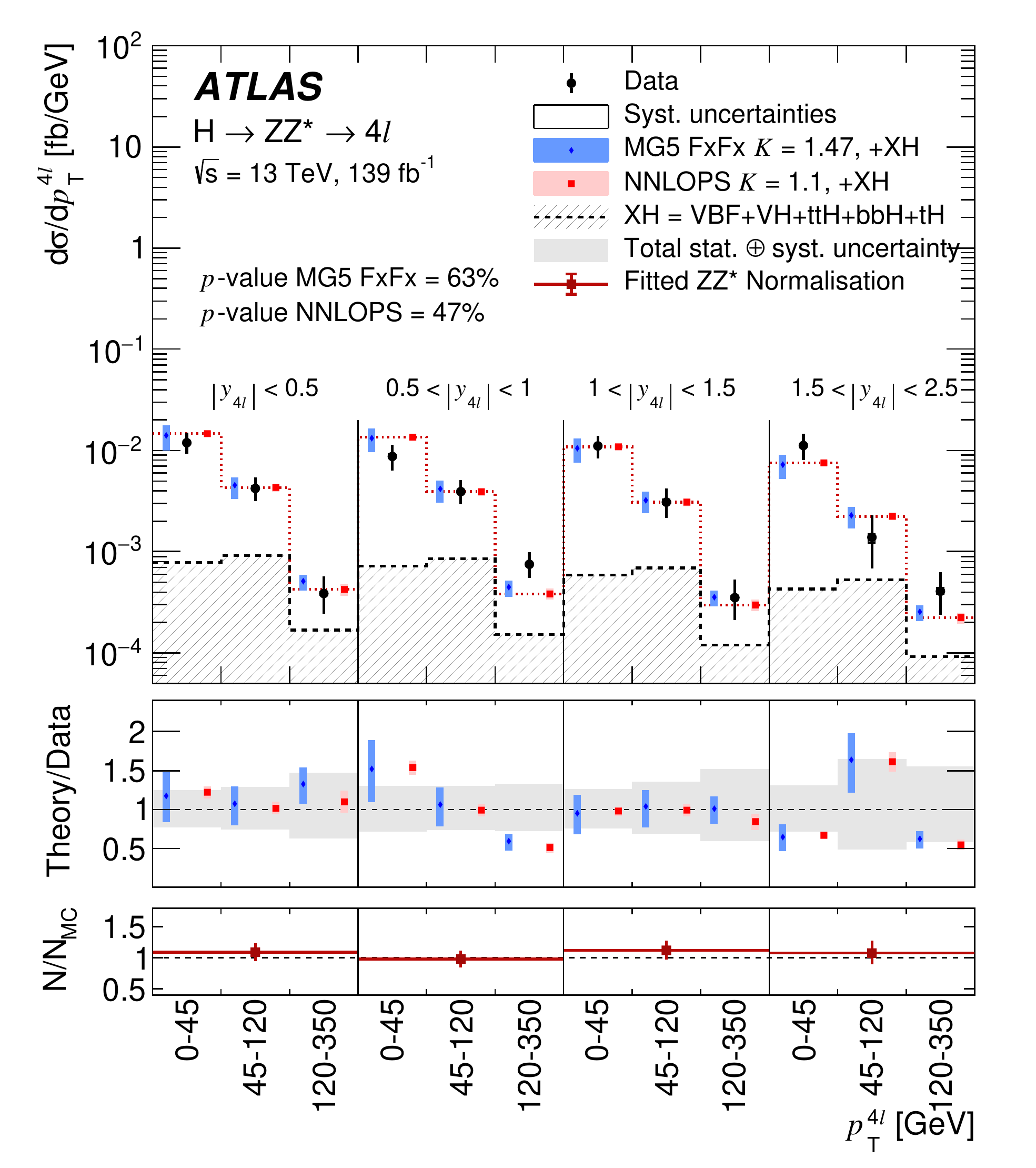}\label{fig:summary_pty}}
\subfloat[]{\includegraphics[width=0.45\linewidth]{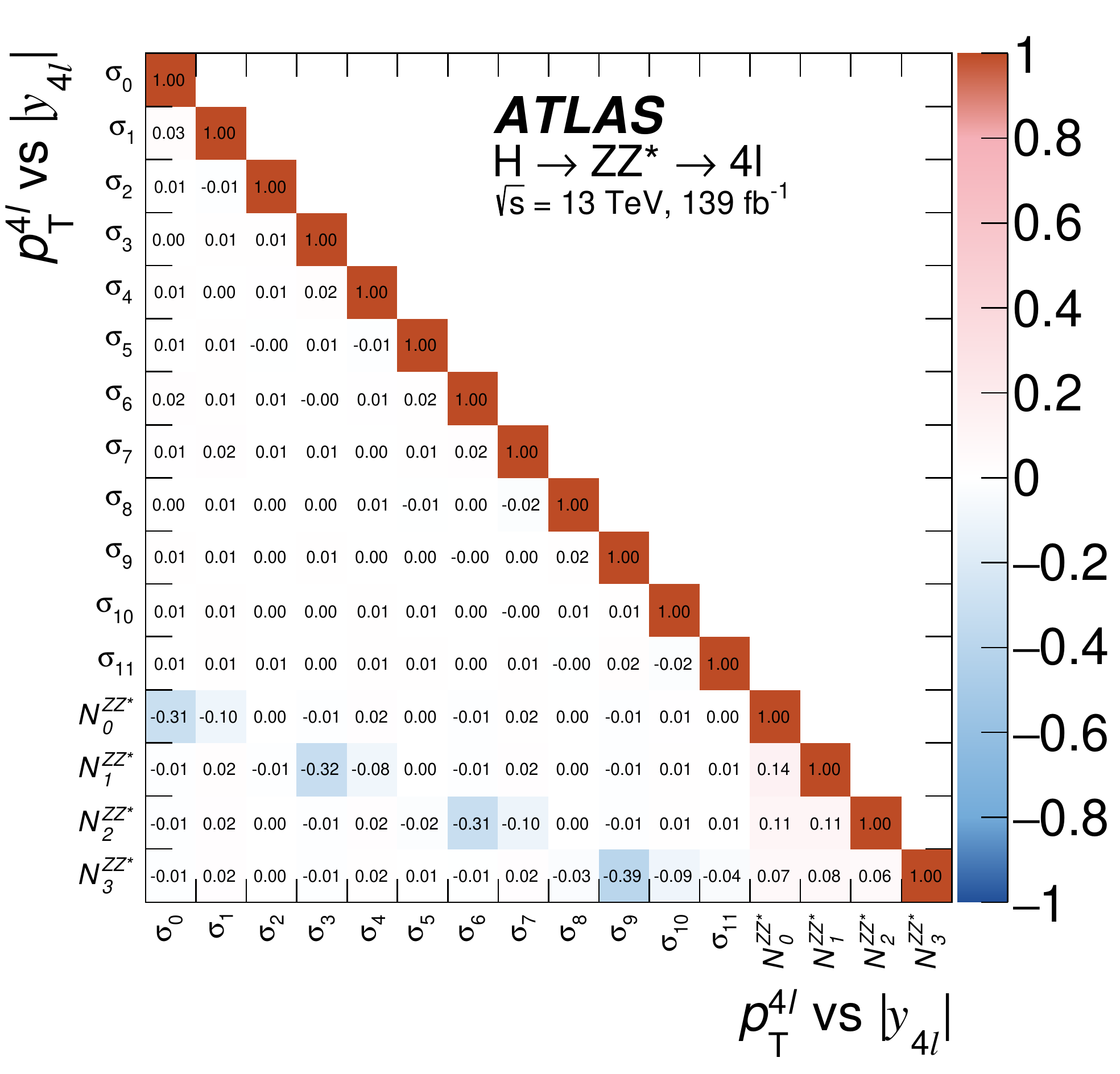}\label{fig:corrMatrix_pty}}\\
\caption{\protect\subref{fig:summary_pty} Double differential fiducial cross sections of the \ptH~distribution in \yH~bins. The corresponding correlation matrix between the measured cross sections and the $ZZ^{\ast}$ background normalisation factors is shown in \protect\subref{fig:corrMatrix_pty}. The $p$-values shown are calculated for all bins across both \ptH\ and \yH\ simultaneously.
\label{fig:pty_unfolded} }
\end{figure}

\begin{figure}[!htbp]
\centering
\subfloat[]{\includegraphics[width=0.45\linewidth]{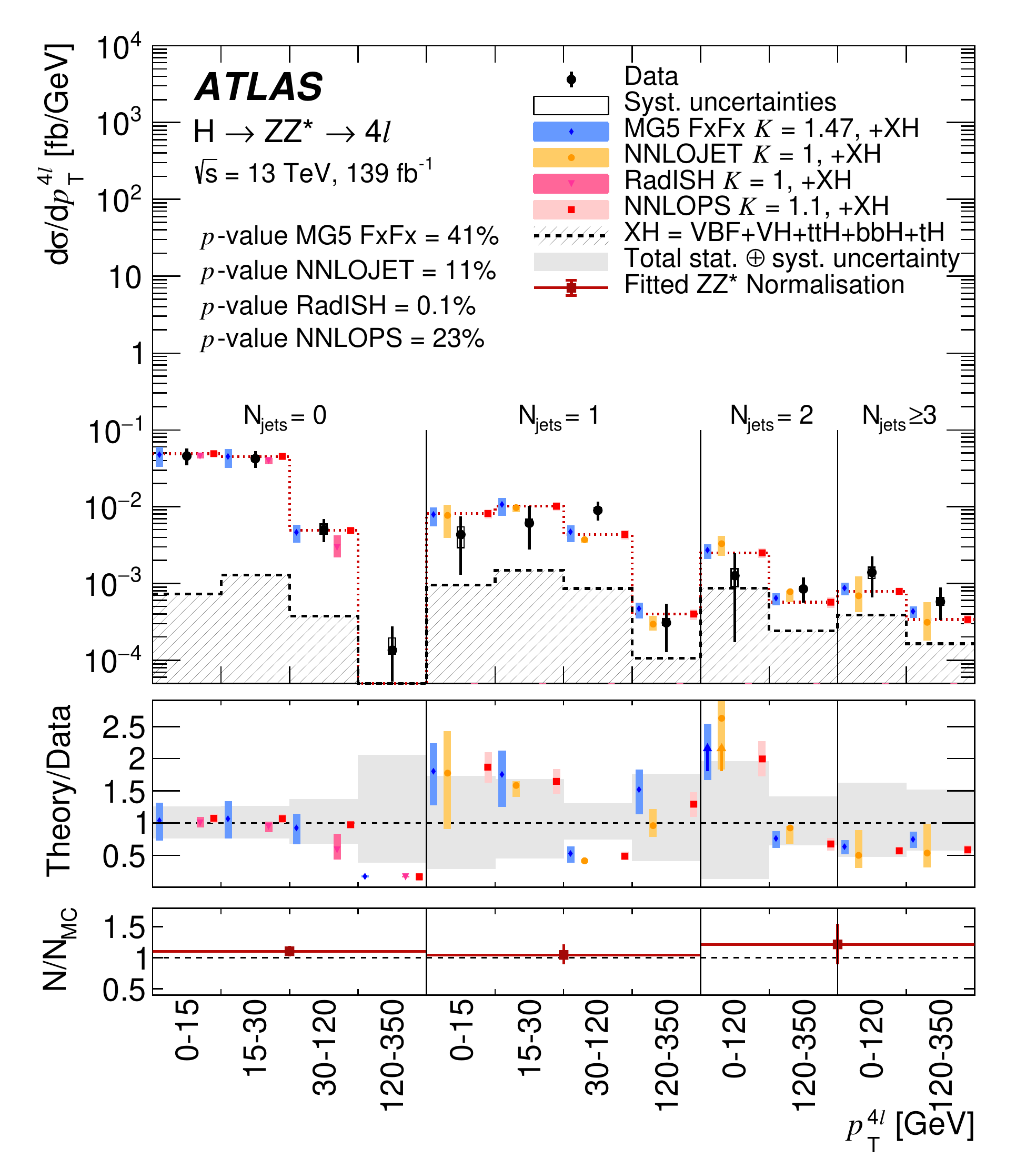}\label{fig:summary_ptnj}}
\subfloat[]{\includegraphics[width=0.45\linewidth]{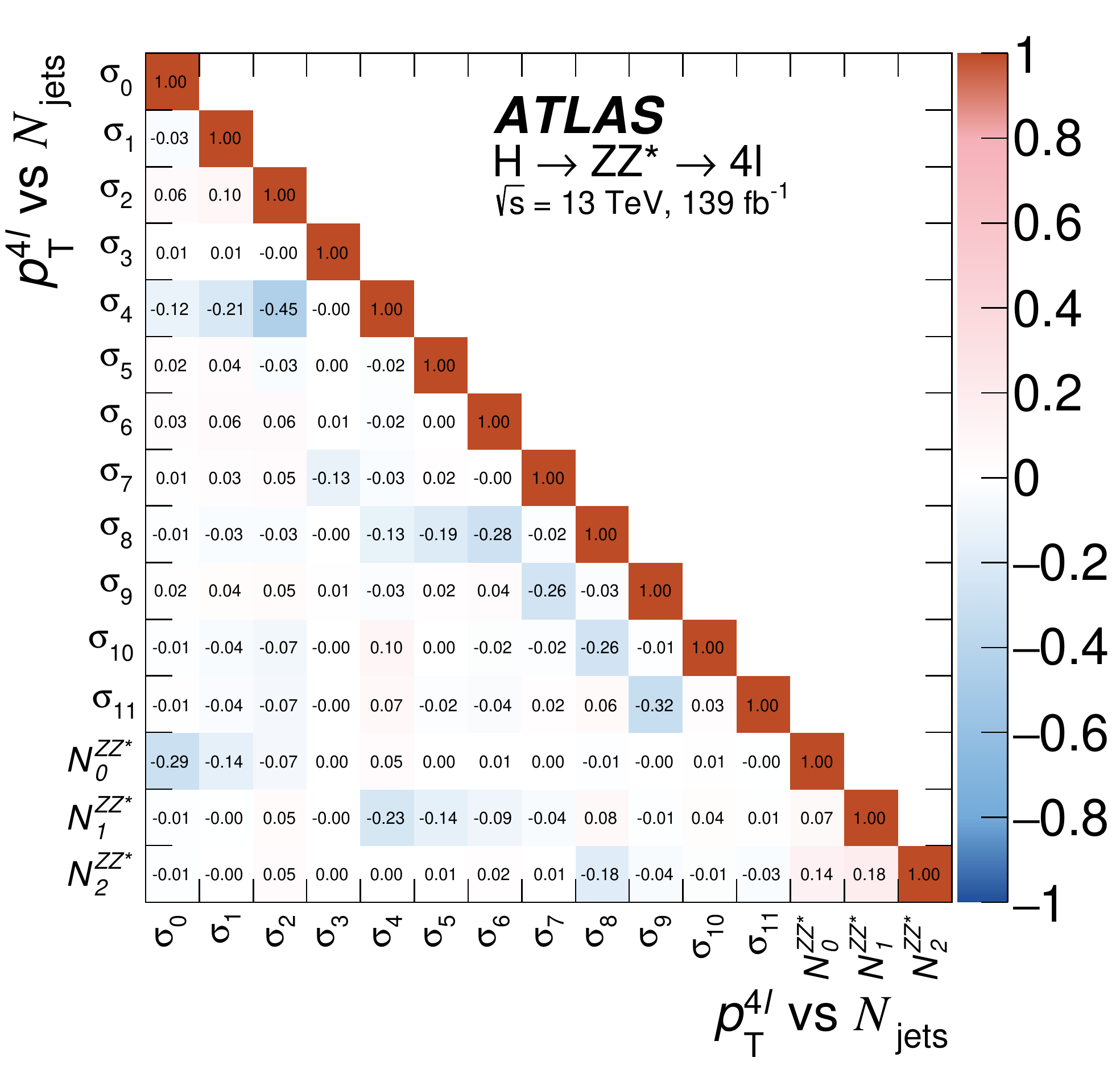}\label{fig:corrMatrix_ptnjet}}\\
\caption{\protect\subref{fig:summary_ptnj} Double differential fiducial cross sections of the \ptH\ distribution in $N_{\mathrm{jets}}$ bins. The corresponding correlation matrix between the measured cross sections and the $ZZ^{\ast}$ background normalisation factors is shown in~\protect\subref{fig:corrMatrix_ptnjet}. The $p$-values shown are calculated for all bins across both \ptH\ and \njets\ simultaneously.
\label{fig:ptnjet_unfolded} }
\end{figure}

\begin{figure}[!htbp]
\centering
\subfloat[]{\includegraphics[width=0.45\linewidth]{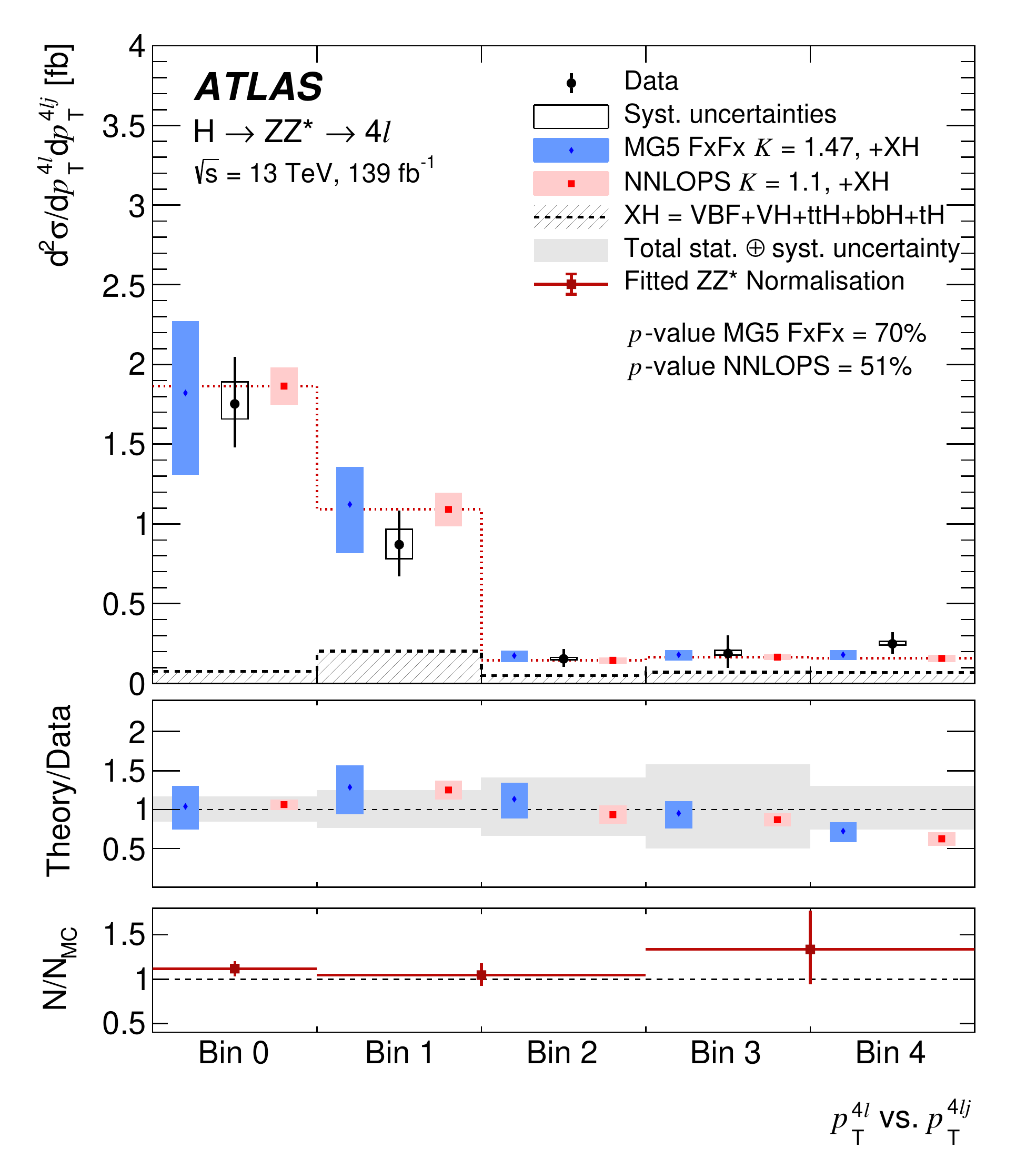}\label{fig:summary_pt4ljpt4lj}}
\subfloat[]{\includegraphics[width=0.45\linewidth]{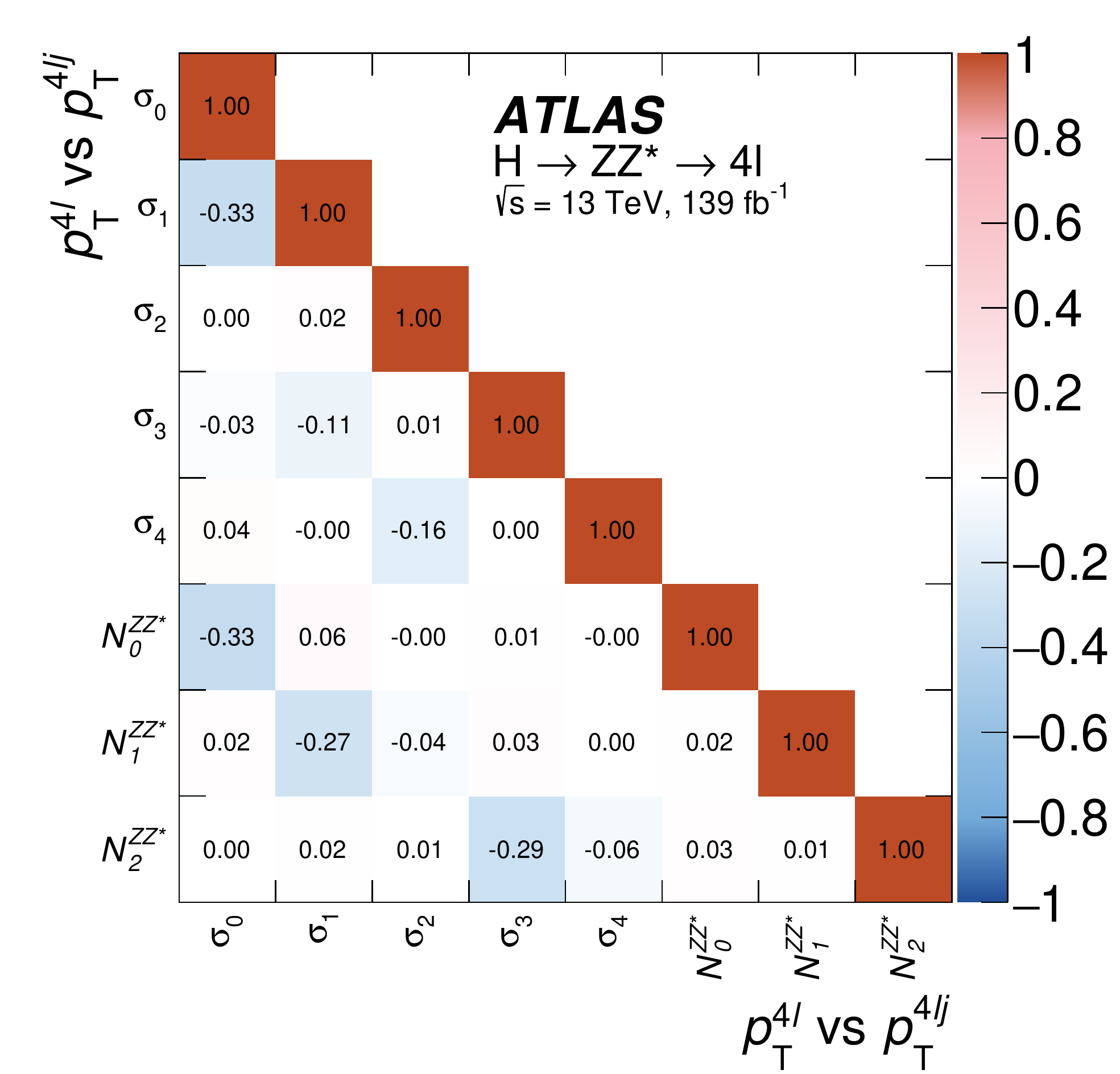}\label{fig:corrMatrix_pt4lpt4lj}}\\
\caption{\protect\subref{fig:summary_pt4ljpt4lj} Differential fiducial cross section for  the transverse momentum of the four-lepton system vs.\ the transverse momentum of the four-lepton plus jet system, \ptH\ vs.\ \ptHj\, and ~\protect\subref{fig:corrMatrix_pt4lpt4lj} the corresponding correlation matrix between the measured cross sections and the $ZZ^{\ast}$ background normalisation factors. The bin boundaries are defined in Figure~\ref{fig:categoryVars9}.
\label{fig:pt4lpt4lj_unfolded} }
\end{figure}

\begin{figure}[!htbp]
\centering
\subfloat[]{\includegraphics[width=0.45\linewidth]{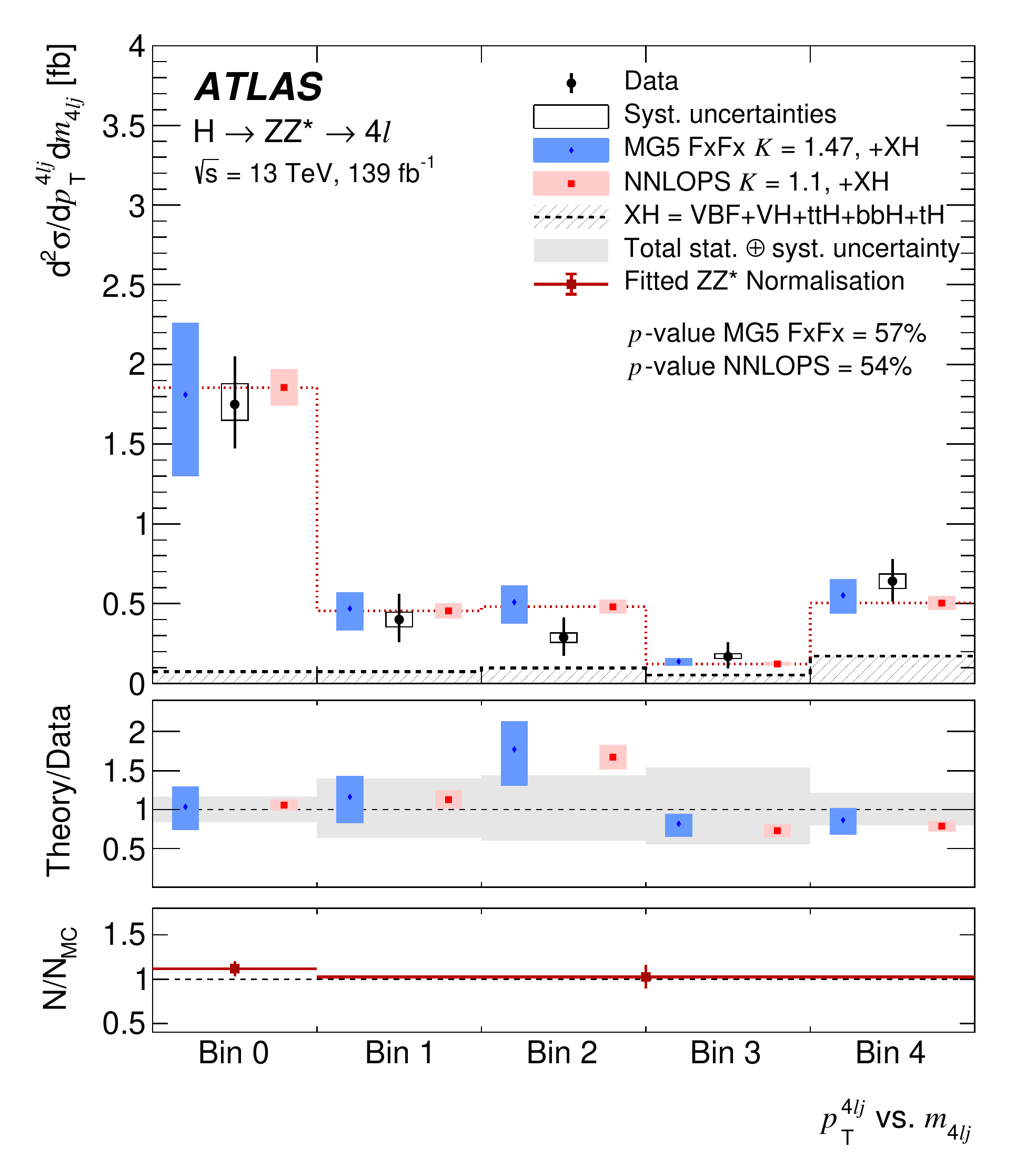}\label{fig:summary_pt4ljm4lj}}
\subfloat[]{\includegraphics[width=0.45\linewidth]{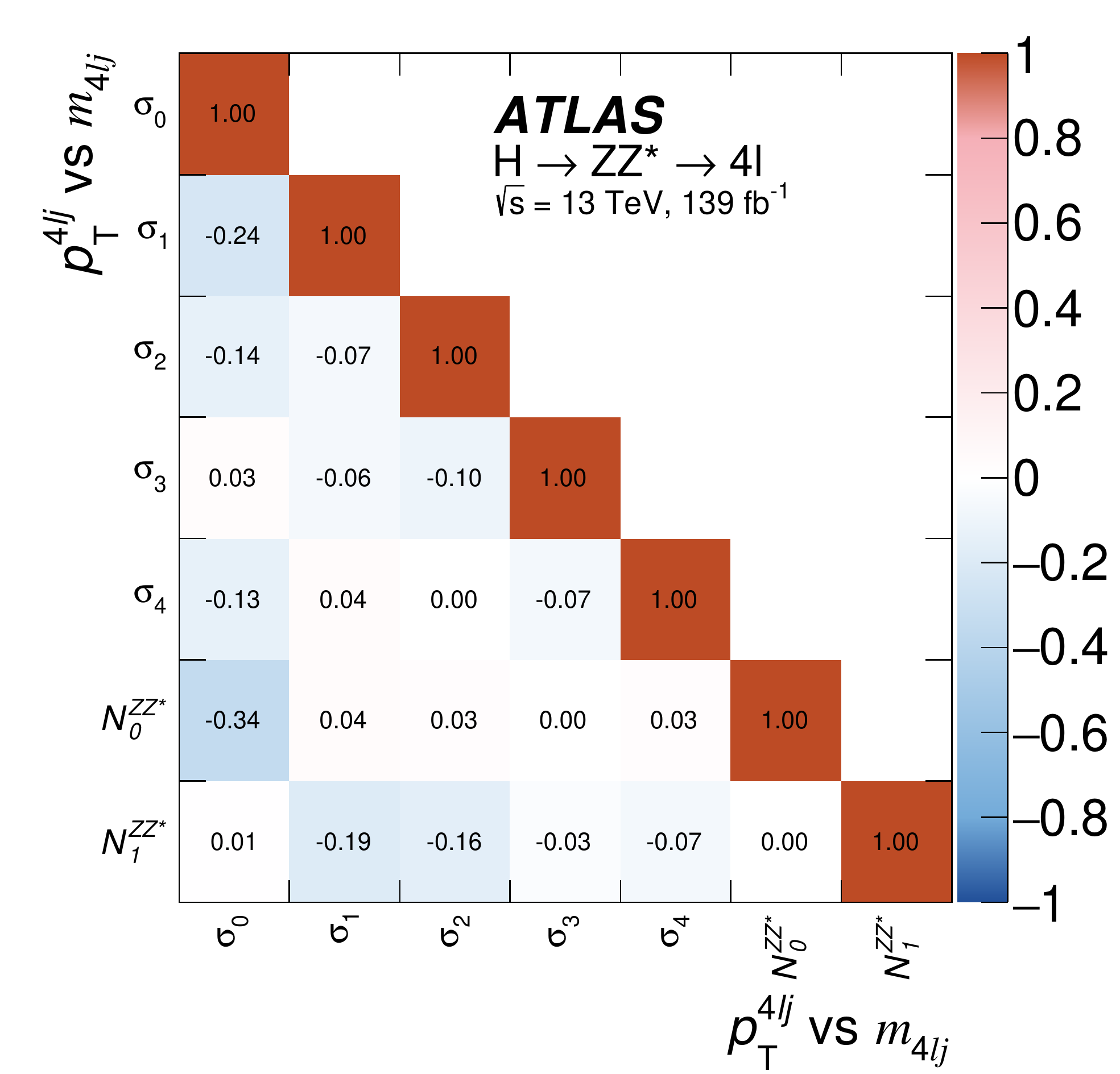}\label{fig:corrMatrix_ptjm4lj}}\\
\caption{\protect\subref{fig:summary_pt4ljm4lj} Double differential fiducial cross section for   the transverse momentum of the four-lepton plus jet system vs.\ the invariant mass of the four-lepton plus jet system, \ptHj\ vs.\ \mflj\, and~\protect\subref{fig:corrMatrix_ptjm4lj}  the corresponding correlation matrix between the measured cross sections and the $ZZ^{\ast}$ background normalisation factors. The bin boundaries are defined in Figure~\ref{fig:categoryVars8}.
\label{fig:pt4ljm4lj_unfolded} }
\end{figure}

\begin{figure}[!htbp]
\centering
\subfloat[]{\includegraphics[width=0.45\linewidth]{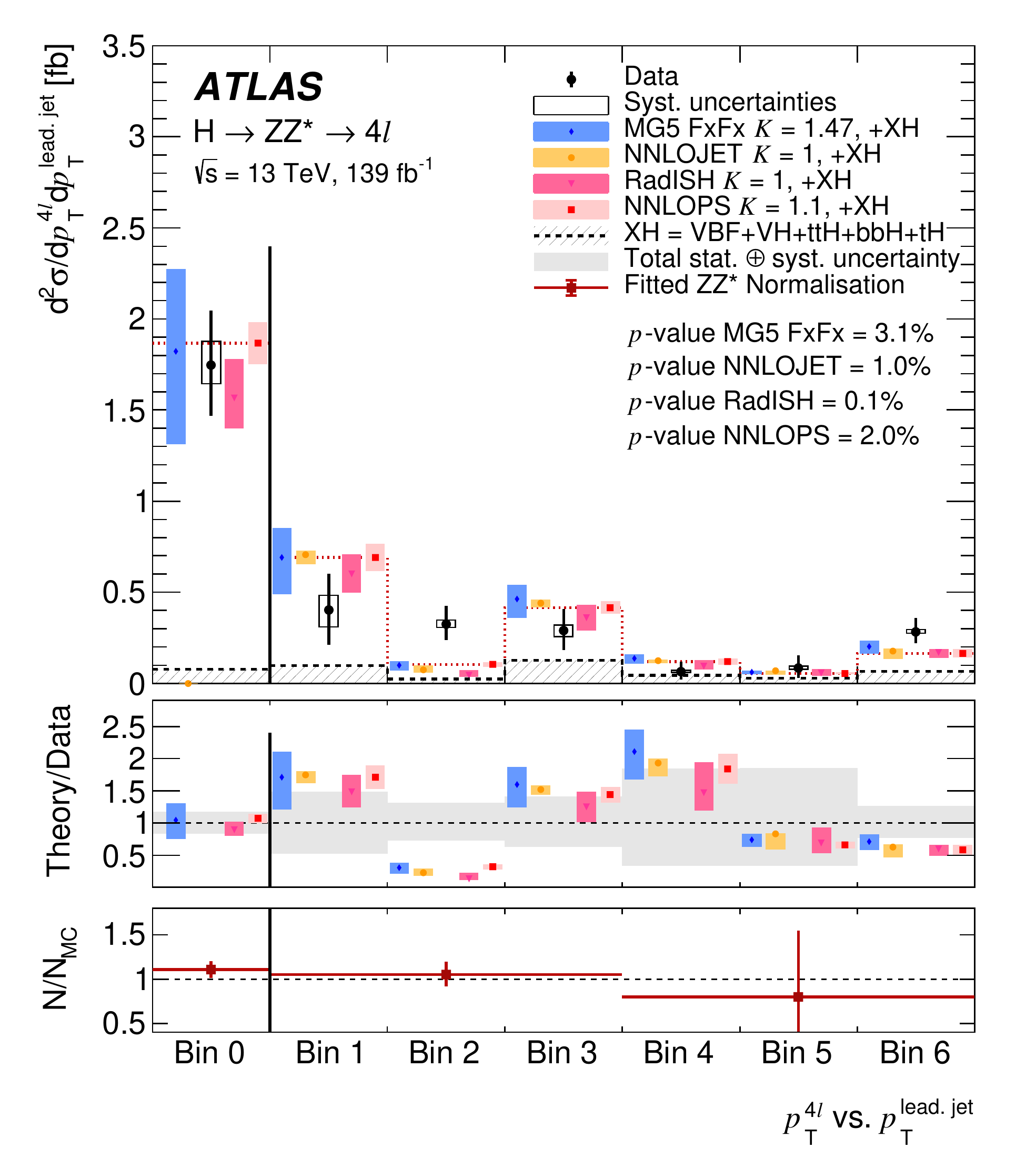}\label{fig:summary_ptljpt}}
\subfloat[]{\includegraphics[width=0.45\linewidth]{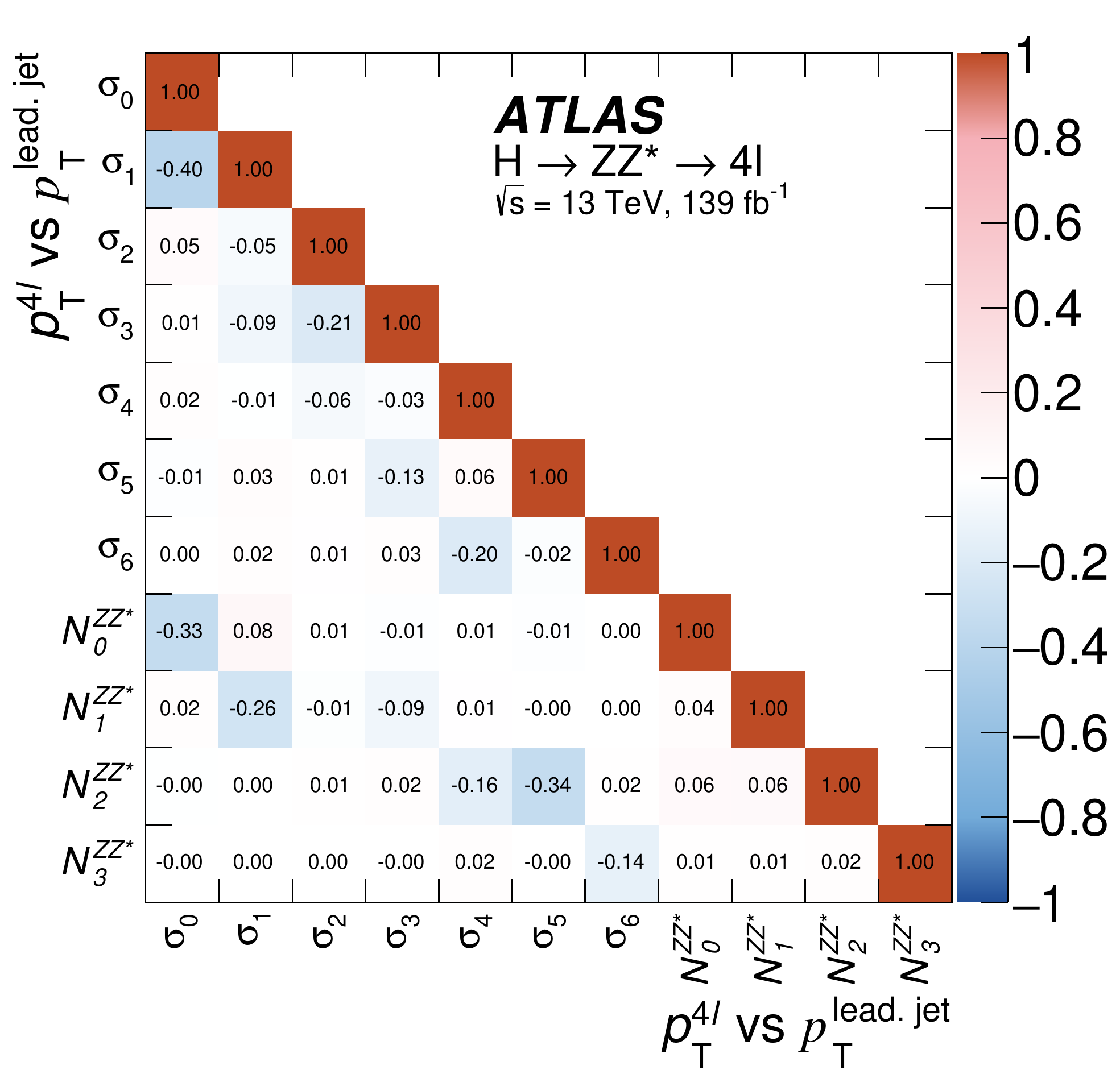}\label{fig:corrMatrix_ptljpt}}\\
\caption{\protect\subref{fig:summary_ptljpt} Double differential fiducial cross section for  the transverse momentum of the four-lepton system vs.\ the transverse momentum of the leading jet, \ptH\ vs.\ \ljpt, and~\protect\subref{fig:corrMatrix_ptljpt}   the corresponding correlation matrix between the measured cross sections and the $ZZ^{\ast}$ background normalisation factors. The bin boundaries are defined in Figure~\ref{fig:categoryVars10}.
\label{fig:ptljpt_unfolded} }
\end{figure}

\begin{figure}[!htbp]
\centering
\subfloat[]{\includegraphics[width=0.45\linewidth]{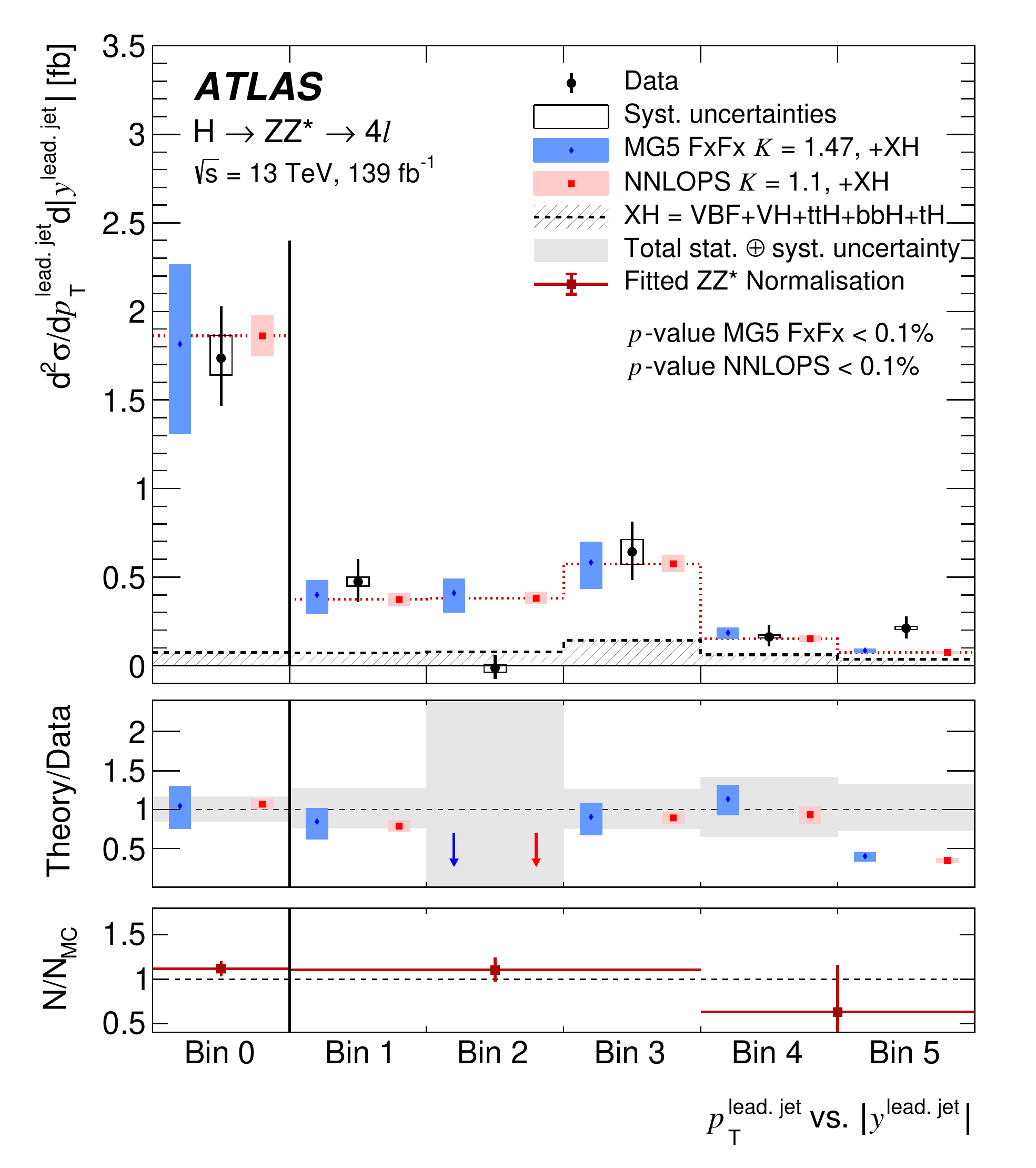}\label{fig:summary_ljptljy}}
\subfloat[]{\includegraphics[width=0.45\linewidth]{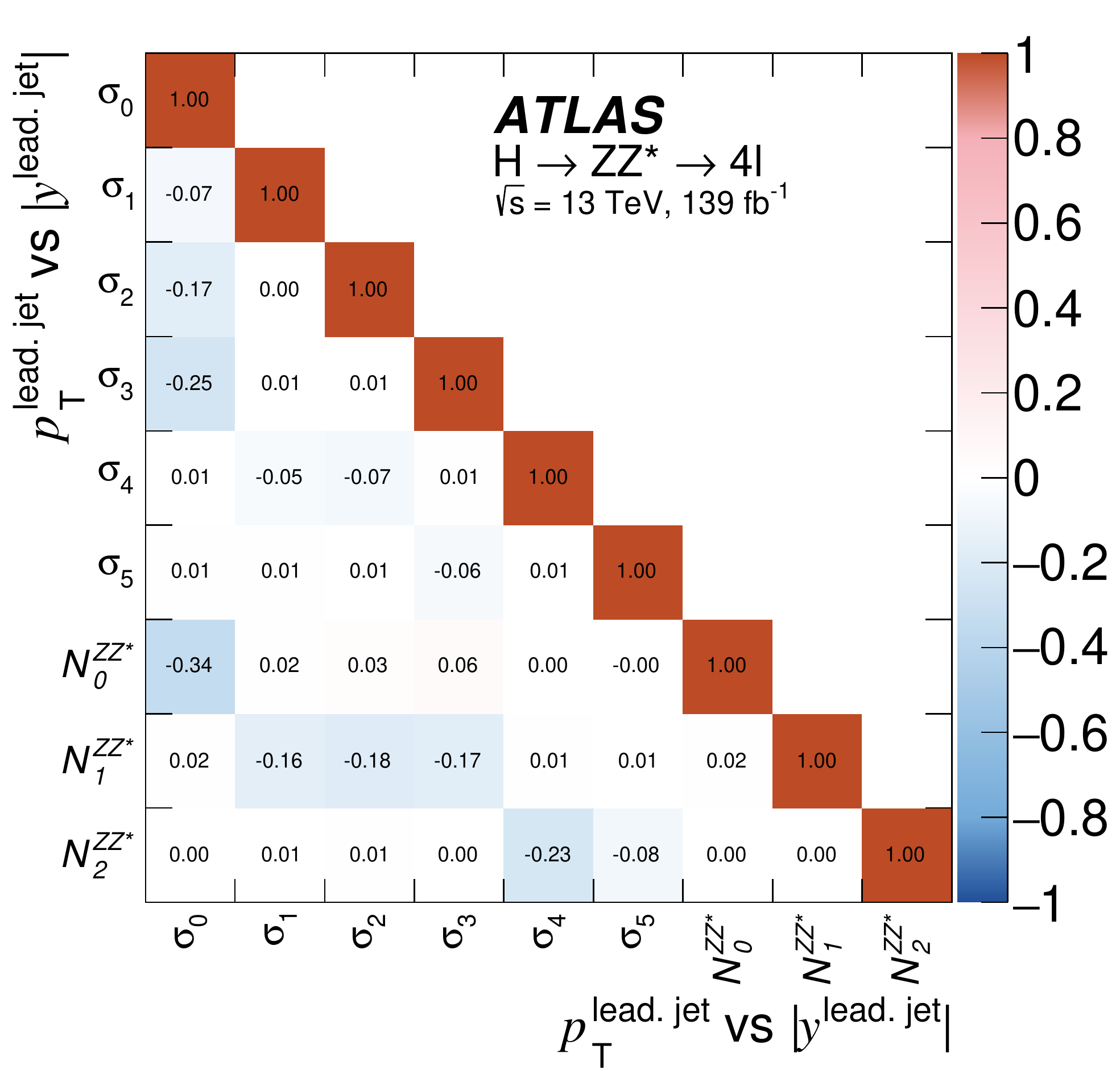}\label{fig:corrMatrix_ljptljy}}\\
\caption{\protect\subref{fig:summary_ljptljy} Double differential fiducial cross section for   the transverse momentum of the leading jet vs.\ the rapidity of the leading jet, \ljpt\ vs.\ \ljy, and~\protect\subref{fig:corrMatrix_ljptljy}  the corresponding correlation matrix between the measured cross sections and the $ZZ^{\ast}$ background normalisation factors. The bin boundaries are defined in Figure~\ref{fig:categoryVars12}.
\label{fig:ljptljy_unfolded} }
\end{figure}

Finally, the double differential measurements in bins of $m_{12}~\mathrm{vs.}~m_{34}$, $\ptH~\mathrm{vs.}~\yH$, $\ptH~\mathrm{vs.}~\njets$, $\ptH~\mathrm{vs.}~\ptHj$, $\ptHj~\mathrm{vs.}~\mflj$, $\ljpt~\mathrm{vs.}~\sjpt$, and $\ljpt$ vs.\ $\ljy$ are provided in Figures~\ref{fig:m12m34_unfolded}--\ref{fig:ljptljy_unfolded}.

The data are compared with SM expectations constructed from the ggF predictions provided by \progname{NNLOPS} and \progname{MadGraph5\_aMC@NLO}-FxFx. Certain distributions related to the production of the Higgs boson also include a comparison with the predictions from   \progname{NNLOJET} and \progname{RadISH} and some of the measurements  related to the Higgs boson decay are  compared also with predictions from \progname{Hto4l} and \progname{Prophecy4f}.
The ggF predictions from \progname{MadGraph5\_aMC@NLO}-FxFx and
\progname{NNLOPS} are
normalised to the  N$^3$LO prediction
while the normalisations for \progname{NNLOJET} and \progname{RadISH} are to their respective predicted cross sections.
All the other Higgs boson production modes are normalised to the most accurate SM predictions, as discussed in Section~\ref{sec:samples}.
The shaded bands on the expected cross sections indicate the PDF and scale uncertainties \footnote{Given the accuracy of some predictions, this procedure may underestimate the associated uncertainties. In particular, \progname{NNLOPS} predictions for $\geq 3$ jets, which are affected in part by additional uncertainties which are not accounted by the procedure described in Section~\ref{subsec:systematics_theory_couplings}.}.
The figures include the $p$-values quantifying the probability of compatibility of the measurements and the SM predictions and show in addition fitted values of the $ZZ^{\ast}$ normalisation factors. Finally, the correlation matrices between the measured cross sections and the $ZZ^{\ast}$ background normalisation factors are shown in all figures along with the cross-section measurements.

\begin{figure}[!htbp]
\centering
\subfloat[]{\includegraphics[width=0.45\linewidth]{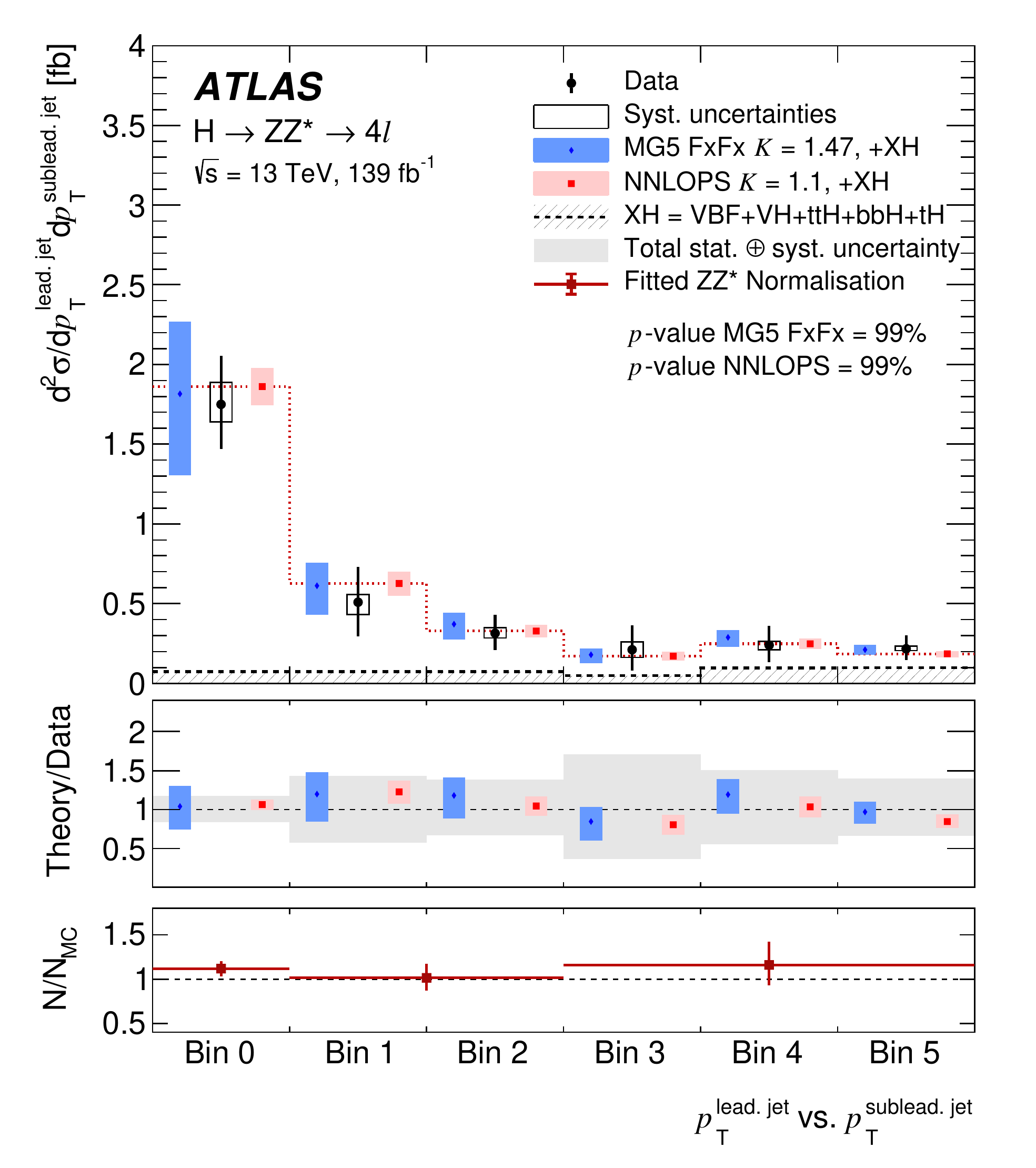}\label{fig:summary_ljptsjpt}}
\subfloat[]{\includegraphics[width=0.45\linewidth]{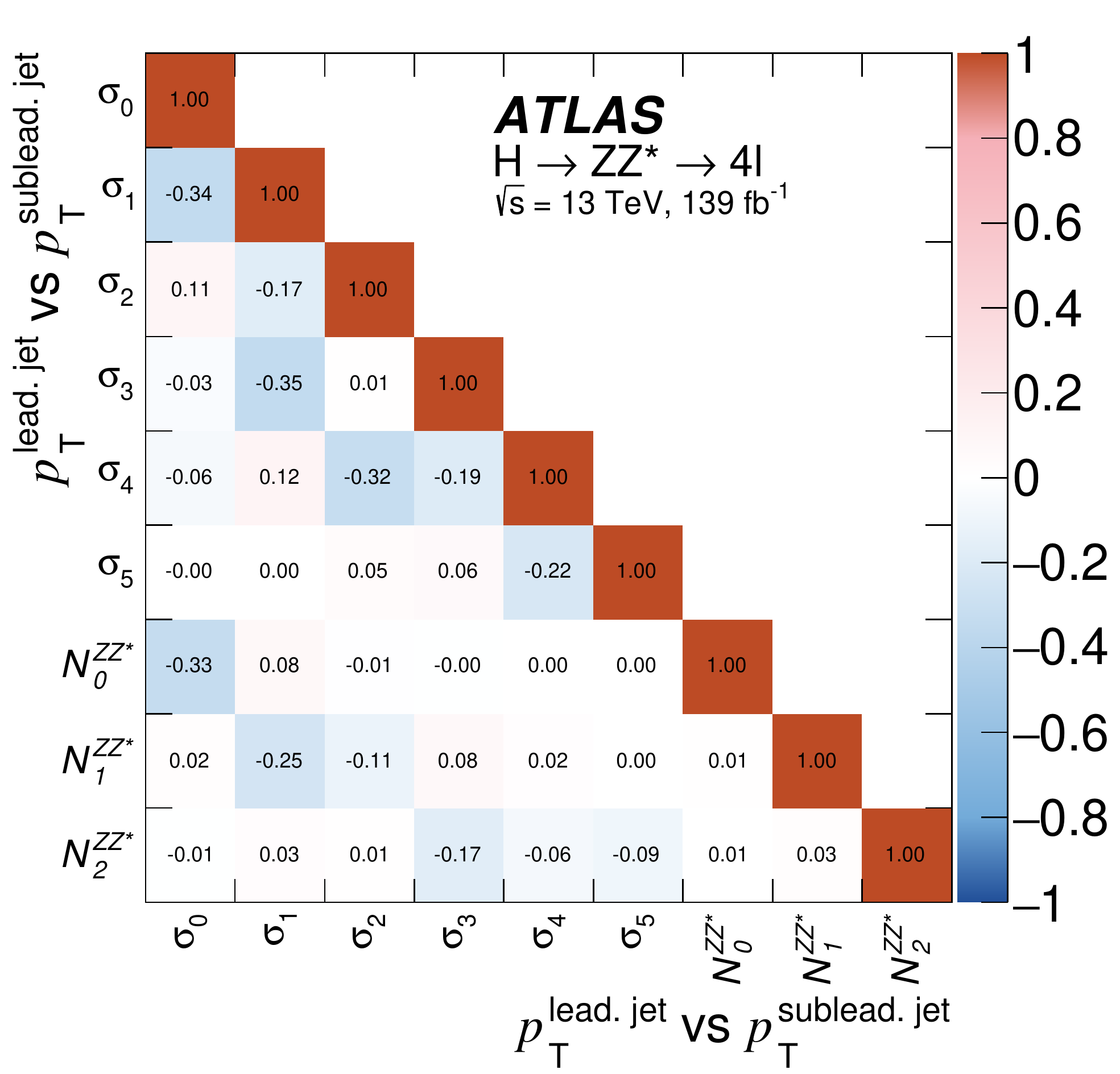}\label{fig:corrMatrix_ljptsjpt}}\\
\caption{\protect\subref{fig:summary_ljptsjpt} Double differential fiducial cross section for   the transverse momentum of leading vs.\ subleading jet, \ljpt\ vs.\ \sjpt, and~\protect\subref{fig:corrMatrix_ljptsjpt}  the corresponding correlation matrix between the measured cross sections and the $ZZ^{\ast}$ background normalisation factor. The bin boundaries are defined in Figure~\ref{fig:categoryVars11}.
\label{fig:ljptsjpt_unfolded} }
\end{figure}
 
\clearpage

Overall, there is good agreement between measured cross sections and predictions. Small differences between measurement and prediction occur in several of the angular observables, as well as in bins of \mfljj, and several of the double differential measurements. For example, the $p$-value for the double differential distribution \ljpt~vs.~\ljy~in Figure~\ref{fig:ljptljy_unfolded} is particularly low due to the downward fluctuation in bin~2. However, when considering the size of the uncertainties these differences are not significant.
Since no events are observed in the highest bin for \ptH\ in Figure~\ref{fig:pt_unfolded}, an upper limit of 27~ab at 95\% confidence level (CL) is set on the cross section using CLs~\cite{Read:2002}. Similarly, a limit of $\sigma < 38$~ab at 95\% CL is also set in the last bin of \ptHjj\ in Figure~\ref{fig:pt4lj_pt4ljj_unfolded}.

\clearpage
 
\section{Interpretation of differential distributions}
\label{sec:interpretations}
 
The measured differential fiducial cross sections can be used to probe possible effects of physics beyond the SM. Two possible interpretations of the results are presented. In Section~\ref{sec:PseudoObs}, the $m_{12}$ vs.\ $m_{34}$ double differential cross section is used to probe several BSM scenarios within the framework of  pseudo-observables~\cite{Gonzalez-Alonso:2014eva}, while in Section~\ref{sec:LightYukawa}, the \ptH\ differential cross section is used to constrain the Yukawa couplings of the Higgs boson with the $b$- and $c$-quarks~\cite{Bishara:2016jga}.

\subsection{Constraints on BSM effects within the pseudo-observables framework}
\label{sec:PseudoObs}
 
In this interpretation, the couplings related to the BSM contact interactions of the Higgs boson decaying into four leptons are considered. As defined in Ref. \cite{Gainer:2018qjm}, the pseudo-observables framework introduces modified contact terms between the Higgs boson, the $Z$ boson, and the left- or right-handed leptons  $\epsilon_{Z,\ell_{L}}$ and $\epsilon_{Z,\ell_{R}}$. In order to reduce the number of independent parameters considered in the pseudo-observables framework for the $H\rightarrow4\ell$ decay amplitudes, specific symmetries are  imposed~\cite{Gainer:2018qjm}. In all the scenarios considered,  the  parameters associated with other pseudo-observables affecting the angular distributions, such as  $\epsilon_{ZZ}^{(CP)}$, $\epsilon_{Z\gamma}^{(CP)}$ and $\epsilon_{\gamma\gamma}^{(CP)}$, are set to zero. Thus, the contact terms considered have the same Lorentz structure as the SM term and only affect the dilepton invariant mass distributions.
 
Four scenarios are investigated~\cite{Gainer:2018qjm}. In the first scenario, referred to as the \textit{flavour-universal contact terms}, the parameters of interest are the $\epsilon_{Z,\ell_{L}}$ and the $\epsilon_{Z,\ell_{R}}$ couplings, where the interactions described by these contact terms have the same strength for electrons and muons. The second scenario considered is \textit{linear EFT-inspired}, where lepton-flavour universality is again imposed and the Higgs boson is assumed to be part of a $SU(2)_L$ doublet. This is reflected in the condition $\epsilon_{R}=0.48\epsilon_{L}$ \cite{Gainer:2018qjm}. The parameters of interest are  $\epsilon_{L}$ and the coupling strength of the Higgs boson to the $Z$ boson, $\kappa_{ZZ}$. In the following two scenarios, lepton-flavour universality can be violated. For the third scenario, referred to as \textit{ flavour non-universal vector contact terms}, the helicity structure of the couplings is fixed to be vector ($\epsilon_{Z,e_{L}}$= $\epsilon_{Z,e_{R}}$, $\epsilon_{Z,\mu_{L}}$= $\epsilon_{Z,\mu_{R}}$) and the independent parameters are the couplings to electrons $\epsilon_{Z,e_{R}}$ and muons $\epsilon_{Z,\mu_{R}}$.  Finally, a fourth scenario with \textit{flavour non-universal axial-vector contact terms} is considered. In this case the helicity structure of the couplings is fixed to be axial-vector, with the parameters of interest being the couplings to electrons $\epsilon_{Z,e_{R}}$ and muons $\epsilon_{Z,\mu_{R}}$ and the condition $\epsilon_{Z,\ell_{L}}=-$ $\epsilon_{Z,\ell_{R}}$ is imposed. Using the $m_{12}$ vs.\ $m_{34}$ double differential cross sections for these interpretations provides sensitivity to distinguish between potential contributions from the contact terms and those from changes to the coupling strength of the Higgs boson to the $Z$ boson.

The variation of the fiducial cross section as a function of the BSM couplings is computed relative to the SM by \progname{MadGraph5\_aMC@NLO} in each of the bins of the measured $m_{12}$ vs.\ $m_{34}$  differential cross section. This is done for a grid of points in the BSM parameter space in each scenario. These relative variations are then fit to a two-dimensional quadratic function. The parameterisation, which also includes any changes in the acceptance, is then encoded into the likelihood and corresponding limits are set for each scenario.
 
\begin{figure}[!htbp]
\centering
\subfloat[]{\includegraphics[width=0.45\linewidth]{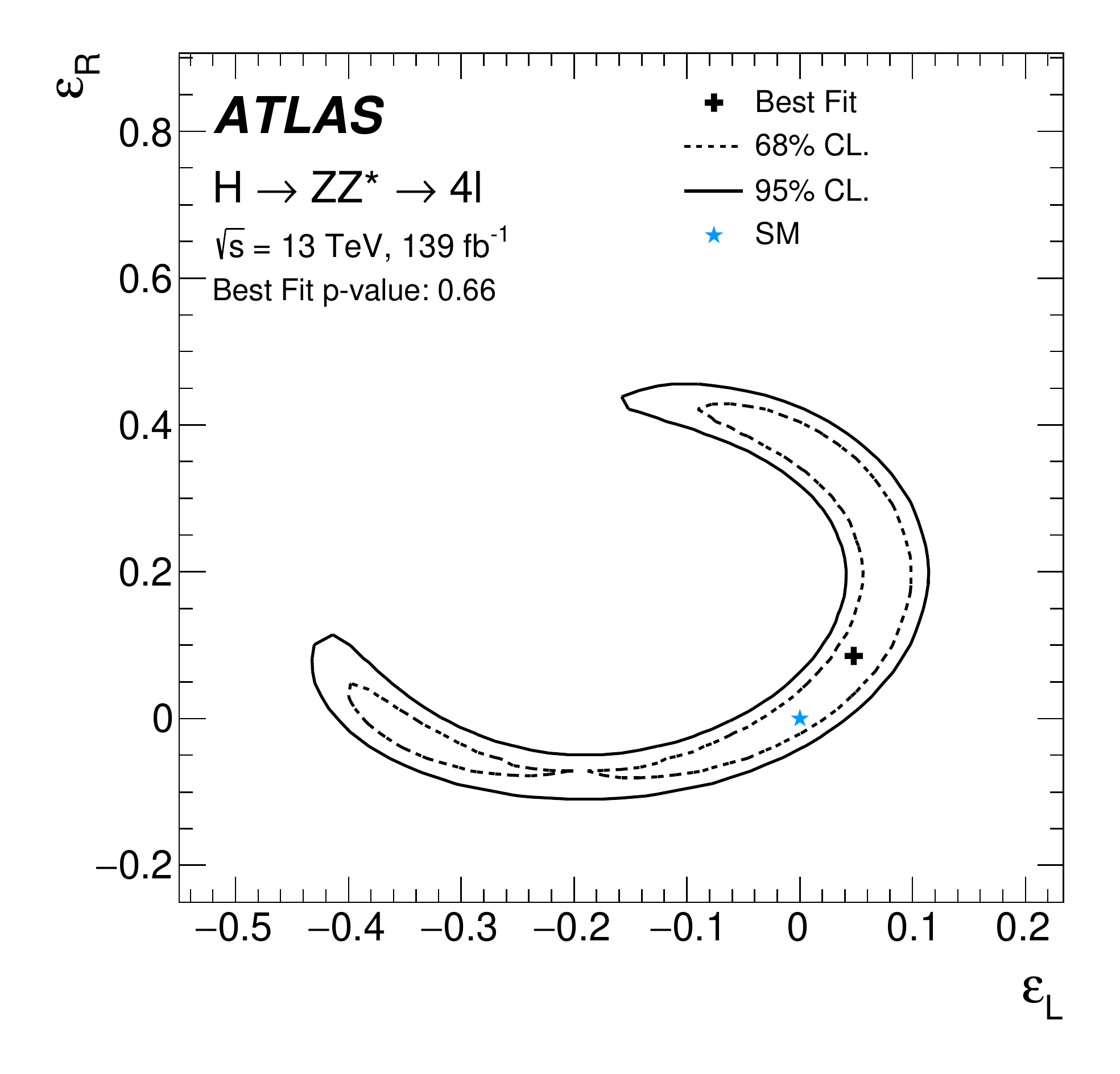}\label{fig:eL_eR}}
\subfloat[]{\includegraphics[width=0.45\linewidth]{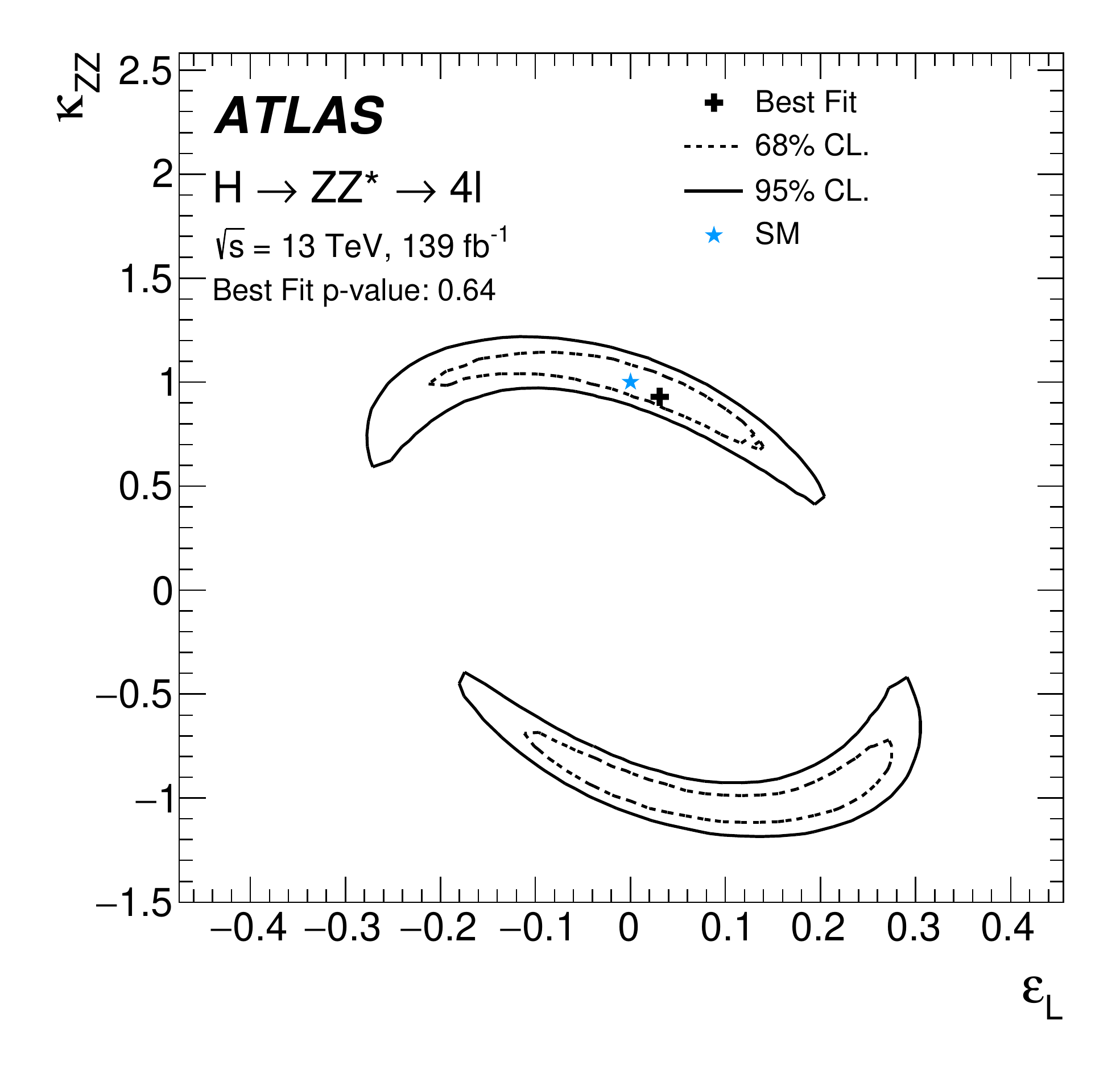}\label{fig:eL_kZZ}}\\
\subfloat[]{\includegraphics[width=0.45\linewidth]{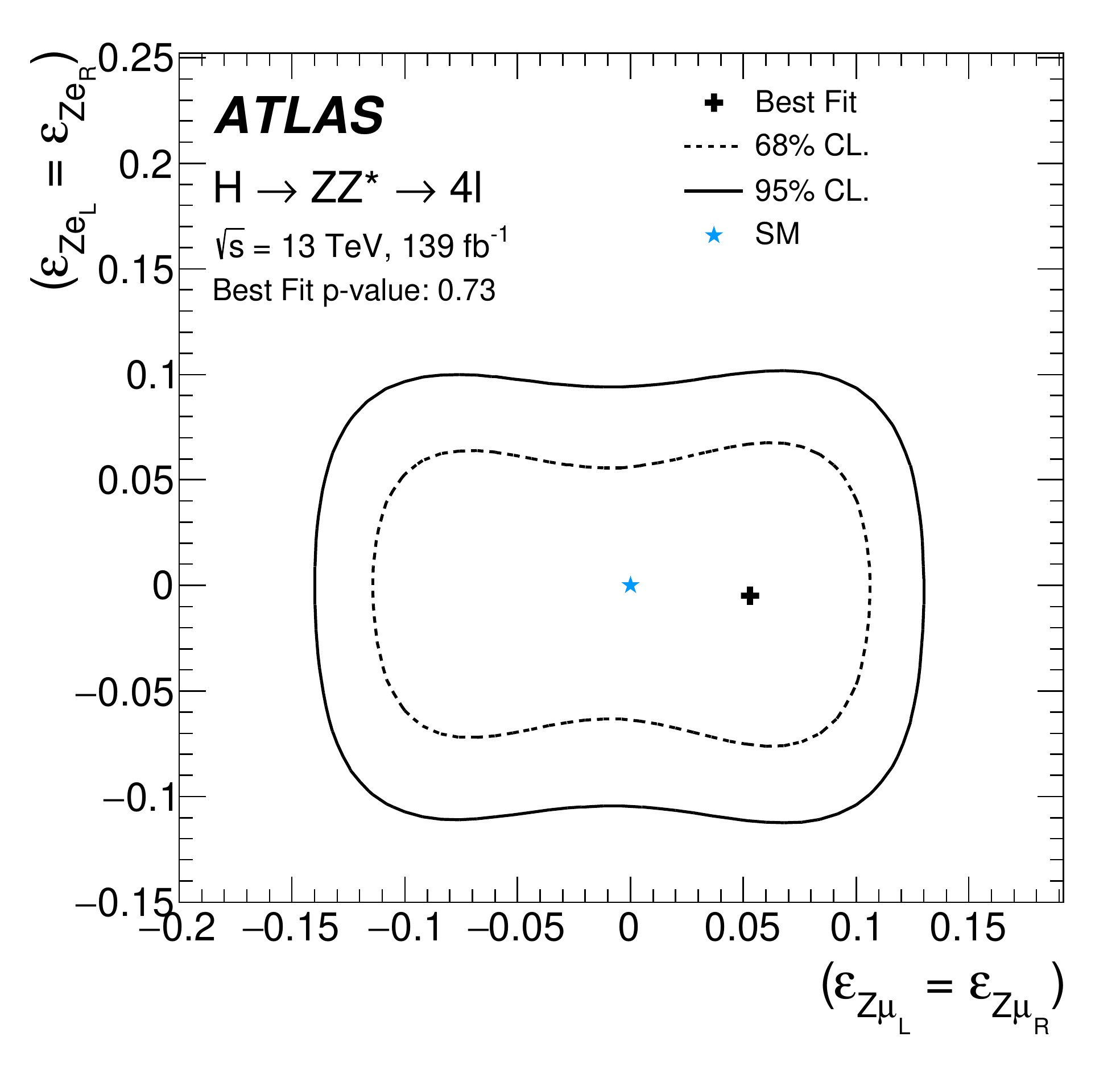}\label{fig:eMu_eEl}}
\subfloat[]{\includegraphics[width=0.45\linewidth]{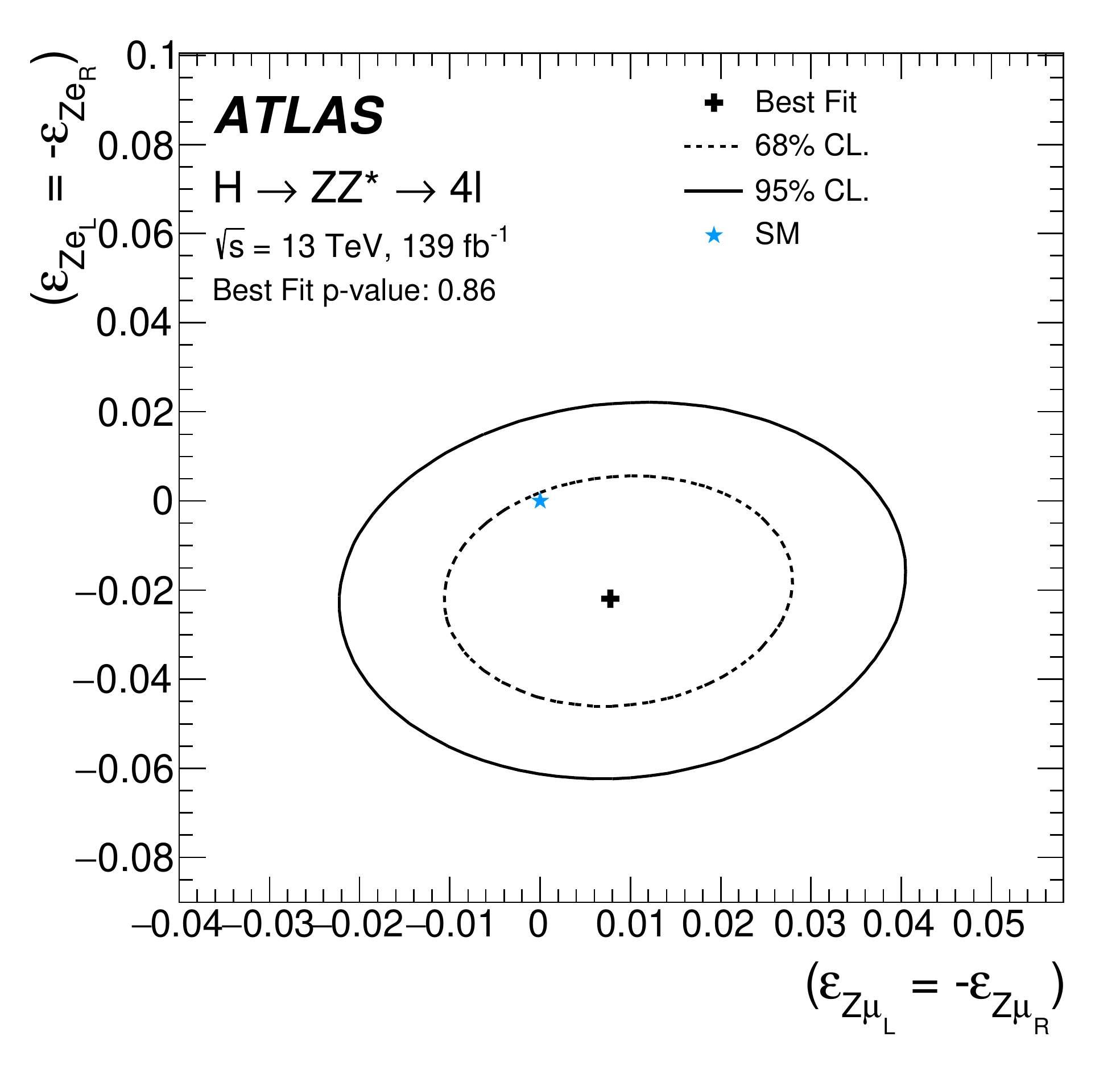}\label{fig:eMuR_eElR}}
\caption{Observed limits at 68\% and  95\% CL on the modified Higgs boson decays within the framework of the pseudo-observables: ~\protect\subref{fig:eL_eR} flavour universal contact terms;~\protect\subref{fig:eL_kZZ} linear EFT-inspired;~\protect\subref{fig:eMu_eEl} flavour non-universal vector contact terms;~\protect\subref{fig:eMuR_eElR} flavour non-universal axial-vector contact terms.  The $p$-values shown represent the probability of compatibility between the data and the  $m_{12}$ vs.\ $m_{34}$ prediction corresponding to the best-fit values of the parameters of interest for each of the four scenarios considered. The SM predictions ($\ast$) and the observed best-fit values (+) are indicated on the plots.
\label{fig:interp_po}}
\end{figure}

Figure~\ref{fig:interp_po}   shows the limits on  BSM interactions of the Higgs boson for the four considered cases. The corresponding 95\% confidence intervals for each of the parameters are listed in Table~\ref{tab:interp_PO_results}.
 
\begin{table*}[htbb!]
\centering
\caption{Confidence intervals for the scenarios considered in the pseudo-observables framework. Based on the observed 2D exclusion contours, 1D exclusion intervals are provided for the EFT-inspired, flavour non-universal vector, and flavour non-universal axial-vector scenarios. The observed limits are calculated while profiling the other parameters of interest. For the EFT-inspired interpretation, the limits are derived assuming $\kappa_{ZZ} \geq 0$. This constraint has no impact on the limit as the analysis is not sensitive to the sign of this parameter.}
\begin{tabular}{ l r l c }
\hline
\hline
Interpretation & \multicolumn{2}{c}{Parameter best-fit value} &  95\% confidence interval \\
\hline
 
\multirow{2}{*}{EFT-inspired}              & $ \epsilon_{L}$ & = ~0.03 & $[-0.25, 0.17]$ \\
& $ \kappa_{ZZ} $ & = ~0.93 & $~~[0.51 , 1.16]$ \\
 
\multirow{2}{*}{Flavour non-universal vector}   & $ \epsilon_{Ze}$    & = -0.005 & $[-0.097, 0.082]$ \\
& $ \epsilon_{Z \mu}$ &  = ~0.054 & $[-0.131, 0.114]$ \\
 
\multirow{2}{*}{Flavour non-universal axial-vector}    & $ \epsilon_{Ze}$ & = -0.022 & $[-0.056, 0.012]$ \\
& $ \epsilon_{Z \mu}$ &  =  ~0.008 & $[-0.016, 0.033]$ \\
 
\hline
\hline
\end{tabular}
\label{tab:interp_PO_results}
\end{table*}

\newpage
\subsection{Constraints on Yukawa couplings}
\label{sec:LightYukawa}
 
Although the couplings of the Higgs boson to the top and bottom quarks have been established recently, obtaining evidence for the coupling of the Higgs boson to the charm quark is more challenging. Direct methods are  limited either by low branching fraction ($H\rightarrow J/\psi\gamma\rightarrow\mu^+\mu^-\gamma$) or by large backgrounds ($H\rightarrow c\bar{c}$). Nevertheless, it has been shown recently that it is possible to indirectly constrain the Yukawa coupling to quarks by analysing the $p_\mathrm{T}^H$ spectrum~\cite{Bishara:2016jga, CMS-HIG-17-028}. In particular, the effects of BSM contributions to the coupling modifiers for the Higgs boson to charm quarks, $\kappa_c$, and for the Higgs boson to bottom quarks,  $\kappa_b$, are investigated.
 
The fiducial cross section is parameterised as a function of the  $\kappa_c$ and $\kappa_b$ values in each measured bin of \ptH. Both the gluon-initiated and quark-initiated components of the prediction show a larger variation, different in size and shape, of the cross section especially at $\ptH < 10$ GeV.
The theoretical uncertainties of these predictions are calculated separately for the gluon-initiated and quark-initiated components by varying the normalisation and factorisation scales by factors of two.
The configuration with largest uncertainty across all the \ptH\ bins across $\kappa_c\in$ [$-10$, 10] and $\kappa_b\in$ [$-2$, 2] ranges is used to define the systematic uncertainty for the predictions. These uncertainties are uncorrelated for each component.
The impact of this uncertainty is about 20\% on the expected limits.
 
\begin{figure}[!htbp]
\centering
\subfloat[]{\includegraphics[width=0.45\linewidth]{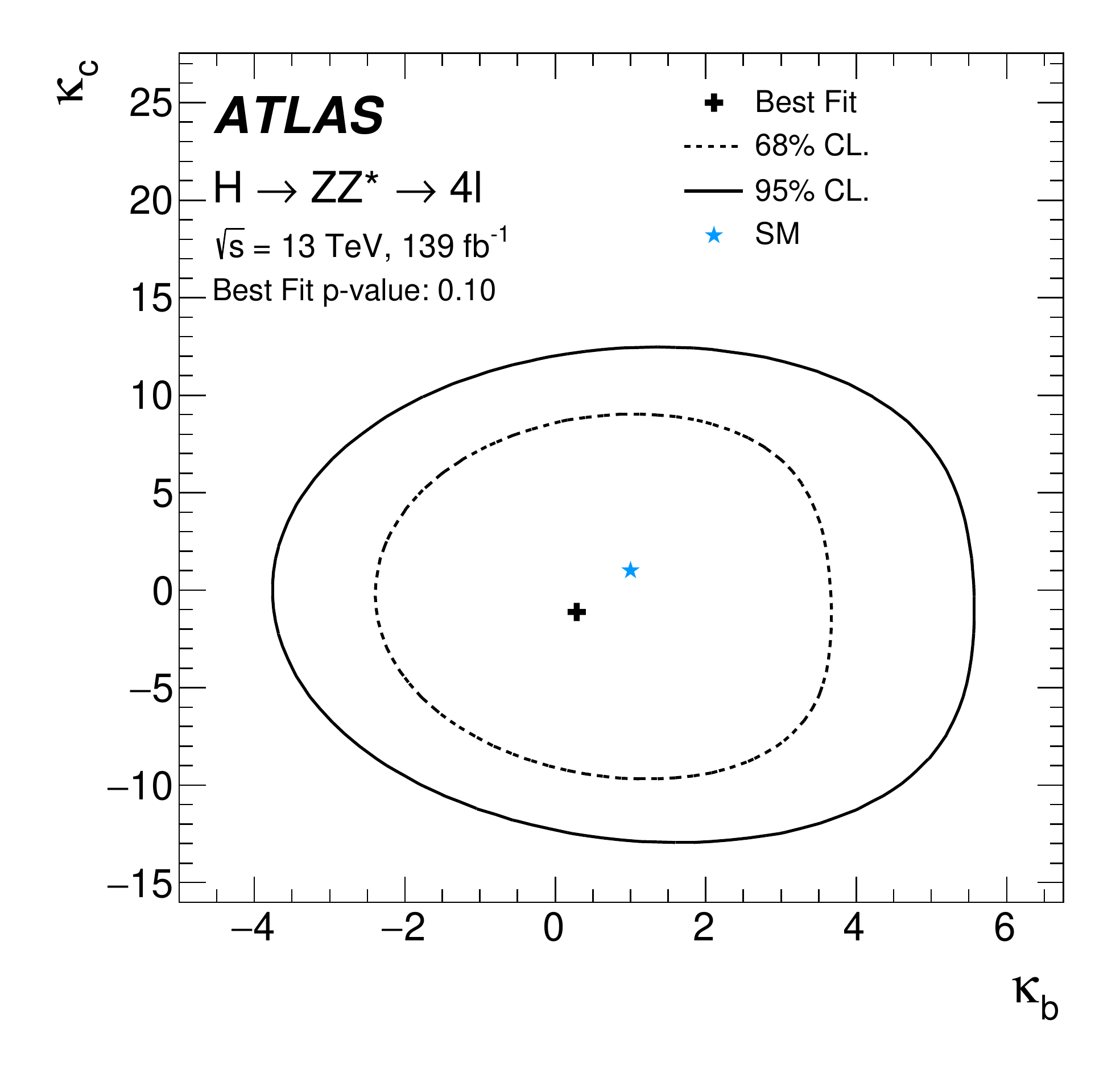}\label{fig:kbkc_SO}}
\subfloat[]{\includegraphics[width=0.45\linewidth]{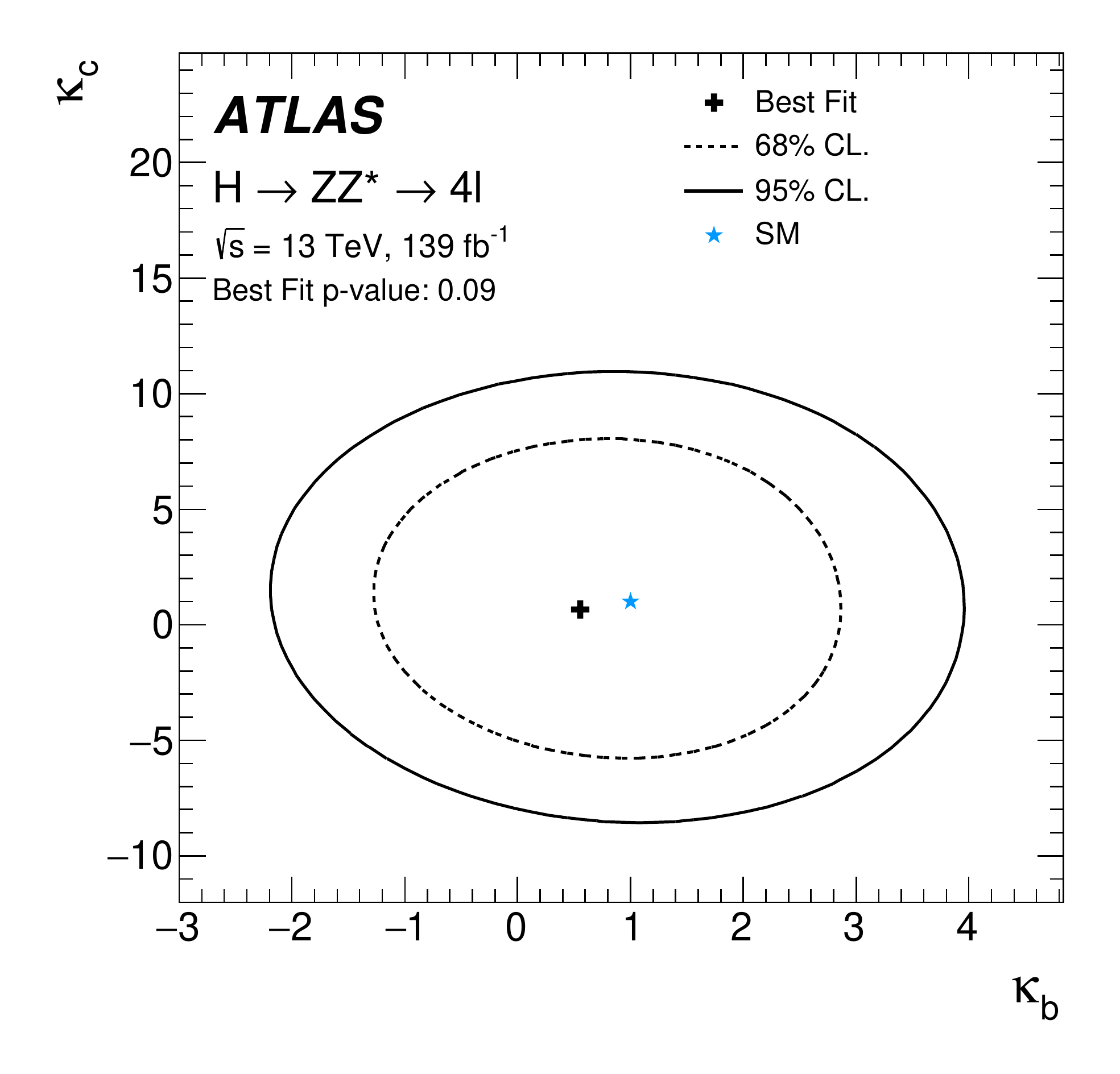}\label{fig:kbkc_XO}}\\
\subfloat[]{\includegraphics[width=0.45\linewidth]{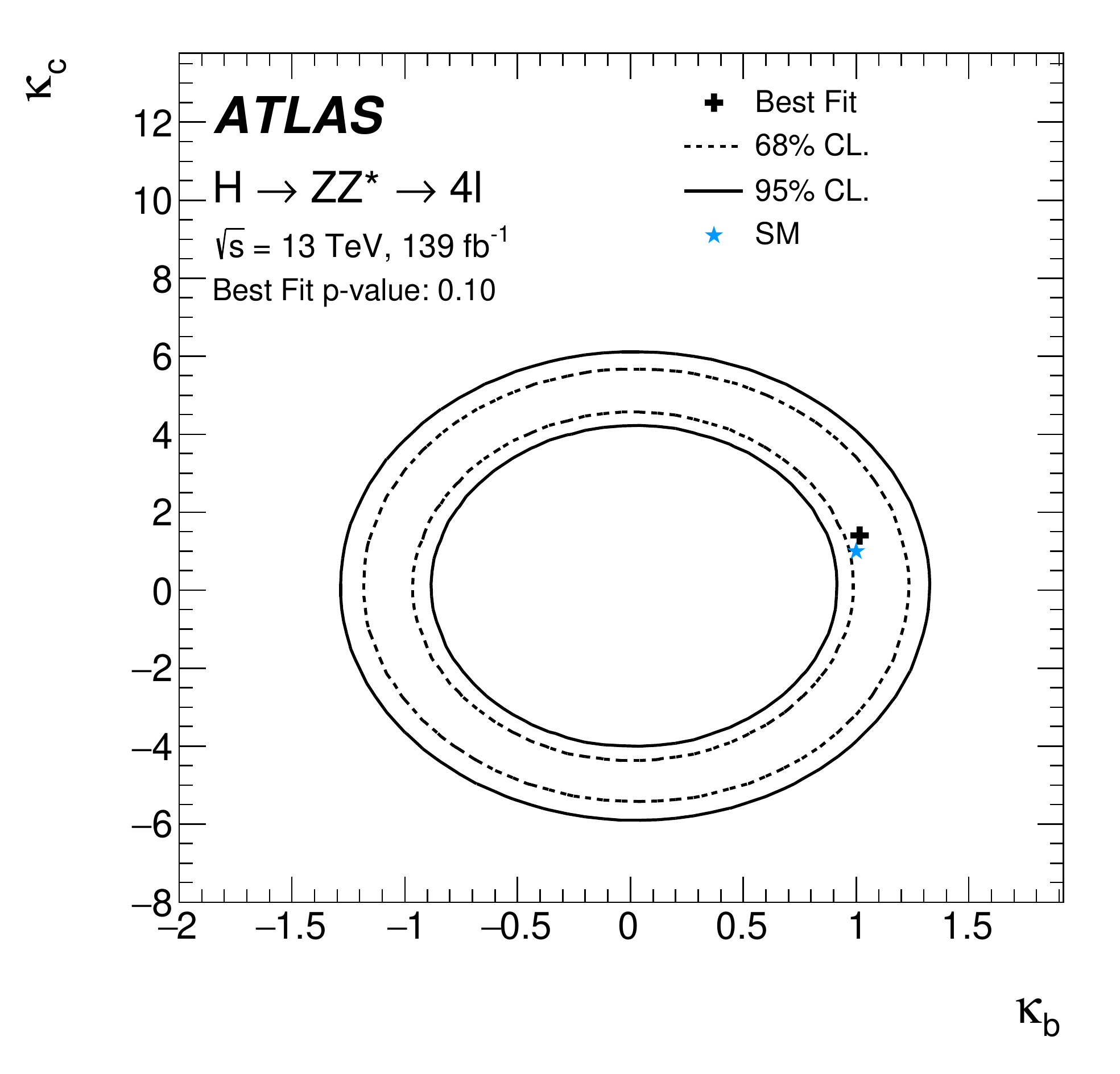}\label{fig:kbkc_XB}}
\caption{Observed limits at 95\% CL on Yukawa couplings
$\kappa_c$ and $\kappa_b$ for the three scenarios considered:~\protect\subref{fig:kbkc_SO} only the \ptH\ shape is used to constrain $\kappa_c$ and $\kappa_b$;~\protect\subref{fig:kbkc_XO} the predicted \ptH\ differential cross section is used;~\protect\subref{fig:kbkc_XB} both the prediction of the \ptH\ differential cross section and the modification to the branching ratio due to the  $\kappa_c$ and $\kappa_b$ values are used. The $p$-values shown represent the probability of compatibility between the data and the \ptH\ prediction corresponding to the best-fit values of  $\kappa_c$ and $\kappa_b$.
The SM predictions ($\ast$) and the observed best-fit values (+) are indicated on the plots.
\label{fig:interp_kbkc}}
\end{figure}

Three different scenarios are considered, with an increasing level of model dependency. In the first case, the modified fiducial cross sections in each bin due to the value of the $b$- and $c$-quark Yukawa couplings  are fit to the data together with a global normalisation factor. The corresponding observed limits on $\kappa_c$ and $\kappa_b$ are shown in Figure~\ref{fig:kbkc_SO}. The sensitivity in this case comes mainly from the modification of the shape induced by $\kappa_c$ and $\kappa_b$, while possible overall normalisation effects are factorised out.
In a second scenario, no additional normalisation factor is introduced in the likelihood and the obtained limits for the Yukawa couplings are shown in Figure~\ref{fig:kbkc_XO}. Finally, in a third scenario, a modification to the total width, and correspondingly to the branching ratio as function of the modified Yukawa couplings, is also encoded in the likelihood and the corresponding limits are shown in Figure~\ref{fig:kbkc_XB}. The 95\% confidence intervals for the first and second scenarios are also listed in Table~\ref{tab:interp_LY_results}.  These are comparable to results from direct searches in $VH, H\rightarrow c\bar{c}$~\cite{Sirunyan:2019qia,HIGG-2017-01}. Constraining $\kappa_b$ to the results from Ref.~\cite{HIGG-2018-57} leads to a  less than 5\%  improvement  in the observed limits for $\kappa_c$ for the scenarios considered.
 
\begin{table*}[htbb!]
\centering
\caption{Confidence intervals for the Yukawa couplings. Based on the observed 2D exclusion contours, 1D exclusion intervals are only provided for interpretations where modification to the \ptH~shape and predictions are considered. The observed limits are calculated while profiling the other parameter of interest.}
\begin{tabular}{ l c c }
\hline
\hline
Interpretation & Parameter best-fit value&  95\% confidence interval \\
\hline
\multirow{2}{*}{Modifications to only \ptH~shape}
& $\kappa_{c} = -1.1$ & $[-11.7,~10.5]$ \\
& $\kappa_{b} = 0.28$ & $[-3.21,~4.50]$ \\
 
\multirow{2}{*}{Modifications to \ptH~predictions}
& $\kappa_{c} = 0.66$ & $[-7.46,~9.27]$ \\
& $\kappa_{b} = 0.55$ & $[-1.82,~3.34]$ \\
\hline
\hline
\end{tabular}
\label{tab:interp_LY_results}
\end{table*}

\FloatBarrier
 
\section{Summary}
\label{sec:conclusion}
Fiducial inclusive and  differential cross-section measurements
of the  Higgs boson in the \hfourl\ decay channel  are presented.
They are based
on  \Lum\ of $\sqrt{s} = 13$~\tev\ proton--proton collisions recorded by the ATLAS detector at the LHC in 2015$-$2018.
The inclusive fiducial cross section in the $H \rightarrow ZZ^* \rightarrow 4\ell$ decay channel is measured to be \Fiducial, in agreement with the Standard Model prediction \FiducialExp. The measurement is about 40\% more precise than the previous ATLAS result. 
The inclusive fiducial cross section is
also extrapolated to the full phase space.
Differential cross sections defined in a fiducial region close to the reconstructed event selection are measured for several variables sensitive to the Higgs boson production and decay such as the transverse momentum of the Higgs boson,  the number of jets produced in association with the Higgs boson, the leading and subleading invariant masses of the lepton pairs.
The measured cross sections are compared with different Standard Model predictions and in general good agreement is found.
The results are also used to set new and more stringent constraints on BSM scenarios where contact term interactions in the $H\rightarrow 4\ell$ amplitudes are introduced. \enlargethispage{\baselineskip}
In addition, the \ptH\ spectrum is used to constrain the $b$- and $c$-quark Yukawa couplings of the Higgs boson. In the scenario with minimal assumptions, values of $\kappa_c$ outside the range $\kappa_c\in [-12, +11]$ are excluded at 95\% CL.
 
 
\clearpage
\appendix
\part*{Appendix}
\addcontentsline{toc}{part}{Appendix}
\section{Results with regularised unfolding}
 
For all the variables investigated in this paper, the unfolding matrix used is well conditioned and no regularisation is required, as discussed in Section~\ref{sec:sig_extraction}. Nevertheless, a Tikhonov regularisation has been tested for the \njets\ and \ljpt\ observables where perceptible off-diagonal terms in the response matrix are observed.
In the Tikhonov regularisation~\cite{tikho}, a prior
assumption about the final result of the measurement is added to the PDF, where the impact of this assumption is controlled by a tunable parameter, $\tau$.
In practice, this method is implemented by adding a penalty term to the negative log-likelihood that is minimised in the fit as
 
\begin{equation}
\tau\cdot\sum_{i=2}^{n-1} \left(\left(\frac{\sigma_{i+1}}{\sigma_{i+1,\mathrm{truth}}}-\frac{\sigma_{i}}{\sigma_{i,\mathrm{truth}}}\right)-\left(\frac{\sigma_{i}}{\sigma_{i,\mathrm{truth}}}-\frac{\sigma_{i-1}}{\sigma_{i-1,\mathrm{truth}}}\right)\right)^2  \,,  \notag
\end{equation}
 
\noindent where $\sigma_i$ is the cross section in bin $i$.
Therefore, a  second-derivative expression  for the curvature is used, with the
parameters normalised  by their expected values from the MC simulation as done in the  SVD unfolding method ~\cite{Hocker:1995kb}. As is done for the main results, only the signal is unfolded.
 
\begin{figure}[!htbp]
\centering
\subfloat[]{\includegraphics[width=0.45\linewidth]{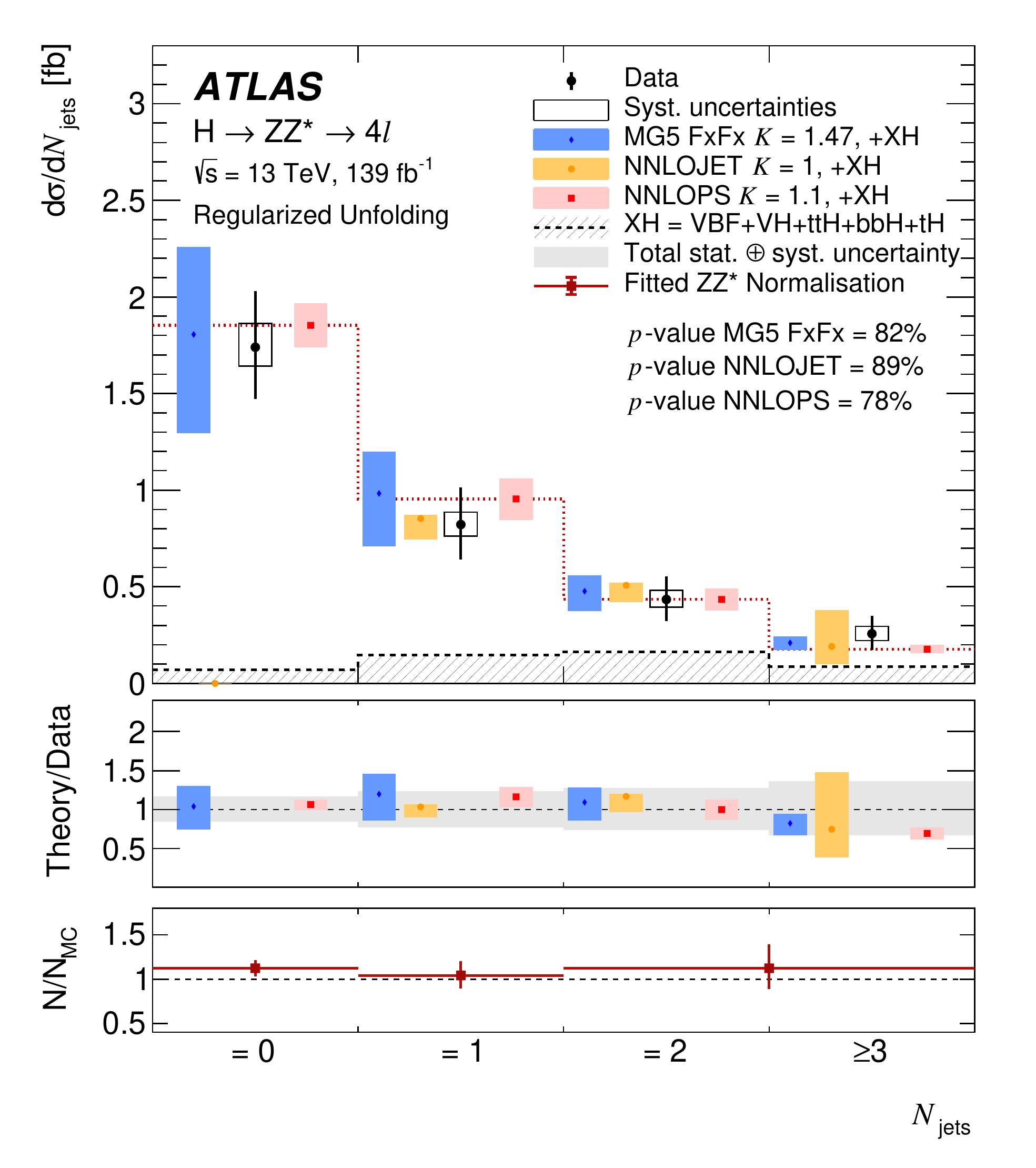}\label{fig:summary_njet_svd}}
\subfloat[]{\includegraphics[width=0.45\linewidth]{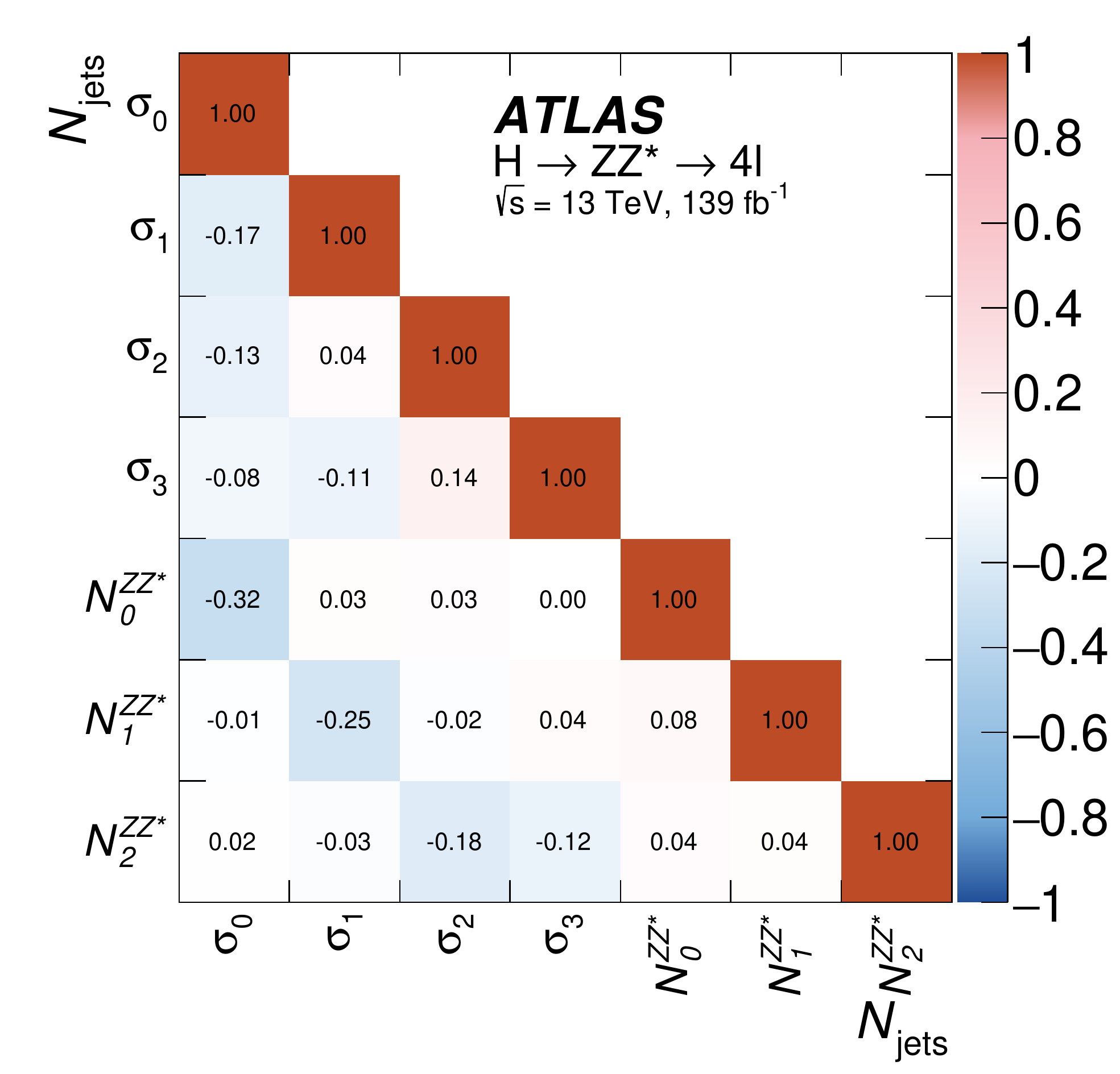}\label{fig:corrMatrix_njet_reg}}\\
\subfloat[]{\includegraphics[width=0.45\linewidth]{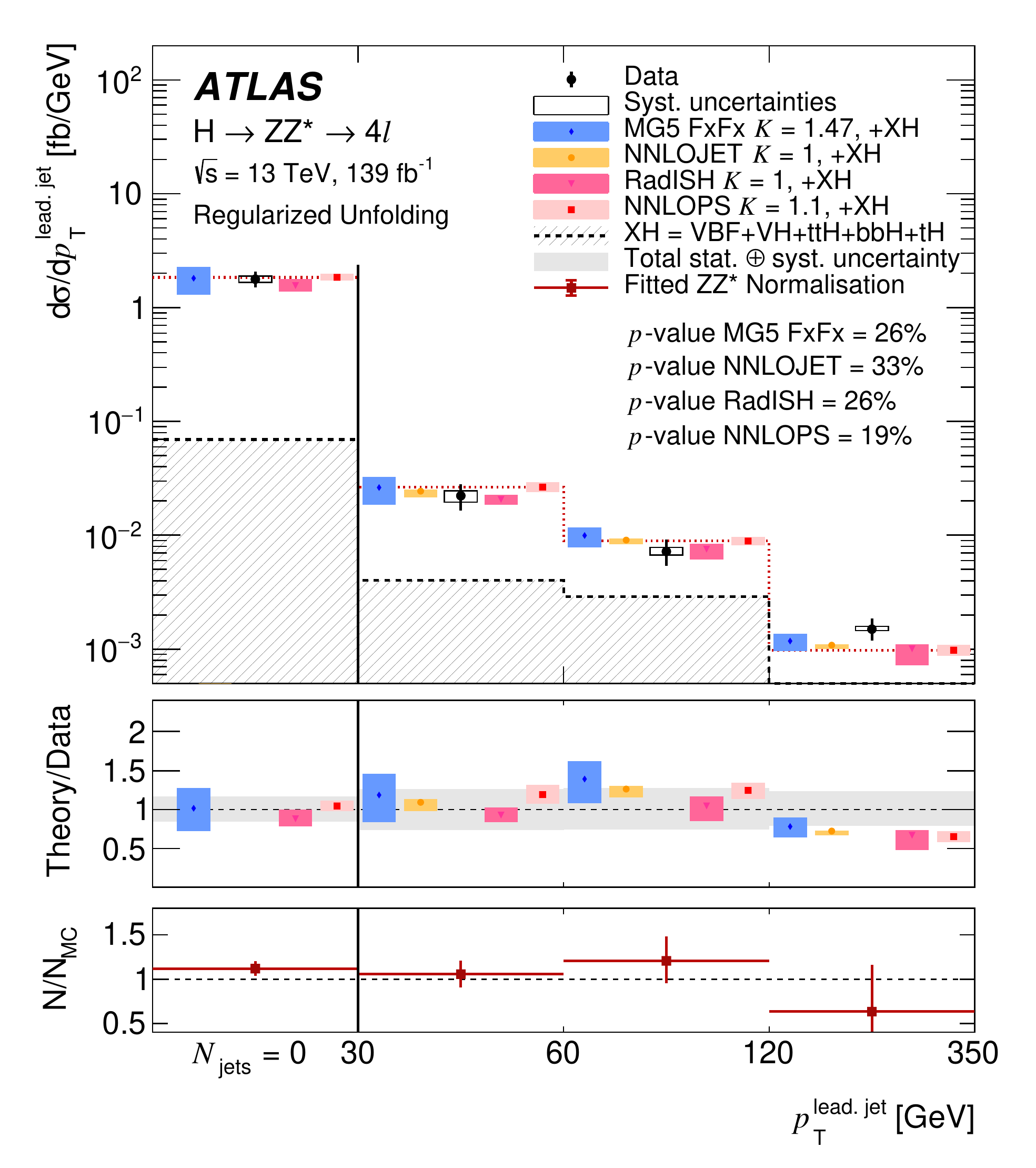}\label{fig:summary_ptlj_svd}}
\subfloat[]{\includegraphics[width=0.45\linewidth]{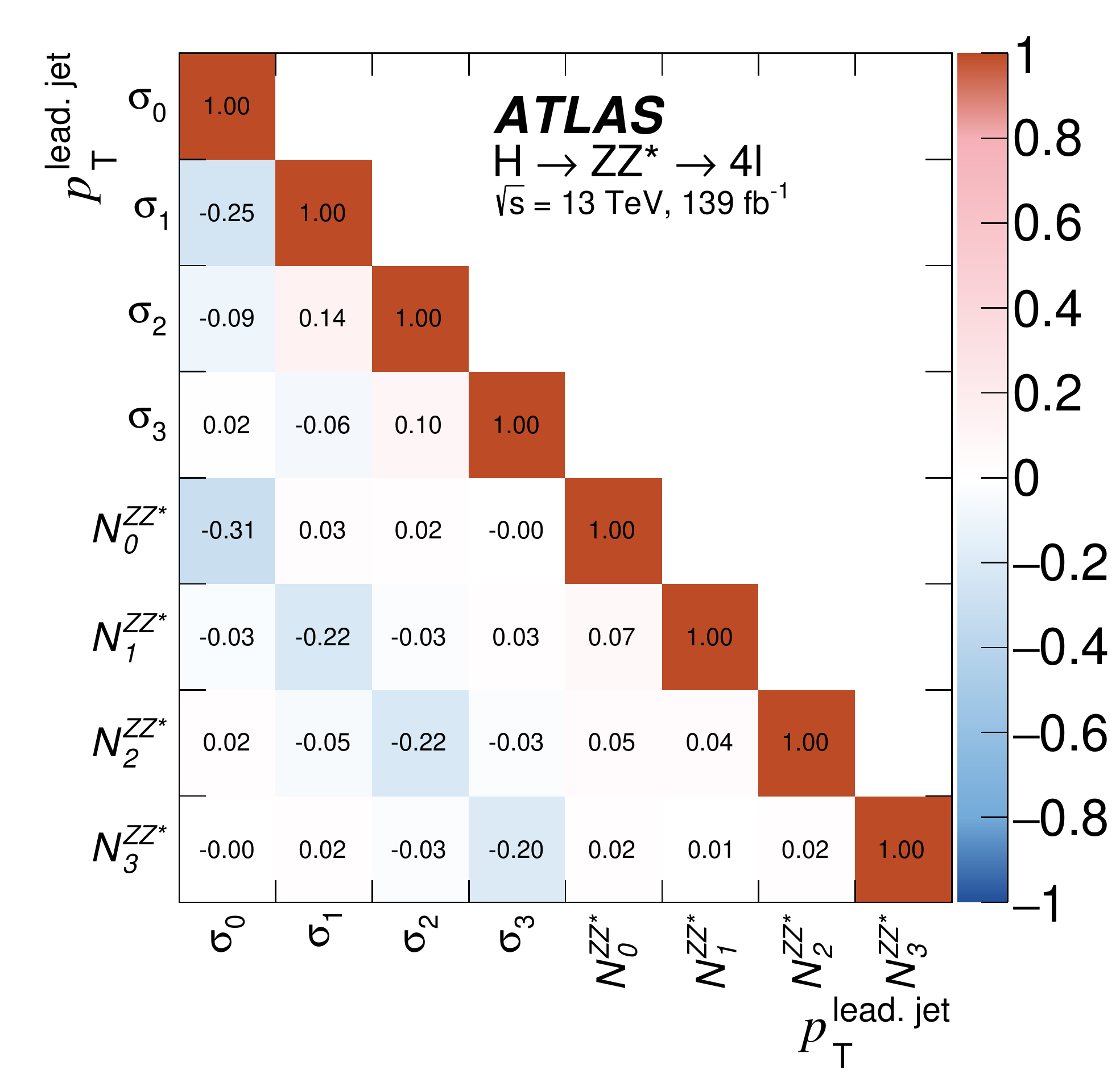}\label{fig:corrMatrix_ptlj_reg}}
\caption{~\protect\subref{fig:summary_njet_svd},~\protect\subref{fig:summary_ptlj_svd} Differential fiducial cross sections as a function of the jet multiplicity, \njets, and leading jet $\pt$ in events with at least one jet, using a regularised matrix unfolding with the  $\tau$ parameters set to 0.6 and 0.7 respectively.  
The corresponding correlation matrix between the measured cross sections and the $ZZ^{\ast}$ background normalisation factors are also shown in ~\protect\subref{fig:corrMatrix_njet_reg} and~\protect\subref{fig:corrMatrix_ptlj_reg}.
\label{fig:njet_unfolded_SVD} }
\end{figure}

The unfolded \njets\ and \ljpt\ distributions using the regularised unfolding with a  $\tau$ parameter set to $\tau=0.6$ and 0.7, respectively, are shown in Figure~\ref{fig:njet_unfolded_SVD}. The  uncertainty which accounts for a possible bias in this regularisation ranges from less than 1\% to about 10\%, depending on the differential bin.
As expected, the comparison of Figures~\ref{fig:summary_njet_svd} with~\ref{fig:summary_njet} and Figures~\ref{fig:summary_ptlj_svd} with~\ref{fig:summary_ljpt} shows that the regularisation tends to reduce the off-diagonal anti-correlation terms of the correlation matrix among the measured cross sections, reducing its uncertainty.
Nevertheless, the $p$-values for the different predictions are close to the ones obtained with the matrix unfolding without any regularisation.
 
\FloatBarrier
\section{Invariant mass of the leading lepton pair in same-flavour and opposite-flavour final states}
 
\begin{figure}[!htbp]
\centering
\subfloat[]{\includegraphics[width=0.45\linewidth]{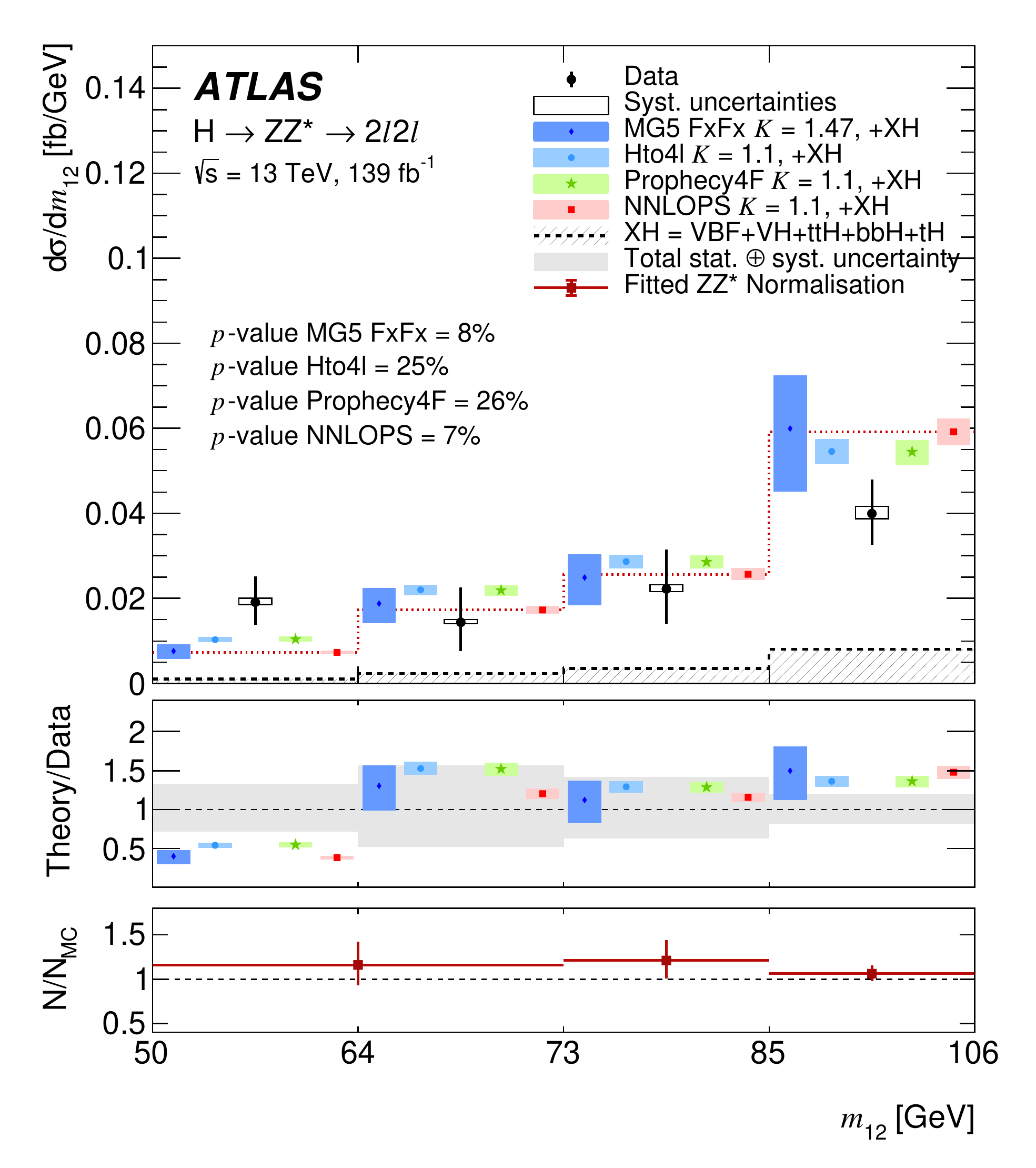}\label{fig:summary_m12_4l}}
\subfloat[]{\includegraphics[width=0.45\linewidth]{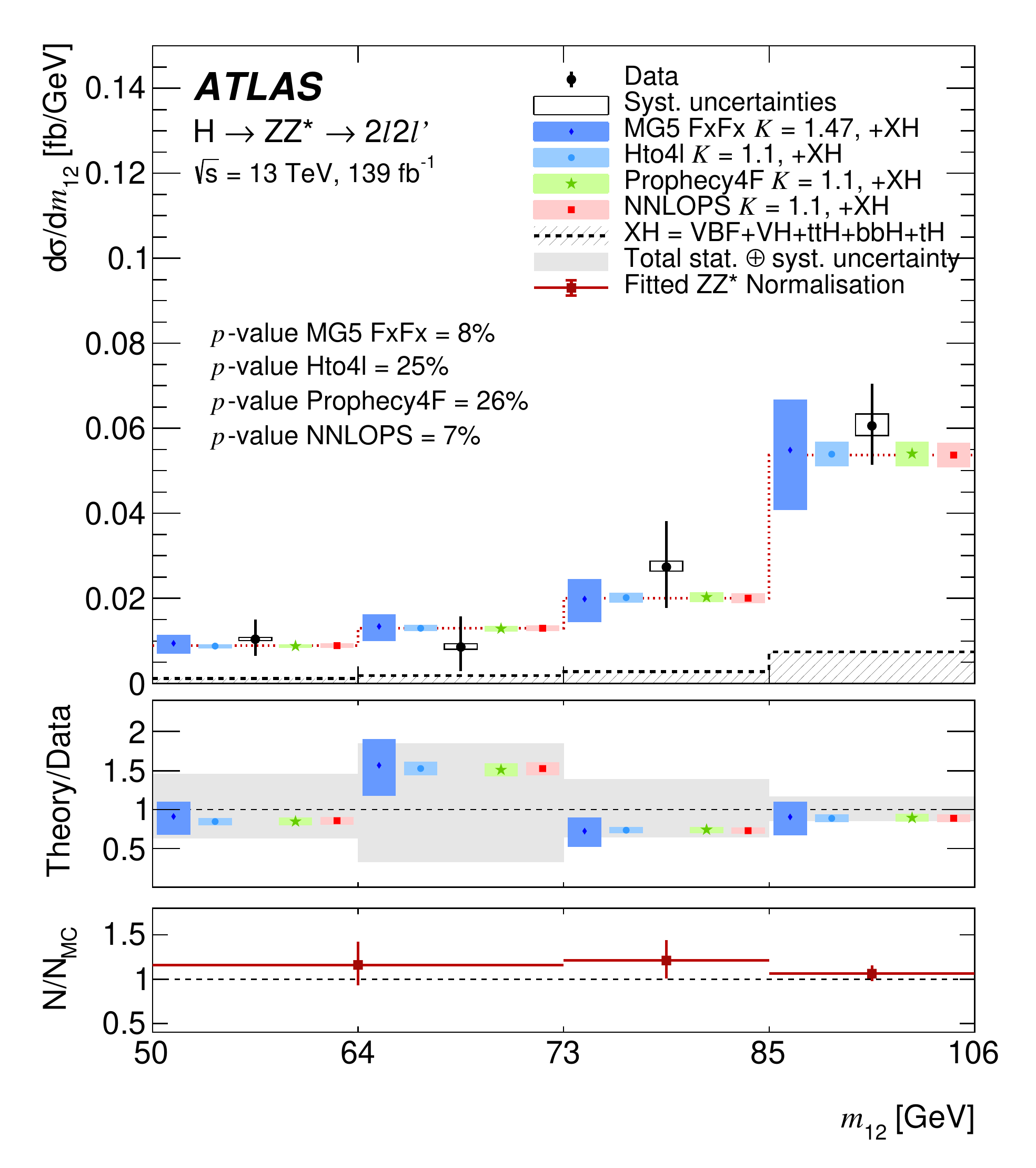}\label{fig:summary_m12_2l2l}} \\
\subfloat[]{\includegraphics[width=0.45\linewidth]{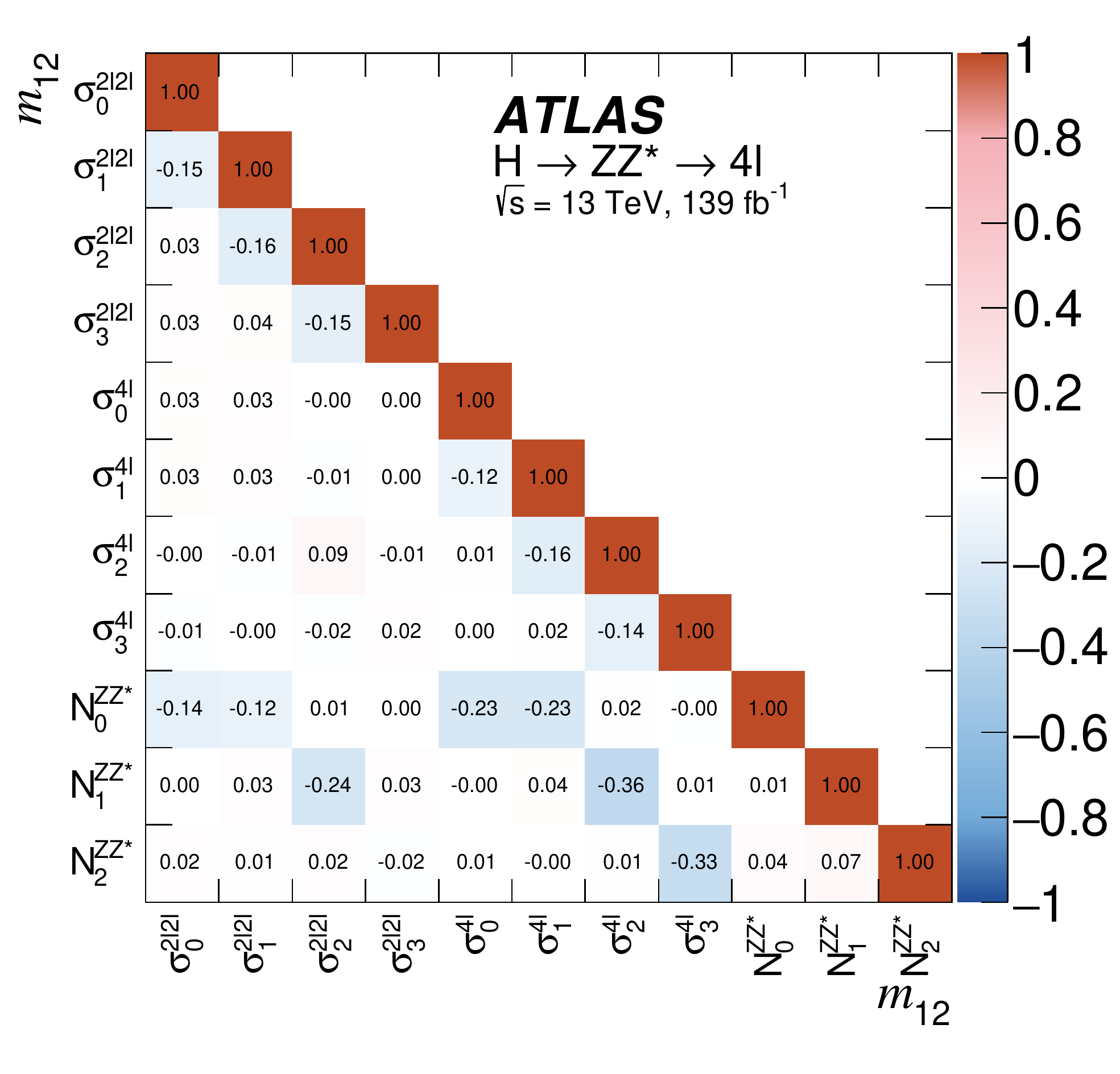}\label{fig:corrHist_m12_splitF}}
\caption{Differential fiducial cross sections for the invariant mass \monetwo~of the leading $Z$ boson in~\protect\subref{fig:summary_m12_4l} the $4\mu$ and $4e$ decay channels and~\protect\subref{fig:summary_m12_2l2l} the $2e2\mu$ and $2\mu2e$ decay channels. The corresponding correlation matrix is shown in~\protect\subref{fig:corrHist_m12_splitF}. }
\label{fig:m12_unfolded_SFOF}
\end{figure}

Figure~\ref{fig:m12_unfolded_SFOF} presents results for the invariant mass of the leading lepton pair in same-flavour and different-flavour final states.
 
\newpage


\section*{Acknowledgements}
 

We thank CERN for the very successful operation of the LHC, as well as the
support staff from our institutions without whom ATLAS could not be
operated efficiently.
 
We acknowledge the support of ANPCyT, Argentina; YerPhI, Armenia; ARC, Australia; BMWFW and FWF, Austria; ANAS, Azerbaijan; SSTC, Belarus; CNPq and FAPESP, Brazil; NSERC, NRC and CFI, Canada; CERN; CONICYT, Chile; CAS, MOST and NSFC, China; COLCIENCIAS, Colombia; MSMT CR, MPO CR and VSC CR, Czech Republic; DNRF and DNSRC, Denmark; IN2P3-CNRS and CEA-DRF/IRFU, France; SRNSFG, Georgia; BMBF, HGF and MPG, Germany; GSRT, Greece; RGC and Hong Kong SAR, China; ISF and Benoziyo Center, Israel; INFN, Italy; MEXT and JSPS, Japan; CNRST, Morocco; NWO, Netherlands; RCN, Norway; MNiSW and NCN, Poland; FCT, Portugal; MNE/IFA, Romania; MES of Russia and NRC KI, Russia Federation; JINR; MESTD, Serbia; MSSR, Slovakia; ARRS and MIZ\v{S}, Slovenia; DST/NRF, South Africa; MINECO, Spain; SRC and Wallenberg Foundation, Sweden; SERI, SNSF and Cantons of Bern and Geneva, Switzerland; MOST, Taiwan; TAEK, Turkey; STFC, United Kingdom; DOE and NSF, United States of America. In addition, individual groups and members have received support from BCKDF, CANARIE, Compute Canada and CRC, Canada; ERC, ERDF, Horizon 2020, Marie Sk{\l}odowska-Curie Actions and COST, European Union; Investissements d'Avenir Labex, Investissements d'Avenir Idex and ANR, France; DFG and AvH Foundation, Germany; Herakleitos, Thales and Aristeia programmes co-financed by EU-ESF and the Greek NSRF, Greece; BSF-NSF and GIF, Israel; CERCA Programme Generalitat de Catalunya and PROMETEO Programme Generalitat Valenciana, Spain; G\"{o}ran Gustafssons Stiftelse, Sweden; The Royal Society and Leverhulme Trust, United Kingdom.
 
The crucial computing support from all WLCG partners is acknowledged gratefully, in particular from CERN, the ATLAS Tier-1 facilities at TRIUMF (Canada), NDGF (Denmark, Norway, Sweden), CC-IN2P3 (France), KIT/GridKA (Germany), INFN-CNAF (Italy), NL-T1 (Netherlands), PIC (Spain), ASGC (Taiwan), RAL (UK) and BNL (USA), the Tier-2 facilities worldwide and large non-WLCG resource providers. Major contributors of computing resources are listed in Ref.~\cite{ATL-GEN-PUB-2016-002}.
 

\printbibliography
\clearpage 
 
\begin{flushleft}
\hypersetup{urlcolor=black}
{\Large The ATLAS Collaboration}

\bigskip

\AtlasOrcid[0000-0002-6665-4934]{G.~Aad}$^\textrm{\scriptsize 102}$,    
\AtlasOrcid[0000-0002-5888-2734]{B.~Abbott}$^\textrm{\scriptsize 128}$,    
\AtlasOrcid{D.C.~Abbott}$^\textrm{\scriptsize 103}$,    
\AtlasOrcid[0000-0002-2788-3822]{A.~Abed~Abud}$^\textrm{\scriptsize 36}$,    
\AtlasOrcid[0000-0002-1002-1652]{K.~Abeling}$^\textrm{\scriptsize 53}$,    
\AtlasOrcid[0000-0002-2987-4006]{D.K.~Abhayasinghe}$^\textrm{\scriptsize 94}$,    
\AtlasOrcid[0000-0002-8496-9294]{S.H.~Abidi}$^\textrm{\scriptsize 166}$,    
\AtlasOrcid[0000-0002-8279-9324]{O.S.~AbouZeid}$^\textrm{\scriptsize 40}$,    
\AtlasOrcid{N.L.~Abraham}$^\textrm{\scriptsize 155}$,    
\AtlasOrcid[0000-0001-5329-6640]{H.~Abramowicz}$^\textrm{\scriptsize 160}$
% ...
\bigskip
\\

% ...
$^{36}$CERN, Geneva; Switzerland.\\
% ...
$^{40}$Niels Bohr Institute, University of Copenhagen, Copenhagen; Denmark.\\
% ...
$^{53}$II. Physikalisches Institut, Georg-August-Universit\"{a}t G\"ottingen, G\"ottingen; Germany.\\
% ...
$^{94}$Department of Physics, Royal Holloway University of London, Egham; United Kingdom.\\
% ...
$^{102}$CPPM, Aix-Marseille Universit\'e, CNRS/IN2P3, Marseille; France.\\
$^{103}$Department of Physics, University of Massachusetts, Amherst MA; United States of America.\\
% ...
$^{128}$Homer L. Dodge Department of Physics and Astronomy, University of Oklahoma, Norman OK; United States of America.\\
% ...
$^{155}$Department of Physics and Astronomy, University of Sussex, Brighton; United Kingdom.\\
% ...
$^{160}$Raymond and Beverly Sackler School of Physics and Astronomy, Tel Aviv University, Tel Aviv; Israel.\\
% ...
$^{166}$Department of Physics, University of Toronto, Toronto ON; Canada.\\
% ...

\end{flushleft}


\end{document}